\documentclass[namedreferences]{solarphysics}

\usepackage[hyperref,optionalrh]{spr-sola-addons} 
\usepackage{graphicx}        
\usepackage[binary-units=true]{siunitx}
\usepackage{color}           
\usepackage{breakurl}        
\usepackage{pdflscape}
\usepackage{caption}        

\usepackage{multirow}
\usepackage{ragged2e}
\usepackage{rotating}

\ifx\urlurl    \undefined
       \def\urlurl#1{\href{http://#1}{\textsf{#1}}}\fi

\newcommand{\Rs}{\(\mathrm{R}_\odot\)}

\chardef\us=`\_

\begin{document}

\begin{article}
\begin{opening}

\title{Defining the Middle Corona\\ {\it Solar Physics}}

\author[addressref={SwRI},corref,email={mwest@boulder.swri.edu}]{\inits{M.J.}\fnm{Matthew~J.}~\lnm{West}\orcid{0000-0002-0631-2393}}
\author[addressref={SwRI},email={dseaton@boulder.swri.edu}]{\inits{D.B.}\fnm{Daniel~B.}~\lnm{Seaton}\orcid{0000-0002-0494-2025}}
\author[addressref={aff3},email={David$\us$Wexler@uml.edu}]{\inits{D.B.}\fnm{David~B.}~\lnm{Wexler}\orcid{0000-0002-5763-6267}}
\author[addressref={cfa},email={jraymond@cfa.harvard.edu}]{\inits{J.C.}\fnm{John~C.}~\lnm{Raymond}\orcid{0000-0002-7868-1622}}
\author[addressref={aff6},email={gd232@cam.ac.uk}]{\inits{G.}\fnm{Giulio}~\lnm{Del Zanna}\orcid{0000-0002-4125-0204}}
\author[addressref={cfa},email={yeimy.rivera@cfa.harvard.edu}]{\inits{Y.J.}\fnm{Yeimy~J.}~\lnm{Rivera}\orcid{0000-0002-8748-2123}}
\author[addressref={msfc},email={adam.kobelski@nasa.gov}]{\inits{A.R.}\fnm{Adam R.}~\lnm{Kobelski}\orcid{0000-0002-4691-1729}}
\author[addressref={NJIT},email={bin.chen@njit.edu}]{\inits{B.}\fnm{Bin}~\lnm{Chen}\orcid{0000-0002-0660-3350}}
\author[addressref={SwRI},email={cdeforest@boulder.swri.edu}]{\inits{C.E.}\fnm{Craig}~\lnm{DeForest}\orcid{0000-0002-7164-2786}}
\author[addressref={cfa},email={lgolub@cfa.harvard.edu}]{\inits{L.}\fnm{Leon}~\lnm{Golub}\orcid{0000-0001-9638-3082}}
\author[addressref={SwRI},email={amir@boulder.swri.edu}]{\inits{A.}\fnm{Amir}~\lnm{Caspi}\orcid{0000-0001-8702-8273}}
\author[addressref={SwRI},email={chris.gilly@swri.org}]{\inits{C.R.}\fnm{Chris~R.}~\lnm{Gilly}\orcid{0000-0003-0021-9056}}
\author[addressref={nrl},email={jason.kooi@nrl.navy.mil}]{\inits{J.E.}\fnm{Jason~E.}~\lnm{Kooi}\orcid{0000-0002-5595-2522}}
\author[addressref={UoD},email={kmeyer001@dundee.ac.uk}]{\inits{K.A.}\fnm{Karen A.}~\lnm{Meyer}\orcid{0000-0001-6046-2811}}
\author[addressref={SwRISAT},email={blalterman@swri.org}]{\inits{B.L.}\fnm{Benjamin~L.}~\lnm{Alterman}\orcid{0000-0001-6673-3432}}
\author[addressref={GSFC,aff8},email={nathalia.alzate@nasa.gov}]{\inits{N.A.}\fnm{Nathalia}~\lnm{Alzate}\orcid{0000-0001-5207-9628}}
\author[addressref={inaf1},email={vincenzo.andretta@inaf.it}]{\inits{V.}\fnm{Vincenzo}~\lnm{Andretta}\orcid{0000-0003-1962-9741}}
\author[addressref={UPS},email={frederic.auchere@universite-paris-saclay.fr}]{\inits{F.}\fnm{Fr\'ed\'eric}~\lnm{Auch\`ere}\orcid{0000-0003-0972-7022}}
\author[addressref={IIAS},email={dipu@aries.res.in}]{\inits{D.}\fnm{Dipankar}~\lnm{Banerjee}\orcid{0000-0003-4653-6823}}
\author[addressref={ROB},email={david.berghmans@oma.be}]{\inits{D.B.}\fnm{David}~\lnm{Berghmans}\orcid{0000-0003-4052-9462}}
\author[addressref={LASP},email={phil.Chamberlin@lasp.colorado.edu}]{\inits{P.}\fnm{Phillip}~\lnm{Chamberlin}\orcid{0000-0003-4372-7405}}
\author[addressref={mps},email={chitta@mps.mpg.de}]{\inits{L.P.}\fnm{Lakshmi~Pradeep}~\lnm{Chitta}\orcid{0000-0002-9270-6785}}
\author[addressref={psi},email={cdowns@predsci.com}]{\inits{C.}\fnm{Cooper}~\lnm{Downs}\orcid{0000-0003-1759-4354}}
\author[addressref={inaf},email={silvio.giordano@inaf.it}]{\inits{S.G.}\fnm{Silvio}~\lnm{Giordano}\orcid{0000-0002-3468-8566}}
\author[addressref={PMOD},email={Louise.Harra@pmodwrc.ch}]{\inits{L.}\fnm{Louise}~\lnm{Harra}\orcid{0000-0001-9457-6200}}
\author[addressref={GSFC},email={aleida.k.higginson@nasa.gov}]{\inits{A.}\fnm{Aleida}~\lnm{Higginson}\orcid{0000-0003-1380-8722}}
\author[addressref={jhuapl},email={Russell.Howard@jhuapl.edu}]{\inits{R.A.}\fnm{Russell~A.}~\lnm{Howard}\orcid{0000-0001-9027-8249}}
\author[addressref={AU,GSFC}, email={pankaj.kumar@nasa.gov}]{\inits{P.}\fnm{Pankaj}~\lnm{Kumar}\orcid{0000-0001-6289-7341}}
\author[addressref={psi},email={emason@predsci.com}]{\inits{E.I.}\fnm{Emily}~\lnm{Mason}\orcid{0000-0002-8767-7182}}
\author[addressref={jhuapl},email={james.mason@jhuapl.edu}]{\inits{J.P.}\fnm{James P.}~\lnm{Mason}\orcid{0000-0002-3783-5509}}
\author[addressref={North},email={richard.morton@northumbria.ac.uk}]{\inits{R.~J.}\fnm{Richard~J.}~\lnm{Morton}\orcid{0000-0001-5678-9002}}
\author[addressref={ERAU},email={nykyrik@erau.edu}]{\inits{K.}\fnm{Katariina}~\lnm{Nykyri}\orcid{0000-0002-6905-9487}}
\author[addressref={SwRI},email={ritesh.patel@swri.org}]{\inits{R.}\fnm{Ritesh}~\lnm{Patel}\orcid{0000-0001-8504-2725}}
\author[addressref={ncei},email={laurel.rachmeler@noaa.gov}]{\inits{L.R.}\fnm{Laurel}~\lnm{Rachmeler}\orcid{0000-0002-3770-009X}}
\author[addressref={NSO},email={kreardon@nso.edu}]{\inits{K.P.}\fnm{Kevin~P.}~\lnm{Reardon}\orcid{0000-0001-8016-0001}}
\author[addressref={cfa},email={kreeves@cfa.harvard.edu}]{\inits{K.K.}\fnm{Katharine~K.}~\lnm{Reeves}\orcid{0000-0002-6903-6832}}
\author[addressref={msfc},email={sabrina.savage@nasa.gov}]{\inits{S.L.}\fnm{Sabrina}~\lnm{Savage}\orcid{0000-0002-6172-0517}}
\author[addressref={GSFC},email={barbara.j.thompson@nasa.gov}]{\inits{B.J.}\fnm{Barbara J.}~\lnm{Thompson}\orcid{0000-0001-6952-7343}}
\author[addressref={SwRI},email={samuel.vankooten@swri.org}]{\inits{S.J.}\fnm{Samuel~J.}~\lnm{Van~Kooten}\orcid{0000-0002-4472-8517}}
\author[addressref={GSFC},email={nicholeen.m.viall@nasa.gov}]{\inits{N.M.}\fnm{Nicholeen~M.}~\lnm{Viall}\orcid{0000-0003-1692-1704}}
\author[addressref={jhuapl},email={angelos.vourlidas@jhuapl.edu}]{\inits{A.V.}\fnm{Angelos}~\lnm{Vourlidas}\orcid{0000-0002-8164-5948}}
\author[addressref={ROB,SINP},email={andrei.zhukov@sidc.be}]{\inits{A.N.}\fnm{Andrei~N.}~\lnm{Zhukov}\orcid{0000-0002-2542-9810}}

\address[id=SwRI]{Southwest Research Institute, 1050 Walnut Street, Suite 300, Boulder, CO 80302, USA}
\address[id=aff3]{Space Science Laboratory, University of Massachusetts Lowell, Lowell, Massachusetts, USA }
\address[id=cfa]{Center for Astrophysics, Harvard~\&~Smithsonian, Cambridge, MA, 02138 USA}
\address[id=aff6]{DAMTP, CMS, University of Cambridge, Wilberforce Road, Cambridge CB3 0WA, UK}
\address[id=msfc]{NASA Marshall Space Flight Center, Huntsville, AL 35812, USA}
\address[id=NJIT]{New Jersey Institute of Technology, 323 Martin Luther King Jr. Blvd., Newark, NJ 07102, USA}
\address[id=nrl]{U.S. Naval Research Laboratory, Code 7213, 4555 Overlook Ave. SW, Washington, DC 20375, USA}
\address[id=UoD]{Mathematics, School of Science \& Engineering, University of Dundee, Nethergate, Dundee, DD1 4HN, UK}
\address[id=SwRISAT]{Southwest Research Institute, 6220 Culebra Road, San Antonio, TX 78238, USA}
\address[id=GSFC]{NASA Goddard Space Flight Center, Code 670, Greenbelt, MD 20771, USA}
\address[id=aff8]{ADNET Systems, Inc., Greenbelt, MD 20771, USA}
\address[id=inaf1]{INAF - Osservatorio Astronomico di Capodimonte, Salita Moiariello 16, I-80131 Naples, Italy}
\address[id=UPS]{Universit\'{e} Paris-Saclay, CNRS, Institut d'Astrophysique Spatiale, 91405 Orsay, France}
\address[id=IIAS]{Indian Institute of Astrophysics, 2nd Block, Koramangala, Bangalore 560034, India}
\address[id=ROB]{Solar-Terrestrial Centre of Excellence -- SIDC, Royal Observatory of Belgium, Ringlaan - 3 - Avenue Circulaire, 1180 Brussels, Belgium}
\address[id=LASP]{Laboratory for Atmospheric and Space Physics, Space Science, 3665 Discovery Dr, Boulder, CO 80303}
\address[id=mps]{Max-Planck-Institut f\"ur Sonnensystemforschung, Justus-von-Liebig-Weg 3, 37077 G\"ottingen, Germany}
\address[id=psi]{Predictive Science Inc., 9990 Mesa Rim Rd, Suite 170,San Diego, CA 92121}
\address[id=inaf]{INAF-Astrophysical Observatory of Torino, via Osservatorio 20, I-10025, Pino Torinese, Italy}
\address[id=PMOD]{ETH-Z\"urich, H\"onggerberg campus, HIT building, Z\"urich, Switzerland}
\address[id=jhuapl]{Applied Physics Laboratory, Johns Hopkins University, 11100 Johns Hopkins Rd., Laurel, MD 20723, USA}
\address[id=AU]{American University, Washington, DC 20016, USA}
\address[id=North]{Department of Maths, Physics and Electrical Engineering, Northumbria University, UK}
\address[id=ERAU]{Embry-Riddle Aeronautical University, 1 Aerospace Blvd., Daytona Beach, FL, 32114}
\address[id=ncei]{NOAA National Centers for Environmental Information, 325 Broadway, Boulder, CO 80305, USA}
\address[id=NSO]{National Solar Observatory, 3665 Discovery Drive, Boulder, CO 80303, USA}
\address[id=SINP]{Skobeltsyn Institute of Nuclear Physics, Moscow State University, 119992 Moscow, Russia}

\runningauthor{M.J. West et al.}
\runningtitle{The Middle Corona}

\begin{abstract}
The middle corona, the region roughly spanning heliocentric altitudes from $1.5$ to $6$ solar radii, encompasses almost all of the influential physical transitions and processes that govern the behavior of coronal outflow into the heliosphere.  The solar wind, eruptions, and flows pass through the region, and are shaped by it.  Importantly, the region also modulates inflow from above that can drive dynamic changes at lower heights in the inner corona.  Consequently, the middle corona is essential for comprehensively connecting the corona to the heliosphere and for developing corresponding global models.  Nonetheless, because it is challenging to observe, the middle corona has been poorly studied by both major solar remote sensing and in-situ missions and instruments, extending back to the \textit{Solar and Heliospheric Observatory} (SOHO) era.  Thanks to recent advances in instrumentation, observational processing techniques, and a realization of the importance of the region, interest in the middle corona has increased.  Although the region cannot be intrinsically separated from other regions of the solar atmosphere, there has emerged a need to define the region in terms of its location and extension in the solar atmosphere, its composition, the physical transitions it covers, and the underlying physics believed to be encapsulated by the region.  This article aims to define the middle corona, its physical characteristics,  and give an overview of the processes that occur there.
\end{abstract}
\end{opening}

\section{Introduction}
\label{sec:Introduction} 

\citet{Parker1958} showed that the hot corona cannot maintain a hydrostatic equilibrium.  Instead, the pressure-gradient force exceeds gravity and produces a radial acceleration of the coronal plasma to supersonic velocities, the \textit{solar wind}.  Early solar wind velocity observations by the Ulysses spacecraft showed that the solar wind was split rather simply between fast and slow components, the fast wind emanating generally from the interiors of (polar) coronal holes and the slow wind originating near the ecliptic plane.  Observations frequently deviate from this traditional fast/slow dichotomous view, so models of coronal heating and solar-wind acceleration must encompass a much more diverse set of conditions and phenomena to truly achieve a realistic description of the physics of the solar wind \citep{Verscharen2019}. 

The solar wind acceleration region was originally thought to originate beyond 10~solar radii (\Rs); however, new observations suggest that this critical region originates closer to the solar surface \citep{Wexler2020, Raouafi2023}. This height is dictated by the interplay between the open and closed magnetic field, their origins and boundaries, as described by open flux corridors and the S-web \citep[][]{Antiochos2011,Titov2011}.  A new system of potential source, release, and acceleration mechanisms for solar wind types characterized beyond the traditional fast-slow wind dichotomy was presented in \citet{Viall2020}.  Several of those mechanisms (e.g. streamer blob release) take place at locations within the \textit{middle corona}.

The middle corona is a critical transition region between the highly disparate physical regimes of the inner and outer solar corona\footnote{Throughout this article we will adopt the common nomenclature of \emph{inner} and \emph{outer} corona, as opposed to \emph{lower} and \emph{upper}, or \emph{extended} corona.}.  Nonetheless, the region remains poorly understood due primarily to historical difficulties in observing it.  The boundaries of the region have been debated for many years. Nevertheless, through a series of open community meetings and extensive discussions we have arrived at a common set of boundaries to define the middle corona.  Our consensus considers both the variation in roles that different physical mechanisms play throughout the corona and the historical observational context of coronal observations. We define the middle corona: $\approx1.5-6$\,\Rs (measured from disk center).

The inner boundary roughly traces the tops of the closed magnetic field structures that dominate the inner corona, below which loops appear and hydrostatic scale heights are often applicable \citep[e.g.][]{Koutchmy1992, Koutchmy1994, Winebarger2002, Koutchmy2004}.  The outer boundary is roughly pinned to where the solar atmosphere is believed to have \emph{fully} transitioned to an outflow regime, and is observed to be fully radial in structure.  This is evident in coronal hole structures which appear to be purely radial beyond 3\,--\,4 \Rs and no longer exhibit super-radial expansion \citep[][]{DeForest1997, DeForest2001b}.  \citet{Schatten1969} chose the source surface height for potential field source surface \citep[PFSS; see also][]{Wang1992} extrapolations based on matching the interplanetary magnetic field to the number of surviving field lines; a value that has typically been located between 3 and 6 \Rs\ \citep[e.g.][]{McGregor2008}.  It is also around this height that ``Sheeley Blobs,'' small-scale density inhomogeneities frequently observed flowing both inwards and outwards in streamers, are believed to be pinched-off through magnetic reconnection \citep[e.g.][]{SanchezDiaz2017}.

The region thus encapsulates several important physical transitions, including the change from predominantly \textit{closed} to \textit{open} magnetic field structures, and the change from low to high plasma $\beta$ in quiet sun regions \citep{Vourlidas2020}.  A list of transitions occurring in this region can be found in Table \ref{tbl:MiddleCoronaTransitions}, several inner coronal transitions are also included for comparison. 

\begin{table}
\caption{A table of transitions, in the inner and middle corona where: FSW = Fast Solar Wind, SSW = Slow Solar Wind.}
\begin{tabular}{p{0.13\linewidth} p{0.26\linewidth} p{0.28\linewidth} p{0.14\linewidth}}
  \hline
{\RaggedRight \textbf{Type of transition}} & \textbf{Inner corona} & \textbf{Middle corona} & \textbf{Context}  \\
  \hline

\multirow{8}{1\linewidth}{\textbf{Structure}} &  & {\RaggedRight Closed-to-open magnetic field configurations} & {\RaggedRight SSW, streamer regions} \\ \cline{2-4}
&  & {\RaggedRight Density structures/ ``blobs'' released into outflow} & SSW, streamer cores \\ \cline{2-4}
& {\RaggedRight Confinement regime with elevated density power law radial dependence} & {\RaggedRight Density radial dependence drops to near inverse-square scaling} & {\RaggedRight SSW, streamer regions} \\
\hline

\multirow{4}{1\linewidth}{\textbf{Dynamics}} &  & Subsonic-to-supersonic solar wind outflow & SSW \\ \cline{2-4}
&  & CME main acceleration and initial shock formation & CME \\
\hline

\multirow{15}{1\linewidth}{\textbf{Plasma physics}}  & Plasma $\beta$ $\ll$\,1 &  & {\RaggedRight Coronal Holes \& FSW} \\ \cline{2-4}

 & {\RaggedRight Plasma $\beta$ $<$\,1 in innermost corona} & {\RaggedRight Broad range of $\beta$ spanning} $<$\,1 to $>$\,1 & SSW, streamer regions \\ \cline{2-4}

 & Charge state freeze-in &  & FSW \\ \cline{2-4}

 &  & {\RaggedRight Stabilization/freeze-in of ionization charge states} & SSW \\ \cline{2-4}

 & {\RaggedRight Gravitational settling affecting FIP abundances} &  & Streamer bases \\ \cline{2-4}
    
 &  & {\RaggedRight Gravitational settling affecting FIP abundances} & Streamer cores \\ \cline{2-4}
  
 & {\RaggedRight Coulomb collisions to kinetic plasma processes} &  & FSW \\ \cline{2-4}
 
 &  & Coulomb collisions to kinetic plasma processes & SSW \\

\hline
\label{tbl:MiddleCoronaTransitions}
\end{tabular}
\end{table}
New observations reported by \citet{Seaton2021} suggest heliospheric solar wind structures not only originate in the inner corona \citep[e.g.][]{DeForest2018}, but can originate from complex dynamics in the middle corona \citep{Chitta2022}.  The region is also believed to influence the inner corona, where downflows have been shown to interact with structures below.  For example, supra-arcade downflows \citep[SADs: e.g.][]{Savage2012, Shen2022} observed in the wake of eruptions correspond to plasma pile up in the inner corona, and smaller or fainter downflows may also be ubiquitous in the less dynamic atmosphere \citep{Sheeley2002}. Such downflows may trigger larger scale eruptive phenomena, or erode magnetic fields that could trigger eruptions through mechanisms such as magnetic breakout \citep[e.g.][]{Antiochos1999}. Thus, the middle corona not only plays an important role in shaping outflow, as the region through which all outflow and eruptions must pass and be modulated, but the middle corona's physics also has important implications for unified coronal-heliospheric models.

\begin{figure}
\centering
\includegraphics[width=1.00\columnwidth]{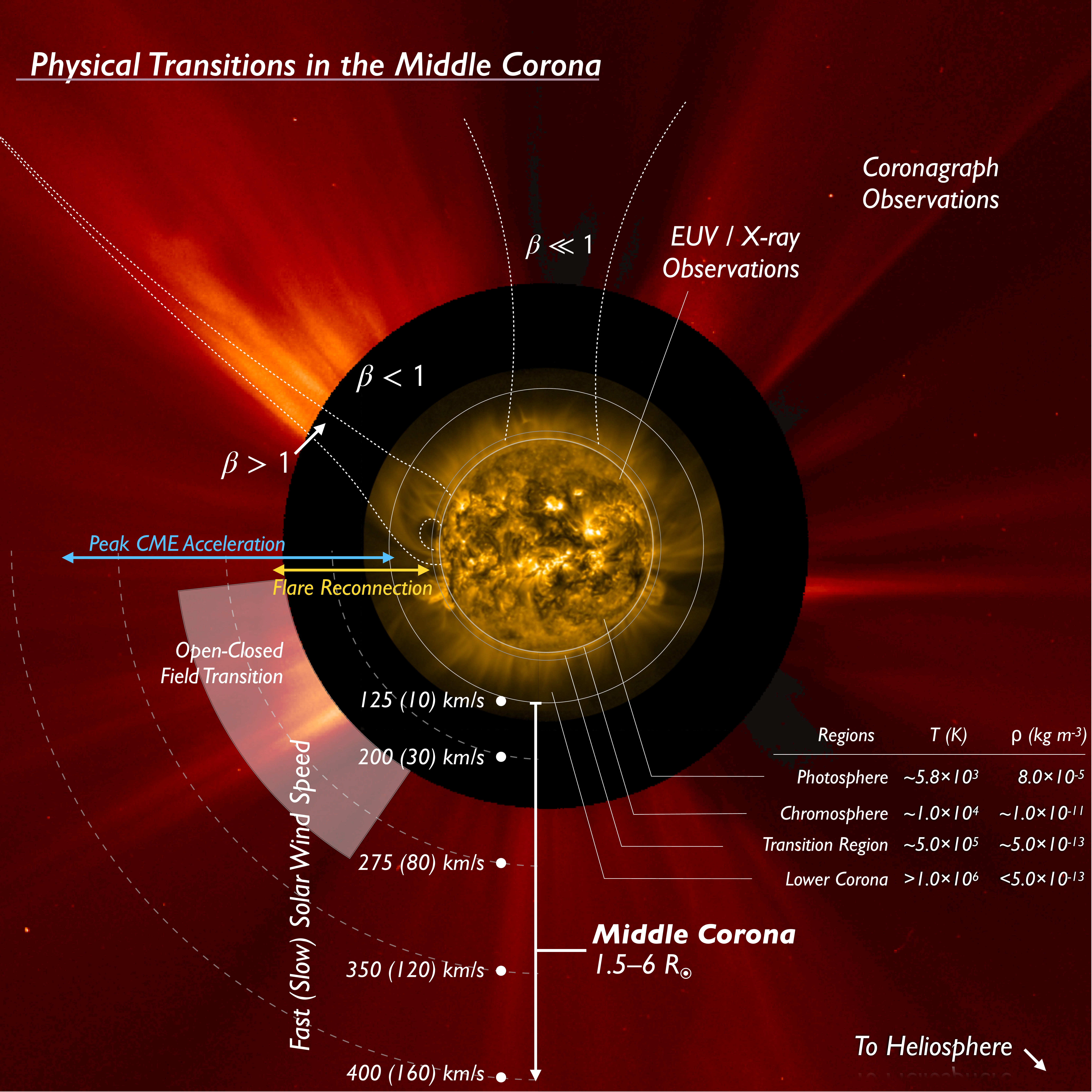}
\caption{A SWAP and LASCO composite image highlighting the middle corona, and the physical transitions that extend through the region. The image also highlights the observational gap between EUV observations of the inner corona and visible-light observations of the outer corona, currently experienced from the Earth perspective \citep[e.g.][]{Byrne2014}. The image is annotated to highlight key heights, coronal characteristics, and physical transitions.} 
\label{fig:CoronalOverview}
\end{figure}

Historically, the solar corona and its continuous evolution has most commonly been studied using a combination of extreme ultra-violet (EUV) and X-ray observations of the inner corona with visible-light coronagraph observations of the outer corona, as shown in Figure \ref{fig:CoronalOverview}.  The observations in the figure, from 2014 when the Sun was near the peak of its activity cycle, include an EUV image in the center (gold false color), from the large field-of-view (FOV) PROBA2/{\it Sun Watcher with Active Pixels and Image Processing} \citep[SWAP:][]{Seaton2013,Halain2013} imager, whose passband is centered on \SI{17.4}{\nano\metre}, and a visible-light image (red false color) from the SOHO/{\it Large Angle and Spectrometric Coronagraph} \citep[LASCO:][]{Brueckner1995} C2 coronagraph, around the edge.

Figure \ref{fig:CoronalOverview} is annotated to highlight atmospheric regions, phenomena, characteristic solar wind speeds, and various coronal transitions. The EUV observations of the inner corona reveal the shape of structures permeating the region -- highlighted by the emitting plasma -- that are constrained by the corona's magnetic field. EUV observations of this region reveal it to be largely dominated by closed magnetic structures.  In contrast, the visible-light observations reveal more striated structures, indicative of open magnetic structures, extending out into the heliosphere. 

Although EUV and visible-light observations have both served as synoptic probes of the corona, these two observational regimes have generally been focused on different regions of the middle corona, and through disparate passbands. Thus, they capture different physical characteristics of the underlying plasma: emission measure within a specific temperature range in the case of EUV and temperature-independent electron density in the case of visible-light. 

The general lack of continuously available overlap between the different methods of observation, especially from the Earth's perspective (highlighted by the observational gap in Figure \ref{fig:CoronalOverview}) can lead to ambiguity, both when tracking structures and inferring plasma properties such as temperatures and densities.  Methods to continuously infer plasma properties include extrapolation and modeling \citep[e.g.][]{Lynch2020, Schlenker2021}; however, even for the relatively simple case of a quiet sun-streamer structure, various different estimates of the densities and temperatures have been published \citep{DelZanna2018}. To fully elucidate the mechanisms affecting the large-scale structural and dynamic changes occurring over the middle corona, complementary observations that overlap adjacent zones are essential. 

In this article, we propose a definition for the region called the middle corona, we review how we observe it and what we know about it, and we present both the open questions concerning the region and a strategy to explore it. In Section\,\ref{sec:how_we_observe} we describe how we currently -- and historically -- observe the middle corona; in Section\,\ref{sec:Properties_and_transitions} we describe the properties and topology of the middle corona; in Section\,\ref{sec:modeling_the_mc} we describe some of the efforts to model and extrapolate properties of the region. Finally, in Section \ref{sec:Discussion} we present a discussion of the region, in the form of open questions pertaining to the region and ways of answering them.

\section{Partitioning the Solar Atmosphere}
\label{sec:partition}

Although the Sun is effectively a continuous ball of plasma with no physical boundaries, the solar interior is typically demarcated into layers based on the dominant physical processes that govern the energy transport in the respective regions.  A similar logic is applied to the solar atmosphere, where the partitions are based on thermal and magnetic properties. These properties not only dictate the emission mechanisms and physical length-scales at play, but ultimately how we observe and model the different regions. 

In general the high magnetic field strengths, plasma conductivity, temperatures, and densities, and the inhomogeneity of these properties within the inner corona make formal calculations of its properties inherently complex, so the average properties of particles are often adopted. This introduces the magnetohydrodynamic (MHD) approach to modelling the region, which treats the plasma as a bulk magnetized fluid \citep[][and references therein]{Gombosi2018}. In the outer corona, where length-scales have increased, kinetic models are both more practical and more commonly used, and the equations of motion for each particle, subject to various forces, are calculated \citep[][and references therein]{Marsch2006}. The middle corona acts as the interface between these two regions, and therefore requires a combination of approaches.

The transitions between the three very distinct physical regimes of the inner, middle, and outer corona are not themselves distinct, largely due to the range in length-scales and scale heights experienced among different coronal regions \citep{Chhiber2022,Malanushenko2022}, and their variation throughout the solar cycle \citep{Badalyan1993,Edwards2022}. However, rapidly advancing observational and data processing techniques have provided new insights into the region, and new proposed missions to explore the region have led to the term ``middle corona'' entering the solar and heliospheric physicists' lexicon in recent years \citep{Koutchmy2004}. There is a clear need to define both the terminology describing this region as well as its properties, which is the goal of this article.

\section{How We Observe the Middle Corona}
\label{sec:how_we_observe} 
There are a variety of reasons that the middle corona has not been as well characterized as other regions of the solar atmosphere. These include limitations on instrumentation and instrumentation capabilities, prioritization of other investigations, and the observation of other regions. Nonetheless, through dedicated observation campaigns and increasingly sophisticated spectroscopic, imaging, and data processing techniques, large portions of the middle corona have been intermittently probed. Figure \ref{fig:MC_Missions} presents a rough overview of many past, present, planned, and proposed observatories that contribute to our knowledge of the region.

\begin{figure}
\begin{center}
\includegraphics[angle=270,
width=0.80\columnwidth]{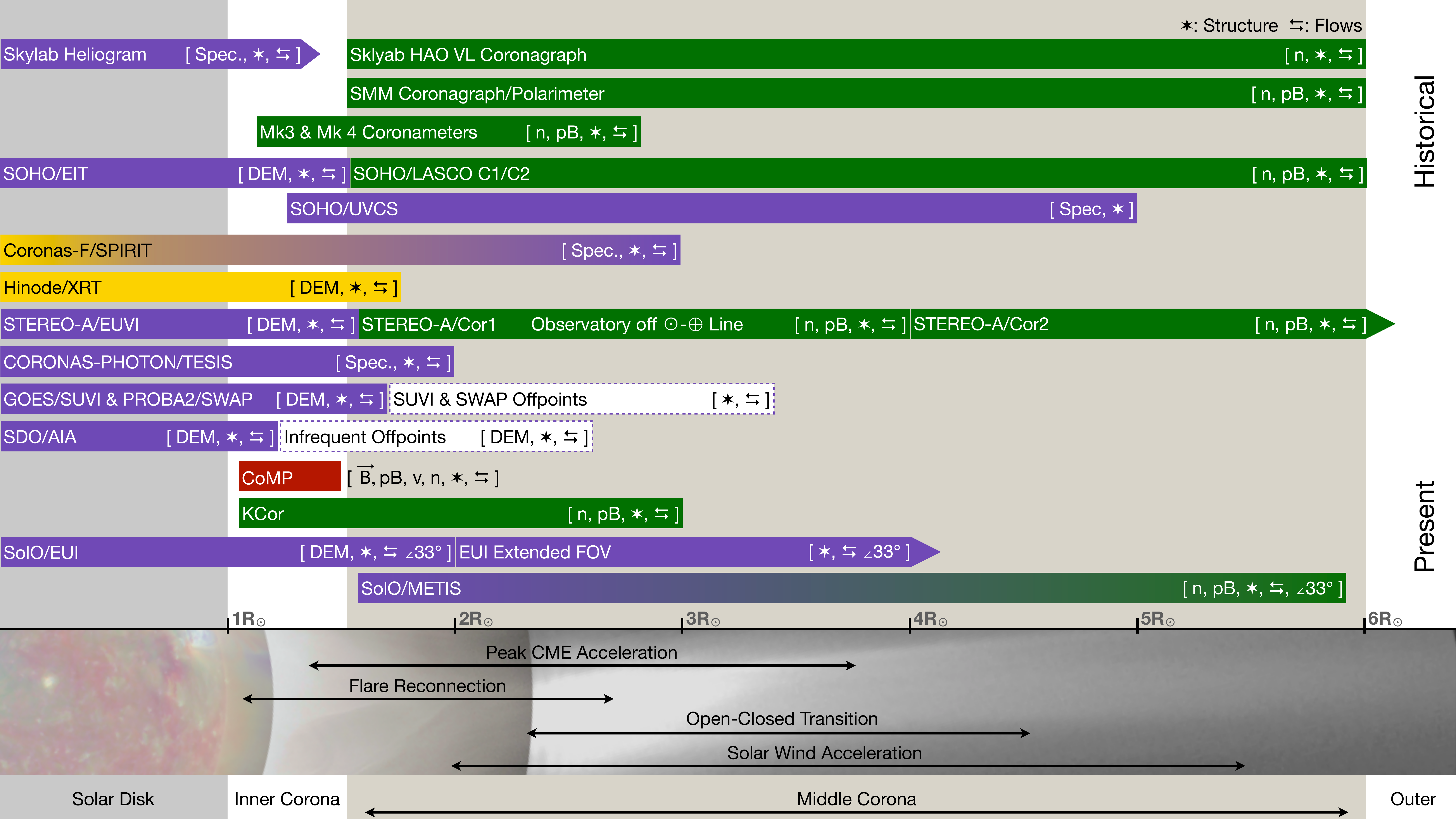}
\caption{\small (\textbf{a}) Summary of past, present, planned, and proposed middle corona observatories. \textit{The type of observation is indicated in the brackets}, with key to symbolic abbreviations in upper right of the figure. \textit{Color corresponds to the wavelength regime of the observation, X-ray (Gold), EUV/UV (Violet), Visible (Green), Infrared (Red), and Radio (Gray)}.}
\label{fig:MC_Missions}
\end{center}
\end{figure}

\begin{figure}
\begin{center}
\ContinuedFloat
\includegraphics[angle=270, width=0.80\columnwidth]{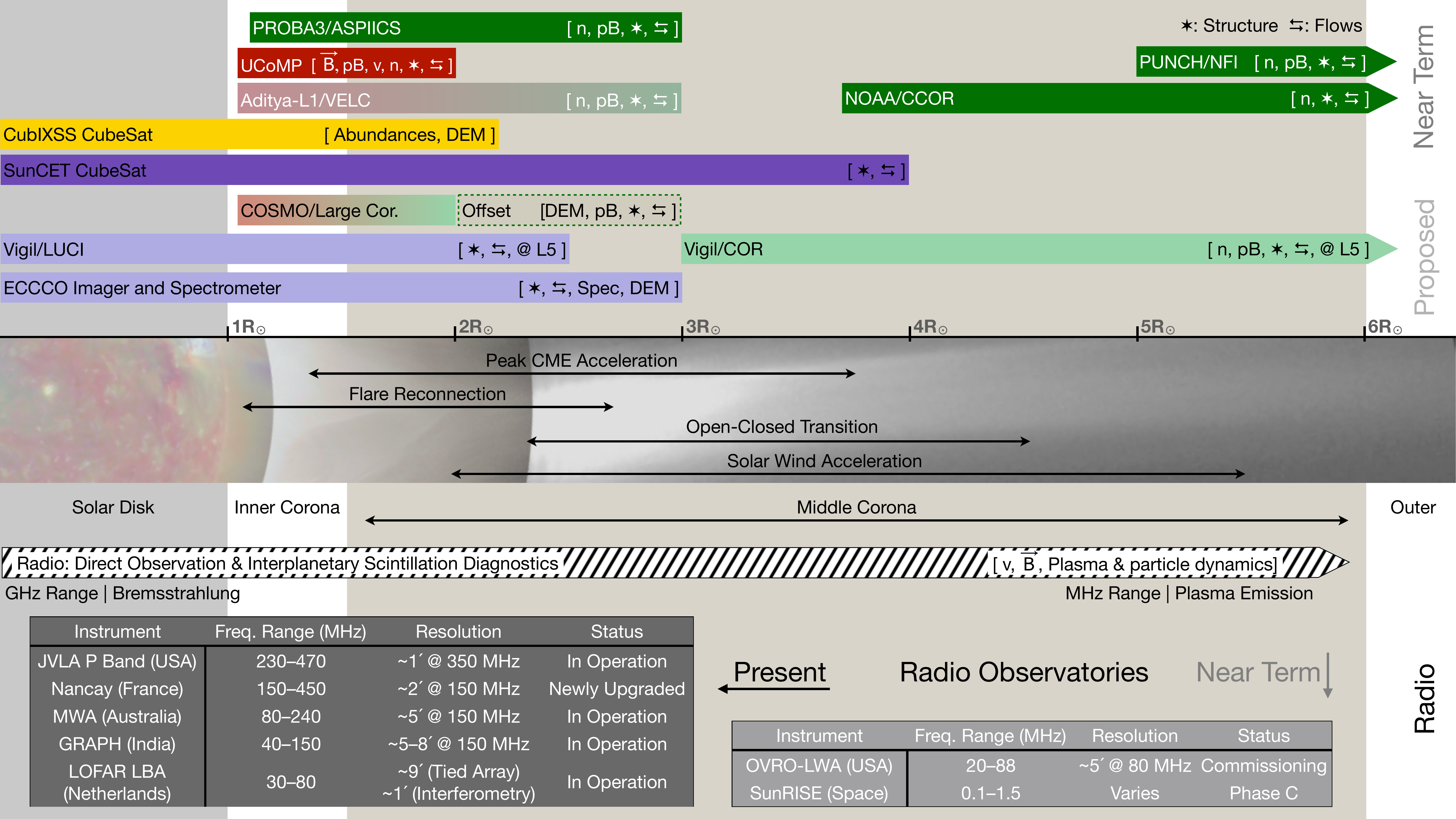}
\caption{\small (\textbf{b}) Continuation of middle-corona observatories.  \textit{The type of observation is indicated in the brackets}, with key to symbolic abbreviations in upper right of the figure. \textit{Color corresponds to the wavelength regime of the observation, X-ray (Gold), EUV/UV (Violet), Visible (Green), Infrared (Red), and Radio (Gray)}.}
\end{center}
\end{figure}

Many of the most prominent observations of the middle corona have been made in wavelengths ranging from X-rays to infrared, but radio imaging and radio measurements of the middle corona have provided important insights into the underlying plasma characteristics. Spectroscopic instruments, particularly in the ultraviolet, have also made important contributions to our understanding of the properties and dynamics of middle corona plasma.

In general, instruments that make continuous observations in visible-light, EUV, and X-ray passbands are located on space-based platforms, where they can observe the corona unencumbered by the Earth's atmosphere and day-night cycles, whereas observations at radio wavelengths are made from ground based sites due to the size of instruments. Both sets of observation utilize different observing techniques, and rely on different emission mechanisms, which we review below. The following section has been divided in to two main subsections: the first examines observations made through IR, visible-light, EUV, and X-ray wavelengths (\ref{ssec:ShortWavelengths}), and the second covers observations made through radio imaging and measurements (\ref{ssec:radioWavelengths}).

\subsection{Short Wavelengths: Infrared, Visible, UV, and X-Rays}
\label{ssec:ShortWavelengths}

\begin{figure}
\centering
\includegraphics[width=0.92\columnwidth]{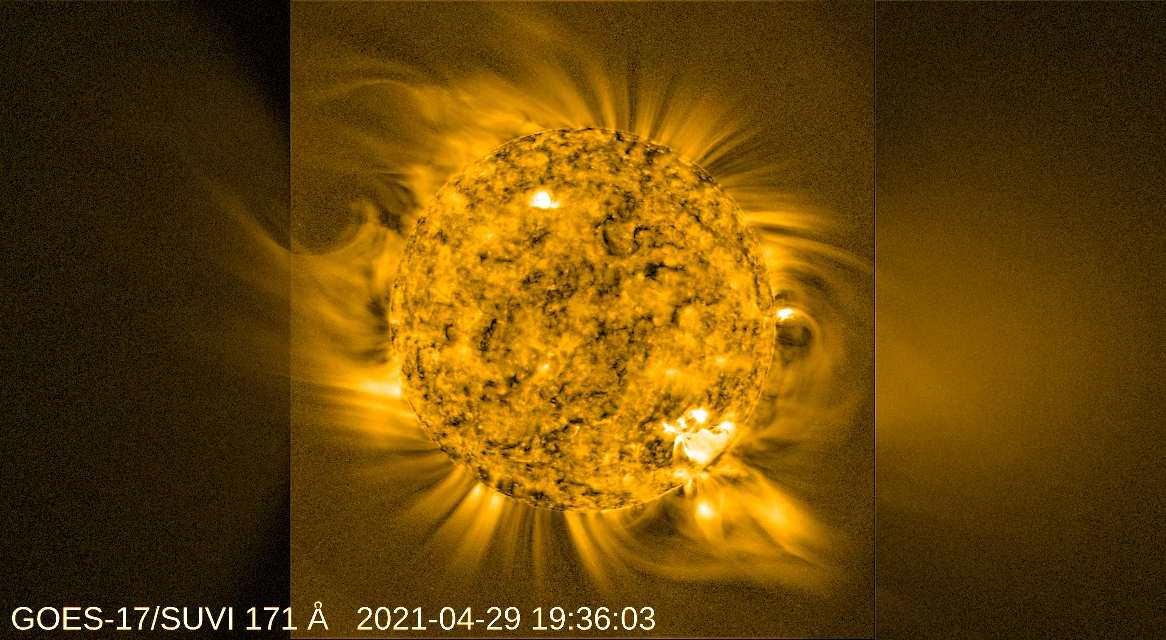}
\includegraphics[width=0.92\columnwidth]{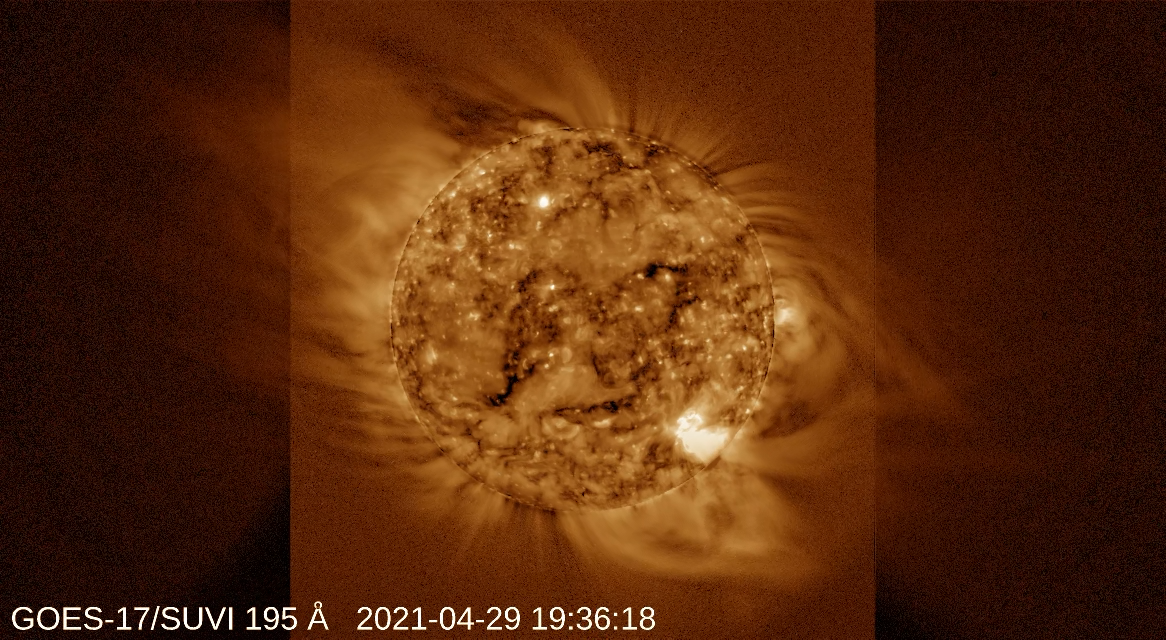}
\includegraphics[width=0.455\columnwidth]{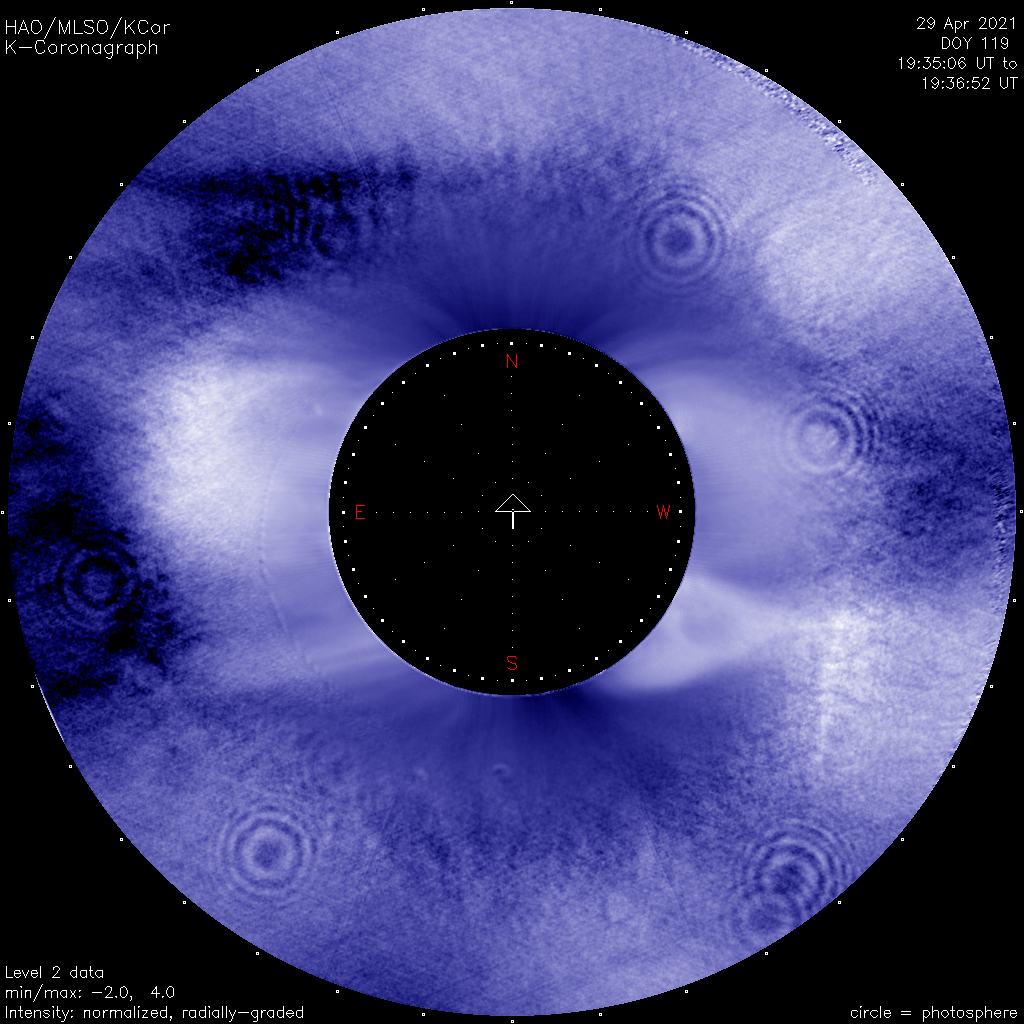}\ \includegraphics[width=0.455\columnwidth]{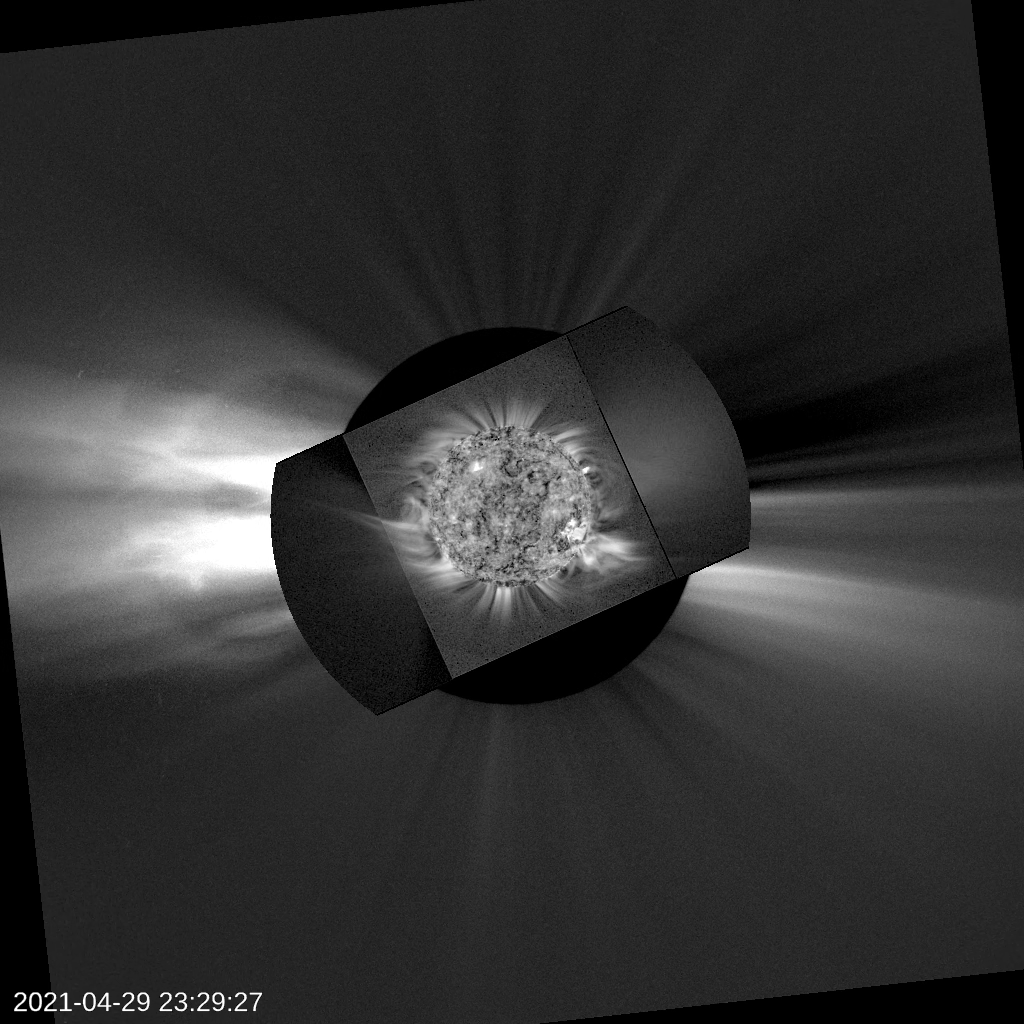}
\caption{Different views of the middle corona, observed on 29 April 2021, in EUV from SUVI (top in \SI{17.1}{\nano\metre}, and middle panel in \SI{19.5}{\nano\metre}) and visible light from K-Cor (bottom left) and LASCO (bottom right; with SUVI superimposed). Images are in camera coordinates and not necessarily co-aligned, though solar north is generally upwards in each frame.} 
\label{fig:MC_Imaging}
\end{figure}

Although there remain persistent observational gaps \citep[][]{Byrne2014}, the middle corona has occasionally been observed by a disparate set of instruments in passbands that range from the infrared to X-ray.  The most extensive observations have been made with visible-light images, both from coronagraphs and eclipses, as well as direct EUV imaging, primarily through dedicated off-point campaigns by imagers designed to observe the inner corona.  Figure\,\ref{fig:MC_Imaging} shows examples of several such observations, from a coordinated campaign during April 2021, that included offpoints by the GOES Solar Ultraviolet Imager \citep[SUVI][]{Darnel2022}, the Mauna Loa Solar Observatory's K-coronameter \citep[K-Cor:][]{Elmore2003}, and LASCO on SOHO.  These coordinated observations allow us to characterize different aspects of the middle corona, leveraging several different mechanisms through which plasma in the region manifests itself.  Here we provide a brief overview of the history of these observations and the variety of phenomena observed here using these approaches.

\subsubsection{Observed Emission Mechanisms}
\label{sssec:VLEUVXRayEmissionMechanisms}

The inner corona exhibits temperatures ranging between $T\approx5\times10^{5}$ and $>$2$\times10^{7}$\,K, and consequently highly ionized atoms, emitting at UV, EUV and X-ray wavelengths, provide a key diagnostic of temperature and are very commonly used to observe this region.  This highly ionised emission is strongly dependent on electron density ($n_{ \mathrm{e} }$).  The dominant emission mechanisms are spontaneous emission following \textit{collisional excitation} and \textit{resonant scattering} of incident light by ions.  The intensity of emission resulting from scattering mechanisms is proportional to number density ($\propto n_{ \mathrm{e} }$) while emission from collisional excitation is proportional to density squared ($\propto n_{ \mathrm{e} }^{2}$). In the innermost corona, collisional excitation dominates all emission mechanisms other than broadband Thomson scattering, and in the absence of large scale structures, its $\propto n_{ \mathrm{e} }^{2}$ relationship gives rise to a rapid drop-off in brightness as density decreases with height.  The belief that this drop-off would limit the viability of EUV observations above $1.5$\,\Rs\ led most past observational efforts in these wavelengths to focus only on the inner corona.

\begin{figure}[ht!]
\centering
\includegraphics[width=0.85\columnwidth]{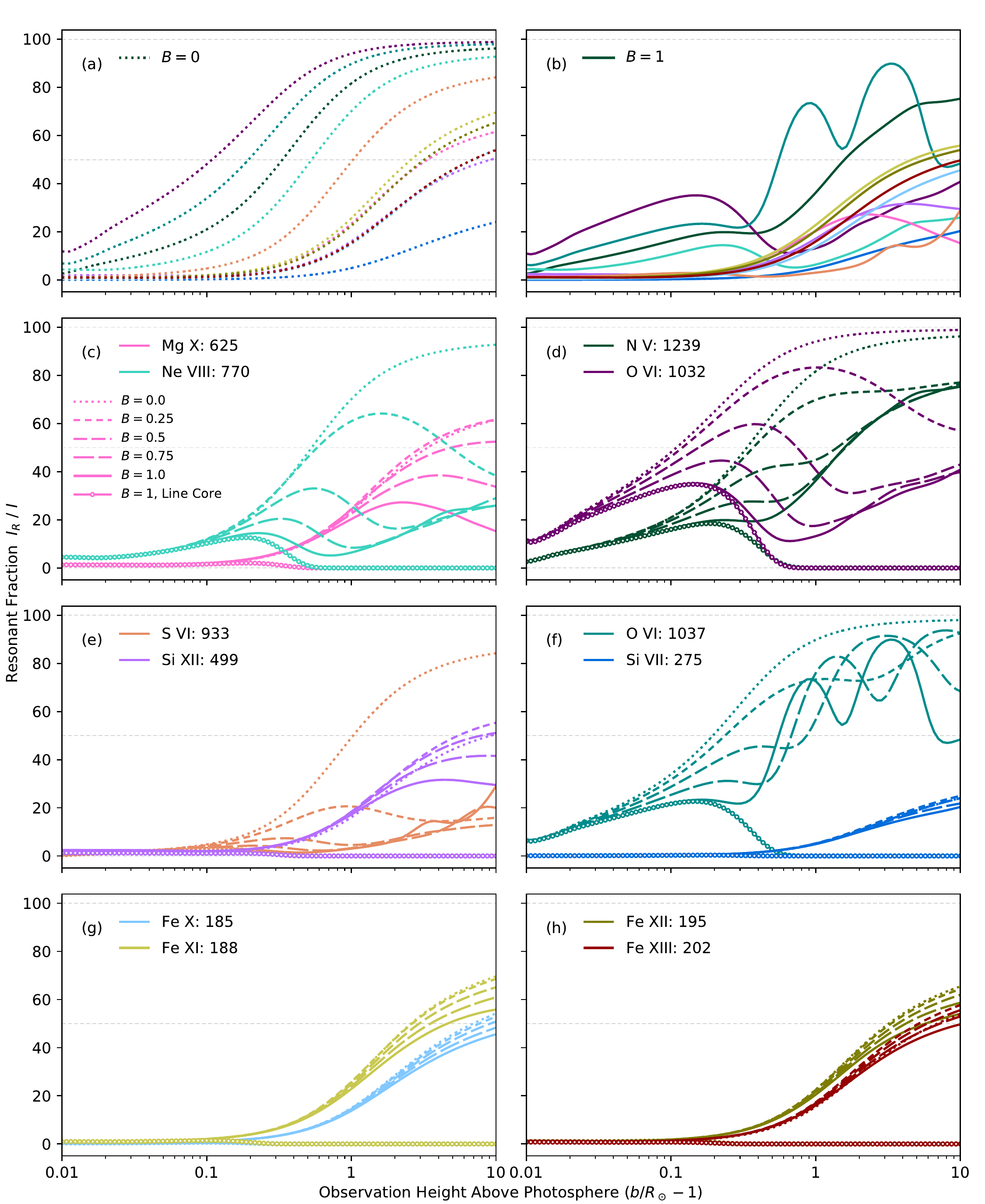}
\caption{The proportion of the total emissivity contributed by resonant scattering as a function of height, for various fractions $B$ of the model value of the solar wind speed. $B$ is a scalar factor applied to the radial wind speed profile, with $B=0$ indicating no wind and $B=1$ indicating wind at nominal modelled values. Ion-line wavelengths are given in units of Angstroms. 
\citep[Figure 17 of][used with permission]{Gilly2020}.}
\label{fig:emissivity-compare}
\end{figure}

At larger heights, resonant scattering can begin to dominate the ion and neutral emission.  The relative contribution of resonant scattering and collisional excitation to the total emissivity of the plasma depends on the local density (both ion and electron), temperature, the collisional excitation rate, and the incident radiation at a given wavelength.  The resonant scattering generally increases the emission, but for some lines, those excited by radiation from particularly strong chromospheric emission lines, Doppler-dimming can lead to a strong decrease of the amplitude of the scattered radiation.  As the solar wind leaves the inner corona, it is accelerated until it reaches such a velocity that the incident light is no longer at the same wavelength as the spectral line at rest, thereby reducing the total amount of photon scattering from that particle. Note that in addition to the bulk outflow velocities, there are large thermal motions of the ions (with a dependence on atomic mass) that effectively smears out the relative velocities with respect to the solar surface and reduces the amplitude of the dimming.  Occasionally, the scattering can lead to Doppler-pumping, where a Doppler shift causes the resonant wavelength of the coronal ions to match the wavelength of a spectrally-adjacent line, such as is the case with O\,\textsc{vi} 103.8\,nm. Figure \ref{fig:emissivity-compare} from \cite{Gilly2020} highlights the relative proportion of resonant scattering to the total emissivity as a function of wind speed, pointing to a potential diagnostic for solar wind acceleration in the middle corona. 

New EUV observations \citep[e.g.][]{Goryaev2014, Seaton2021} have shown that resonant scattering of emission from EUV-bright inner-corona features ($\propto n_{ \mathrm{e} }$) occurs in many structures. Thus the brightness of the EUV corona declines less precipitously than anticipated for purely collisionally excited emission.  This resonantly scattered emission can enhance the visibility of large-scale features in the middle corona for a new class of coronal observatories (see Section\,\ref{sssec:VLEUVXRaySpatiallyResolvedObservations}). %

In contrast to emission line diagnostics, broadband visible-light observations from coronagraphs and eclipses reveal Thomson scattered emission \citep{Inhester2015}, which is only sensitive to electron density.  Differences in the nature of complex 3D structures that are manifested in these different emission mechanisms cannot always be reconciled in multi-wavelength studies.  Thus visual confusion in the fine structure of this region has led to additional barriers to resolving key questions about the middle corona.

In narrowband visible and near infrared, because of the much greater flux from the photospheric radiation, resonant scattering begins to dominate the emission processes for lines in this spectral interval already at relatively low heights above the surface. Because of the scattering of that same photospheric emission in the Earth's atmosphere, and the near lack of space-borne coronagraphs capable of observing coronal emission lines, it is more typical to study the visible-light corona at these wavelengths, both at eclipses and using a polarization-discriminating coronagraph \citep[but see][for emission line measurements at 2\,\Rs]{Ding_Habbal_2017}. 

\subsubsection{Optical Observations}
\label{sssec:VLEUVXRaySpatiallyResolvedObservations}

Images of the middle corona are primarily produced in visible and infrared light using coronagraphic instruments or during eclipses, or in UV, EUV and X-ray passbands using telescopes that can directly image the solar disk.  High quality eclipse observations that include the middle corona date to the nineteenth century \citep{Holden1894}, while coronagraphic observations extend the pioneering work of \citet{Lyot1939}.

The space age opened the door to both higher quality coronagraphic observations and exploration in EUV and X-Rays.  An important milestone was the \textit{Skylab} mission, which carried both the High Altitude Observatory (HAO) visible-light coronagraph \citep{MacQueen1974}, whose FOV covered the middle corona more or less exactly -- 1.5 to 6.0\,\Rs -- and the Naval Research Lab's \textit{Extreme Ultraviolet Spectroheliograph}, which made spectrally dispersed images of the inner and, occasionally, middle corona over a wide range of EUV wavelengths \citep{Tousey1973, Tousey1977}. An even more significant breakthrough in middle corona studies came with the \textit{Solar Maximum Mission} (SMM) coronagraph and polarimeter instrument \citep{MacQueen1980}, which shared significant heritage and its FOV with the Skylab coronagraph but made many more systematic observations.

These space-based visible-light observations were augmented by a more sporadic set of ground-based observations: numerous eclipses, coronagraphs such as HAO's several instruments, first in Climax, Colorado, \citep{Wlerick1957}, and then on Mauna Loa \citep[including, notably, the Mark III K-coronameter][and several subsequent improved designs]{Fisher1981}.  These visible-light instruments and their space-based counterparts exploit the scattering of photospheric light by electrons in the corona (Thomson scattering) to image the corona, and must contend with the challenge of eliminating light from the photosphere, which is nearly $10^7$ times brighter than the corona at 1.5\,\Rs. Coronagraphic imaging therefore requires very efficient stray light suppression, which must overcome both scattering of light and diffraction at the edges of optical components.  This is generally easier to achieve with instruments having large separations between occulter and primary objective, in the case of externally occulted instruments, though the specifics of the designs of these instruments differ considerably.  The challenge of observing close to the solar limb is particularly acute, and as a result, coronagraphic observations of the innermost middle corona are generally affected by stray light. 

Nonetheless, in the more than 25 years since the beginning of the SOHO mission, most of the middle corona has been observed in visible-light by the LASCO suite \citep{Brueckner1995} and in the UV/EUV by the Ultraviolet Coronagraph Spectrometer \citep[UVCS:][]{Kohl1995},in EUV in the far edges and corners of images from the Extreme-ultraviolet Imaging Telescope \citep[EIT:][]{EIT1995}, and subsequently by a fleet of instruments that observed in visible-light, EUV, and X-Rays, including from multiple perspectives.  These include the EUV/Visible Sun-Earth Connection Coronal and Heliospheric Investigation (SECCHI) on the twin STEREO spacecraft \citep{Howard2008}, the SPectrographIc X-Ray Imaging Telescope-spectroheliograph \citep[SPIRIT:][]{Zhitnik2002}, the TElescopic Spectroheligraphic Imaging System telescope \citep[TESIS:][]{Kuzin2009}, the SWAP EUV Imager on PROBA2 \citep{Seaton2013, Halain2013}, the GOES Solar Ultraviolet Imager \citep[SUVI:][]{Darnel2022}, the GOES Soft X-Ray Imager \citep[SXI:][]{Hill2005, Pizzo2005}, the \textit{Hinode}/X-Ray Telescope \citep[XRT:][]{Golub2007}, the \textit{Solar Orbiter} Extreme Ultraviolet Imager \citep[EUI:][]{Rochus2020} and the Metis coronagraph \citep{Antonucci2020}.  Additional planned missions will soon push the boundaries of observations of the middle corona both farther outwards (in EUV) and inwards (for coronagraphs).

\begin{figure*}
\centering
\includegraphics[width=0.99\textwidth]{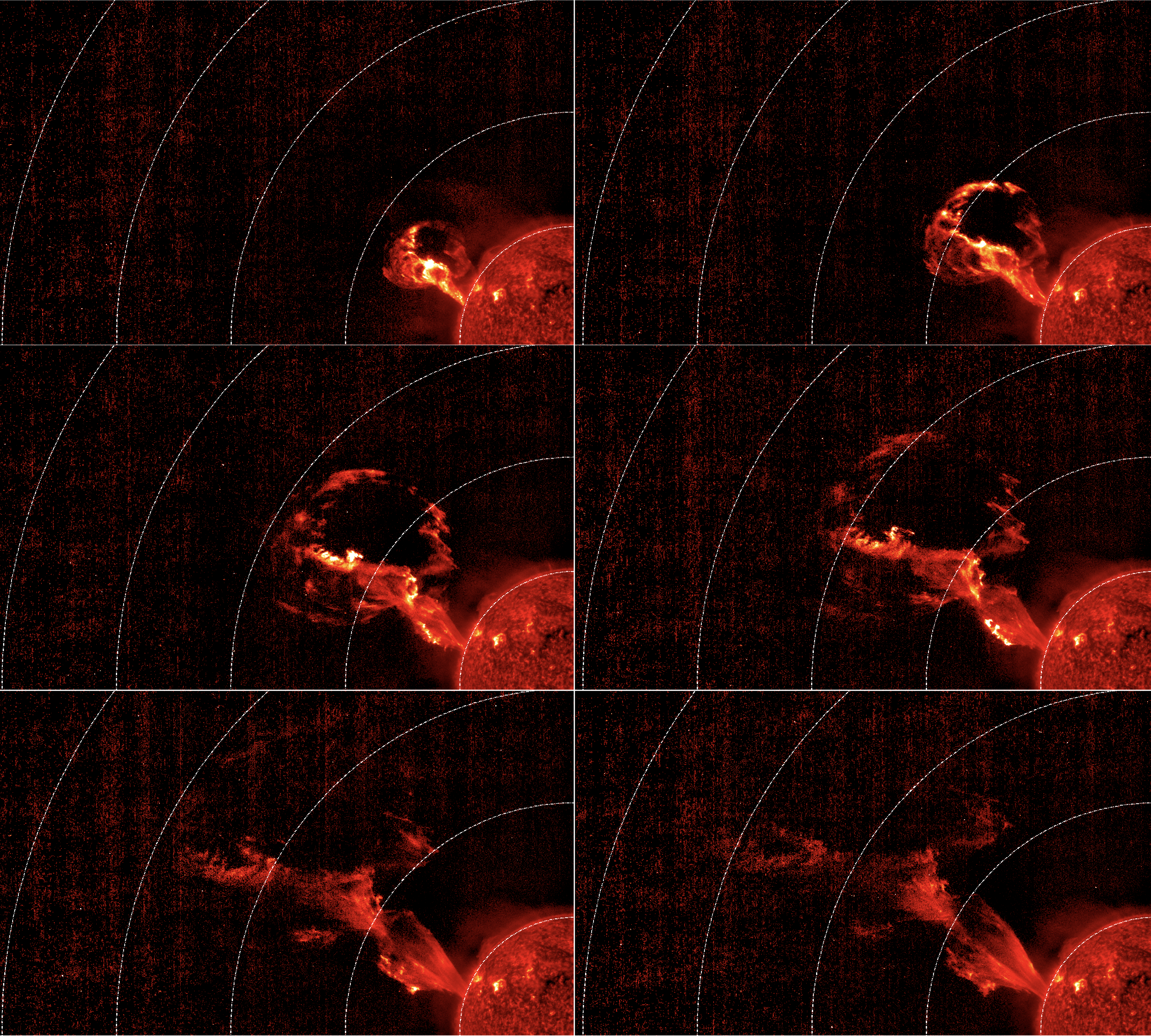}
\caption{A prominence eruption observed through the \SI{30.4}{\nano\metre} passband of the EUI Full Sun Imager on 15 Feb 2022, when the Solar Orbiter satelite was located at 0.73\,AU from the Sun, at 22:00\,UT (top left), 22:04\,UT (top right), 22:10\,UT (middle left), 22:14\,UT (middle right), 22:20\,UT (bottom left), and 22:24\,UT (bottom right).  The observations have been processed using the radial filtering technique described in \citet{Seaton2023} to enhance the off limb signal, allowing the eruption to be tracked out to 5\,\Rs.  See \citet{Mierla2022} for further details about this event.}
\label{fig:EUI_example}
\end{figure*}

Arguably the most important innovation in middle corona studies of the last decade has been a series of exploratory campaigns using off-pointed EUV images. These include both short-term campaigns with the SWAP imager \citep{OHara2019, Goryaev2014} and long-term campaigns using SUVI \citep{Seaton2021, Chitta2022}.  Such observations, using instruments with medium fields of view in novel ways to extend their observational range -- along with a handful of reports from less well known instruments with dedicated larger fields of view \citep[e.g.][]{Reva2017} -- definitively proved the feasibility of middle corona observations with dedicated EUV instruments.  The most recent and prominent of these instruments is the Full-Sun Imager (FSI) in Solar Orbiter's EUI suite, with a varying instantaneous FOV due to its highly variable distance from the Sun.  Preliminary FSI observations have already demonstrated its ability to track erupting prominences from their genesis to the outer edge of the middle corona \citep{Mierla2022}. Figure~\ref{fig:EUI_example} shows the propagation of such a prominence observed by the EUI Full Sun Imager on 15 Feb 2022, the observations have been processed using the radial filtering technique described in \citet{Seaton2023} to enhance the off limb signal, allowing the eruption to be tracked out to 5\,\Rs.


These pioneering EUV instruments have paved the way for a new generation of EUV instruments and techniques that focus specifically on the middle corona, including the Sun Coronal Ejection Tracker (SunCET) CubeSat \citep{JMason2021, Mason2022}, in development now, and proposed EUV CME and Coronal Connectivity Observatory \citep[ECCCO; previously referred to as the COronal Spectrographic Imager in the EUV or COSIE:][]{Golub2020} and a potential successor to the Lagrange eUv Coronal Imager \citep[LUCI:][]{West2020} on the Vigil mission.

Likewise, pioneering visible and near-IR observations both from coronagraphs and eclipses, have paved the way for a new generation of coronagraph instruments with improved imaging capabilities in the inner and middle corona.  These include the Coronal Solar Magnetism Observatory \citep[COSMO:][]{Tomczyk2016}, a suite of ground-based coronagraphic instruments, and Association of Spacecraft for Polarimetric and Imaging Investigation of the Corona of the Sun \citep[ASPIICS:][]{Lamy2010, Galano2018, Shestov2021}, the visible coronagraph on the PROBA-3 formation-flying space mission. See Figure\,\ref{fig:MC_Missions} for a summary of notable historical, active, and planned and proposed middle corona observations.

Middle corona studies have also benefited from the development of advanced image processing techniques during the past two decades.  The steep gradient in intensity as a function of height in the corona, both in visible and shorter wavelength observations, means that the dynamic range of solar images is far greater than can be captured in a single exposure by typical scientific cameras or displayed on a computer screen.  Therefore, techniques that can overcome this to generate high-quality, large-FOV images, which still preserve fine details on many scales have been developed. There are over 20 separate methods in the literature that process solar imagery to draw out hidden detail \citep[e.g.][]{Druckmullerova2011, Seaton2023, Auchere2023}.

\begin{figure*}
\centering
\includegraphics[width=0.49\textwidth]{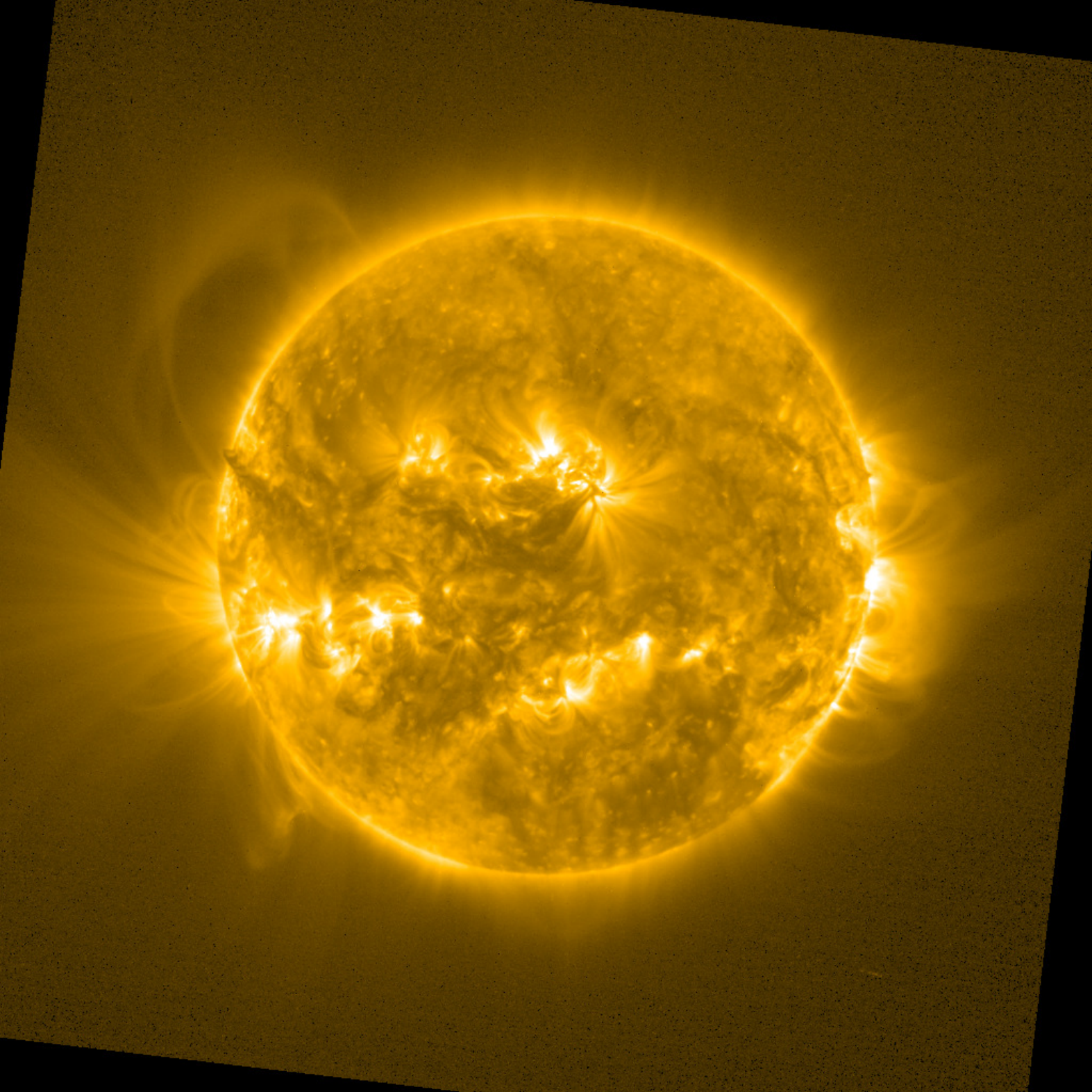}
\includegraphics[width=0.49\textwidth]{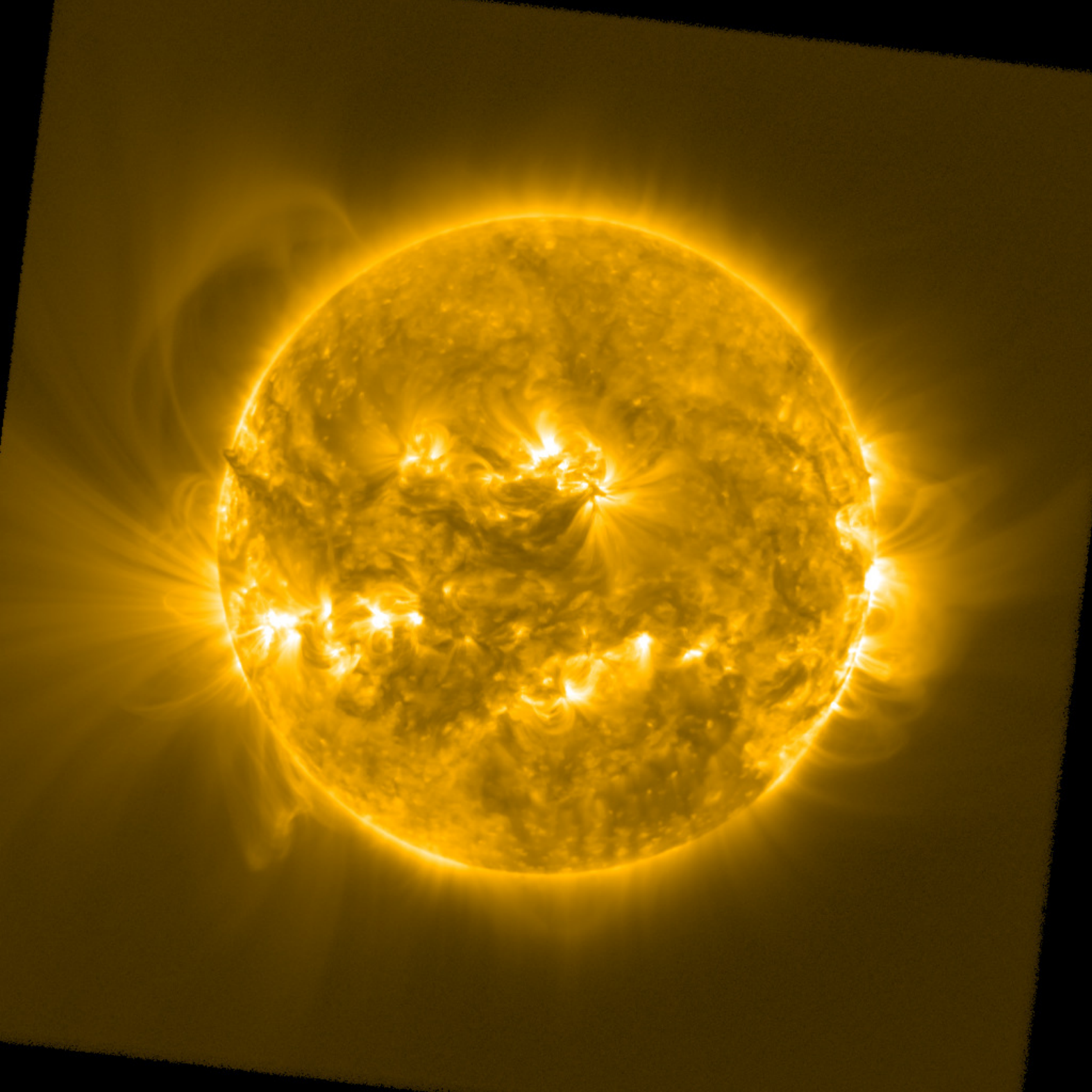}
\includegraphics[width=0.60\textwidth]{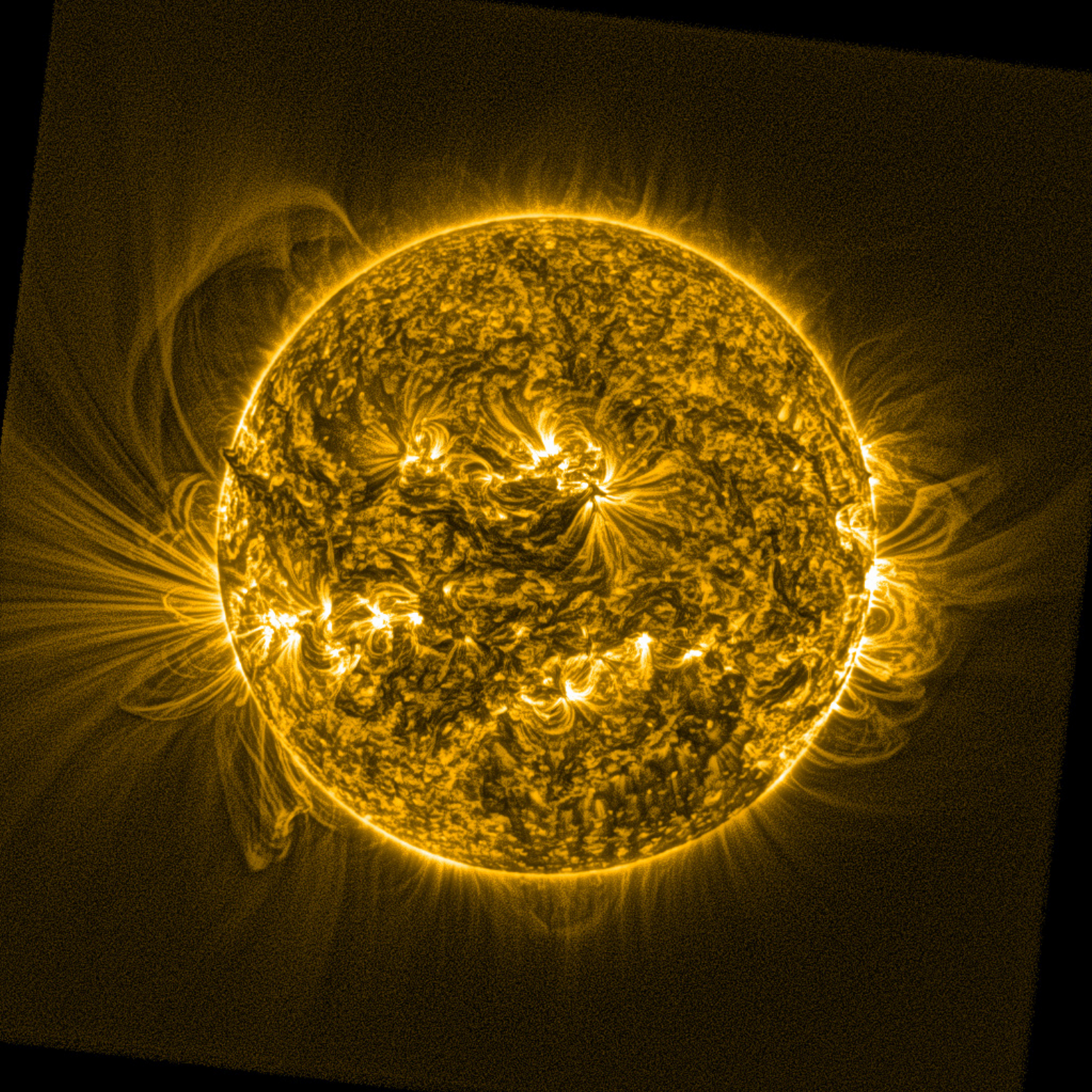}
\caption{An example of how large FOV images can be processed to reveal structures extending into the middle corona. The three images show the same SWAP (\SI{17.4}{\nano\metre}) observation from 10 November 2014, processed nominally (top left), using a stacking technique (top right, see \citet{West2022} for further details), and using the MGN technique \citep[bottom;][]{Morgan2014}.} 
\label{fig:swap_stack_mgn}
\end{figure*}

Historically, such dynamic range challenges were addressed with radially varying optical filters \citep{Eddy1989, Newkirk1970}.  Contemporary imaging techniques include the stacking of multiple short exposure observations to approximate a long exposure \citep[e.g.][]{West2022} and the use of detectors with locally variable exposure times \citep{Mason2022}.  Post-processing techniques, which improve the display of these high-dynamic-range images, include computational radial graded filters \citep[e.g.][]{Martinez1978, Seaton2023}, wavelet-based techniques \citep[]{Stenborg2008} and Multiscale Gaussian Normalization (MGN) \citep[e.g.][]{Morgan2014}. Figure\,\ref{fig:swap_stack_mgn} shows a SWAP EUV \SI{17.4}{\nano\metre} image from 10 November 2014 (top left) and a high-dynamic-range stacked image from the same time (top right) with improved noise characteristics in the outer FOV. The bottom image shows how image processing with the MGN technique can improve the visibility of finer structures.

\subsubsection{Spectroscopy}
\label{sssec:VLEUVXRaySpectroscopicObservations}

\begin{figure*}[t!]
\centering
\includegraphics[width=2.3in]{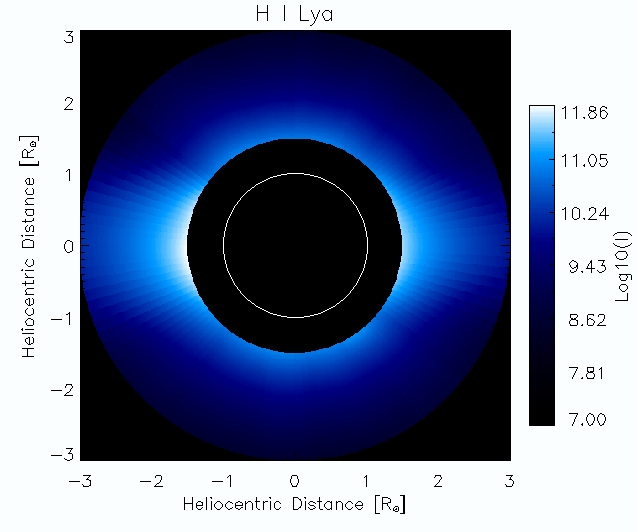}
\includegraphics[width=2.3in]{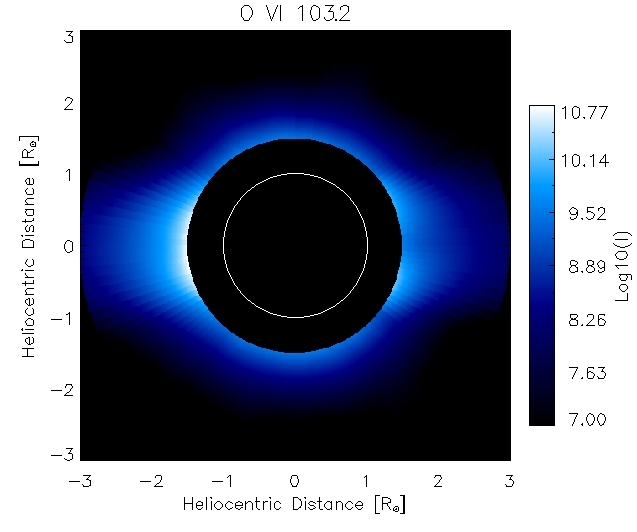}
\caption{Intensity images of H\,\textsc{i} Ly$\alpha$ and O\,\textsc{vi} (103.2\,nm) reconstructed from the sets of UVCS synoptic images during 1 June 1996 through 3 June. Note the different morphologies above the west limb. Note that the units in the Figure are given in Angstroms.} 
\label{fig:uvcs_synoptic}
\end{figure*}

Extensive UV spectroscopy of the middle corona was obtained by the \textit{Ultraviolet Coronagraph Spectrometer} (UVCS) onboard SOHO.  UVCS generally observed heights above 1.5\,\Rs\ (often extending out to 5\,\Rs) in a wavelength range from \SI{50.0} to \SI{135.0}{\nano\metre}.  Its spatial and spectral resolutions were about $7''$ and 30\,km\,s$^{-1}$ per pixel, but for most observations the pixels were binned due to telemetry limitations.  A review is given in \cite{Kohl2006}.  Daily synoptic observations covered a range of heights at eight position angles around the Sun, allowing the reconstruction of intensity images such as those shown in Figure\,\ref{fig:uvcs_synoptic}.

A wide variety of plasma parameters were measured from the UVCS spectra.  Line intensity ratios among different ions of a single element yield the ionization state, which directly gives the electron temperature at low heights where the plasma is in ionization equilibrium.  Once the ionization state has been established, intensity ratios of lines of different elements give their abundances; absolute abundances can be computed using the the Lyman lines of H\,\textsc{i}.  The column density along the LOS is obtained from the intensity of any line, such as Ly\,$\alpha$, that is produced by scattering of photons from the disk.  Like the densities obtained from visible-light images, these measurements yield the average density.  A second approach to estimating density estimate is to use collisionally excited lines such as Mg\,\textsc{x}, whose intensity is proportional to the density squared.  In some cases density-sensitive line ratios such as O\,\textsc{v} 121.8/121.3\,nm are also available. 

The spectral line widths give effective temperatures, which include both the kinetic temperatures of the ions and bulk motions due to turbulence or expansion.  A unique diagnostic method using UV lines is the Doppler dimming measurement of the velocity component away from the Sun.  For example, the O\,\textsc{vi} doublet has both collisional and radiative scattering components, and their intensity ratio depends on the Doppler shift of the absorption profile away from the emission profile of disk photons.  Analysis of line widths and intensities O\,\textsc{vi} combined with Ly\,$\alpha$ or visible-light data makes it possible to infer temperature anisotropy.  It is also possible to use sungrazing comets as probes to measure density, proton temperature and wind speed at points along the trajectory, rather than integrated LOS averages \citep{Bemporad2007, Jones2018}.

Some important results of these methods have been strong preferential heating of O and Mg ions compared to H in coronal holes and at heights above 3\,\Rs\ in streamers \citep{Cranmer1999, Strachan2002, Frazin2003}. Strong oxygen temperature anisotropies in the coronal hole plasma were also indicated.  Outflow speeds increase from around 20\,km\,s$^{-1}$  at 1.5\,\Rs\ to around 550\,km\,s$^{-1}$ at 6~\Rs\ in coronal holes \citep{Cranmer2008, Raymond2018}, while reaching speeds of about 100\,km\,s$^{-1}$ or more by 6\,\Rs\ in streamers \citep{Sheeley1997, Wexler2020}.  Elemental abundances in streamers show a first ionization potential fractionation (FIP fractionation; see Section\,\ref{ssec:Elemental_Composition}) similar to that seen in the slow solar wind, but the absolute abundances in streamer cores are reduced by at least a factor or 3, probably by gravitational settling \citep{Raymond1997, Feldman1998, Uzzo2006}.

Ultraviolet observations of CMEs in the middle corona have also determined the temperatures, thicknesses and turbulent velocities in current sheets \citep{Ciaravella2008, Bemporad2008}, as well as the Mach numbers and electron-ion equilibration in CME shocks \citep{Frassati2020}.  Studies of the energy budgets of CME ejecta have shown that they continue to be heated after leaving the solar surface, and that the cumulative heating is comparable to the kinetic energy \citep{Murphy2011, Wilson2022}.

Recent technical advances have enabled great strides in UV spectroscopy of the corona under a variety of launch platforms that can provide fundamental observations of coronal plasma that are inaccessible with other means \citep{Ko_etal._2016, Laming_etal._2019,Strachan2017}, including the recently launched Ultraviolet Spectro-Coronagraph (UVSC) Pathfinder instrument, which has a thirty-fold increase in sensitivity compared with UVCS and a multi-slit design to simultaneously observe two heights. Improved spatial and spectral resolution and increased spectral range are also feasible.

\subsubsection{Phenomena Observed}
\label{sssec:VLEUVXRayPhenomenaObserved}

Structures that pervade the middle corona can roughly be divided into long-lived and dynamic phenomena.  The long-lived structures are generally those that make up the background coronal environment; the dynamic phenomena are more transient, often passing through the region, and are often influenced by the background structures. 

\paragraph*{Long-lived structures.} Long-lived structures are generally larger structures that persist for weeks to months -- and in certain cases even years -- and make up the background coronal environment.  These include streamers and pseudostreamers \citep[e.g.][]{Pneuman1971, Wang2007}, both of which are observed in the outer corona as bright radial features extending outwards.  The inner and middle coronal magnetic topology cannot be discerned from such observations, but large-FOV EUV observations allow the magnetic topology to be traced from the inner corona out into visible-light observations \citep{Zhukov2008}.  Several studies have focused on the extended streamer structures: \citet{Rachmeler2014} used SWAP with Coronal Multichannel Polarimeter \citep[CoMP;][]{Tomczyk2008} (1074.7\,nm), and Chromospheric Telescope \citep[ChroTel;][]{Bethge2011} (H\,$\alpha$ 656.3\,nm) observations to investigate the long-term evolution of a streamer-pseudostreamer structure extending into the middle corona.  \citet{Guennou2016} also used SWAP data to investigate a pseudostreamer/cavity system, determining its large-scale three-dimensional structure, temperature, and density, and its associated cavity.  Separately, \citet{Pasachoff2011} used ground-based eclipse observations combined with EUV observations of a streamer structure to draw comparisons between the observations in the different passbands.

Coronal fans are another example of an extended large-scale structure, observed as fan like structures extending off the solar limb \citep[see e.g.][]{Koutchmy2002, Morgan2007}. They often overlie polar crown filaments, bending over before extending outwards and tracing out the edges of boundaries between distinct topological magnetic field regions, and are often observed to extend far out into the heliosphere.  \citet{Seaton2013b} showed fans are the single largest source of brightness at heights above 1.3\,\Rs\ in SWAP \SI{17.4}{\nano\metre} observations, and can persist for multiple rotations.  \citet{Mierla2020} extended this study and showed some fans can persist for over a year, and can be observed extending out to at least $1.6$\,\Rs\ in EUV observations.

\paragraph*{Dynamic phenomena.} Dynamic phenomena come in many forms, unfold over minutes to days, and include all structures that pass through the middle corona, traveling both inwards and outwards \citep{Seaton2021, Chitta2022}. The most prominent and energetic structures to pass through the middle corona are CMEs \citep[e.g.][]{Zhang2021}.  CMEs come in a range of sizes \citep{Robbrecht2009}, ranging from halo CMEs to eruptions whose angular widths are barely wider than their smaller counterparts, coronal jets \citep[e.g.][]{Sterling2015}.  These structures also have a range of speeds, from a few hundred to thousands of km\,s$^{-1}$ \citep[e.g.][]{Yashiro2004}. The faster eruptions develop a shock front ahead of the ejecta front \citep[e.g.][]{Zhang2006}, which in turn can produce solar energetic particles \citep[SEPs:][]{Reames1999}. CME-generated shocks  can also trigger transverse waves in solar helmet streamers, which have also been observed in the middle corona \citep{Decraemer2020}.

Beyond their impulsive drivers, eruptions are mainly influenced by the background corona/solar wind \citep[e.g.][]{Mierla2013, Schrijver2008}, especially in the dense inner and middle coronal regions. \citet{Sieyra2020} used wide-field EUV imagers to assess where CMEs can become deflected, and found deflections often occur in the inner or middle corona, during their acceleration phase.  \citet{Majumdar2020ApJM} studied the deflection of CMEs and drew comparisons to the 3D graduated cylindrical shell (GCS) model \citep{Thernisien2006,Thernisien2009}.  It is reported that the velocity and width of the CMEs become constant at heights around $\approx$3\,\Rs. 

The evolution of eruptions through the middle corona has been studied by many authors.  Many discuss the difficulties linking structure in the EUV and visible-light passbands \citep[e.g.][]{Byrne2014}. \citet{OHara2019} used unique SWAP EUV (\SI{17.4}{\nano\metre}) off-point observations to directly trace an eruption from EUV observations (up to $\approx$2.5\,\Rs) into surrounding visible-light LASCO coronagraph observations.  Although the overarching kinematics could be matched, exact features were difficult to reconcile due to the emitting plasma and differences in the observing passbands.

While not observed extensively, the middle corona should also be full of MHD wave phenomena that acts on a range of timescales.  Observations from the inner corona with CoMP have revealed the presence of ubiquitous propagating Alfv\'enic waves \citep{Tomczyk2007,Morton2015,Morton2016ApJ, Morton2019}.  The waves are present along the closed loops at the base of streamers and are also seen to leave the FOV ($\approx$1.3 R$_\odot$) along near radially oriented structures, suggesting they propagate directly into the middle corona. The Alfv\'enic fluctuations have also been long reported in the heliosphere, where they constitute an integral part of the fast wind streams \citep{BrunoCarbone2013}. The Alfv\'enic waves are thought to play a key role in heating the extended corona and adding momentum to the solar wind streams \citep[e.g.][]{CranmerVanB2005,Cranmer2007,Shoda2018}.  While there is some suggestion of in-situ wave generation, the majority of fluctuations observed in the heliosphere are believed to originate at the Sun. However, it is unknown how their journey is impacted as they passage through the dynamic and structured middle corona and is unaccounted for in current wave-driven models of the corona and heliosphere.

\subsection{Radio Wavelengths}
\label{ssec:radioWavelengths}

\begin{figure}
\centering
\includegraphics[width=0.95\textwidth]{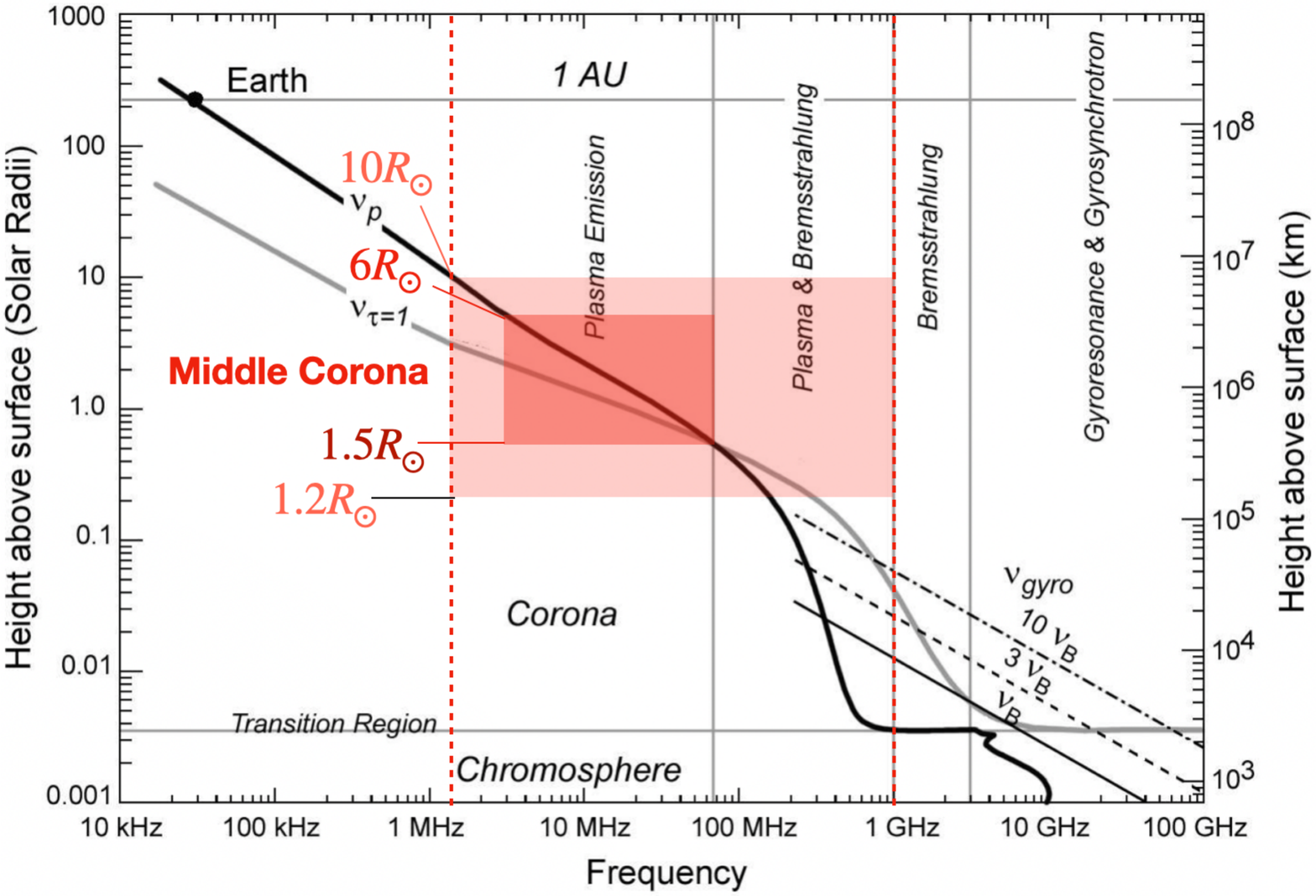}
\caption{Characteristic radio frequencies in the solar atmosphere. The middle corona includes a critical region where the transition of radio emission mechanisms occurs. The dark pink box marks the nominal range of the middle corona ($\approx$1.5\,--\,6\,\Rs) and the light pink box marks an extended range taking into account the highly structured and dynamic nature of the corona. The corresponding frequencies that are relevant to radio observations of the middle corona range from $<$10 MHz to $\approx$1 GHz. (Adapted from Figure 4.1 in \citealt{Gary2004} with permission.)}
    \label{fig:radio_observations}
\end{figure}

From the perspective of radio observers, the middle corona includes the coronal heights where the key transition from incoherent radio emission to coherent radio emission occurs \citep{Chen2023}. Figure\,\ref{fig:radio_observations} \citep[adapted from][]{Gary2004}, shows the variation of plasma frequency ($\nu_p$; thick black curve), gyrofrequency ($\nu_{\rm B}$; thin black curve), and the frequency of the free-free opacity $\approx$1 ($\nu_\tau$=1) layer as a function of coronal height under typical quiescent coronal conditions. 

The transition region and the innermost inner corona ($\lesssim$ 1.1\,\Rs\ from the center of the Sun) are dominated by incoherent gyromagnetic emission and free-free emission. At around 1.5\,\Rs, the plasma frequency $\nu_p$ layer takes over and becomes higher (closer to the observer) than both the $\nu_\tau$=1 curve and the curves of $v_{\rm B}$ and its harmonics. Such a transition has a profound implication on radio observations: the quiescent free-free radio corona is no longer playing a dominant role due to the strong refraction near the plasma frequency. Meanwhile, bright coherent radio bursts, due to plasma radiation occurring near $\nu_{\rm p}$ and its second harmonic, start to be important among the observed radio phenomena. Of course, even in the region where the coherent plasma radiation dominates, incoherent radio emission from transients (e.g. CMEs) can still be observed, providing crucial diagnostics for these coronal transients, including the magnetic field and nonthermal electrons trapped in the CME or accelerated by the CME-driven shock \citep[e.g.][]{bastian2001,mondal2020,chhabra2021}. Therefore, at radio wavelengths, a broad frequency range of $<$10\,MHz to $\approx$300\,MHz is relevant to the highly dynamic and structured middle corona (light pink box in Figure\,\ref{fig:radio_observations}). It is worth emphasizing that the magnetic field and non-thermal electron distribution diagnostics in the middle corona are unique to the radio techniques, and they are otherwise difficult to achieve (if not unavailable) for remote sensing at any other wavelengths.

\subsubsection{Observed Emission Mechanisms}
\label{ssec:RadioEmissionMechanisms}

There are numerous radio emission mechanisms relevant to the solar corona, which include gyroresonance (thermal electrons gyrating in the coronal magnetic field), gyrosynchrotron (nonthermal electrons gyrating in the coronal magnetic field), bremsstrahlung (or free-free; electrons interacting with ions), as well as a variety of coherent emissions such as plasma radiation (e.g. the nonlinear growth of Langmuir waves) and electron cyclotron masers (i.e. the nonlinear growth of plasma waves at harmonics of the electron cyclotron frequency). These emission mechanisms co-exist, but because the physical parameters differ in various coronal locations/conditions, the importance of each emission mechanism also varies. In particular, the plasma density $n_{ \mathrm{e} }$ and magnetic field $B$ vary dynamically throughout the corona, and hence the corresponding plasma frequency $\nu_p$ and gyrofrequency $\nu_{\rm B}$, thus the dominant radio emission mechanism varies over the corona, and can change due to local conditions.

\subsubsection{Radio Observations}
\label{sssec:RadioSpatiallyResolvedRadioImaging}

Observing the middle corona at radio wavelengths requires a wide frequency coverage from $<$10 MHz to $\approx$300\,MHz (c.f. Figure\,\ref{fig:radio_observations}). The $>$20\,MHz range is generally accessible from the ground, but the lowest frequencies can only be observed from space due to the ionospheric cutoff.  Currently, multiple ground-based instruments have been available to observe in the frequency range relevant to the middle corona.  In space, new missions, such as the the Sun Radio Interferometer Space Experiment (SunRISE), are being designed to locate radio bursts. Figure\,\ref{fig:MC_Missions} summarizes the currently operating and upcoming radio facilities that provide imaging capabilities in the frequency range relevant to middle corona studies. This list is representative, as there are, of course, a large number of additional radio instruments that provide total-power (full-Sun integrated) dynamic spectral measurements.

Over the past decade, new advances have been made with radio facilities equipped with \textit{broadband dynamic imaging spectroscopy}. This exciting new technique allows simultaneous imaging and spectroscopy to be performed over a broad frequency range and at a high time cadence. In other words, a detailed spectrum can be derived from \textit{each pixel} in the radio image for spectral analysis. First realized by the Karl G. Jansky Very Large Array at the decimetric wavelengths \citep{chen2013} and followed by the commissioning of LOFAR, MWA, EOVSA, and MUSER, this technique is just beginning to reach the full potential of radio studies using the rich diagnostics tools available \citep[e.g.][]{Carley_etal._2020}.

\subsubsection{Phenomena Observed}
\label{sssec:RadioPhenomenaObserved}

\begin{figure}[htb!]
	\begin{center}
	\includegraphics[width=\textwidth, trim = {20mm 30mm 20mm 30mm}, clip]{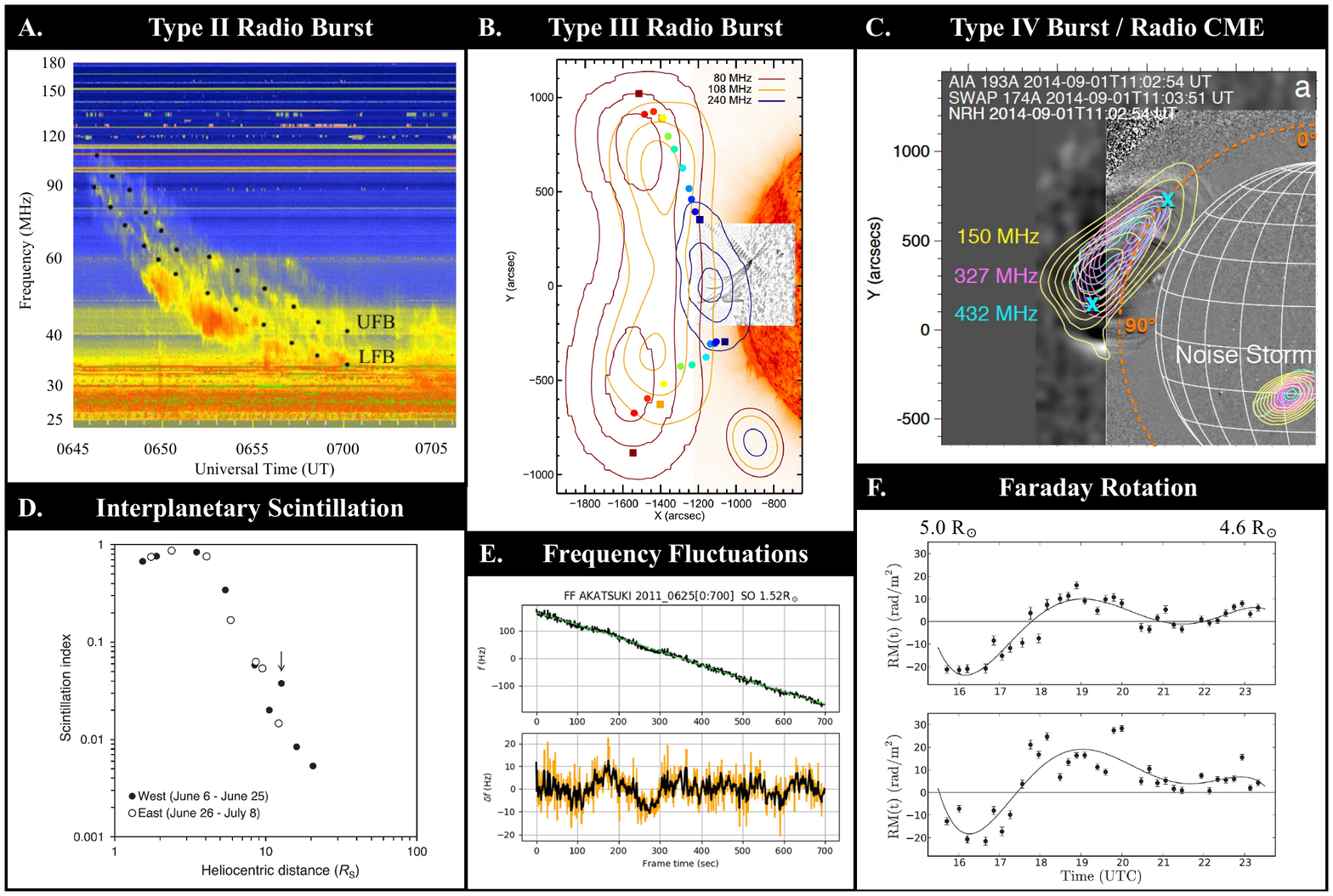}
	\caption{Overview of radio phenomena in the middle corona (A) Type II burst with a well-defined split-band feature into an upper and lower frequency branch (UFB and LFB, respectively), which, if interpreted as plasma radiation from the shock upstream and downstream, can be used to estimate the shock  compression ratio and Mach number (Figure 2 in \citealt{Mahrous2018}, used with permission; see also \citealt{Zimovets2012}). (B) Type III burst contours overlaid on an {\it SDO}/AIA \SI{30.4}{\nano\metre},image. Tracking the radio burst over several frequencies illustrates an evolution from a single source in the inner corona to two separate sources split between two separate flux tubes in the middle corona \citep[Figure 14 in][used with permission]{McCauley2017}. (C) Type IV burst associated with a radio CME resulting from trapped non-thermal electrons emitting gyrosynchrotron radiation, which can be used to determine the CME's magnetic field strength \citep[Figure 2 in][used with permission]{Carley2017}. (D) The scintillation index (representing the magnitude of the intensity fluctuations) as a function of heliocentric distance; intensity scintillation provides information on the plasma density and solar wind speed \citep[Figure 3 in][used with permission]{Imamura2014}. (E) Frequency fluctuations provide information on plasma density fluctuations and solar wind speed. Upper panel shows raw frequency data dominated by a Doppler shift and the bottom panel shows the frequency fluctuations with the Doppler shift removed \citep[Figure 2 in][used with permission]{Wexler2020}. (F) Faraday rotation provides information on the plasma density and magnetic field component along the line of sight. Differences between measurements along two closely-spaced lines of sight (provided here by a background radio galaxy) can be used to probe coronal electric currents \citep[Figure 5 in][used with permission]{Kooi2014}.
	}
	\label{fig:radio_obs_examples}
	\end{center}
\end{figure}

\paragraph*{Type II Bursts and Coronal Shocks}
Type II radio bursts are seen from metric to kilometric wavelengths (a few times 100 MHz to tens of kHz) and are notable for their relatively slow drift to lower frequencies compared to type III radio bursts \citep[see, e.g., the empirical expression of their drift rate in][]{Aguilar-Rodriguez2005}. They result from coherent plasma radiation of energetic electrons accelerated at or near the shock front propagating outward at super-Alfv{\'e}nic speeds. Therefore, they bear important diagnostics for both the shock parameters and shock-accelerated electrons. Figure\,\ref{fig:radio_obs_examples}(A) shows an example of a metric type II burst that shows a split-band feature in the time-frequency domain. This feature is interpreted as plasma radiation at the shock upstream and downstream regions, which, in turn, can be used to estimate the shock compression ratio and Mach number. Recently, thanks to the imaging spectroscopy capability provided by instruments such as LOFAR, new insights have been provided into their source region at the CME-driven shock front. For example, \citet{morosan2019} found shock-accelerated electrons ``beaming out" from multiple acceleration sites located at the nose and flank of the shock.

\paragraph*{Type III Bursts and Electron Beams.} Type III radio bursts are produced by fast electron beams ($\approx$0.1--0.5$c$) escaping along open magnetic field lines (see, e.g., \citealt{reid2014, Reid2020} for recent reviews). Observations of type III bursts span an extremely wide frequency range from $>$GHz to kHz and exhibit a much greater frequency drift than that of type II bursts. 

In the middle corona, these bursts are predominantly associated with open field lines.  With imaging spectroscopy provided by general purpose facilities such as LOFAR and MWA, new advances have been made in tracing the trajectories of the electron beams (e.g. Panel B in Figure\,\ref{fig:radio_obs_examples}) which, in turn, outline the electron-beam-conducting magnetic field lines in the middle corona \citep[e.g.][]{McCauley2017,Mann2018}. The emission frequencies and fine structures in the dynamic spectra have been used to derive the coronal density variation in height and properties of the coronal turbulence \citep{Kontar2017,McCauley2018b,Mann2018,Reid2021}.

\paragraph*{Type IV Bursts and Trapped Electrons.} Type IV radio bursts are broadband bursts characterized by their slow- or non-drifting appearance in the radio dynamic spectrum. Typically observed after the flare peak, they are thought to be produced by nonthermal electrons trapped in closed coronal structures, emitting coherent (plasma or electron cyclotron maser) radiation and, in some cases, incoherent gyrosynchrotron radiation. Depending on the underlying emission mechanism, type IV bursts can, on one hand, trace and outline the closed magnetic structure of interest, and on the other hand, provide diagnostics of the source region \citep[see, e.g., review by][and references therein]{Carley2020b}.  First detected and coined in the 1950s \citep{Boischot1957}, type IV radio bursts have been generally sub-categorized into stationary and moving type IV bursts. The latter, by virtue of their close association with CMEs, are of particular interest because of their diagnostic potential for CME magnetic fields and energetic electrons. 

\paragraph*{Radio CMEs.} Faint radio emissions that closely resemble their visible-light CME counterparts are dubbed ``radio'' CMEs because of their similar appearance (e.g. Panel C in Figure\,\ref{fig:radio_obs_examples}). In fact, they were discovered around the same period as LASCO's start of science operations in 1996 \citep[see recent review by][]{Vourlidas_Carley_Vilmer_2020}. Since the emission occurs at large harmonics of the electron gyrofrequency, this emission can be found at frequencies above the local plasma frequency, thereby being less affected by the scattering effects.  Thanks to their incoherent nature, when imaged at multiple frequencies, they can be used to map the coronal magnetic field and non-thermal electron distribution associated with the CMEs \citep[see, e.g.][]{bastian2001, Maia2007, mondal2020}. 

\paragraph*{Propagation Effects.} The propagation effects of radio waves provide other means for studying the middle corona. These observations utilize a known, point-like background radio source (e.g. a spacecraft transmitter or a natural celestial source such as a pulsar or radio galaxy) to ``shine through" the corona. The observed radio signatures can be used to probe the structure and dynamics of the middle corona.  Importantly, these trans-coronal radio sensing methods are applicable in all solar activity states and do not rely on observations of specific episodic outburst phenomena. Signal delays at different frequencies (i.e. dispersion measure) can be used to constrain the coronal density. Signal broadening and scintillation provide information on the density inhomogeneities in the turbulent coronal plasma \citep[][]{Rickett1990}. Analysis of radio scintillation and frequency fluctuations (Panels D and E in Figure\,\ref{fig:radio_obs_examples}) can provide estimates of solar wind speed \citep{Imamura2014,Wexler2019,Wexler2020}. In addition, modulations of the signal polarization due to Faraday rotation (Panel F in Figure\,\ref{fig:radio_obs_examples}) can be used to constrain the coronal magnetic field and its fluctuations \citep[see, e.g.][and references therein]{Wexler2021a,Wexler2017,Kooi2022}.  

\section{Properties and Transitions in the Middle Corona}
\label{sec:Properties_and_transitions} 

The inner corona exhibits a broad range of temperatures, which can exceed $T>10^7\,K$ in case of flares, and electron densities of $n_{ \mathrm{e} }\sim10^{15}\,$m$^{-3}$ in closed structure regions. These closed magnetic field regions are generally associated with the relative confinement of plasma, with subsonic flow speeds and increased elemental abundances. The open-field configurations associated with coronal holes are known to produce fast solar winds and relatively low scale-heights, and exhibit FIP elemental abundances close to those of the photosphere.  In contrast to the inner corona, the outer corona is generally a region of supersonic solar wind outflow, with an open magnetic field pattern and stabilized ionization charge states.

The characteristics of the middle corona straddle those of the inner and outer corona, and accordingly the region hosts a number of structural, dynamic and plasma physics transitions, as described in Table \ref{tbl:MiddleCoronaTransitions}.  The most important structural change is the transition from a mix of open and closed magnetic configuration to almost exclusively open-field structures.  

Due to instrumental limitations, the middle corona has not been continuously or comprehensively probed by instruments that can provide self-consistent plasma parameters. As a result, the multiple physical transitions that occur here have not been fully characterized, and methods to study plasma properties must include extrapolation and modeling, often drawn from measurements of surrounding regions  \citep[e.g.][]{Lynch2020,Schlenker2021}.  Table \ref{tbl:Plasma_properties} presents a list of canonical plasma properties measured/derived on either side of the middle corona illustrating the transitions that occur within the region.

\begin{table}
\caption{Representative middle corona properties in fast and slow solar wind regions. The top portion includes representative measured and modelled quantities, the bottom portion includes derived quantities}
\begin{tabular}{l c c c c l}
\hline
\textbf{Symbol}&   \multicolumn{2}{c}{\textbf{1.5\,\Rs}} & \multicolumn{2}{c}{\textbf{6.0\,\Rs}} & \textbf{Units: Definition} \\
               &     \textbf{Fast}   &  \textbf{Slow}  & \textbf{Fast}  & \textbf{Slow} & \\
\hline
$n_e$$^a$& $1\times 10^{12}$  & $7\times 10^{12}$ & $6\times 10^9$  & $3\times 10^{10}$ & m$^{-3}$: electron no. density \\
$T_{p,\parallel}$$^b$ & 1.6    & 2.0  & 1.9    & 0.85 & MK: proton $\parallel$ temperature \\
$T_{p,\perp}$$^b$   & 2.0      & 2.6  &  ---   & 1.1 & MK: proton $\perp$ temperature \\
$T_e$$^c$         & 1.4    &  1.8    &  0.8   & ---    & MK: electron temperature \\
$T_{O,\parallel}$$^d$ & 2   & $>$1   & 60    & $>$5 & MK: oxygen $\parallel$ temperature \\
$T_{O,\perp}$$^d$ & 10   & 20     & 200  &  20  & MK: oxygen $\perp$ temperature \\
$V_{SW}$$^e$             & $>$100 & $<$25  & 550 & 150 & km/s: outflow speed \\
He/H$^f$  &  ---   & 8\% &  ---   &   ---   & --- : Helium/Hydrogen ratio   \\
\multirow{3}*{FIP$_{bias}$$^g$}   &     \multirow{3}*{1.5\,--\,2.5}   &  \multirow{3}*{4\,--\,6}  &   \multirow{3}*{---}  &   \multirow{3}*{---}  & --- : Elemental composition \\
  & &   &     &     & \,\, compared to photospheric\\ 
  & &   &     &     & \,\, composition\\
$B^h$             & 1.3$\times10^5$& $7\times10^4$& 4$\times$10$^3$    & 4$\times$10$^3$  & nT: magnetic field \\
\hline
$C_S$         & 150 & 170 & 160    & 100  & km/s: sound speed \\
$V_A$$^i$         & 3000   & 600 &  1100   & 500  & km/s: Alfv\'{e}n speed \\
$\omega_{pe}$    &  $5.6\times10^7$  &  $1.5\times10^8$   &  $4.4\times10^6$   & $9.8\times10^6$    & Hz: e$^-$  plasma frequency \\
$\beta^j$       & $<0.01$   &   $\geq$0.08  &  $<0.1$   & $\geq$0.04   & plasma $\beta$, P$_{gas}$/P$_{mag}$ \\
\hline
\end{tabular}
$^a$ \cite{bird1990, guhathakurta1999, Raymond2018, Wexler2019}  \\
$^b$ \cite{Strachan2002,Frazin2003, Cranmer2008, cranmer2020} \\
$^c$ \cite{Raymond1997,Cranmer2009} \\
$^d$ \cite{Strachan2002,Frazin2003,Cranmer2008} \\
$^e$ \cite{woo1978, strachan1993, Raymond2018, Wexler2020, romoli2021} \\
$^f$ \cite{moses2020} \\
$^g$ \cite{Feldman1998, young1999, Raymond1997, uzzo2004} \\
$^h$ \cite{Kooi2022, yang2020, Wexler2021a, Alissandrakis&Gary2021, hofmeister2017} \\
$^i$ \cite{evans2008}\\
$^j$ \cite{gary2001} for slow SW; note $\beta=\frac{C_s^2}{V_A^2}$ \\
Note:  1 gauss (cgs) =  10$^5$ nT = 10$^{-4} $T (mks, S.I.). 
\label{tbl:Plasma_properties}
\end{table}

The parameters in Table \ref{tbl:Plasma_properties} are also subdivided into categories of fast and slow solar wind, to represent the range of values that are present in the different regions. Coronal holes are considered the source of the fast solar wind, streamers and pseudostreamers contain slow solar wind, and the remaining regions are predominantly slow interspersed with some fast regions. In slow solar wind and streamer regions the supersonic solar wind outflow is achieved by approximately 5\,--\,6\,\Rs\ \citep{Sheeley1997,Wexler2020}.

As discussed in Sections~\ref{ssec:ShortWavelengths} and \ref{ssec:radioWavelengths}, several instruments including UVCS and various radio arrays have sporadically provided direct diagnostics of specific middle corona properties, yielding estimates of density, proton temperature, ion temperatures, temperature anisotropy, outflow speed, ionization state and elemental composition. However, even for the relatively simple case of a quiet-sun streamer various different estimates of the densities and temperatures have been published \citep{DelZanna2018, Seaton2021}. This might be due to the large amplitude density contrasts on small scales \citep{Raymond2014} and estimates based on scattered light (average density) or emission lines (average density squared).

To a good approximation the magnetic field in the inner corona is force-free since the plasma $\beta$ is much smaller than unity. Throughout the middle corona the magnetic control is only partial.  The confinement of plasma by closed fields diminishes. At the same time, the stabilization or ``freeze-in'' of the ionization charge states occurs. This provides the basis for source region diagnostics based on measurements far from the middle corona. From the global heliospheric magnetic field modeling point of view, the middle corona is critical; the PFSS is nominally placed between 2.5 and 3.0\,\Rs, but actually may be more suitably placed at different middle corona region altitudes \citep{Lee2011}. With these several key transitions occurring over a relatively small radial distance range, intensive cross-disciplinary analysis is necessary to create internally-consistent models of the complex processes.

\subsection{Elemental Composition}
\label{ssec:Elemental_Composition}

It is now well-established that the chemical composition of the corona varies depending on the structures observed and differs from the solar photospheric composition, although both recent revisions of older data and new analyses indicate that, at least up to 1\,MK, the composition of the quiet solar corona is also close to photospheric \citep{delzanna2018b,DelZanna2018,madsen2019}. 
The variability in chemical abundances depends, among other factors, on the First Ionization Potential (FIP) of the element and gravitational settling effects.

The FIP effect is a process in which elements with neutral atoms with ionization potentials below 10\,eV (e.g. Fe, Si) are preferentially enhanced by a factor of 2 to 4 relative to those with higher FIP values (e.g. O, Ne). The FIP effect is most prominent in active regions and helmet streamers at the Sun and is also reflected in the in-situ observations of the slow solar wind and SEPs that originate from those structures \citep{Geiss1995, vonSteiger2000, Uzzo2006, Baker2013, Reames1999}.  In coronal holes and the fast wind the FIP enhancement is small or non-existent \citep{Feldman1993}.

Gravitational stratification (settling) of higher mass elements (compared to lighter ones) can appear in large, long-lived coronal structures, such as the cores of helmet streamers, which are observed throughout the middle corona \citep{Raymond1997}. Spectral observations of helmet streamers from UVCS have shown a significant depletion of both low (FIP $< 10$\,eV) and high (FIP $> 10$\,eV) FIP elements (O, Si, Mg) in accordance to particle mass that is thought to be caused by gravitational settling taking place high in the corona \citep{uzzo2003, uzzo2004}.

A similar effect was observed by the Solar Ultraviolet Measurements of Emitted Radiation \citep[SUMER;][]{Wilhelm1995} instrument on SOHO.  This phenomenon results in mass fractionated coronal plasma where the loop apex becomes depleted of the heaviest elements as they sink towards the footpoints faster compared to lighter elements.

The gravitational settling shows strong spatial dependence, such that it becomes less pronounced between the helmet streamer core and legs. This variation is attributed to the transition between closed (core) and open/closed field (streamer edge) where plasma confined to the streamer core resides in the corona long enough for notable gravitational settling to take place, $\approx 1$ day, while plasma on the open/closed field boundary is released on a faster timescale \citep{lenz1998}. These observations indicate that gravitational settling can be important in regulating the plasma's chemical composition in large coronal loops, and can be a distinctive compositional signature of helmet streamer plasma observed in the form of heavy element dropouts in the solar wind and CMEs at 1\,AU \citep{Weberg2012, Weberg2015, Rivera2022}. These are believed to correspond to pulses of gas released from the cusps of helmet streamers by reconnection.  However, further examination of gravitational settling and variability in the chemical composition across the middle corona is necessary to further characterize solar wind origin and the pathways to its formation and connection to heliospheric structures.

\subsection{Charge State Evolution and Freeze-in Distances}
\label{ssec:Ionization} 

\begin{figure}
\centering
\includegraphics[width=1.00\columnwidth]{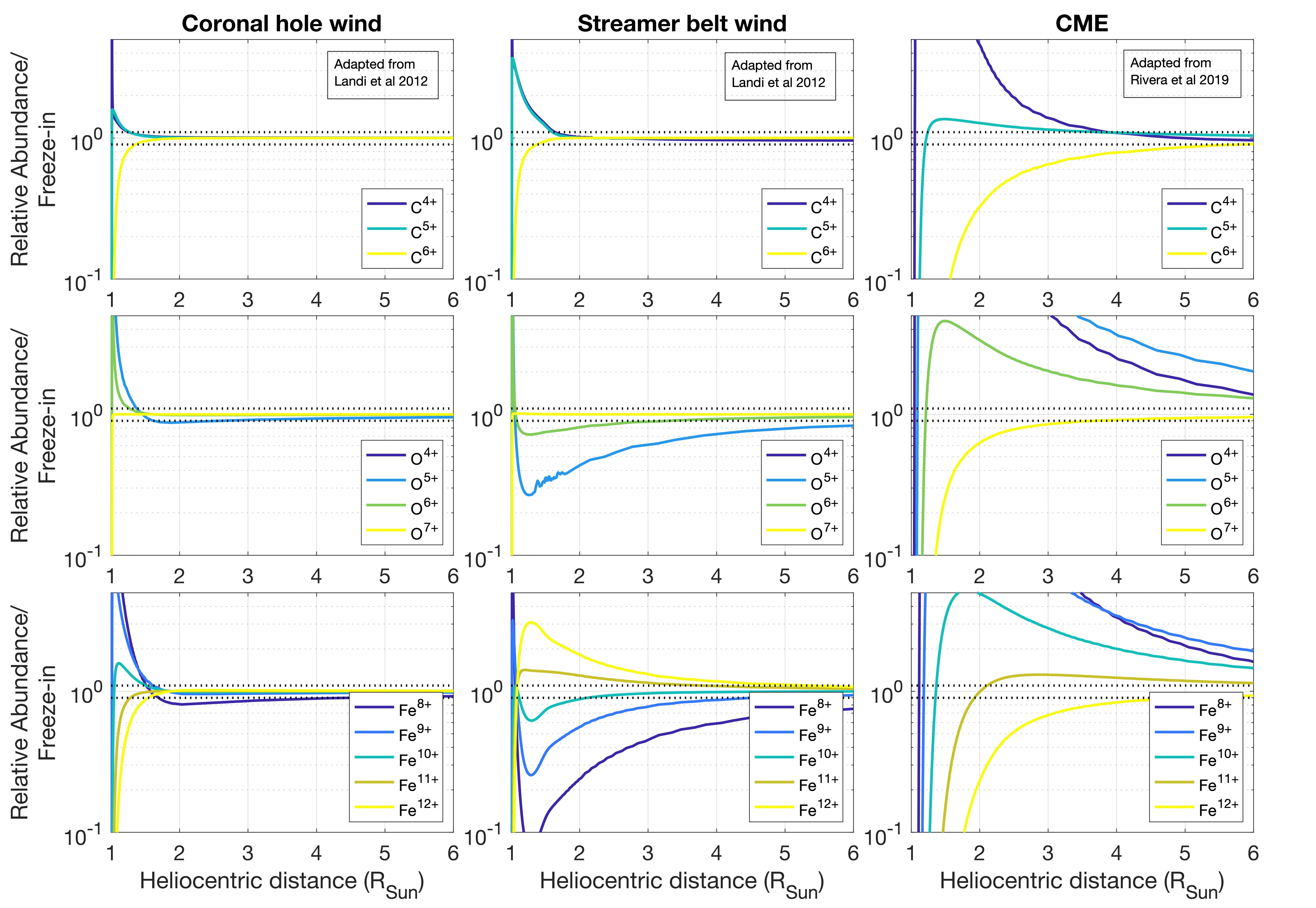}
\caption{\small Radial evolution for selected C, O, Fe ions within simulated coronal hole wind, equatorial streamer belt solar wind, adapted from \cite[][used with permission]{landi2012}, and a CME adapted from \cite{rivera2019}, and used with permission. The horizontal dashed lines represent ions reaching 10\% of its freeze-in value.} 
\label{fig:solwind_chargestates}
\end{figure}

One important transition point occurring in the middle corona is height at which heavy ion abundances in the solar wind and CMEs reach their ``freeze-in'' altitude \citep[]{hundhausen1968,owocki1983}. The so-called freeze-in process takes place as charge states become fixed at some radial distance from the Sun where the plasma becomes too tenuous to sustain ionization and recombination processes any further. After this transition, ions become uncoupled from thermodynamic changes in the plasma and remain fixed throughout the solar wind's radial evolution. 

In-situ measurements of ions can be tied to their sources in the corona by comparing frozen-in populations. As a result freeze-in states can be used to probe the heating and cooling in the nascent solar wind prior to freeze-in.

Freeze-in distances are governed by the plasma's electron temperature, density, and outflow speed, which result in large ranges of freeze-in distances among solar structures throughout the middle corona. \citet{boe2018} used the regions where the resonant scattering dominates in visible-light observations of Fe$^{10,13+}$ as a proxy to estimate freeze-in distances.  They found in coronal holes and helmet streamers distances of 1.25\,--\,2\,\Rs\ and 1.45\,--\,2.2\,\Rs, respectively. However, theoretical freeze-in heights can be considerably larger in pseudostreamers \citep{shen2018}. Similarly, simulations of the solar wind have shown that freeze-in distances for other ions of C, O, and Fe in coronal hole wind can range between 1\,--\,2\,\Rs, while in equatorial streamer belt wind ions may evolve beyond 5\,\Rs, as shown in the left and middle column of Figure \ref{fig:solwind_chargestates} \citep{ko1997,landi2012, Gilly2020}. 

In CME plasma, ion freeze-in distances are predicted to reach beyond 6\,\Rs\ in the dense prominence core, as shown in the right column of Figure\,\ref{fig:solwind_chargestates} \citep{rivera2019}. Also, the higher velocities in CMEs can be an important factor in the higher freeze-in heights \citep{rakowski2007}. Simulations of the freeze-in process using nonequilbrium ionzation (NEI) conditions have enabled studies of the corona's thermodynamic state using heavy ion composition that reflect the plasma's early stages of ionization evolution \cite[see, e.g.][]{landi2012, Gilly2020}. Multi-wavelength observations throughout the middle corona will place more stringent constraints on the ion evolution at these critical freeze-in heights to strengthen the connection made between the Sun and in-situ observations taken by orbiting spacecraft. 

Anomalous charge states observed at 1 AU are believed to arise in the middle corona, where CMEs show very high (Fe$^{+17}$) and very low (Fe$^{+3,+4}$) charge states.  Anomalous dropouts of fully stripped ions such as C$^{+6}$ are also observed. However, there is no clear interpretation of how these bare ion dropouts occur \citep{Kocher2017, Zhao2017, Rivera2021}.

\subsection{Magnetic Topology of the Middle Corona}
\label{ssec:Topology} 

There are three types of large scale features that dominate the middle corona associated with distinct magnetic topologies: the closed streamers, including their cusps; rays of various types; and open field regions.  The three have unique characteristic speeds, densities, plasma betas, compositional and FIP values. (An additional, transient, topological feature is large-scale closed loop systems, or giant arches, formed by magnetic reconnection during large eruptions, which can reach well into the middle corona and persist in active regions for days to weeks; \citealt{West2015}.)

Coronal holes undergo several changes in the middle corona region: the magnetic field expands super-radially and fast solar wind acceleration occurs, generally in the lower reaches of the middle corona \citep[][and references therein]{Cranmer2009a}. The cause of this acceleration is still a topic of some debate, and is one of the middle corona's most important open questions. In contrast, the slow solar wind is organized in the middle corona and initial acceleration to the supersonic threshold occurs.  The wind is believed to become super-Alfv\'{e}nic at varying distances between 10\,--\,25\,\Rs\ into the extended corona and heliosphere \citep{Wexler2021b}; where the kinetic energy dominates over the magnetic energy, regardless of the value of the plasma $\beta$.  The middle corona is important to mediating the overall morphology of coronal holes: the high rate of forced reconnection in the ``magnetic carpet'' of the solar photosphere \citep{SimonTitleWeiss2001} induces a high diffusion rate of small-scale magnetic flux \citep{Hagenaar1999}, which should break up large-scale coronal holes on a time scale of days; this is not observed \citep{Cranmer2009}, implying that the structure of the open flux is somehow communicated downward from or through the middle corona, to affect reconnection patterns near the surface.

Helmet streamer cusps lie in the middle corona region; these form the heliospheric current sheet, as well as some secondary topological surfaces \citep[e.g. above polar crown filaments:][]{Rachmeler2014}. Here high-$\beta$ plasma and magnetic field fluctuations near the magnetic y-points pinch off to form plasmoids or ``blobs'' \citep{Wang2018}. While helmet streamers are generally fairly quiescent and contain the only magnetic field not directly connected to the solar wind in this region, they are also the source of streamer blow-outs, some of the largest and most internally-coherent coronal mass ejections in the heliosphere \citep{Lynch2010,Vourlidas2018}. 

Rays are a term that can be applied to any of various structures of the same basic null point topology. Plumes and jets, which generally lie in open field regions, have extensive collimated columns of enhanced-density plasma extending above their domes in the low corona, which have been observed to extend into the heliosphere, with direct imaging as high as 40 solar radii \citep{DeForest1997,DelZanna2003,Raouafi2016a,Karpen2017,Kumar2019,Uritsky2021}. They have long been postulated as a small but relatively stable source of contributions to the solar wind, and some middle corona observations show outflows into such smaller-scale features \citep{Seaton2021}. Recent observations have revealed quasi-periodic energy release and jetlets (period=5 min) at the base of plumes which are important to understand the coronal heating and origin of solar wind in plumes \citep{Kumar2022}.

Pseudostreamers, which are similar to streamers in appearance, but topologically more complex, also have outer spines that are often seen in coronagraph imagery, potentially appearing as miniature low-lying streamer cusps, as the narrow spines themselves, or rarely as dim fans curving away from the dome surface, depending upon the height and angle of viewing. Evidence shows that coherent magnetic structures attempting to escape the low corona can be destroyed by reconnection in these null-point topologies, leading to large streams of unstructured plasma being ejected into the solar wind from these narrow rays \citep{Vourlidas2017,Kumar2021,EMason2021,Wyper2021}. In addition, pseudostreamers can produce unstructured slow CMEs \citep{Wang2015} as well as ``bubble-shaped'' fast CME ($>$ 1000\,km\,s$^{-1}$) via interchange/breakout reconnection at 3D null \citep{Kumar2021}. 

This collection of qualities is described by the S-web, a map of separatrices and quasi-separatrix layers (QSLs) in the heliosphere \citep[e.g.][]{Antiochos2011}. The major separatrix lines show the HCS, while the QSLs are smaller arcs corresponding to pseudo-streamers, jets, etc. Taken together, the topological picture of this region is diverse and rich; the closed but dynamic streamer belt regularly extrudes blobs of closed field and relatively dense plasma into the otherwise narrow and well-structured heliospheric current sheet. Much of the remaining volume is filled in by the expanding field and tenuous plasma of the coronal holes, occasionally punctuated by tight spears of condensed field and plasma introduced by null-point topologies. New large-field-of-view EUV observations have recently provided direct imaging of the S-web and its complex dynamic behavior in the middle corona \citep{Chitta2022}, validating models that predicted its importance in governing the topological and dynamic transitions that occur in here.

\section{Modeling the Middle Corona}
\label{sec:modeling_the_mc}

Because of the multiple physical transitions within the middle corona -- and the instrumental limitations that have hampered a complete characterization of them -- a unified model of middle corona physics does not yet exist. The lack of continuous, comprehensive measurements of the region as a whole has also limited the availability of high-quality, data-based model boundary and initial condition parameters. However, a limited number of direct measurements from UVCS and various radio arrays have provided estimates of density, proton temperature, ion temperatures, temperature anisotropy, outflow speed, ionization state and elemental composition. (See Section\,\ref{sec:Properties_and_transitions} for a thorough discussion.)

\subsection{Spectral Diagnostics and Implications for Forward Modeling}
\label{ssec:spectral-forward}

A general description of the underlying atomic data needed to model the coronal emission and obtain information about the plasma state appears in the Living Review by \cite{delzanna2018b}.  Modeling the visible/IR continuum emission resulting from the disk radiation being Thomson-scattered by the free electrons is relatively simple, although a knowledge of the spatial distribution of the electron density is required.  Following \cite{vandehulst:1950}, in most cases the modeling assumes a homogeneous distribution with spherical/cylindrical symmetry. This is routinely used to infer the radial density profile from measurements of the polarized Brightness (pB).  However, this is an over-simplification, as the corona is known to be finely structured \citep[e.g. the images in][and many other similar solar eclipse observations]{Habbal2014}. More sophisticated approaches can provide density diagnostics using broadband visible light imaging without the simplifying assumptions of homogeneous distributions and spherical symmetry (as in the van de Hulst inversion), for example, \cite{Decraemer2019}, who provided diagnostics using a more elaborate geometric approach.

As the middle quiescent corona appears to have an electron temperature around 1\,MK \citep{Seaton2021, Boe2020}, it emits a range of coronal lines from the X-rays (above \SI{2.0}{\nano\metre}) to the near infrared, mostly from highly ionized atoms. The strongest coronal lines are allowed transitions in the EUV/UV, between \SI{17.0} and \SI{110.0}{\nano\metre}, and forbidden transitions in the visible and near infrared. The modeling of most of the allowed transitions is relatively simple, as their emissivity mainly depends on the local electron density and temperature, as well as chemical abundances and ionization states.  The main populating mechanism is excitation by electron collisions (collisional excitation), and the observed radiance is proportional to the square of the electron density. However, there are cases where photo-excitation by the disk radiation in the visible/near infrared affects the ion populations, as in the case of Fe XIII \citep[see, e.g.][]{dudik2021}, which somewhat changes the predicted emission of the allowed transitions. Measurements of the density from line ratios are available for the inner corona. 

There is a range of allowed transitions from neutrals or ionized atoms which become very strong by direct resonant photo-excitation from the disk radiation. Examples are lines from H\,\textsc{i}, He\,\textsc{i}, He\,\textsc{ii}, and O\,\textsc{vi}, which are the strongest in the lower and middle corona. Such atoms produce a very strong solar disk emission from the chromosphere/transition region, and naturally produce little emission at coronal temperatures/densities. Hence, a large fraction of their coronal emission is produced by resonant photo-excitation. Their modeling is complex as it depends strongly on the distribution of the disk radiation, which is highly variable and, in the case of He, also controls the charge states via photo-ionization.   Modeling the Helium emission has several extra complications, as discussed in the first coronal model by \cite{delzanna2020}. Lines from these ions offer several diagnostics, the most widely used one being Doppler Dimming, to measure the outflow velocity
\citep[see, e.g.][]{noci1987}.

\begin{figure}
\centering
\includegraphics[width=0.75\columnwidth]{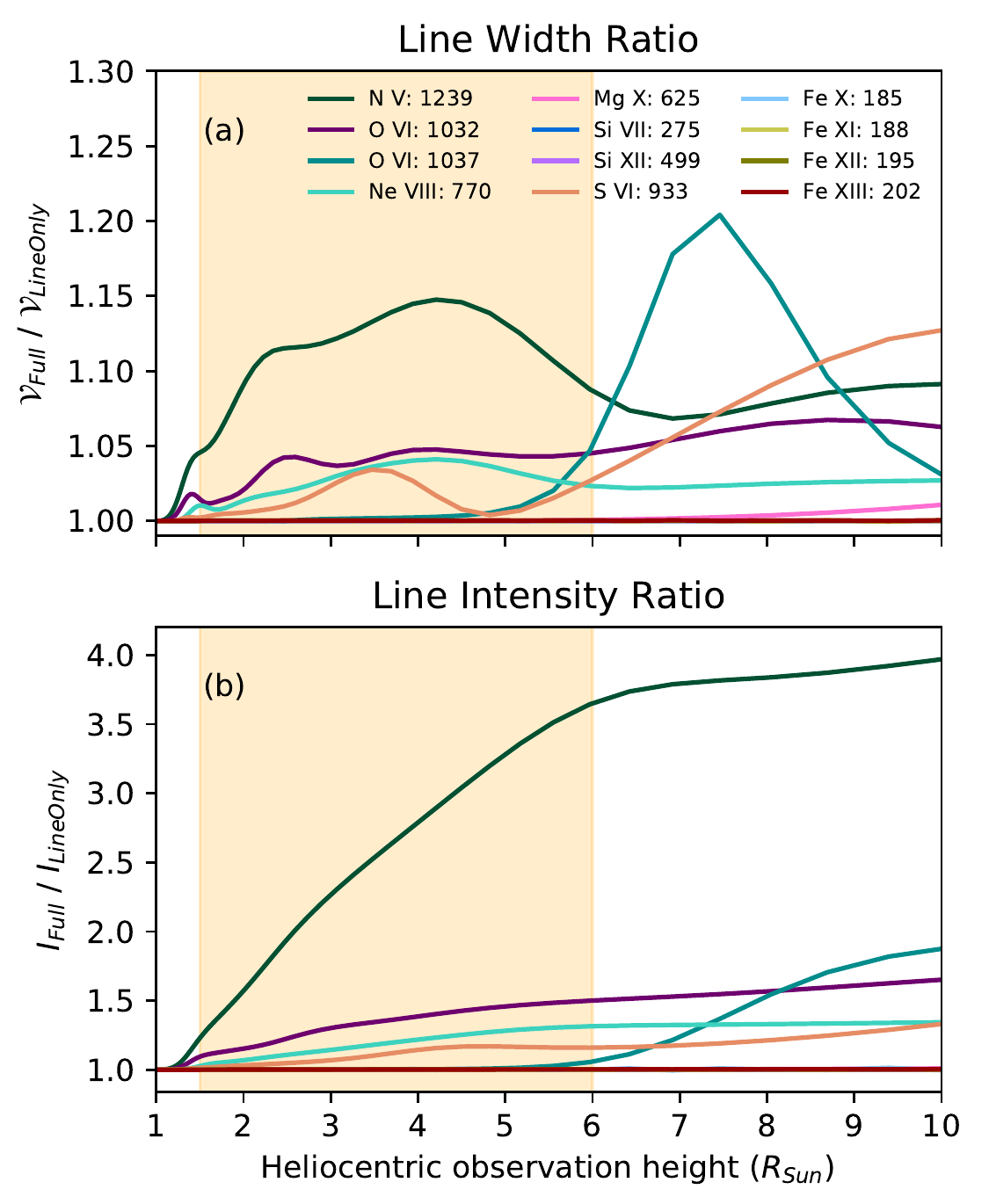}
\caption{Ratio of modelled ion-line properties as a function of height, with and without including a modelled continuum in the resonantly scattered light. Panel (a) indicates excess line-width caused by including the continuum, while Panel (b) shows excess intensity.  The shaded orange area indicates the middle coronal region. Ion-line wavelengths are given in units of Angstroms.  Adapted from Figure 14 of \citet{Gilly2020}, and used with permission.}
\label{fig:LineRatios}
\end{figure}

Doppler effects must be considered when forward modelling the corona, or the calculation will be truncated. Figure \ref{fig:LineRatios} shows the difference between two runs of the GHOSTS code \citep{Gilly2020}, which simulates both collisional excitation and resonant scattering along lines of sight over the North solar pole. The only change between the two runs is the choice of incident light profile in the resonant scattering calculation: The ``Full" case uses a window that contains realistic continuum out to the edge of the Doppler-shifted scattering window, while the ``Line Only" case simply uses a model Gaussian spectral line. Panel \ref{fig:LineRatios}(a) shows the change in the full width half maximum of the lines produced in each case, which can be as much as 15\% in the middle corona. The effect on line intensities is much greater: Panel \ref{fig:LineRatios}(b) demonstrates that some lines can be brightened by a factor of 1.5\,--\,4.  Because of these effects, it is important to include a sufficient range into and out of the plane of the sky along the line of sight, and the incident light profile used in the scattering calculations must be wide enough and include a realistic continuum component such that the scattered light profile doesn't artificially truncate.

Calculations of the collisional excitation rates, which started in the 1970s, have now reached, for a few key ions, an accuracy of the order of 10\,--\,20\%. (For a recent review of a series of calculations for astrophysical ions see \citealt{badnell_etal:2016}). Calculations of the decay rates for spontaneous emission now have  an even better accuracy \citep[see e.g. the review by][]{jonsson_etal:2017}. The atomic rates for the coronal ions in the EUV/UV are relatively complete and accurate, as a series of benchmark studies has shown \citep[see][and references therein]{delzanna2020b,delzanna2019,delzanna2018b}, although significant improvements in the soft X-rays are still needed \citep{delzanna:2012_sxr1}. 
The latest set of atomic rates made available to the community are included in CHIANTI version 10 \citep{chianti_v10}. 
The database also includes some approximate treatment of resonant scattering.

The ionization state is controlled by collisions with free electrons. The ionization and recombination rates, which are needed to calculate the ion abundances, either in equilibrium or not, are somewhat more uncertain.  Fortunately, the modeling of the ionization state is relatively simple as most of the ion populations are in their ground state, hence it only depends on the electron temperature. Note, however, that there can be cases when photo-ionization from the disk can affect the ion balance in the corona and the solar wind \citep{Landi2015}.

In summary, to model the radiances of the allowed transitions not affected by resonant photo-excitation, knowledge of the electron density and temperature is needed. Estimates of the averaged density in the middle corona are widely available via the pB measurements and the \cite{vandehulst:1950} inversion. However, direct measurements of the electron temperature have been lacking \citep{delzanna2018b}. This is one of the major problems when modeling the middle corona.

Therefore, modeling usually relies on the temperature obtained from line ratios assuming that ionization equilibrium holds i.e. the ionization temperature. That is usually a reasonable assumption in the low quiescent corona but not necessarily in the middle corona, where the ionization state needs to be calculated taking into account estimates of local flows, densities and temperatures. 

Strong emission in the middle corona is also produced by forbidden lines in the visible/near infrared by highly-ionized atoms, see the review by \cite{delzanna_deluca:2018}. There are also many weaker forbidden lines in the UV. As in the case of the allowed transitions by neutral or low-charge ions, these lines are photo-pumped by the disk radiation. The advantage of such transitions is that they are visible out to great distances \citep[cf.][]{Habbal2011}. However, they are also complex to model and use for diagnostic purposes.

Firstly, accurate collisional excitation rates for these forbidden lines are difficult to obtain as they require large-scale scattering calculations. Such calculations for iron ions have shown significant increases (50\,--\,100\%) in the predicted  emissivities of some key transitions \citep[cf references in][]{delzanna2018b}.  However, not all ions have accurate atomic rates available. Also, large-scale models that are not currently available in CHIANTI are needed to account for all the cascading effects from high-lying states. 

Secondly, as any atom affected by resonant photo-excitation, accurate estimates of the disk radiation are needed. For the visible and near infrared lines this is achievable as the disk radiation has little variability, but is more challenging for UV lines, where disk radiance is both variable and inhomogeneous \citep{Vernazza1978}. Thirdly, an accurate knowledge of the local density is needed, to calculate the relative contribution of the collisional excitation and resonant photo-excitation processes. Obtaining densities from e.g. line ratios of forbidden lines is not trivial:  the plane of the sky approximation is reasonable for the allowed transitions not affected by resonant photo-excitation, but the lines affected have a significant long-range contribution \citep[see, e.g.][]{yang2020, DelZanna2023}. As a consequence, measurements of the ionization temperature from the resonant photo-excitation lines becomes strongly dependent on the distributions of the electron densities. The same issues apply when measuring chemical abundances. 

Detailed knowledge of the emission mechanisms that act in the middle corona has led to the development of a variety of forward models and modeling frameworks \citep[e.g. FORWARD:][]{Gibson2016}. However, as knowledge of the nature of emission from the middle corona is quickly evolving, the terrain for modeling of this region is also shifting rapidly. Such forward models have been used both to characterize middle corona structure and improve understanding of the emission sources themselves, which is critical for developing more robust plasma diagnostics.

\begin{figure}[t!]
\centering
\includegraphics[width=0.99\columnwidth]{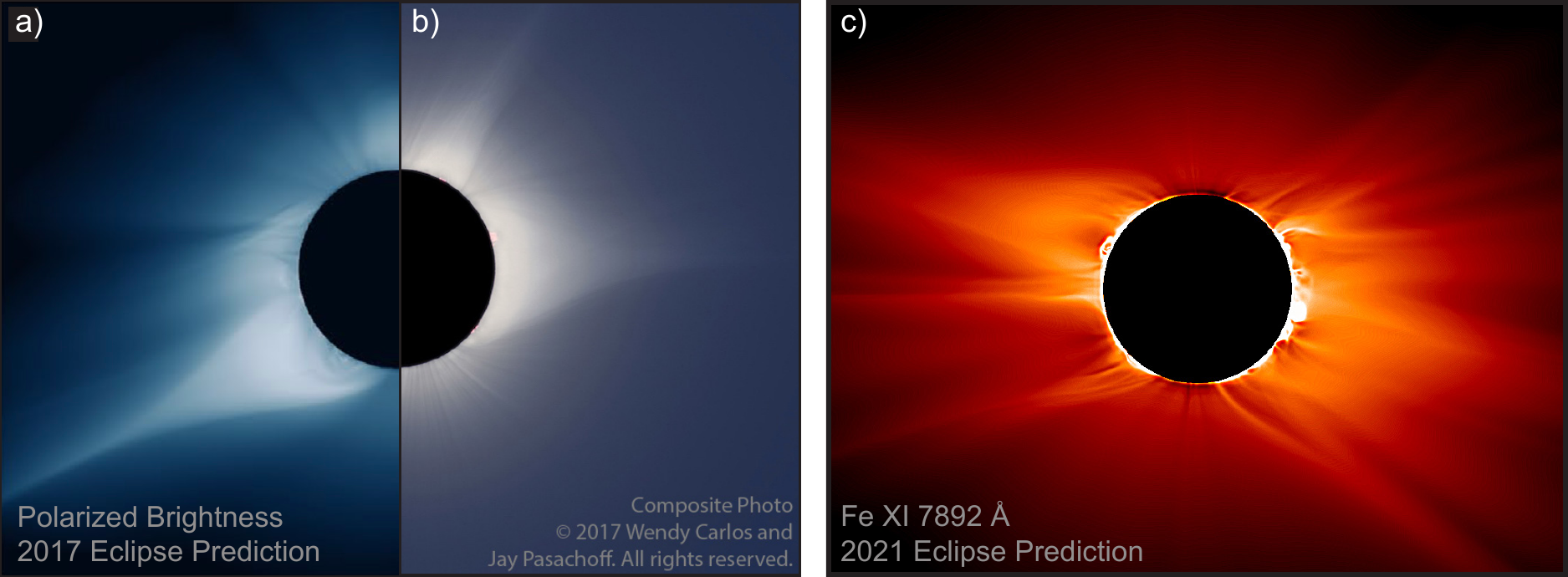}
\caption{Two examples of forward modeling from the PSI eclipse predictions. Left panels: a merged image comparing the polarized brightness prediction for the 21 August 2017 total solar eclipse (a) to a processed eclipse photo \citep[b, images adapted from][]{Mikic2018}. Panel (c) shows radially filtered, sharpened radiances for the photoexcited Fe XI \SI{789.2}{\nano\metre} emission line for the 14 December 2021 eclipse prediction \citep[\urlurl{predsci.com/eclipse2021}, see][for details on the method]{Boe2022_accepted}.}
\label{fig:PSI_eclipses}
\end{figure}

The best known examples of forward models that include the middle corona are probably the \textit{Predictive Science Inc.} eclipse predictions, which capture global coronal structure extending out through the middle corona, in an attempt to predict the corona's appearance across a variety of wavelengths, prior to a total eclipse \citep{Mikic2018}\footnote{\urlurl{www.predsci.com/corona}}. These predictions leverage the Magnetohydrodynamic Algorithm outside a Sphere (MHD-MAS; see additional discussion in Section\,\ref{ssec:Global_Coronal_Models}) global coronal model, which has also been used to extensively characterize the topology and thermodynamics of the corona, addressing a number of open questions about the nature of the corona's large-scale magnetic structure, including within the middle corona \citep[e.g.][]{Riley2019}. Examples of forward modeling from these simulations are shown in Figure\,\ref{fig:PSI_eclipses}, which highlight how both broadband K-corona signatures as well as photo-excited coronal emission lines can be synthesized from MHD models. Such diagnostics can be used to tease out information about the K and F corona from eclipse observations \citep{Boe2021} as well as benchmark the temperature, densities, and charge-state distributions predicted by the MHD models through comparison to narrowband emission lines \citep{Boe2022_accepted}.

Other forward modeling efforts specifically focused on structures within the middle corona include those of \cite{Goryaev2014}, who developed a forward model to simulate the coronal emission of a streamer in EUV and visible-light, using assumed distributions of the electron density and temperature. The distribution parameters were determined by the solution that best fit EUV observations from SWAP and \textit{Hinode}/EIS, and visible-light observations from the Mauna Loa Mk4 Coronagraph. The streamer plasma temperature near the solar limb was found to be nearly isothermal from 1.2\,--\,2\,\Rs, at $1.43\pm0.08$\,MK. They estimated the hydrostatic scale-height temperature from the determined density distribution and found it to be significantly higher, at $1.72\pm0.08$\,MK. They suggested that an outward plasma flow along the streamer could be the cause of the discrepancy. They estimated that more than 90\,\% of the observed EUV emission from the streamer was due to collisional excitation, whereas in the background corona above $\approx$ 2\,\Rs, resonant scattering may become comparable to collisional excitation in its contribution. 
    
\cite{DelZanna2018} developed a cylindrical-symmetry model which reproduced SOHO/UVCS observations of the H\,\textsc{i} Ly$\alpha$ and coronal lines between 1.4\,--\,3\,\Rs\ in quiescent streamers. The radial profile of the electron density was close to what was obtained from pB measurements, and the ionization temperature was constant at 1.4\,MK. The extrapolated densities at lower heights and the same temperature were successful in predicting the signal of the inner corona in near-infrared lines as measured during two solar eclipses in 2017 and 2019 by AIR-Spec, an Airborne Infrared Spectrometer, see \cite{madsen2019,Samra2021}.

\subsection{Modeling the Energetic Events}
\label{ssec:modeling-energetics}

The basic picture of slow energy build-up through magnetic field contortions and rapid energy release through magnetic reconnection is well established. However, the details of \textit{how} that energy is released remain an area of active research. For CME energy storage and release alone, there are at least 26 review articles and $\geq$75 model articles spanning 18 physical mechanisms over the past 2 decades (\citealt{Green2018} and references therein). A significant portion of that energy release is most clear in the middle corona, where CMEs experience the bulk of their acceleration (e.g. \citealt{Bein2011, DHuys2014}). 

Each of the numerous CME mechanism models can produce predicted kinematic profiles for the resultant CME (height--time, speed--time, acceleration--time), which have characteristic shapes that can be altered by varying the dependencies in the model. For example, the torus instability model \citep{Kliem2006} can be modified with an upward velocity perturbation whose duration can be modified -- an approximation for continued energy release powering the acceleration -- and not only does the acceleration-time profile peak at earlier times with a longer velocity perturbation, it can change from a single acceleration peak to having two acceleration peaks \citep{Schrijver2008, Majumdar2022}.

As another example, the helical kink instability tends to produce acceleration profiles with very strong jerks resulting from the magnetic flux rope twist exceeding a critical threshold of 448$^{\circ}$ \citep[e.g.][]{Fan2016}. In these two cases, and many others, the acceleration profiles differentiate themselves in the middle corona. A comprehensive summary of all of the physical mechanisms is beyond the scope of this article, but it has already been well covered by, e.g. \citet{Green2018} and \citet{Chen2011}. 

Most of these models require magnetic reconnection to liberate the stored energy needed to accelerate CMEs and power their companion solar flares. Many reconnection models and observations also predict the formation of a large-scale plasma sheet associated with reconnection, extending from the low to middle corona. Though some models predict that reconnection itself occurs primarily in the low corona \citep{Forbes2018}, both these models and observations of real events predict that upward-directed reconnection jets will dominate the dynamics of the middle corona in the wake of a CME \citep{SYu2020}. Sophisticated numerical models now appear to capture the dynamics of eruptive CME reconnection itself and the supra-arcade downflows (SADs) which often accompany this process \citep{Shen2022}. Since SADs often appear to originate from the middle corona \citep{Savage2011} these new models help to explain one of the most important manifestations of energy release in this region in the wake of large eruptive events.

Importantly, both the limited observations we already have (Section \ref{sec:how_we_observe}) and the numerous models suggest that the middle corona is a key region for developing comprehensive understanding of CME energy release and acceleration. However, only with the next generation of high-sensitivity middle corona observatories are we likely to obtain sufficient observations to develop comprehensive, data-constrained models of eruptions, that include the global-scale processes that govern the early evolution of these events.

\subsection{Global Coronal Models}
\label{ssec:Global_Coronal_Models}

Many studies have used PFSS extrapolations \citep{Schatten1969,Altschuler1969} to estimate the topology of the global coronal magnetic field, and hence consider the magnetic field of structures within the middle corona. For example, \cite{Goryaev2014} in estimating the magnetic structure of a coronal streamer and \cite{Seaton2013b} considering a coronal fan.

One parameter in a PFSS is the source surface height [$R_\mathrm{ss}$] which is the height at which magnetic field lines become radial and are considered open. Many take the ``default'' value of the source surface to be $R_\mathrm{ss}$=2\,\Rs, although some studies have shown that a lower source surface height may give a better fit to observations \citep[e.g.][]{Asvestari2019}. \citet{Sarkar2019} combined observations from the large FOV of SWAP and the LASCO-C2/C3, to cover the whole middle corona region, and by tracking the evolution of a cavity (EUV) into the three-part structure of the associated coronal mass ejection (visible-light), observed on 13~June~2010, captured the kinematics of the eruption. By applying successive geometrical fits they found that the cavity exhibited non-self-similar expansion in the low and middle corona, below $2.2\pm0.2$\,\Rs, indicating a spatial scale for the radius of the source surface.

Recent work by \cite{badman2022} have used different parameters to constrain global models. The aim was to have different global models fit the coronal holes on the disk and the neutral line topology. Different observational datasets were used to then determine the accuracy. These included visible-light Carrington maps, EUV imaging and Parker Solar Probe magnetic structures. The three parameters could not be optimised simultaneously which means there is a trade off between measuring the coronal holes or the streamer belt topology.  

A non-potential magnetic field model is a step up in complexity from a PFSS, allowing for free magnetic energy and electric currents within the volume. \cite{Meyer2020} investigated large-scale structures in the middle corona by comparing a data-driven, non-potential, global coronal magnetic field model with EUV observations from SWAP.  The lower boundary condition for the model was a global photospheric magnetic flux transport simulation which incorporated observed active region magnetic field data derived from the SDO/HMI-driven AFT model \citep{Upton2014}, from September 2014 to March 2015. The initial condition for the model was a PFSS extrapolation on 1 September. The global coronal magnetic field was then evolved in time using a magnetofrictional relaxation method \citep{vanBallegooijen2000,Mackay2006}, which produced a continuous series of non-potential equilibria in response to lower boundary motions from the flux transport simulation. 

\citeauthor{Meyer2020} considered the simulated coronal magnetic field from October 2014 onward, to allow sufficient time for the field to evolve away from its initially potential state. The model was found to reproduce the general structure of the global corona with a good degree of accuracy. Discrepancies between the observed and model corona typically occurred off the east solar limb, caused by active regions having emerged on the far side of the Sun, which cannot be incorporated into the model until they are observed on the near-side. The simulated corona was found to self-correct within a few days, but only after ``late'' active region emergences were incorporated into the model.  They followed the evolution of a particular coronal fan that was observed by SWAP over four Carrington rotations, from 2014~October to 2015~January.  The model was able to reproduce the observed structure of the fan, particularly when observed off the west limb. The model indicated that the magnetic structure underlying the fan changed from a streamer to a pseudostreamer configuration during its evolution.

\cite{Yeates2018} compared seven different global non-potential coronal magnetic field models, which were all used to model the solar corona during the 20 March 2015 total eclipse. Included in the comparison were a magnetohydrostatic model \citep{Bogdan1986}; non-linear force-free field models, including optimization \citep{Wiegelmann2007}, Grad--Rubin \citep{Amari2013}, force-free electrodynamics \citep{Contopoulos2011} and the time-evolving magnetofrictional method \citep{Mackay2006}; and MHD models including the AMR SIP--CESE Solar Wind model \citep{Feng2012} and MHD-MAS \citep{Mikic1999}. All models produced static extrapolations of the corona on the day of the eclipse, with the exception of the magneto-frictional model, which simulated a continuous time-evolution of the global corona from 1 September 2014 to 20 March 2015. 

Filament channel locations based on this magnetofrictional simulation were used to energize the MHD-MAS model. To evaluate their success, the plane-of-sky coronal structure of each model was compared with a stacked EUV image from SWAP and an Fe\,\textsc{xiv} \SI{530.5}{\nano\metre} image of the corona during the eclipse, and sheared magnetic field structures in the models were compared with filaments observed in an H$\alpha$ image from Big Bear Solar Observatory. \cite{Yeates2018} found that the models showed general agreement in magnetic topology and the ratio of total to potential magnetic energy, but showed significant differences in electric current distributions. Static extrapolations were found to best reproduce active regions, whilst the time evolving simulation could successfully recover filament channel fields. The authors recommended overall that a hybrid approach may be most suitable, using static extrapolations that are energized by a simplified evolution model, such as the MHD-MAS/magnetofrictional hybrid example they presented.

Indeed, \cite{Mikic2018} produced a prediction of the global corona for the 21 August 2017 solar eclipse using the MHD-MAS model, energized by filament channel information from a time-evolving magnetofrictional simulation in the months leading up to the eclipse. They compared the simulated corona with visible-light and EUV observations of the corona during the eclipse, finding that discrepancies between the model and simulations arose due to limitations in our current ability to observe the solar magnetic field, such as new active regions having emerged on the far side of the Sun. 

Two observational constraints for global magnetic field modeling are that the open field regions in the model should approximately correspond to coronal holes observed in emission, and that the magnitude of open flux from the model should match that determined from in-situ spacecraft. \cite{Linker2017} computed MHD and PFSS models from five different types of observatory magnetograms around 2010~July. They found that for all combinations of maps and models, the models which had open flux areas consistent with the observed coronal holes underestimated the interplanetary magnetic flux, and the models that matched the interplanetary magnetic flux had larger open flux areas than the observed coronal holes, hence raising an open flux problem. 

\cite{Riley2019} investigated whether the ``missing'' open flux could be explained by adding flux to the polar regions, at latitudes too high to be resolved by ground-based observatories or Earth-based satellites. Through PFSS and MHD magnetic field modeling, they showed that this additional polar flux could partially address the open flux problem. These models were constructed to coincide with the 11 July 2010 total eclipse, so that plane of sky coronal structures could also be compared between the models and visible-light observations of the corona during the eclipse, where the global structure of the magnetic field becomes clear predominantly in the middle corona. Through this comparison, they concluded that the additional polar flux did not generate any new observational discrepancies, and indirectly demonstrated the value of middle corona observations as important constraints on global coronal models.

\section{Open Questions and a Strategy to Answer Them}
\label{sec:Discussion}

The middle corona is a region of critical transitions straddling the inner and outer corona, to the point where it has occasionally been labelled \textit{the transition corona} \citep[e.g.][]{Masson2014,Vourlidas2020,Golub2020}. These transitions include the change from predominantly closed to open magnetic field structures, and the change from low to high plasma $\beta$ in specific regions. The middle corona is implicitly connected to both the inner and outer corona (and heliosphere by extension) through the continuation of the medium, and the bulk plasma and kinetic motions that pass from one to the other. Processes that occur within the middle corona can drive important effects in these regions and farther afield, including at the Earth and other celestial bodies, especially as the result of its modulation of solar wind outflow and CME kinematics.

The global-scale transitions that occur in the middle corona are neither tidy nor monotonic, and they depend strongly on the structures in which they occur. Plasma $\beta$, for example, varies widely within the middle corona. In general, however, $\beta \ll 1$ in the lower corona, and magnetic field dominates plasma dynamics almost everywhere, while in the outer corona, $\beta$ can be variably above or below one depending on local conditions. The location at which this transition occurs depends strongly on the type of structure observed and, in particular, these structures' embedded magnetic field. Some observations \citep{Seaton2021} suggest that large-scale dynamic processes in the middle corona can be driven by the gas dynamics of plasma flows, particularly in streamers.

Ultimately the plasma kinetic energy dominates the magnetic field structure in the super-Alfv\'enic flow regime beyond 10\,--\,20\,\Rs, as seen in Parker Solar Probe \citep{Fox2016} in-situ data, even as plasma $\beta$ varies across the unity threshold  \citep{Wexler2021b}.  In \textit{open} field structures such as plumes \citep[e.g.][]{DeForest2001} or streamers it is believed that the release from low-$\beta$ dominance occurs mainly in the outer corona, where plasma can flow freely outwards. However, the quiet sun and active region inner corona is dominated by closed field structures with more complex topology, where under certain conditions the plasma pressure can overwhelm the magnetic pressure, such as along magnetic neutral lines in streamers \citep{Vasquez2003}.  The dominance of a particular force can have significant consequences for dynamic events. The dominance of magnetic pressure in the inner corona allows for the build up of magnetic energy, and field aligned currents, which under certain conditions can be released as eruptions and flares.  The relative magnetic easing from very low $\beta$ to a higher and variable role of gas pressure occurs largely in the middle corona. Understanding where the transitions occurs will help us better understand the plasma dynamics and how flows and eruptions are influenced. 

In spite of these important transitions, however, remote sensing observations in both radio and shorter wavelengths have been insufficient to definitively characterize its global properties (see Section\,\ref{sec:how_we_observe}). Occasional instrumental off-points, eclipses (see Section \ref{sssec:VLEUVXRaySpatiallyResolvedObservations}), and radio imaging have helped bridge the gap, but only intermittently.  

Importantly, the primary methods used to observe the inner and outer corona already create an artificial boundary between the these regions \citep{Byrne2014}. From an imaging perspective, the differing X-ray, EUV, and narrowband visible observations primarily sample line-of-sight \textit{emission measure}, the amount of emitting material at the temperature the passband samples, while broad-band visible coronagraph observations sample electron density (at all temperatures) along the line-of-sight. 

Reconciling large-scale, multi-thermal, three-dimensional bright structures, such as solar eruptions, across multiple passbands is difficult \citep[see e.g.][]{OHara2019}. The absence of continuous and self-consistent observations, and an incomplete understanding of the underlying plasma properties, (see Section \ref{sec:modeling_the_mc}), has exacerbated the challenge of developing a deep understanding of the middle corona and its properties and behavior.  We are left with several important questions that must be addressed to close this knowledge gap. In the following section we discuss a few of these key questions before presenting a broad strategy that could help to address them in Section\,\ref{ssec:strategy}.

\subsection{Open Questions Relating to the Middle Corona}
\label{ssec:open_questions}

\subsubsection{Questions Concerning Transitions}

\textbf{What is the nature of middle corona plasma, and how does its nature change from its inner to outer boundary?}

The many transitions that occur within the middle corona include the change from predominantly \textit{closed} to \textit{open} magnetic field structures and the change from low to high plasma $\beta$ in quiet sun regions.  These changes will vary throughout the region depending on the underlying coronal structures and plasma properties. The lack of comprehensive, systematic, and self-consistent observations through the region, in particular those that can provide density, temperature, and magnetic field estimates, has impeded progress in determining where and how these transitions occur. Developing this understanding is critical to determining where and how processes such as solar wind acceleration, ionization state freezing-in, supersonic flows, and eruption and flow kinematic shaping occur.\\

\noindent\textbf{Where does freeze-in occur in the middle corona? What can it tell us about the origins of solar wind accelerated within middle corona structures?}

The ``freeze-in'' altitude is the height at which charge states become fixed due to the plasma becoming too tenuous to sustain ionization and recombination processes any further (Section\,\ref{ssec:Ionization}). After this transition, ions become uncoupled from thermodynamic changes in the plasma and remain fixed. As a consequence, charge states are directly related to the heating and cooling experienced prior to freeze-in, making them an indirect diagnostic of coronal conditions.  The height of freezing-in is still debated and can occur throughout the middle corona.  New observations are required to constrain modelled plasma properties which are used to derive freeze-in heights \citep{Rivera2022b}.\\

\noindent\textbf{How does the magnetic topology of the corona transition from mostly closed to almost entirely open in the middle corona? What is the role of topology in determining dynamics within the region?}

Outside of coronal holes, the inner corona is composed primarily of closed magnetic structures, while the outer corona is almost entirely radial, open magnetic field. The transition between these two regimes occurs entirely within the middle corona, but neither existing observations nor models have been sufficient to fully characterize how this transition occurs or the important role it plays in determining  the dynamics that occur here. Increasingly, simulations \citep{Higginson2016} and observations \citep{Chitta2022} have revealed the ways in which the complex topology of this region and the interactions that occur in the S-web dictate structure embedded throughout the heliosphere, but much more work is needed.

\subsubsection{Questions Concerning Outflow and Inflows}

\textbf{How does the evolving structure of the middle corona drive the structures that shape outflow into the solar wind?}

Central to the understanding of solar wind formation is the knowledge of the connection between the solar corona and the heliosphere \citep{Viall2020}. The heliospheric magnetic field is composed of an open field anchored in the photosphere, while lower in the corona the field is dominated by closed structures. The boundary between the open and closed field, situated in the middle corona around 2\,--\,3\,\Rs, fluctuates and is distorted by physical processes on a broad range of scales \citep[e.g. magnetic reconnection, eruptions, and continual flux emergence,][]{Abbo2016}. The feedback between these processes and the open/closed transition boundary is poorly understood, largely due to a lack of sensitivity and coverage in the middle corona region; though the existence of such feedback may be inferred from the longevity of coronal holes, compared to the small-scale magnetic diffusion timescale at the photosphere. Understanding this feedback is critical for heliospheric studies since it determines how hot magnetized plasma enters interplanetary space.  Furthermore, the open magnetic field and associated plasma are diverted from a purely radial direction by currents that produce a complex magnetic topology determined by photospheric evolution, prior dynamic events, and the field's global structure \citep{Wang1996,Newkirk1968, McComas2007, Yeates2008}.  These deviations from the radial field have implications for the large-scale energy storage in the corona.  These implications have yet to be fully explored because of the lack of observations in the middle corona region.\\

\noindent\textbf{What is the role of fine-scale plasma inhomogeneity perpendicular to the magnetic field?}

The sonic point is considered an important benchmark for energy deposition within the corona, and is believed to lie within the middle corona, potentially around 2\,R$_\odot$ \citep[e.g.][]{Cranmer2007,Telloni2019}. Energy deposition below or above this region is known to influence the properties of the outflowing solar wind, i.e., density, flow speed, temperature \citep[e.g.][]{Leer1980}. This has been confirmed in wave-driven solar wind models where amplifying the influence of different dissipation mechanisms, which predominantly act at different heights in the corona, leads to winds with different characteristics \citep{Shoda2018}. Although, knowledge of Alfv\'enic wave propagation from the photosphere out into the heliosphere has long suffered from a lack of wave observations in the inner and middle corona that are able to provide meaningful constraints.

To this end, it has often sufficed to assume that the plasma throughout the corona has no plasma inhomogeneity perpendicular to the magnetic field, leading to simulations of coronal heating and wind acceleration focusing on the evolution of pure Alfv\'en waves. However, recent observations have demonstrated that the inner corona is highly structured, with over-dense magnetised plasma structures present in the quiet Sun and coronal holes \citep[e.g.][]{Thurgood2014, Morton2019, Uritsky2021}. This perpendicular inhomogeneity has been found to remain present out until at least to 14 R$_\odot$ \citep{DeForest2018}, implying it must also be present in the middle corona. The presence of the inhomogeneities plays a critical role in wave propagation, making pure Alfv\'en modes impermissible. In their place are surface Alfv\'en waves \citep{Goossens2012}, which are subject to resonances and enhanced phase mixing that pure Alfv\'en modes would not \citep{Terradas2010, Pascoe2010, Soler2019}. Such phenomena concentrate wave energy to scales associated with the density structuring \citep{Magyar2022}. Previously such mechanisms for wave dissipation were dismissed as unimportant for the wave heating and acceleration. Hence there are reawakened questions as to whether the structure in the inner and middle corona enables wave dissipation through resonances and phase mixing, and whether such physics is efficient enough to dissipate a meaningful fraction of energy before the sonic point. \\

\noindent\textbf{What is the nature of the interface between the middle and outer corona? How do changes within the middle and outer corona propagate back to the Sun?}

Inflows have been shown to interact with structures in the inner corona, including large scale flows seen in EUV observations \citep{Seaton2021}, SADs seen in the wake of solar eruptions \citep[][]{Savage2012}, weaker inflows on many scales \citep{Sheeley2014}. Smaller or fainter downflows may also be ubiquitous in the less dynamic atmosphere, but could trigger eruptions through mechanisms such as magnetic breakout \citep[e.g.][]{Antiochos1999}. The exact nature of this interaction and the frequency of downflows are not fully known due to the weak signal in the far-field of EUV observations, and consequently a lack of observations. The lack of understanding of this feedback also leaves gaps in unified coronal-heliospheric models.

\subsubsection{Questions Concerning Impulsive Events}

\textbf{What role does the middle corona play in CME acceleration? How does the middle corona influence the overall evolution of CMEs?}

Impulsive CME acceleration is known to occur in the middle corona \citep{Bein2011}. Likewise, interactions within the middle corona can sometimes alter the trajectories of mature CMEs \citep{DHuys2017, Reva2017}, potentially under the influence of the structure of magnetic field in the vicinity of the eruption \citep{OHara2019} even to the point of preventing the CME from escaping at all \citep{Thalmann2015, Alvarado-Gomez2018}. However, the lack of self-consistent observations of the region presents a barrier to comprehensive understanding of how the forces and structures that emerge here manifest to shape the evolution of solar eruptions. By fully characterizing CME kinematics from the inner corona through the middle corona, we can ascertain how the background solar atmosphere interacts with the eruption, which forces are dominating, and perhaps understand the background solar conditions.\\

\noindent\textbf{How does magnetic reconnection in the middle corona release stored magnetic energy to accelerate CMEs and heat the surrounding environment? What determines where this occurs?}

Theoretical predictions suggest the magnetic reconnection that powers eruptive solar flares should occur relatively low in the corona \citep{Forbes2018}, but only a few observations, such as those by \citet{SYu2020, patel2020A&A}, have successfully isolated this location. Other manifestations of reconnection, such as SADs, can originate much higher -- well into the middle corona -- posing a mystery: what is the relationship between SADs and reconnection, and what do they have to teach us about one another? Likewise, other types of reconnection, like magnetic breakout, may occur high above pre-eruptive structures \citep{Lynch2013}, potentially within the middle corona, but such processes have only rarely been observed. Recent observations of low coronal pseudostreamers have revealed the onset of CMEs via breakout reconnection at the 3D null \citep{Kumar2021}. The similar mechanism is expected for the larger pseudostremers in the middle corona, which requires further investigations.
 Still other types of CMEs, including so-called ``stealth CMEs,'' originate from unknown processes even higher in the middle corona \citep{DHuys2014}, but may be driven by reconnection in streamers. Better observations of the middle corona are needed to provide important insight into the role of magnetic reconnection in all of these disparate situations.\\

\noindent\textbf{How do CME-driven waves and shocks influence the middle corona, particularly to accelerate particles? What can these tell us about CMEs themselves?}

The interaction between CMEs, and their associated shocks, with the ambient middle corona is often studied from the viewpoint of how the CME kinematics are modulated by the ambient plasma conditions. However, the CME can also have important effects on the local surroundings. This can be manifest in many ways, and includes: through through the aforementioned downflows generated in the wake of eruptions (SADs); the movement of surrounding structures, which can in turn force remote restructuring of the coronal magnetic structure and potentially generate sympathetic eruptions \citep{Torok2011}; and the generation of SEPs from CME-driven shocks interacting with surrounding structures, such as streamers \citep{Kong2017,Frassati2020}. Understanding these interactions are particularly important for the space-weather community.

\subsection{A Strategy to Maximize our Understanding of the Middle Corona}
\label{ssec:strategy}

Figure \ref{fig:MC_Missions} presents a summary overview of the past, present, and near-term future of middle corona observations, highlighting the patchwork nature of our coverage of this important region.  The systematic observations of the solar disk and inner corona over the past few decades have been extraordinarily successful in addressing the longstanding questions that the instruments were optimized for. However, while middle-corona-optimized missions presently in development or proposed for the future are likely to lead to some progress towards more systematic observations of the region, we remain a long way from the structured, well-coordinated observations needed to resolve the questions outlined in Section\,\ref{ssec:open_questions}.

In particular, numerous studies, such as those of \citet{Byrne2014} and \citet{OHara2019}, have demonstrated how difficult it is to associate the complex, 3D features of the middle corona that are observed in EUV with those observed in visible-light. This is complicated by the huge disparity in coronal brightness across the region, necessitating complex image processing to coherently reveal structures and dynamic events that span the region. This observational gap must be closed. 

To bridge this gap, UV and X-ray observations must be extended to greater heights, which can only be achieved through the development of high-sensitivity instrumentation, incorporating both low noise detectors and strategies to obtain higher dynamic range observations.  Missions like SunCET (in development) and ECCCO (proposed) can serve as important pathfinders, so the technologies and strategies that follow them can lead to a generation of imagers that can fully connect the inner, middle, and outer coronae in a single FOV. 

In contrast, the inner edge of the occulters required for visible-light and infrared coronagraphs  must be reduced to lower heights. This can only be achieved with instrumentation incorporating improved stray-light rejection. The PROBA-3, Aditya-L1 Visible Emission Line Coronagraph \citep[VELC:][]{Prasad2017}, UCOMP and COSMO coronagraphs again serve as key pathfinders, but require complementary observations to extend the FOV to the outer edge of the middle corona.  Future strategic planning is required to ensure the availability of co-temporal observations from all types of instruments discussed above.  Additional targeted opportunities using low-cost platforms such as rockets, balloons, and eclipse observations can also fill important observational gaps.

Spectral observations are a crucial part of any middle corona observation program, and they are required to help derive detailed understanding of the plasma properties of features captured in traditional images. Optical, UV and EUV spectra provide unique diagnostics for densities, electron temperatures, ionization states, elemental compositions, kinetic temperatures and temperature anisotropies, Doppler shift velocities along the line of sight, and velocities radially away from the Sun.  Except for optical observations during eclipses, spectral observations of the middle corona have been largely limited to the UV spectra from  UVCS, which operated from 1996 to 2013.  These UVCS observations were severely hampered by SOHO's low telemetry rate as well as a limited instantaneous spectral range and low sensitivity, all of which can easily be overcome by modern instruments and spacecraft. Closer integration of spectral and imaging observations, as designed for  the \textit{Large Optimized Coronagraphs for KeY Emission line Research} (LOCKYER) mission concept \citep{2019AGUFMSH31B..15L} would greatly enhance the effectiveness of both.

While multiple radio facilities are available around the globe, there is no solar-dedicated radio instrument that provides true broadband dynamic imaging spectroscopy in the $\approx$0.4\,--\,1 GHz spectral range, which is critical to producing observations in support of key open questions about the middle corona, including CME initiation and acceleration, understanding the CME-accelerated electrons, and perhaps most importantly, to provide unique measurements of the evolving magnetic field of CMEs in the lower portion of the middle corona ($\approx$\,1.2\,--\,2\,\Rs).

CMEs are faint and diffuse structures, which necessitate radio interferometers with a large number of antennas (several 10s to 100) to achieve sufficiently high-dynamic-range imaging ($>$10$^3$:1) and high surface-brightness sensitivity. In fact, these requirements for advancing radio studies of the middle corona science toward the next stage already comprise one of the core objectives of the Frequency Agile Solar Radiotelescope (FASR) concept, which is envisioned to provide high resolution, high dynamic range, and high fidelity dynamic imaging spectroscopy over a wide frequency range from 0.2\,--\,20\,GHz.

Structures throughout the corona are defined by the underlying magnetic fields; however, very few instruments can probe coronal magnetic fields at all, and only the Upgraded Coronal Multi-channel Polarimeter (UCoMP; \citealt{Landi2016}, still under development) will be able to measure them anywhere close to the middle corona. Techniques to ascertain the coronal magnetic field are restricted to extrapolating magnetic fields from photospheric magnetograms and inferring them from density/temperature models. However, future instruments that leverage the Hanle effect, particularly in Lyman-$\alpha$ measurements \citep{Raouafi2016b}, will be able to much more directly ascertain the strength and orientation of the coronal magnetic field. Coupled with radio measurements, these observations can provide strong constraints on global models. Such new observations will significantly improve our ability to understand the topology, evolution, and global scale effects of the middle corona's complex magnetic field.

All of these individual measurements are important in their own right, but the middle corona in particular is a dynamic, three-dimensional environment that cannot be fully understood if only observed from a single, Earth-bound, perspective. Thus, developing true understanding requires 360$^{\circ}/4\pi$ views of the Sun, including both the photospheric magnetic field and multiple lines of sight through coronal features and magnetic field structures. Such observations, coupled to global magnetic field models and advanced 3D reconstruction techniques \citep[e.g.][]{Plowman2021} would facilitate comprehensive understanding of the entire middle (and inner) corona. Such multi-perspective observations are required especially to characterize the highly structured and complex interfaces between middle corona structures and the inner and outer coronae. Given the exotic solar orbits required to achieve these multi-perspective views, it is especially important to prioritize development of miniaturized instrumentation for multi-platform, deep space constellations. The community must expedite the development such efforts via, e.g., expanded opportunities within NASA's CubeSat and LCAS programs.

Trade studies are needed to prioritize limited resources in a coherent observing framework, balancing cost, risk, and criticality of observed physical parameters across the wide range of conditions in the middle corona. Some measurements can be made with distributed ground-based instrument networks, and some with miniaturized space-borne instruments, while others require significantly larger space-based investments. $360^{\circ}/4\pi$ observations -- including out-of-ecliptic perspectives -- should prioritize measurements that cannot be made from the Earth or ecliptic perspective, or that facilitate significant research or space weather forecasting progress using additional vantage points.

Coupling all of these new observations to global models is a major challenge, particularly determining how new magnetic field and three-dimensional observations can be assimilated to provide model constraints. Advanced 3D reconstruction techniques and robust forward-modeling frameworks \citep{Gibson2016} provide promising pathways to achieve better model/data integration. Further investments in models and model data assimilation are required, and important lessons could be drawn from Earth and atmospheric science communities, which already have extensive data--model integration capabilities \citep{Lahoz2014}.

\section{Conclusion} 
\label{sec:Conclusion} 
The middle corona, the region roughly spanning heliocentric heights from $1.5$ to $6$\,\Rs, encompasses almost all of the influential physical transitions and processes that govern the behavior of coronal outflow. These transitions include the change from predominantly closed to open magnetic field structures, and the change from low to high plasma $\beta$ in specific regions. As a consequence of these transitions, the region is generally the location of primary solar wind acceleration, ionization state freezing-in, composition anomalies in long-lived structures, supersonic flow where the dynamical pressure exceeds the thermal pressure, and eruption and flow (and associated shock) kinematic shaping. 

In spite of the important transitions that occur here, the middle corona remains poorly understood compared to both the inner and outer corona, primarily because it has been much more poorly observed than these regions. Remote sensing observations, both in radio and shorter wavelengths, have been insufficient to definitively characterize its global properties. Occasional imaging opportunities, along with radio imaging and spectroscopic observations, have helped bridge the gap, but only intermittently and not self-consistently.  Developing deep understanding of the large-scale multi-thermal structures from which the middle corona is predominantly composed has therefore proved difficult.  We are left with numerous important questions that must be addressed to close this knowledge gap. 

The object of this article has not been simply to invent another naming convention, but rather to help define the new discovery space that is the middle corona. In particular, we aimed to highlight the deficiencies in our sporadic observations of the region and  the numerous important open questions that follow from this. It is our hope the article will serve as a valuable summary and reference for what we know about important middle coronal properties at the present time.

\begin{acks}
The need to define the middle corona, and consequently the need for this article, emerged from an ever increasing community interested in exploring the region. The seeds of the community were planted in a series of AGU and other conference sessions that began in Washington, DC, in 2018. The conveners and presenters of those sessions, and the conversations they sparked within our informal middle corona working group, have strongly contributed to and influenced the narrative of this article, for which we are extremely thankful.

During the COVID-disrupted AGU conference of 2020 the middle corona session was driven online in a video-conference format, and in this remote format we discovered that meaningful, wide-ranging discussion was much easier. From these early conversations, a monthly, free for all, community led meeting was born. Through this series of meetings and the broad and inspiring discussions at the online Heliophysics 2050 Workshop, we decided that a clearer, more focused definition of this region was needed to help motivate studies that focused on the region and highlight its importance to the Sun-Heliosphere system. This article is the outcome of that decision. Many people contributed to its genesis, development, and writing -- more than just those who appear on the author list, and more than we can name briefly here. Nonetheless, we thank everyone who has engaged with this community over the years, from session attendees to discussion group members to participants in discussions hosted on our Slack workspace.

To get involved in the middle corona community, please reach out via our web page, where you will find additional information on our meetings and other opportunities to interact: \urlurl{middlecorona.com}.
\end{acks}

\begin{authorcontribution}
M.J. West led the consortium, writing, ordering, reviewing, soliciting and editing contributions throughout the manuscript.  D. B. Seaton led the writing of the Short Wavelengths section of the manuscript, as well as writing, ordering, reviewing, soliciting and editing contributions throughout the manuscript where required.  D.B. Wexler led the writing of the Radio Wavelengths section of the manuscript, as well as writing, ordering, reviewing, soliciting and editing contributions throughout the manuscript where required.  J.C. Raymond led the writing of the Spectroscopy Section, and made significant contributions to the Properties and Transitions section, as well as contributing throughout the remainder of the manuscript.  G. Del Zanna led the sections focusing on Emission Mechanisms and the underlying atomic physics, as well as reviewing throughout.  Y. Rivera led the Elemental Composition section of the manuscript.  K.A. Meyer led the Global Coronal Model section and contributed significantly throughout the Modelling section of the manuscript.  B. Chen, A. Kobelski, and J. Kooi contributed significantly to the Radio Wavelengths section of the manuscript, as well as writing, ordering, and reviewing throughout.  C.R. Gilly contributed to the emission mechanisms and Charge state sections of the manuscript, as well as contributing throughout. R.J. Morton contributed expertise on wave phenomena in several Sections. All authors contributed material pertaining to their expertise, were involved in discussions about the structure and style of the manuscript.
\end{authorcontribution}

\begin{fundinginformation}
D.B. Seaton and M. J. West acknowledge support from NASA Grant 80NSSC22K0523.  G. Del Zanna acknowledges support from STFC (UK) via the consolidated grant ST/T000481/1. Y.J. Rivera acknowledges support from the Future Faculty Leaders postdoctoral fellowship at Harvard University. J.E. Kooi acknowledges support by 6.1 Base funding for basic research at the US Naval Research Laboratory (NRL). L.P. Chitta gratefully acknowledges funding by the European Union. Views and opinions expressed are however those of the author(s) only and do not necessarily reflect those of the European Union or the European Research Council (grant agreement No 101039844). Neither the European Union nor the granting authority can be held responsible for them. N.M. Viall acknowledges support from the NASA/GSFC Internal Scientist Funding Model competitive work package program ``Connecting the corona to solar wind structure and magnetospheric impact using modeling and remote and in-situ observations''. A. Vourlidas is supported by NASA grants 80NSSC21K1860. A.N. Zhukov thanks the Belgian Federal Science Policy Office (BELSPO) for the provision of financial support in the framework of the PRODEX Programme of the European Space Agency (ESA) under contract number 4000136424.
\end{fundinginformation}

\begin{dataavailability}
Data used in the production of images in this manuscript are freely available from the instrument teams and data distribution nodes.
\end{dataavailability}

\begin{ethics}
\begin{conflict}
The authors declare no competing interests.
\end{conflict}
\end{ethics}

\bibliographystyle{spr-mp-sola}
\bibliography{references.bib}

\begin{thebibliography}{307}
\ifx\bisbn     \undefined \def\bisbn  #1{ISBN #1}\fi
\ifx\binits    \undefined \def\binits#1{#1}\fi
\ifx\bauthor   \undefined \def\bauthor#1{#1}\fi
\ifx\batitle   \undefined \def\batitle#1{#1}\fi
\ifx\bjtitle   \undefined \def\bjtitle#1{\textit{#1}}\fi
\ifx\bvolume   \undefined \def\bvolume#1{\textbf{#1}}\fi
\ifx\byear     \undefined \def\byear#1{#1}\fi
\ifx\bissue    \undefined \def\bissue#1{#1}\fi
\ifx\bfpage    \undefined \def\bfpage#1{#1}\fi
\ifx\blpage    \undefined \def\blpage #1{#1}\fi
\ifx\burl      \undefined \def\burl#1{#1}\fi
\ifx\href      \undefined \def\href#1#2{#2}\fi
\ifx\betal     \undefined \def\betal{et al.}\fi
\ifx\bctitle   \undefined \def\bctitle#1{#1}\fi
\ifx\beditor   \undefined \def\beditor#1{#1}\fi
\ifx\bbtitle   \undefined \def\bbtitle#1{\textit{#1}}\fi
\ifx\bedition  \undefined \def\bedition#1{#1}\fi
\ifx\bseriesno \undefined \def\bseriesno#1{\textbf{#1}}\fi
\ifx\blocation \undefined \def\blocation#1{#1}\fi
\ifx\bsertitle \undefined \def\bsertitle#1{\textit{#1}}\fi
\ifx\bsnm      \undefined \def\bsnm#1{#1}\fi
\ifx\bsuffix   \undefined \def\bsuffix#1{#1}\fi
\ifx\bparticle \undefined \def\bparticle#1{#1}\fi
\ifx\barticle  \undefined \def\barticle#1{}\fi
\ifx\binstitute  \undefined \def\binstitute#1{#1}\fi
\ifx\bpublisher  \undefined \def\bpublisher#1{#1}\fi
\ifx\doiurl    \undefined \def\doiurl#1{\href{#1}{DOI}}\fi
\makeatletter
\def\safeHref#1#2#3{\in@{http}{#2}\ifin@\href{#2}{#3}\else\href{#1#2}{#3}\fi}
\makeatother
\ifx\adsurl    \undefined
  \def\adsurl#1{\safeHref{https://ui.adsabs.harvard.edu/abs/}{#1}{ADS}}\fi
\ifx\arxivurl  \undefined
  \def\arxivurl#1{\safeHref{http://arxiv.org/abs/}{#1}{arXiv}}\fi
\ifx\botherref \undefined \def\botherref#1{}\fi
\ifx\url       \undefined \def\url#1{#1}\fi
\ifx\bchapter  \undefined \def\bchapter#1{}\fi
\ifx\bbook     \undefined \def\bbook#1{}\fi
\ifx\bcomment  \undefined \def\bcomment#1{#1}\fi
\ifx\oauthor   \undefined \def\oauthor#1{#1}\fi
\ifx\citeauthoryear \undefined\def \citeauthoryear#1{#1}\fi
\def\endbibitem {}
\ifx\bconflocation  \undefined \def\bconflocation#1{#1} \fi

\bibitem[\protect\citeauthoryear{{Abbo} et~al.}{2016}]{Abbo2016}
\begin{barticle}
\bauthor{\bsnm{{Abbo}}, \binits{L.}},
\bauthor{\bsnm{{Ofman}}, \binits{L.}},
\bauthor{\bsnm{{Antiochos}}, \binits{S.K.}},
\bauthor{\bsnm{{Hansteen}}, \binits{V.H.}},
\bauthor{\bsnm{{Harra}}, \binits{L.}},
\bauthor{\bsnm{{Ko}}, \binits{Y.-K.}},
\bauthor{\bsnm{{Lapenta}}, \binits{G.}},
\bauthor{\bsnm{{Li}}, \binits{B.}},
\bauthor{\bsnm{{Riley}}, \binits{P.}},
\bauthor{\bsnm{{Strachan}}, \binits{L.}},
\bauthor{\bsnm{{von Steiger}}, \binits{R.}},
\bauthor{\bsnm{{Wang}}, \binits{Y.-M.}}:
\byear{2016},
\batitle{{Slow Solar Wind: Observations and Modeling}}.
\bjtitle{\ssr}
\bvolume{201},
\bfpage{55}.
\doiurl{https://doi.org/10.1007/s11214-016-0264-1}.
\adsurl{2016SSRv..201...55A}.
\end{barticle}
\endbibitem

\bibitem[\protect\citeauthoryear{{Aguilar-Rodriguez}
  et~al.}{2005}]{Aguilar-Rodriguez2005}
\begin{bchapter}
\bauthor{\bsnm{{Aguilar-Rodriguez}}, \binits{E.}},
\bauthor{\bsnm{{Gopalswamy}}, \binits{N.}},
\bauthor{\bsnm{{MacDowall}}, \binits{R.}},
\bauthor{\bsnm{{Yashiro}}, \binits{S.}},
\bauthor{\bsnm{{Kaiser}}, \binits{M.I.}}:
\byear{2005},
\bctitle{{A Study of the Drift Rate of Type II Radio Bursts at Different
  Wavelengths}}.
In: \beditor{\bsnm{{Fleck}}, \binits{B.}},
\beditor{\bsnm{{Zurbuchen}}, \binits{T.H.}},
\beditor{\bsnm{{Lacoste}}, \binits{H.}} (eds.)
\bbtitle{Solar Wind 11/SOHO 16, Connecting Sun and Heliosphere}
\bseriesno{SP-592},
\bpublisher{ESA, Noordwijk},
\bfpage{393}.
\adsurl{2005ESASP.592..393A}.
\end{bchapter}
\endbibitem

\bibitem[\protect\citeauthoryear{{Alissandrakis} and
  {Gary}}{2021}]{Alissandrakis&Gary2021}
\begin{barticle}
\bauthor{\bsnm{{Alissandrakis}}, \binits{C.E.}},
\bauthor{\bsnm{{Gary}}, \binits{D.E.}}:
\byear{2021},
\batitle{{Radio Measurements of the Magnetic field in the Solar Chromosphere
  and the Corona}}.
\bjtitle{Front. Astron. Space Sci.}
\bvolume{7},
\bfpage{77}.
\doiurl{https://doi.org/10.3389/fspas.2020.591075}.
\adsurl{2021FrASS...7...77A}.
\end{barticle}
\endbibitem

\bibitem[\protect\citeauthoryear{{Altschuler} and
  {Newkirk}}{1969}]{Altschuler1969}
\begin{barticle}
\bauthor{\bsnm{{Altschuler}}, \binits{M.D.}},
\bauthor{\bsnm{{Newkirk}}, \binits{G.}}:
\byear{1969},
\batitle{{Magnetic Fields and the Structure of the Solar Corona. I: Methods of
  Calculating Coronal Fields}}.
\bjtitle{\solphys}
\bvolume{9},
\bfpage{131}.
\doiurl{https://doi.org/10.1007/BF00145734}.
\adsurl{1969SoPh....9..131A}.
\end{barticle}
\endbibitem

\bibitem[\protect\citeauthoryear{Alvarado-G{\'{o}}mez
  et~al.}{2018}]{Alvarado-Gomez2018}
\begin{barticle}
\bauthor{\bsnm{Alvarado-G{\'{o}}mez}, \binits{J.D.}},
\bauthor{\bsnm{Drake}, \binits{J.J.}},
\bauthor{\bsnm{Cohen}, \binits{O.}},
\bauthor{\bsnm{Moschou}, \binits{S.P.}},
\bauthor{\bsnm{Garraffo}, \binits{C.}}:
\byear{2018},
\batitle{{Suppression of Coronal Mass Ejections in Active Stars by an Overlying
  Large-scale Magnetic Field: A Numerical Study}}.
\bjtitle{\apj}
\bvolume{862},
\bfpage{93}.
\doiurl{https://doi.org/10.3847/1538-4357/aacb7f}.
\end{barticle}
\endbibitem

\bibitem[\protect\citeauthoryear{{Amari} et~al.}{2013}]{Amari2013}
\begin{barticle}
\bauthor{\bsnm{{Amari}}, \binits{T.}},
\bauthor{\bsnm{{Aly}}, \binits{J.-J.}},
\bauthor{\bsnm{{Canou}}, \binits{A.}},
\bauthor{\bsnm{{Mikic}}, \binits{Z.}}:
\byear{2013},
\batitle{{Reconstruction of the solar coronal magnetic field in spherical
  geometry}}.
\bjtitle{\aap}
\bvolume{553},
\bfpage{A43}.
\doiurl{https://doi.org/10.1051/0004-6361/201220787}.
\adsurl{2013A&A...553A..43A}.
\end{barticle}
\endbibitem

\bibitem[\protect\citeauthoryear{{Antiochos}, {DeVore}, and
  {Klimchuk}}{1999}]{Antiochos1999}
\begin{barticle}
\bauthor{\bsnm{{Antiochos}}, \binits{S.K.}},
\bauthor{\bsnm{{DeVore}}, \binits{C.R.}},
\bauthor{\bsnm{{Klimchuk}}, \binits{J.A.}}:
\byear{1999},
\batitle{{A Model for Solar Coronal Mass Ejections}}.
\bjtitle{\apj}
\bvolume{510},
\bfpage{485}.
\doiurl{https://doi.org/10.1086/306563}.
\adsurl{1999ApJ...510..485A}.
\end{barticle}
\endbibitem

\bibitem[\protect\citeauthoryear{{Antiochos} et~al.}{2011}]{Antiochos2011}
\begin{barticle}
\bauthor{\bsnm{{Antiochos}}, \binits{S.K.}},
\bauthor{\bsnm{{Miki{\'c}}}, \binits{Z.}},
\bauthor{\bsnm{{Titov}}, \binits{V.S.}},
\bauthor{\bsnm{{Lionello}}, \binits{R.}},
\bauthor{\bsnm{{Linker}}, \binits{J.A.}}:
\byear{2011},
\batitle{{A Model for the Sources of the Slow Solar Wind}}.
\bjtitle{\apj}
\bvolume{731},
\bfpage{112}.
\doiurl{https://doi.org/10.1088/0004-637X/731/2/112}.
\adsurl{2011ApJ...731..112A}.
\end{barticle}
\endbibitem

\bibitem[\protect\citeauthoryear{{Antonucci} et~al.}{2020}]{Antonucci2020}
\begin{barticle}
\bauthor{\bsnm{{Antonucci}}, \binits{E.}},
\bauthor{\bsnm{{Romoli}}, \binits{M.}},
\bauthor{\bsnm{{Andretta}}, \binits{V.}},
\bauthor{\bsnm{{Fineschi}}, \binits{S.}},
\bauthor{\bsnm{{Heinzel}}, \binits{P.}},
\bauthor{\bsnm{{Moses}}, \binits{J.D.}},
\bauthor{\bsnm{{Naletto}}, \binits{G.}},
\bauthor{\bsnm{{Nicolini}}, \binits{G.}},
\bauthor{\bsnm{{Spadaro}}, \binits{D.}},
\bauthor{\bsnm{{Teriaca}}, \binits{L.}},
\bauthor{\bsnm{{Berlicki}}, \binits{A.}},
\bauthor{\bsnm{{Capobianco}}, \binits{G.}},
\bauthor{\bsnm{{Crescenzio}}, \binits{G.}},
\bauthor{\bsnm{{Da Deppo}}, \binits{V.}},
\bauthor{\bsnm{{Focardi}}, \binits{M.}},
\bauthor{\bsnm{{Frassetto}}, \binits{F.}},
\bauthor{\bsnm{{Heerlein}}, \binits{K.}},
\bauthor{\bsnm{{Landini}}, \binits{F.}},
\bauthor{\bsnm{{Magli}}, \binits{E.}},
\bauthor{\bsnm{{Marco Malvezzi}}, \binits{A.}},
\bauthor{\bsnm{{Massone}}, \binits{G.}},
\bauthor{\bsnm{{Melich}}, \binits{R.}},
\bauthor{\bsnm{{Nicolosi}}, \binits{P.}},
\bauthor{\bsnm{{Noci}}, \binits{G.}},
\bauthor{\bsnm{{Pancrazzi}}, \binits{M.}},
\bauthor{\bsnm{{Pelizzo}}, \binits{M.G.}},
\bauthor{\bsnm{{Poletto}}, \binits{L.}},
\bauthor{\bsnm{{Sasso}}, \binits{C.}},
\bauthor{\bsnm{{Sch{\"u}hle}}, \binits{U.}},
\bauthor{\bsnm{{Solanki}}, \binits{S.K.}},
\bauthor{\bsnm{{Strachan}}, \binits{L.}},
\bauthor{\bsnm{{Susino}}, \binits{R.}},
\bauthor{\bsnm{{Tondello}}, \binits{G.}},
\bauthor{\bsnm{{Uslenghi}}, \binits{M.}},
\bauthor{\bsnm{{Woch}}, \binits{J.}},
\bauthor{\bsnm{{Abbo}}, \binits{L.}},
\bauthor{\bsnm{{Bemporad}}, \binits{A.}},
\bauthor{\bsnm{{Casti}}, \binits{M.}},
\bauthor{\bsnm{{Dolei}}, \binits{S.}},
\bauthor{\bsnm{{Grimani}}, \binits{C.}},
\bauthor{\bsnm{{Messerotti}}, \binits{M.}},
\bauthor{\bsnm{{Ricci}}, \binits{M.}},
\bauthor{\bsnm{{Straus}}, \binits{T.}},
\bauthor{\bsnm{{Telloni}}, \binits{D.}},
\bauthor{\bsnm{{Zuppella}}, \binits{P.}},
\bauthor{\bsnm{{Auch{\`e}re}}, \binits{F.}},
\bauthor{\bsnm{{Bruno}}, \binits{R.}},
\bauthor{\bsnm{{Ciaravella}}, \binits{A.}},
\bauthor{\bsnm{{Corso}}, \binits{A.J.}},
\bauthor{\bsnm{{Alvarez Copano}}, \binits{M.}},
\bauthor{\bsnm{{Aznar Cuadrado}}, \binits{R.}},
\bauthor{\bsnm{{D'Amicis}}, \binits{R.}},
\bauthor{\bsnm{{Enge}}, \binits{R.}},
\bauthor{\bsnm{{Gravina}}, \binits{A.}},
\bauthor{\bsnm{{Jej{\v{c}}i{\v{c}}}}, \binits{S.}},
\bauthor{\bsnm{{Lamy}}, \binits{P.}},
\bauthor{\bsnm{{Lanzafame}}, \binits{A.}},
\bauthor{\bsnm{{Meierdierks}}, \binits{T.}},
\bauthor{\bsnm{{Papagiannaki}}, \binits{I.}},
\bauthor{\bsnm{{Peter}}, \binits{H.}},
\bauthor{\bsnm{{Fernandez Rico}}, \binits{G.}},
\bauthor{\bsnm{{Giday Sertsu}}, \binits{M.}},
\bauthor{\bsnm{{Staub}}, \binits{J.}},
\bauthor{\bsnm{{Tsinganos}}, \binits{K.}},
\bauthor{\bsnm{{Velli}}, \binits{M.}},
\bauthor{\bsnm{{Ventura}}, \binits{R.}},
\bauthor{\bsnm{{Verroi}}, \binits{E.}},
\bauthor{\bsnm{{Vial}}, \binits{J.-C.}},
\bauthor{\bsnm{{Vives}}, \binits{S.}},
\bauthor{\bsnm{{Volpicelli}}, \binits{A.}},
\bauthor{\bsnm{{Werner}}, \binits{S.}},
\bauthor{\bsnm{{Zerr}}, \binits{A.}},
\bauthor{\bsnm{{Negri}}, \binits{B.}},
\bauthor{\bsnm{{Castronuovo}}, \binits{M.}},
\bauthor{\bsnm{{Gabrielli}}, \binits{A.}},
\bauthor{\bsnm{{Bertacin}}, \binits{R.}},
\bauthor{\bsnm{{Carpentiero}}, \binits{R.}},
\bauthor{\bsnm{{Natalucci}}, \binits{S.}},
\bauthor{\bsnm{{Marliani}}, \binits{F.}},
\bauthor{\bsnm{{Cesa}}, \binits{M.}},
\bauthor{\bsnm{{Laget}}, \binits{P.}},
\bauthor{\bsnm{{Morea}}, \binits{D.}},
\bauthor{\bsnm{{Pieraccini}}, \binits{S.}},
\bauthor{\bsnm{{Radaelli}}, \binits{P.}},
\bauthor{\bsnm{{Sandri}}, \binits{P.}},
\bauthor{\bsnm{{Sarra}}, \binits{P.}},
\bauthor{\bsnm{{Cesare}}, \binits{S.}},
\bauthor{\bsnm{{Del Forno}}, \binits{F.}},
\bauthor{\bsnm{{Massa}}, \binits{E.}},
\bauthor{\bsnm{{Montabone}}, \binits{M.}},
\bauthor{\bsnm{{Mottini}}, \binits{S.}},
\bauthor{\bsnm{{Quattropani}}, \binits{D.}},
\bauthor{\bsnm{{Schillaci}}, \binits{T.}},
\bauthor{\bsnm{{Boccardo}}, \binits{R.}},
\bauthor{\bsnm{{Brando}}, \binits{R.}},
\bauthor{\bsnm{{Pandi}}, \binits{A.}},
\bauthor{\bsnm{{Baietto}}, \binits{C.}},
\bauthor{\bsnm{{Bertone}}, \binits{R.}},
\bauthor{\bsnm{{Alvarez-Herrero}}, \binits{A.}},
\bauthor{\bsnm{{Garc{\'\i}a Parejo}}, \binits{P.}},
\bauthor{\bsnm{{Cebollero}}, \binits{M.}},
\bauthor{\bsnm{{Amoruso}}, \binits{M.}},
\bauthor{\bsnm{{Centonze}}, \binits{V.}}:
\byear{2020},
\batitle{{Metis: the Solar Orbiter visible light and ultraviolet coronal
  imager}}.
\bjtitle{\aap}
\bvolume{642},
\bfpage{A10}.
\doiurl{https://doi.org/10.1051/0004-6361/201935338}.
\adsurl{2020A&A...642A..10A}.
\end{barticle}
\endbibitem

\bibitem[\protect\citeauthoryear{{Asvestari} et~al.}{2019}]{Asvestari2019}
\begin{barticle}
\bauthor{\bsnm{{Asvestari}}, \binits{E.}},
\bauthor{\bsnm{{Heinemann}}, \binits{S.G.}},
\bauthor{\bsnm{{Temmer}}, \binits{M.}},
\bauthor{\bsnm{{Pomoell}}, \binits{J.}},
\bauthor{\bsnm{{Kilpua}}, \binits{E.}},
\bauthor{\bsnm{{Magdalenic}}, \binits{J.}},
\bauthor{\bsnm{{Poedts}}, \binits{S.}}:
\byear{2019},
\batitle{{Reconstructing Coronal Hole Areas With EUHFORIA and Adapted WSA
  Model: Optimizing the Model Parameters}}.
\bjtitle{J. Geophys. Res. (Space Phys.)}
\bvolume{124},
\bfpage{8280}.
\doiurl{https://doi.org/10.1029/2019JA027173}.
\adsurl{2019JGRA..124.8280A}.
\end{barticle}
\endbibitem

\bibitem[\protect\citeauthoryear{{Auch{\`e}re} et~al.}{2023}]{Auchere2023}
\begin{barticle}
\bauthor{\bsnm{{Auch{\`e}re}}, \binits{F.}},
\bauthor{\bsnm{{Soubri{\'e}}}, \binits{E.}},
\bauthor{\bsnm{{Pelouze}}, \binits{G.}},
\bauthor{\bsnm{{Buchlin}}, \binits{{\'E}.}}:
\byear{2023},
\batitle{{Image enhancement with wavelet-optimized whitening}}.
\bjtitle{\aap}
\bvolume{670},
\bfpage{A66}.
\doiurl{https://doi.org/10.1051/0004-6361/202245345}.
\adsurl{2022arXiv221210134A}.
\end{barticle}
\endbibitem

\bibitem[\protect\citeauthoryear{Badalyan, Livshits, and
  Sykora}{1993}]{Badalyan1993}
\begin{barticle}
\bauthor{\bsnm{Badalyan}, \binits{O.G.}},
\bauthor{\bsnm{Livshits}, \binits{M.A.}},
\bauthor{\bsnm{Sykora}, \binits{J.}}:
\byear{1993},
\batitle{{Polarization of the white-light corona and its large-scale structure
  in the period of solar cycle maximum}}.
\bjtitle{\solphys}
\bvolume{145},
\bfpage{279}.
\doiurl{https://doi.org/10.1007/BF00690656}.
\end{barticle}
\endbibitem

\bibitem[\protect\citeauthoryear{{Badman} et~al.}{2022}]{badman2022}
\begin{barticle}
\bauthor{\bsnm{{Badman}}, \binits{S.T.}},
\bauthor{\bsnm{{Brooks}}, \binits{D.H.}},
\bauthor{\bsnm{{Poirier}}, \binits{N.}},
\bauthor{\bsnm{{Warren}}, \binits{H.P.}},
\bauthor{\bsnm{{Petrie}}, \binits{G.}},
\bauthor{\bsnm{{Rouillard}}, \binits{A.P.}},
\bauthor{\bsnm{{Nick Arge}}, \binits{C.}},
\bauthor{\bsnm{{Bale}}, \binits{S.D.}},
\bauthor{\bsnm{{de Pablos Ag{\"u}ero}}, \binits{D.}},
\bauthor{\bsnm{{Harra}}, \binits{L.}},
\bauthor{\bsnm{{Jones}}, \binits{S.I.}},
\bauthor{\bsnm{{Kouloumvakos}}, \binits{A.}},
\bauthor{\bsnm{{Riley}}, \binits{P.}},
\bauthor{\bsnm{{Panasenco}}, \binits{O.}},
\bauthor{\bsnm{{Velli}}, \binits{M.}},
\bauthor{\bsnm{{Wallace}}, \binits{S.}}:
\byear{2022},
\batitle{{Constraining Global Coronal Models with Multiple Independent
  Observables}}.
\bjtitle{\apj}
\bvolume{932},
\bfpage{135}.
\doiurl{https://doi.org/10.3847/1538-4357/ac6610}.
\adsurl{2022ApJ...932..135B}.
\end{barticle}
\endbibitem

\bibitem[\protect\citeauthoryear{{Badnell} et~al.}{2016}]{badnell_etal:2016}
\begin{barticle}
\bauthor{\bsnm{{Badnell}}, \binits{N.R.}},
\bauthor{\bsnm{{Del Zanna}}, \binits{G.}},
\bauthor{\bsnm{{Fern{\'a}ndez-Menchero}}, \binits{L.}},
\bauthor{\bsnm{{Giunta}}, \binits{A.S.}},
\bauthor{\bsnm{{Liang}}, \binits{G.Y.}},
\bauthor{\bsnm{{Mason}}, \binits{H.E.}},
\bauthor{\bsnm{{Storey}}, \binits{P.J.}}:
\byear{2016},
\batitle{{Atomic processes for astrophysical plasmas}}.
\bjtitle{J. Phys. B Atom. Mol. Phys.}
\bvolume{49},
\bfpage{094001}.
\doiurl{https://doi.org/10.1088/0953-4075/49/9/094001}.
\adsurl{2016JPhB...49i4001B}.
\end{barticle}
\endbibitem

\bibitem[\protect\citeauthoryear{{Baker} et~al.}{2013}]{Baker2013}
\begin{barticle}
\bauthor{\bsnm{{Baker}}, \binits{D.}},
\bauthor{\bsnm{{Brooks}}, \binits{D.H.}},
\bauthor{\bsnm{{D{\'e}moulin}}, \binits{P.}},
\bauthor{\bsnm{{van Driel-Gesztelyi}}, \binits{L.}},
\bauthor{\bsnm{{Green}}, \binits{L.M.}},
\bauthor{\bsnm{{Steed}}, \binits{K.}},
\bauthor{\bsnm{{Carlyle}}, \binits{J.}}:
\byear{2013},
\batitle{{Plasma Composition in a Sigmoidal Anemone Active Region}}.
\bjtitle{\apj}
\bvolume{778},
\bfpage{69}.
\doiurl{https://doi.org/10.1088/0004-637X/778/1/69}.
\adsurl{2013ApJ...778...69B}.
\end{barticle}
\endbibitem

\bibitem[\protect\citeauthoryear{Bastian et~al.}{2001}]{bastian2001}
\begin{barticle}
\bauthor{\bsnm{Bastian}, \binits{T.S.}},
\bauthor{\bsnm{Pick}, \binits{M.}},
\bauthor{\bsnm{Kerdraon}, \binits{A.}},
\bauthor{\bsnm{Maia}, \binits{D.}},
\bauthor{\bsnm{Vourlidas}, \binits{A.}}:
\byear{2001},
\batitle{The {{Coronal Mass Ejection}} of 1998 {{April}} 20: {{Direct Imaging}}
  at {{Radio Wavelengths}}}.
\bjtitle{\apj}
\bvolume{558},
\bfpage{L65}.
\doiurl{https://doi.org/10.1086/323421}.
\end{barticle}
\endbibitem

\bibitem[\protect\citeauthoryear{{Bein} et~al.}{2011}]{Bein2011}
\begin{barticle}
\bauthor{\bsnm{{Bein}}, \binits{B.M.}},
\bauthor{\bsnm{{Berkebile-Stoiser}}, \binits{S.}},
\bauthor{\bsnm{{Veronig}}, \binits{A.M.}},
\bauthor{\bsnm{{Temmer}}, \binits{M.}},
\bauthor{\bsnm{{Muhr}}, \binits{N.}},
\bauthor{\bsnm{{Kienreich}}, \binits{I.}},
\bauthor{\bsnm{{Utz}}, \binits{D.}},
\bauthor{\bsnm{{Vr{\v{s}}nak}}, \binits{B.}}:
\byear{2011},
\batitle{{Impulsive Acceleration of Coronal Mass Ejections. I. Statistics and
  Coronal Mass Ejection Source Region Characteristics}}.
\bjtitle{\apj}
\bvolume{738},
\bfpage{191}.
\doiurl{https://doi.org/10.1088/0004-637X/738/2/191}.
\adsurl{2011ApJ...738..191B}.
\end{barticle}
\endbibitem

\bibitem[\protect\citeauthoryear{{Bemporad}}{2008}]{Bemporad2008}
\begin{barticle}
\bauthor{\bsnm{{Bemporad}}, \binits{A.}}:
\byear{2008},
\batitle{{Spectroscopic Detection of Turbulence in Post-CME Current Sheets}}.
\bjtitle{\apj}
\bvolume{689},
\bfpage{572}.
\doiurl{https://doi.org/10.1086/592377}.
\adsurl{2008ApJ...689..572B}.
\end{barticle}
\endbibitem

\bibitem[\protect\citeauthoryear{{Bemporad} et~al.}{2007}]{Bemporad2007}
\begin{barticle}
\bauthor{\bsnm{{Bemporad}}, \binits{A.}},
\bauthor{\bsnm{{Poletto}}, \binits{G.}},
\bauthor{\bsnm{{Raymond}}, \binits{J.}},
\bauthor{\bsnm{{Giordano}}, \binits{S.}}:
\byear{2007},
\batitle{{A review of SOHO/UVCS observations of sungrazing comets}}.
\bjtitle{\planss}
\bvolume{55},
\bfpage{1021}.
\doiurl{https://doi.org/10.1016/j.pss.2006.11.013}.
\adsurl{2007P&SS...55.1021B}.
\end{barticle}
\endbibitem

\bibitem[\protect\citeauthoryear{{Bethge} et~al.}{2011}]{Bethge2011}
\begin{barticle}
\bauthor{\bsnm{{Bethge}}, \binits{C.}},
\bauthor{\bsnm{{Peter}}, \binits{H.}},
\bauthor{\bsnm{{Kentischer}}, \binits{T.J.}},
\bauthor{\bsnm{{Halbgewachs}}, \binits{C.}},
\bauthor{\bsnm{{Elmore}}, \binits{D.F.}},
\bauthor{\bsnm{{Beck}}, \binits{C.}}:
\byear{2011},
\batitle{{The Chromospheric Telescope}}.
\bjtitle{\aap}
\bvolume{534},
\bfpage{A105}.
\doiurl{https://doi.org/10.1051/0004-6361/201117456}.
\adsurl{2011A&A...534A.105B}.
\end{barticle}
\endbibitem

\bibitem[\protect\citeauthoryear{{Bird} and {Edenhofer}}{1990}]{bird1990}
\begin{bchapter}
\bauthor{\bsnm{{Bird}}, \binits{M.K.}},
\bauthor{\bsnm{{Edenhofer}}, \binits{P.}}:
\byear{1990},
\bctitle{{Remote Sensing Observations of the Solar Corona}}.
In: \beditor{\bsnm{{Schwenn}}, \binits{R.}},
\beditor{\bsnm{{Marsch}}, \binits{E.}} (eds.)
\bbtitle{Physics of the Inner Heliosphere I, Physics and Chemistry in Space,
  Space and Solar Phycics}
\bseriesno{20},
\bpublisher{Springer},
\blocation{Berlin},
\bfpage{13}.
\doiurl{https://doi.org/10.1007/978-3-642-75361-9\_2}.
\adsurl{1990pihl.book...13B}.
\end{bchapter}
\endbibitem

\bibitem[\protect\citeauthoryear{{Boe} et~al.}{2018}]{boe2018}
\begin{barticle}
\bauthor{\bsnm{{Boe}}, \binits{B.}},
\bauthor{\bsnm{{Habbal}}, \binits{S.}},
\bauthor{\bsnm{{Druckm{\"u}ller}}, \binits{M.}},
\bauthor{\bsnm{{Landi}}, \binits{E.}},
\bauthor{\bsnm{{Kourkchi}}, \binits{E.}},
\bauthor{\bsnm{{Ding}}, \binits{A.}},
\bauthor{\bsnm{{Starha}}, \binits{P.}},
\bauthor{\bsnm{{Hutton}}, \binits{J.}}:
\byear{2018},
\batitle{{The First Empirical Determination of the Fe$^{10+}$ and Fe$^{13+}$
  Freeze-in Distances in the Solar Corona}}.
\bjtitle{\apj}
\bvolume{859},
\bfpage{155}.
\doiurl{https://doi.org/10.3847/1538-4357/aabfb7}.
\adsurl{2018ApJ...859..155B}.
\end{barticle}
\endbibitem

\bibitem[\protect\citeauthoryear{{Boe} et~al.}{2020}]{Boe2020}
\begin{barticle}
\bauthor{\bsnm{{Boe}}, \binits{B.}},
\bauthor{\bsnm{{Habbal}}, \binits{S.}},
\bauthor{\bsnm{{Druckm{\"u}ller}}, \binits{M.}},
\bauthor{\bsnm{{Ding}}, \binits{A.}},
\bauthor{\bsnm{{Hod{\'e}rova}}, \binits{J.}},
\bauthor{\bsnm{{{\v{S}}tarha}}, \binits{P.}}:
\byear{2020},
\batitle{{CME-induced Thermodynamic Changes in the Corona as Inferred from Fe
  XI and Fe XIV Emission Observations during the 2017 August 21 Total Solar
  Eclipse}}.
\bjtitle{\apj}
\bvolume{888},
\bfpage{100}.
\doiurl{https://doi.org/10.3847/1538-4357/ab5e34}.
\adsurl{2020ApJ...888..100B}.
\end{barticle}
\endbibitem

\bibitem[\protect\citeauthoryear{{Boe} et~al.}{2021}]{Boe2021}
\begin{barticle}
\bauthor{\bsnm{{Boe}}, \binits{B.}},
\bauthor{\bsnm{{Habbal}}, \binits{S.}},
\bauthor{\bsnm{{Downs}}, \binits{C.}},
\bauthor{\bsnm{{Druckm{\"u}ller}}, \binits{M.}}:
\byear{2021},
\batitle{{The Color and Brightness of the F-corona Inferred from the 2019 July
  2 Total Solar Eclipse}}.
\bjtitle{\apj}
\bvolume{912},
\bfpage{44}.
\doiurl{https://doi.org/10.3847/1538-4357/abea79}.
\adsurl{2021ApJ...912...44B}.
\end{barticle}
\endbibitem

\bibitem[\protect\citeauthoryear{{Boe} et~al.}{2022}]{Boe2022_accepted}
\begin{barticle}
\bauthor{\bsnm{{Boe}}, \binits{B.}},
\bauthor{\bsnm{{Habbal}}, \binits{S.}},
\bauthor{\bsnm{{Downs}}, \binits{C.}},
\bauthor{\bsnm{{Druckm{\"u}ller}}, \binits{M.}}:
\byear{2022},
\batitle{{The Solar Minimum Eclipse of 2019 July 2: II. The First Absolute
  Brightness Measurements and MHD Model Predictions of Fe X, XI and XIV out to
  3.4 Rs}}.
\bjtitle{\apj}
\bvolume{935},
\bfpage{173}.
\doiurl{https://doi.org/10.3847/1538-4357/ac8101}.
\adsurl{2022arXiv220610106B}.
\end{barticle}
\endbibitem

\bibitem[\protect\citeauthoryear{{Bogdan} and {Low}}{1986}]{Bogdan1986}
\begin{barticle}
\bauthor{\bsnm{{Bogdan}}, \binits{T.J.}},
\bauthor{\bsnm{{Low}}, \binits{B.C.}}:
\byear{1986},
\batitle{{The Three-dimensional Structure of Magnetostatic Atmospheres. II.
  Modeling the Large-Scale Corona}}.
\bjtitle{\apj}
\bvolume{306},
\bfpage{271}.
\doiurl{https://doi.org/10.1086/164341}.
\adsurl{1986ApJ...306..271B}.
\end{barticle}
\endbibitem

\bibitem[\protect\citeauthoryear{{Boischot}}{1957}]{Boischot1957}
\begin{barticle}
\bauthor{\bsnm{{Boischot}}, \binits{A.}}:
\byear{1957},
\batitle{{Caract{\`e}res d'un type d'{\'e}mission hertzienne associ{\'e} {\`a}
  certaines {\'e}ruptions chromosph{\'e}riques}}.
\bjtitle{Acad. Sci. Paris Comp. Rend.}
\bvolume{244},
\bfpage{1326}.
\adsurl{1957CRAS..244.1326B}.
\end{barticle}
\endbibitem

\bibitem[\protect\citeauthoryear{{Brueckner} et~al.}{1995}]{Brueckner1995}
\begin{barticle}
\bauthor{\bsnm{{Brueckner}}, \binits{G.E.}},
\bauthor{\bsnm{{Howard}}, \binits{R.A.}},
\bauthor{\bsnm{{Koomen}}, \binits{M.J.}},
\bauthor{\bsnm{{Korendyke}}, \binits{C.M.}},
\bauthor{\bsnm{{Michels}}, \binits{D.J.}},
\bauthor{\bsnm{{Moses}}, \binits{J.D.}},
\bauthor{\bsnm{{Socker}}, \binits{D.G.}},
\bauthor{\bsnm{{Dere}}, \binits{K.P.}},
\bauthor{\bsnm{{Lamy}}, \binits{P.L.}},
\bauthor{\bsnm{{Llebaria}}, \binits{A.}},
\bauthor{\bsnm{{Bout}}, \binits{M.V.}},
\bauthor{\bsnm{{Schwenn}}, \binits{R.}},
\bauthor{\bsnm{{Simnett}}, \binits{G.M.}},
\bauthor{\bsnm{{Bedford}}, \binits{D.K.}},
\bauthor{\bsnm{{Eyles}}, \binits{C.J.}}:
\byear{1995},
\batitle{{The Large Angle Spectroscopic Coronagraph (LASCO)}}.
\bjtitle{\solphys}
\bvolume{162},
\bfpage{357}.
\doiurl{https://doi.org/10.1007/BF00733434}.
\adsurl{1995SoPh..162..357B}.
\end{barticle}
\endbibitem

\bibitem[\protect\citeauthoryear{{Bruno} and
  {Carbone}}{2013}]{BrunoCarbone2013}
\begin{barticle}
\bauthor{\bsnm{{Bruno}}, \binits{R.}},
\bauthor{\bsnm{{Carbone}}, \binits{V.}}:
\byear{2013},
\batitle{{The Solar Wind as a Turbulence Laboratory}}.
\bjtitle{Liv. Rev. Solar Phys.}
\bvolume{10},
\bfpage{2}.
\doiurl{https://doi.org/10.12942/lrsp-2013-2}.
\adsurl{2013LRSP...10....2B}.
\end{barticle}
\endbibitem

\bibitem[\protect\citeauthoryear{{Byrne} et~al.}{2014}]{Byrne2014}
\begin{barticle}
\bauthor{\bsnm{{Byrne}}, \binits{J.P.}},
\bauthor{\bsnm{{Morgan}}, \binits{H.}},
\bauthor{\bsnm{{Seaton}}, \binits{D.B.}},
\bauthor{\bsnm{{Bain}}, \binits{H.M.}},
\bauthor{\bsnm{{Habbal}}, \binits{S.R.}}:
\byear{2014},
\batitle{{Bridging EUV and White-Light Observations to Inspect the Initiation
  Phase of a ``Two-Stage'' Solar Eruptive Event}}.
\bjtitle{\solphys}
\bvolume{289},
\bfpage{4545}.
\doiurl{https://doi.org/10.1007/s11207-014-0585-8}.
\adsurl{2014SoPh..289.4545B}.
\end{barticle}
\endbibitem

\bibitem[\protect\citeauthoryear{{Carley}, {Vilmer}, and
  {Vourlidas}}{2020}]{Carley2020b}
\begin{barticle}
\bauthor{\bsnm{{Carley}}, \binits{E.P.}},
\bauthor{\bsnm{{Vilmer}}, \binits{N.}},
\bauthor{\bsnm{{Vourlidas}}, \binits{A.}}:
\byear{2020},
\batitle{{Radio observations of coronal mass ejection initiation and
  development in the low solar corona}}.
\bjtitle{Front. Astron. Space Sci.}
\bvolume{7},
\bfpage{79}.
\doiurl{https://doi.org/10.3389/fspas.2020.551558}.
\adsurl{2020FrASS...7...79C}.
\end{barticle}
\endbibitem

\bibitem[\protect\citeauthoryear{{Carley} et~al.}{2017}]{Carley2017}
\begin{barticle}
\bauthor{\bsnm{{Carley}}, \binits{E.P.}},
\bauthor{\bsnm{{Vilmer}}, \binits{N.}},
\bauthor{\bsnm{{Sim{\~o}es}}, \binits{P.J.A.}},
\bauthor{\bsnm{{{\'O} Fearraigh}}, \binits{B.}}:
\byear{2017},
\batitle{{Estimation of a coronal mass ejection magnetic field strength using
  radio observations of gyrosynchrotron radiation}}.
\bjtitle{\aap}
\bvolume{608},
\bfpage{A137}.
\doiurl{https://doi.org/10.1051/0004-6361/201731368}.
\adsurl{2017A&A...608A.137C}.
\end{barticle}
\endbibitem

\bibitem[\protect\citeauthoryear{Carley et~al.}{2020}]{Carley_etal._2020}
\begin{barticle}
\bauthor{\bsnm{Carley}, \binits{E.P.}},
\bauthor{\bsnm{Baldovin}, \binits{C.}},
\bauthor{\bsnm{Benthem}, \binits{P.}},
\bauthor{\bsnm{Bisi}, \binits{M.M.}},
\bauthor{\bsnm{Fallows}, \binits{R.A.}},
\bauthor{\bsnm{Gallagher}, \binits{P.T.}},
\bauthor{\bsnm{Olberg}, \binits{M.}},
\bauthor{\bsnm{Rothkaehl}, \binits{H.}},
\bauthor{\bsnm{Vermeulen}, \binits{R.}},
\bauthor{\bsnm{Vilmer}, \binits{N.}},
\bauthor{\bsnm{Barnes}, \binits{D.}}:
\byear{2020},
\batitle{Radio observatories and instrumentation used in space weather science
  and operations}.
\bjtitle{J. Space Weather Space Clim.}
\bvolume{10},
\bfpage{7}.
\doiurl{10/gh88nh}.
\end{barticle}
\endbibitem

\bibitem[\protect\citeauthoryear{Chen et~al.}{2013}]{chen2013}
\begin{barticle}
\bauthor{\bsnm{Chen}, \binits{B.}},
\bauthor{\bsnm{Bastian}, \binits{T.S.}},
\bauthor{\bsnm{White}, \binits{S.M.}},
\bauthor{\bsnm{Gary}, \binits{D.E.}},
\bauthor{\bsnm{Perley}, \binits{R.}},
\bauthor{\bsnm{Rupen}, \binits{M.}},
\bauthor{\bsnm{Carlson}, \binits{B.}}:
\byear{2013},
\batitle{Tracing Electron Beams in the Sun's Corona with Radio Dynamic Imaging
  Spectroscopy}.
\bjtitle{\apj}
\bvolume{763},
\bfpage{L21}.
\doiurl{https://doi.org/10.1088/2041-8205/763/1/L21}.
\end{barticle}
\endbibitem

\bibitem[\protect\citeauthoryear{{Chen} et~al.}{2023}]{Chen2023}
\begin{botherref}
\oauthor{\bsnm{{Chen}}, \binits{B.}},
\oauthor{\bsnm{{Kooi}}, \binits{J.E.}},
\oauthor{\bsnm{{Wexler}}, \binits{D.B.}},
\oauthor{\bsnm{{Gary}}, \binits{D.E.}},
\oauthor{\bsnm{{Yu}}, \binits{S.}},
\oauthor{\bsnm{{Mondal}}, \binits{S.}},
\oauthor{\bsnm{{Kobelski}}, \binits{A.R.}},
\oauthor{\bsnm{{Seaton}}, \binits{D.B.}},
\oauthor{\bsnm{{West}}, \binits{M.J.}},
\oauthor{\bsnm{{White}}, \binits{S.M.}},
\oauthor{\bsnm{{Fleishman}}, \binits{G.D.}},
\oauthor{\bsnm{{Saint-Hilaire}}, \binits{P.}},
\oauthor{\bsnm{{Zhang}}, \binits{P.}},
\oauthor{\bsnm{{Gilly}}, \binits{C.R.}},
\oauthor{\bsnm{{Mason}}, \binits{J.P.}},
\oauthor{\bsnm{{Reid}}, \binits{H.}}:
2023,
{Radio Studies of the Middle Corona: Current State and New Prospects in the
  Next Decade}.
\textit{arXiv e-prints},
arXiv:2301.12183.
\doiurl{https://doi.org/10.48550/arXiv.2301.12183}.
\adsurl{2023arXiv230112183C}.
\end{botherref}
\endbibitem

\bibitem[\protect\citeauthoryear{Chen}{2011}]{Chen2011}
\begin{barticle}
\bauthor{\bsnm{Chen}, \binits{P.F.}}:
\byear{2011},
\batitle{{Coronal Mass Ejections: Models and Their Observational Basis}}.
\bjtitle{Liv. Rev. Solar Phys,}
\bvolume{8}.
\doiurl{https://doi.org/10.12942/lrsp-2011-1}.
\end{barticle}
\endbibitem

\bibitem[\protect\citeauthoryear{Chhabra et~al.}{2021}]{chhabra2021}
\begin{barticle}
\bauthor{\bsnm{Chhabra}, \binits{S.}},
\bauthor{\bsnm{Gary}, \binits{D.E.}},
\bauthor{\bsnm{Hallinan}, \binits{G.}},
\bauthor{\bsnm{Anderson}, \binits{M.M.}},
\bauthor{\bsnm{Chen}, \binits{B.}},
\bauthor{\bsnm{Greenhill}, \binits{L.J.}},
\bauthor{\bsnm{Price}, \binits{D.C.}}:
\byear{2021},
\batitle{Imaging {{Spectroscopy}} of {{CME-associated Solar Radio Bursts}}
  Using {{OVRO-LWA}}}.
\bjtitle{\apj}
\bvolume{906},
\bfpage{132}.
\doiurl{https://doi.org/10.3847/1538-4357/abc94b}.
\end{barticle}
\endbibitem

\bibitem[\protect\citeauthoryear{Chhiber et~al.}{2022}]{Chhiber2022}
\begin{barticle}
\bauthor{\bsnm{Chhiber}, \binits{R.}},
\bauthor{\bsnm{Matthaeus}, \binits{W.H.}},
\bauthor{\bsnm{Usmanov}, \binits{A.V.}},
\bauthor{\bsnm{Bandyopadhyay}, \binits{R.}},
\bauthor{\bsnm{Goldstein}, \binits{M.L.}}:
\byear{2022},
\batitle{{An extended and fragmented Alfv{\'{e}}n zone in the Young Solar
  Wind}}.
\bjtitle{\mnras}
\bvolume{513},
\bfpage{159}.
\doiurl{https://doi.org/10.1093/mnras/stac779}.
\end{barticle}
\endbibitem

\bibitem[\protect\citeauthoryear{{Chitta} et~al.}{2022}]{Chitta2022}
\begin{botherref}
\oauthor{\bsnm{{Chitta}}, \binits{L.P.}},
\oauthor{\bsnm{{Seaton}}, \binits{D.B.}},
\oauthor{\bsnm{{Downs}}, \binits{C.}},
\oauthor{\bsnm{{DeForest}}, \binits{C.E.}},
\oauthor{\bsnm{{Higginson}}, \binits{A.K.}}:
2022,
{Direct observations of a complex coronal web driving highly structured slow
  solar wind}.
\textit{Nature Astronomy}.
\doiurl{https://doi.org/10.1038/s41550-022-01834-5}.
\adsurl{2022NatAs.tmp..255C}.
\end{botherref}
\endbibitem

\bibitem[\protect\citeauthoryear{{Ciaravella} and
  {Raymond}}{2008}]{Ciaravella2008}
\begin{barticle}
\bauthor{\bsnm{{Ciaravella}}, \binits{A.}},
\bauthor{\bsnm{{Raymond}}, \binits{J.C.}}:
\byear{2008},
\batitle{{The Current Sheet Associated with the 2003 November 4 Coronal Mass
  Ejection: Density, Temperature, Thickness, and Line Width}}.
\bjtitle{\apj}
\bvolume{686},
\bfpage{1372}.
\doiurl{https://doi.org/10.1086/590655}.
\adsurl{2008ApJ...686.1372C}.
\end{barticle}
\endbibitem

\bibitem[\protect\citeauthoryear{Contopoulos, Kalapotharakos, and
  Georgoulis}{2011}]{Contopoulos2011}
\begin{barticle}
\bauthor{\bsnm{Contopoulos}, \binits{I.}},
\bauthor{\bsnm{Kalapotharakos}, \binits{C.}},
\bauthor{\bsnm{Georgoulis}, \binits{M.K.}}:
\byear{2011},
\batitle{Nonlinear Force-Free Reconstruction of the Global Solar Magnetic
  Field: Methodology}.
\bjtitle{\solphys}
\bvolume{269},
\bfpage{351}.
\doiurl{https://doi.org/10.1007/s11207-011-9713-x}.
\end{barticle}
\endbibitem

\bibitem[\protect\citeauthoryear{{Cranmer}}{2009}]{Cranmer2009a}
\begin{barticle}
\bauthor{\bsnm{{Cranmer}}, \binits{S.R.}}:
\byear{2009},
\batitle{{Coronal Holes}}.
\bjtitle{Liv. Rev. Solar Phys,}
\bvolume{6},
\bfpage{3}.
\doiurl{https://doi.org/10.12942/lrsp-2009-3}.
\adsurl{2009LRSP....6....3C}.
\end{barticle}
\endbibitem

\bibitem[\protect\citeauthoryear{{Cranmer}}{2020}]{cranmer2020}
\begin{barticle}
\bauthor{\bsnm{{Cranmer}}, \binits{S.R.}}:
\byear{2020},
\batitle{{Heating Rates for Protons and Electrons in Polar Coronal Holes:
  Empirical Constraints from the Ultraviolet Coronagraph Spectrometer}}.
\bjtitle{\apj}
\bvolume{900},
\bfpage{105}.
\doiurl{https://doi.org/10.3847/1538-4357/abab04}.
\adsurl{2020ApJ...900..105C}.
\end{barticle}
\endbibitem

\bibitem[\protect\citeauthoryear{{Cranmer} and {van
  Ballegooijen}}{2005}]{CranmerVanB2005}
\begin{barticle}
\bauthor{\bsnm{{Cranmer}}, \binits{S.R.}},
\bauthor{\bsnm{{van Ballegooijen}}, \binits{A.A.}}:
\byear{2005},
\batitle{{On the Generation, Propagation, and Reflection of Alfv{\'e}n Waves
  from the Solar Photosphere to the Distant Heliosphere}}.
\bjtitle{\apjs}
\bvolume{156},
\bfpage{265}.
\doiurl{https://doi.org/10.1086/426507}.
\adsurl{2005ApJS..156..265C}.
\end{barticle}
\endbibitem

\bibitem[\protect\citeauthoryear{{Cranmer}, {Panasyuk}, and
  {Kohl}}{2008}]{Cranmer2008}
\begin{barticle}
\bauthor{\bsnm{{Cranmer}}, \binits{S.R.}},
\bauthor{\bsnm{{Panasyuk}}, \binits{A.V.}},
\bauthor{\bsnm{{Kohl}}, \binits{J.L.}}:
\byear{2008},
\batitle{{Improved Constraints on the Preferential Heating and Acceleration of
  Oxygen Ions in the Extended Solar Corona}}.
\bjtitle{\apj}
\bvolume{678},
\bfpage{1480}.
\doiurl{https://doi.org/10.1086/586890}.
\adsurl{2008ApJ...678.1480C}.
\end{barticle}
\endbibitem

\bibitem[\protect\citeauthoryear{{Cranmer}, {van Ballegooijen}, and
  {Edgar}}{2007}]{Cranmer2007}
\begin{barticle}
\bauthor{\bsnm{{Cranmer}}, \binits{S.R.}},
\bauthor{\bsnm{{van Ballegooijen}}, \binits{A.A.}},
\bauthor{\bsnm{{Edgar}}, \binits{R.J.}}:
\byear{2007},
\batitle{{Self-consistent Coronal Heating and Solar Wind Acceleration from
  Anisotropic Magnetohydrodynamic Turbulence}}.
\bjtitle{\apjs}
\bvolume{171},
\bfpage{520}.
\doiurl{https://doi.org/10.1086/518001}.
\adsurl{2007ApJS..171..520C}.
\end{barticle}
\endbibitem

\bibitem[\protect\citeauthoryear{{Cranmer} et~al.}{1999}]{Cranmer1999}
\begin{barticle}
\bauthor{\bsnm{{Cranmer}}, \binits{S.R.}},
\bauthor{\bsnm{{Kohl}}, \binits{J.L.}},
\bauthor{\bsnm{{Noci}}, \binits{G.}},
\bauthor{\bsnm{{Antonucci}}, \binits{E.}},
\bauthor{\bsnm{{Tondello}}, \binits{G.}},
\bauthor{\bsnm{{Huber}}, \binits{M.C.E.}},
\bauthor{\bsnm{{Strachan}}, \binits{L.}},
\bauthor{\bsnm{{Panasyuk}}, \binits{A.V.}},
\bauthor{\bsnm{{Gardner}}, \binits{L.D.}},
\bauthor{\bsnm{{Romoli}}, \binits{M.}},
\bauthor{\bsnm{{Fineschi}}, \binits{S.}},
\bauthor{\bsnm{{Dobrzycka}}, \binits{D.}},
\bauthor{\bsnm{{Raymond}}, \binits{J.C.}},
\bauthor{\bsnm{{Nicolosi}}, \binits{P.}},
\bauthor{\bsnm{{Siegmund}}, \binits{O.H.W.}},
\bauthor{\bsnm{{Spadaro}}, \binits{D.}},
\bauthor{\bsnm{{Benna}}, \binits{C.}},
\bauthor{\bsnm{{Ciaravella}}, \binits{A.}},
\bauthor{\bsnm{{Giordano}}, \binits{S.}},
\bauthor{\bsnm{{Habbal}}, \binits{S.R.}},
\bauthor{\bsnm{{Karovska}}, \binits{M.}},
\bauthor{\bsnm{{Li}}, \binits{X.}},
\bauthor{\bsnm{{Martin}}, \binits{R.}},
\bauthor{\bsnm{{Michels}}, \binits{J.G.}},
\bauthor{\bsnm{{Modigliani}}, \binits{A.}},
\bauthor{\bsnm{{Naletto}}, \binits{G.}},
\bauthor{\bsnm{{O'Neal}}, \binits{R.H.}},
\bauthor{\bsnm{{Pernechele}}, \binits{C.}},
\bauthor{\bsnm{{Poletto}}, \binits{G.}},
\bauthor{\bsnm{{Smith}}, \binits{P.L.}},
\bauthor{\bsnm{{Suleiman}}, \binits{R.M.}}:
\byear{1999},
\batitle{{An Empirical Model of a Polar Coronal Hole at Solar Minimum}}.
\bjtitle{\apj}
\bvolume{511},
\bfpage{481}.
\doiurl{https://doi.org/10.1086/306675}.
\adsurl{1999ApJ...511..481C}.
\end{barticle}
\endbibitem

\bibitem[\protect\citeauthoryear{{Cranmer} et~al.}{2009}]{Cranmer2009}
\begin{barticle}
\bauthor{\bsnm{{Cranmer}}, \binits{S.R.}},
\bauthor{\bsnm{{Matthaeus}}, \binits{W.H.}},
\bauthor{\bsnm{{Breech}}, \binits{B.A.}},
\bauthor{\bsnm{{Kasper}}, \binits{J.C.}}:
\byear{2009},
\batitle{{Empirical Constraints on Proton and Electron Heating in the Fast
  Solar Wind}}.
\bjtitle{\apj}
\bvolume{702},
\bfpage{1604}.
\doiurl{https://doi.org/10.1088/0004-637X/702/2/1604}.
\adsurl{2009ApJ...702.1604C}.
\end{barticle}
\endbibitem

\bibitem[\protect\citeauthoryear{{Darnel} et~al.}{2022}]{Darnel2022}
\begin{barticle}
\bauthor{\bsnm{{Darnel}}, \binits{J.M.}},
\bauthor{\bsnm{{Seaton}}, \binits{D.B.}},
\bauthor{\bsnm{{Bethge}}, \binits{C.}},
\bauthor{\bsnm{{Rachmeler}}, \binits{L.}},
\bauthor{\bsnm{{Jarvis}}, \binits{A.}},
\bauthor{\bsnm{{Hill}}, \binits{S.M.}},
\bauthor{\bsnm{{Peck}}, \binits{C.L.}},
\bauthor{\bsnm{{Hughes}}, \binits{J.M.}},
\bauthor{\bsnm{{Shapiro}}, \binits{J.}},
\bauthor{\bsnm{{Riley}}, \binits{A.}},
\bauthor{\bsnm{{Vasudevan}}, \binits{G.}},
\bauthor{\bsnm{{Shing}}, \binits{L.}},
\bauthor{\bsnm{{Koener}}, \binits{G.}},
\bauthor{\bsnm{{Edwards}}, \binits{C.}},
\bauthor{\bsnm{{Mathur}}, \binits{D.}},
\bauthor{\bsnm{{Timothy}}, \binits{S.}}:
\byear{2022},
\batitle{{The GOES-R Solar UltraViolet Imager}}.
\bjtitle{Space Weather}
\bvolume{20},
\bfpage{e2022SW003044}.
\doiurl{https://doi.org/10.1029/2022SW003044}.
\adsurl{2022SpWea..2003044D}.
\end{barticle}
\endbibitem

\bibitem[\protect\citeauthoryear{{Decraemer}, {Zhukov}, and {Van
  Doorsselaere}}{2019}]{Decraemer2019}
\begin{barticle}
\bauthor{\bsnm{{Decraemer}}, \binits{B.}},
\bauthor{\bsnm{{Zhukov}}, \binits{A.N.}},
\bauthor{\bsnm{{Van Doorsselaere}}, \binits{T.}}:
\byear{2019},
\batitle{{Three-dimensional Density Structure of a Solar Coronal Streamer
  Observed by SOHO/LASCO and STEREO/COR2 in Quadrature}}.
\bjtitle{\apj}
\bvolume{883},
\bfpage{152}.
\doiurl{https://doi.org/10.3847/1538-4357/ab3b58}.
\adsurl{2019ApJ...883..152D}.
\end{barticle}
\endbibitem

\bibitem[\protect\citeauthoryear{{Decraemer}, {Zhukov}, and {Van
  Doorsselaere}}{2020}]{Decraemer2020}
\begin{barticle}
\bauthor{\bsnm{{Decraemer}}, \binits{B.}},
\bauthor{\bsnm{{Zhukov}}, \binits{A.N.}},
\bauthor{\bsnm{{Van Doorsselaere}}, \binits{T.}}:
\byear{2020},
\batitle{{Properties of Streamer Wave Events Observed during the STEREO Era}}.
\bjtitle{\apj}
\bvolume{893},
\bfpage{78}.
\doiurl{https://doi.org/10.3847/1538-4357/ab8194}.
\adsurl{2020ApJ...893...78D}.
\end{barticle}
\endbibitem

\bibitem[\protect\citeauthoryear{{DeForest}, {Lamy}, and
  {Llebaria}}{2001}]{DeForest2001b}
\begin{barticle}
\bauthor{\bsnm{{DeForest}}, \binits{C.E.}},
\bauthor{\bsnm{{Lamy}}, \binits{P.L.}},
\bauthor{\bsnm{{Llebaria}}, \binits{A.}}:
\byear{2001},
\batitle{{Solar Polar Plume Lifetime and Coronal Hole Expansion: Determination
  from Long-Term Observations}}.
\bjtitle{\apj}
\bvolume{560},
\bfpage{490}.
\doiurl{https://doi.org/10.1086/322497}.
\adsurl{2001ApJ...560..490D}.
\end{barticle}
\endbibitem

\bibitem[\protect\citeauthoryear{{DeForest}, {Plunkett}, and
  {Andrews}}{2001}]{DeForest2001}
\begin{barticle}
\bauthor{\bsnm{{DeForest}}, \binits{C.E.}},
\bauthor{\bsnm{{Plunkett}}, \binits{S.P.}},
\bauthor{\bsnm{{Andrews}}, \binits{M.D.}}:
\byear{2001},
\batitle{{Observation of Polar Plumes at High Solar Altitudes}}.
\bjtitle{\apj}
\bvolume{546},
\bfpage{569}.
\doiurl{https://doi.org/10.1086/318221}.
\adsurl{2001ApJ...546..569D}.
\end{barticle}
\endbibitem

\bibitem[\protect\citeauthoryear{{DeForest} et~al.}{1997}]{DeForest1997}
\begin{barticle}
\bauthor{\bsnm{{DeForest}}, \binits{C.E.}},
\bauthor{\bsnm{{Hoeksema}}, \binits{J.T.}},
\bauthor{\bsnm{{Gurman}}, \binits{J.B.}},
\bauthor{\bsnm{{Thompson}}, \binits{B.J.}},
\bauthor{\bsnm{{Plunkett}}, \binits{S.P.}},
\bauthor{\bsnm{{Howard}}, \binits{R.}},
\bauthor{\bsnm{{Harrison}}, \binits{R.C.}},
\bauthor{\bsnm{{Hassler}}, \binits{D.M.}}:
\byear{1997},
\batitle{{Polar Plume Anatomy: Results of a Coordinated Observation}}.
\bjtitle{\solphys}
\bvolume{175},
\bfpage{393}.
\doiurl{https://doi.org/10.1023/A:1004955223306}.
\adsurl{1997SoPh..175..393D}.
\end{barticle}
\endbibitem

\bibitem[\protect\citeauthoryear{{DeForest} et~al.}{2018}]{DeForest2018}
\begin{barticle}
\bauthor{\bsnm{{DeForest}}, \binits{C.E.}},
\bauthor{\bsnm{{Howard}}, \binits{R.A.}},
\bauthor{\bsnm{{Velli}}, \binits{M.}},
\bauthor{\bsnm{{Viall}}, \binits{N.}},
\bauthor{\bsnm{{Vourlidas}}, \binits{A.}}:
\byear{2018},
\batitle{{The Highly Structured Outer Solar Corona}}.
\bjtitle{\apj}
\bvolume{862},
\bfpage{18}.
\doiurl{https://doi.org/10.3847/1538-4357/aac8e3}.
\adsurl{2018ApJ...862...18D}.
\end{barticle}
\endbibitem

\bibitem[\protect\citeauthoryear{{Del Zanna}}{2012}]{delzanna:2012_sxr1}
\begin{barticle}
\bauthor{\bsnm{{Del Zanna}}, \binits{G.}}:
\byear{2012},
\batitle{{Benchmarking atomic data for astrophysics: a first look at the soft
  X-ray lines}}.
\bjtitle{\aap}
\bvolume{546},
\bfpage{A97}.
\doiurl{https://doi.org/10.1051/0004-6361/201219923}.
\adsurl{2012A\%26A...546A..97D}.
\end{barticle}
\endbibitem

\bibitem[\protect\citeauthoryear{{Del Zanna}}{2019}]{delzanna2019}
\begin{barticle}
\bauthor{\bsnm{{Del Zanna}}, \binits{G.}}:
\byear{2019},
\batitle{{The EUV spectrum of the Sun: Quiet- and active-Sun irradiances and
  chemical composition}}.
\bjtitle{\aap}
\bvolume{624},
\bfpage{A36}.
\doiurl{https://doi.org/10.1051/0004-6361/201834842}.
\adsurl{2019A&A...624A..36D}.
\end{barticle}
\endbibitem

\bibitem[\protect\citeauthoryear{{Del Zanna}}{2020}]{delzanna2020b}
\begin{bchapter}
\bauthor{\bsnm{{Del Zanna}}, \binits{G.}}:
\byear{2020},
\bctitle{{Benchmarked atomic data for astrophysics}}.
In: \beditor{\bsnm{{Salama}}, \binits{F.}},
\beditor{\bsnm{{Linnartz}}, \binits{H.}} (eds.)
\bbtitle{Laboratory Astrophysics: From Observations to Interpretation, Proc.
  IAU Symp.}
\bseriesno{350},
\bpublisher{Cambridge Univ. Press},
\blocation{Cambridge UK},
\bfpage{341}.
\doiurl{https://doi.org/10.1017/S1743921319008329}.
\adsurl{2020IAUS..350..341D}.
\end{bchapter}
\endbibitem

\bibitem[\protect\citeauthoryear{{Del Zanna} and
  {DeLuca}}{2018}]{delzanna_deluca:2018}
\begin{barticle}
\bauthor{\bsnm{{Del Zanna}}, \binits{G.}},
\bauthor{\bsnm{{DeLuca}}, \binits{E.E.}}:
\byear{2018},
\batitle{{Solar Coronal Lines in the Visible and Infrared: A Rough Guide}}.
\bjtitle{\apj}
\bvolume{852},
\bfpage{52}.
\doiurl{https://doi.org/10.3847/1538-4357/aa9edf}.
\end{barticle}
\endbibitem

\bibitem[\protect\citeauthoryear{{Del Zanna} and {Mason}}{2018}]{delzanna2018b}
\begin{botherref}
\oauthor{\bsnm{{Del Zanna}}, \binits{G.}},
\oauthor{\bsnm{{Mason}}, \binits{H.E.}}:
2018,
XUV Spectroscopy.
\textit{Liv. Rev. Solar Phys,}
\textbf{15}.
\end{botherref}
\endbibitem

\bibitem[\protect\citeauthoryear{{Del Zanna}, Bromage, and
  Mason}{2003}]{DelZanna2003}
\begin{barticle}
\bauthor{\bsnm{{Del Zanna}}, \binits{G.}},
\bauthor{\bsnm{Bromage}, \binits{B.J.I.}},
\bauthor{\bsnm{Mason}, \binits{H.E.}}:
\byear{2003},
\batitle{{Spectrosopic characteristics of polar plumes}}.
\bjtitle{\apj}
\bvolume{761},
\bfpage{743}.
\doiurl{https://doi.org/10.1051/0004-6361:20021628}.
\end{barticle}
\endbibitem

\bibitem[\protect\citeauthoryear{{Del Zanna} et~al.}{2018}]{DelZanna2018}
\begin{barticle}
\bauthor{\bsnm{{Del Zanna}}, \binits{G.}},
\bauthor{\bsnm{{Raymond}}, \binits{J.}},
\bauthor{\bsnm{{Andretta}}, \binits{V.}},
\bauthor{\bsnm{{Telloni}}, \binits{D.}},
\bauthor{\bsnm{{Golub}}, \binits{L.}}:
\byear{2018},
\batitle{{Predicting the COSIE-C Signal from the Outer Corona up to 3 Solar
  Radii}}.
\bjtitle{\apj}
\bvolume{865},
\bfpage{132}.
\doiurl{https://doi.org/10.3847/1538-4357/aadcf1}.
\adsurl{2018ApJ...865..132D}.
\end{barticle}
\endbibitem

\bibitem[\protect\citeauthoryear{{Del Zanna} et~al.}{2020}]{delzanna2020}
\begin{barticle}
\bauthor{\bsnm{{Del Zanna}}, \binits{G.}},
\bauthor{\bsnm{{Storey}}, \binits{P.J.}},
\bauthor{\bsnm{{Badnell}}, \binits{N.R.}},
\bauthor{\bsnm{{Andretta}}, \binits{V.}}:
\byear{2020},
\batitle{{Helium Line Emissivities in the Solar Corona}}.
\bjtitle{\apj}
\bvolume{898},
\bfpage{72}.
\doiurl{https://doi.org/10.3847/1538-4357/ab9d84}.
\end{barticle}
\endbibitem

\bibitem[\protect\citeauthoryear{{Del Zanna} et~al.}{2021}]{chianti_v10}
\begin{barticle}
\bauthor{\bsnm{{Del Zanna}}, \binits{G.}},
\bauthor{\bsnm{{Dere}}, \binits{K.P.}},
\bauthor{\bsnm{{Young}}, \binits{P.R.}},
\bauthor{\bsnm{{Landi}}, \binits{E.}}:
\byear{2021},
\batitle{{CHIANTI{\textemdash}An Atomic Database for Emission Lines. XVI.
  Version 10, Further Extensions}}.
\bjtitle{\apj}
\bvolume{909},
\bfpage{38}.
\doiurl{https://doi.org/10.3847/1538-4357/abd8ce}.
\end{barticle}
\endbibitem

\bibitem[\protect\citeauthoryear{{Del Zanna} et~al.}{2023}]{DelZanna2023}
\begin{barticle}
\bauthor{\bsnm{{Del Zanna}}, \binits{G.}},
\bauthor{\bsnm{{Samra}}, \binits{J.}},
\bauthor{\bsnm{{Monaghan}}, \binits{A.}},
\bauthor{\bsnm{{Madsen}}, \binits{C.}},
\bauthor{\bsnm{{Bryans}}, \binits{P.}},
\bauthor{\bsnm{{DeLuca}}, \binits{E.}},
\bauthor{\bsnm{{Mason}}, \binits{H.}},
\bauthor{\bsnm{{Berkey}}, \binits{B.}},
\bauthor{\bsnm{{de Wijn}}, \binits{A.}},
\bauthor{\bsnm{{Rivera}}, \binits{Y.}}:
\byear{2023},
\batitle{{Coronal Densities, Temperatures and Abundances During the 2019 Total
  Solar Eclipse: The Role of Multi-Wavelength Observations in Coronal Plasma
  Characterization}}.
\bjtitle{\apjs}
\bvolume{265},
\bfpage{11}.
\doiurl{https://doi.org/10.3847/1538-4365/acad68}.
\adsurl{2022arXiv221211889D}.
\end{barticle}
\endbibitem

\bibitem[\protect\citeauthoryear{{Delaboudini{\`e}re} et~al.}{1995}]{EIT1995}
\begin{barticle}
\bauthor{\bsnm{{Delaboudini{\`e}re}}, \binits{J.-P.}},
\bauthor{\bsnm{{Artzner}}, \binits{G.E.}},
\bauthor{\bsnm{{Brunaud}}, \binits{J.}},
\bauthor{\bsnm{{Gabriel}}, \binits{A.H.}},
\bauthor{\bsnm{{Hochedez}}, \binits{J.F.}},
\bauthor{\bsnm{{Millier}}, \binits{F.}},
\bauthor{\bsnm{{Song}}, \binits{X.Y.}},
\bauthor{\bsnm{{Au}}, \binits{B.}},
\bauthor{\bsnm{{Dere}}, \binits{K.P.}},
\bauthor{\bsnm{{Howard}}, \binits{R.A.}},
\bauthor{\bsnm{{Kreplin}}, \binits{R.}},
\bauthor{\bsnm{{Michels}}, \binits{D.J.}},
\bauthor{\bsnm{{Moses}}, \binits{J.D.}},
\bauthor{\bsnm{{Defise}}, \binits{J.M.}},
\bauthor{\bsnm{{Jamar}}, \binits{C.}},
\bauthor{\bsnm{{Rochus}}, \binits{P.}},
\bauthor{\bsnm{{Chauvineau}}, \binits{J.P.}},
\bauthor{\bsnm{{Marioge}}, \binits{J.P.}},
\bauthor{\bsnm{{Catura}}, \binits{R.C.}},
\bauthor{\bsnm{{Lemen}}, \binits{J.R.}},
\bauthor{\bsnm{{Shing}}, \binits{L.}},
\bauthor{\bsnm{{Stern}}, \binits{R.A.}},
\bauthor{\bsnm{{Gurman}}, \binits{J.B.}},
\bauthor{\bsnm{{Neupert}}, \binits{W.M.}},
\bauthor{\bsnm{{Maucherat}}, \binits{A.}},
\bauthor{\bsnm{{Clette}}, \binits{F.}},
\bauthor{\bsnm{{Cugnon}}, \binits{P.}},
\bauthor{\bsnm{{van Dessel}}, \binits{E.L.}}:
\byear{1995},
\batitle{{EIT: Extreme-Ultraviolet Imaging Telescope for the SOHO Mission}}.
\bjtitle{\solphys}
\bvolume{162},
\bfpage{291}.
\doiurl{https://doi.org/10.1007/BF00733432}.
\adsurl{1995SoPh..162..291D}.
\end{barticle}
\endbibitem

\bibitem[\protect\citeauthoryear{D'Huys et~al.}{2014}]{DHuys2014}
\begin{barticle}
\bauthor{\bsnm{D'Huys}, \binits{E.}},
\bauthor{\bsnm{Seaton}, \binits{D.B.}},
\bauthor{\bsnm{Poedts}, \binits{S.}},
\bauthor{\bsnm{Berghmans}, \binits{D.}}:
\byear{2014},
\batitle{{Observational characteristics of coronal mass ejections without
  low-coronal signatures}}.
\bjtitle{\apj}
\bvolume{795},
\bfpage{49}.
\doiurl{https://doi.org/10.1088/0004-637X/795/1/49}.
\end{barticle}
\endbibitem

\bibitem[\protect\citeauthoryear{{D'Huys} et~al.}{2017}]{DHuys2017}
\begin{barticle}
\bauthor{\bsnm{{D'Huys}}, \binits{E.}},
\bauthor{\bsnm{{Seaton}}, \binits{D.B.}},
\bauthor{\bsnm{{De Groof}}, \binits{A.}},
\bauthor{\bsnm{{Berghmans}}, \binits{D.}},
\bauthor{\bsnm{{Poedts}}, \binits{S.}}:
\byear{2017},
\batitle{{Solar signatures and eruption mechanism of the August 14, 2010
  coronal mass ejection (CME)}}.
\bjtitle{J. Space Weather Space Clim.}
\bvolume{7},
\bfpage{A7}.
\doiurl{https://doi.org/10.1051/swsc/2017006}.
\adsurl{2017JSWSC...7A...7D}.
\end{barticle}
\endbibitem

\bibitem[\protect\citeauthoryear{Ding and Habbal}{2017}]{Ding_Habbal_2017}
\begin{barticle}
\bauthor{\bsnm{Ding}, \binits{A.}},
\bauthor{\bsnm{Habbal}, \binits{S.R.}}:
\byear{2017},
\batitle{First Detection of Prominence Material Embedded within a 2 × 106 K
  CME Front Streaming away at 100-1500 km s-1 in the Solar Corona}.
\bjtitle{\apjl}
\bvolume{842},
\bfpage{L7}.
\doiurl{https://doi.org/10.3847/2041-8213/aa7460}.
\end{barticle}
\endbibitem

\bibitem[\protect\citeauthoryear{{Druckm{\"u}llerov{\'a}}, {Morgan}, and
  {Habbal}}{2011}]{Druckmullerova2011}
\begin{barticle}
\bauthor{\bsnm{{Druckm{\"u}llerov{\'a}}}, \binits{H.}},
\bauthor{\bsnm{{Morgan}}, \binits{H.}},
\bauthor{\bsnm{{Habbal}}, \binits{S.R.}}:
\byear{2011},
\batitle{{Enhancing Coronal Structures with the Fourier
  Normalizing-radial-graded Filter}}.
\bjtitle{\apj}
\bvolume{737},
\bfpage{88}.
\doiurl{https://doi.org/10.1088/0004-637X/737/2/88}.
\adsurl{2011ApJ...737...88D}.
\end{barticle}
\endbibitem

\bibitem[\protect\citeauthoryear{{Dud{\'\i}k} et~al.}{2021}]{dudik2021}
\begin{barticle}
\bauthor{\bsnm{{Dud{\'\i}k}}, \binits{J.}},
\bauthor{\bsnm{{Del Zanna}}, \binits{G.}},
\bauthor{\bsnm{{Ryb{\'a}k}}, \binits{J.}},
\bauthor{\bsnm{{L{\"o}rin{\v{c}}{\'\i}k}}, \binits{J.}},
\bauthor{\bsnm{{Dzif{\v{c}}{\'a}kov{\'a}}}, \binits{E.}},
\bauthor{\bsnm{{Mason}}, \binits{H.E.}},
\bauthor{\bsnm{{Tomczyk}}, \binits{S.}},
\bauthor{\bsnm{{Galloy}}, \binits{M.}}:
\byear{2021},
\batitle{{Electron Densities in the Solar Corona Measured Simultaneously in the
  Extreme Ultraviolet and Infrared}}.
\bjtitle{\apj}
\bvolume{906},
\bfpage{118}.
\doiurl{https://doi.org/10.3847/1538-4357/abcd91}.
\end{barticle}
\endbibitem

\bibitem[\protect\citeauthoryear{Eddy}{1989}]{Eddy1989}
\begin{bchapter}
\bauthor{\bsnm{Eddy}, \binits{J.A.}}:
\byear{1989},
\bctitle{Gordon Newkirk's Contributions to Coronal Studies}.
In: \beditor{\bsnm{McNally}, \binits{D.}} (ed.)
\bbtitle{Highlights of Astronomy: As presented at the XXth General Assembly of
  the IAU, 1988},
\bpublisher{Springer},
\blocation{Dordrecht},
\bfpage{503}.
\doiurl{https://doi.org/10.1007/978-94-009-0977-9_72}.
\end{bchapter}
\endbibitem

\bibitem[\protect\citeauthoryear{Edwards et~al.}{2022}]{Edwards2022}
\begin{barticle}
\bauthor{\bsnm{Edwards}, \binits{L.}},
\bauthor{\bsnm{Kuridze}, \binits{D.}},
\bauthor{\bsnm{Williams}, \binits{T.}},
\bauthor{\bsnm{Morgan}, \binits{H.}}:
\byear{2022},
\batitle{{A Solar-cycle Study of Coronal Rotation: Large Variations, Rapid
  Changes, and Implications for Solar-wind Models}}.
\bjtitle{\apj}
\bvolume{928},
\bfpage{42}.
\doiurl{https://doi.org/10.3847/1538-4357/ac54ba}.
\end{barticle}
\endbibitem

\bibitem[\protect\citeauthoryear{{Elmore} et~al.}{2003}]{Elmore2003}
\begin{bchapter}
\bauthor{\bsnm{{Elmore}}, \binits{D.F.}},
\bauthor{\bsnm{{Burkepile}}, \binits{J.T.}},
\bauthor{\bsnm{{Darnell}}, \binits{J.A.}},
\bauthor{\bsnm{{Lecinski}}, \binits{A.R.}},
\bauthor{\bsnm{{Stanger}}, \binits{A.L.}}:
\byear{2003},
\bctitle{{Calibration of a ground-based solar coronal polarimeter}}.
In: \beditor{\bsnm{{Fineschi}}, \binits{S.}} (ed.)
\bbtitle{Polarimetry in Astronomy},
\bsertitle{\procspie}
\bseriesno{CS-4843},
\bfpage{66}.
\doiurl{https://doi.org/10.1117/12.459279}.
\adsurl{2003SPIE.4843...66E}.
\end{bchapter}
\endbibitem

\bibitem[\protect\citeauthoryear{{Evans} et~al.}{2008}]{evans2008}
\begin{barticle}
\bauthor{\bsnm{{Evans}}, \binits{R.M.}},
\bauthor{\bsnm{{Opher}}, \binits{M.}},
\bauthor{\bsnm{{Manchester}}, \binits{I.} \bsuffix{W.~B.}},
\bauthor{\bsnm{{Gombosi}}, \binits{T.I.}}:
\byear{2008},
\batitle{{Alfv{\'e}n Profile in the Lower Corona: Implications for Shock
  Formation}}.
\bjtitle{\apj}
\bvolume{687},
\bfpage{1355}.
\doiurl{https://doi.org/10.1086/592016}.
\adsurl{2008ApJ...687.1355E}.
\end{barticle}
\endbibitem

\bibitem[\protect\citeauthoryear{Fan}{2016}]{Fan2016}
\begin{barticle}
\bauthor{\bsnm{Fan}, \binits{Y.}}:
\byear{2016},
\batitle{{Modeling the Initiation of the 2006 December 13 Coronal Mass Ejection
  in AR 10930: The Structure and Dynamics of the Erupting Flux Rope}}.
\bjtitle{\apj}
\bvolume{824},
\bfpage{12}.
\doiurl{https://doi.org/10.3847/0004-637x/824/2/93}.
\end{barticle}
\endbibitem

\bibitem[\protect\citeauthoryear{{Feldman} and {Widing}}{1993}]{Feldman1993}
\begin{barticle}
\bauthor{\bsnm{{Feldman}}, \binits{U.}},
\bauthor{\bsnm{{Widing}}, \binits{K.G.}}:
\byear{1993},
\batitle{{Elemental Abundances in the Upper Solar Atmosphere of Quiet and
  Coronal Hole Regions (T E 4.3 X 10 5 K)}}.
\bjtitle{\apj}
\bvolume{414},
\bfpage{381}.
\doiurl{https://doi.org/10.1086/173084}.
\adsurl{1993ApJ...414..381F}.
\end{barticle}
\endbibitem

\bibitem[\protect\citeauthoryear{{Feldman} et~al.}{1998}]{Feldman1998}
\begin{barticle}
\bauthor{\bsnm{{Feldman}}, \binits{U.}},
\bauthor{\bsnm{{Sch{\"u}hle}}, \binits{U.}},
\bauthor{\bsnm{{Widing}}, \binits{K.G.}},
\bauthor{\bsnm{{Laming}}, \binits{J.M.}}:
\byear{1998},
\batitle{{Coronal Composition above the Solar Equator and the North Pole as
  Determined from Spectra Acquired by the SUMER Instrument on SOHO}}.
\bjtitle{\apj}
\bvolume{505},
\bfpage{999}.
\doiurl{https://doi.org/10.1086/306195}.
\adsurl{1998ApJ...505..999F}.
\end{barticle}
\endbibitem

\bibitem[\protect\citeauthoryear{Feng et~al.}{2012}]{Feng2012}
\begin{barticle}
\bauthor{\bsnm{Feng}, \binits{X.}},
\bauthor{\bsnm{Yang}, \binits{L.}},
\bauthor{\bsnm{Xiang}, \binits{C.}},
\bauthor{\bsnm{Jiang}, \binits{C.}},
\bauthor{\bsnm{Ma}, \binits{X.}},
\bauthor{\bsnm{Wu}, \binits{S.T.}},
\bauthor{\bsnm{Zhong}, \binits{D.}},
\bauthor{\bsnm{Zhou}, \binits{Y.}}:
\byear{2012},
\batitle{Validation of the 3D AMR SIP--CESE Solar Wind Model for Four
  Carrington Rotations}.
\bjtitle{\solphys}
\bvolume{279},
\bfpage{207}.
\bisbn{1573-093X}.
\doiurl{https://doi.org/10.1007/s11207-012-9969-9}.
\burl{https://doi.org/10.1007/s11207-012-9969-9}.
\end{barticle}
\endbibitem

\bibitem[\protect\citeauthoryear{{Fisher} et~al.}{1981}]{Fisher1981}
\begin{barticle}
\bauthor{\bsnm{{Fisher}}, \binits{R.R.}},
\bauthor{\bsnm{{Lee}}, \binits{R.H.}},
\bauthor{\bsnm{{MacQueen}}, \binits{R.M.}},
\bauthor{\bsnm{{Poland}}, \binits{A.I.}}:
\byear{1981},
\batitle{{New Mauna Loa coronagraph systems}}.
\bjtitle{\applopt}
\bvolume{20},
\bfpage{1094}.
\doiurl{https://doi.org/10.1364/AO.20.001094}.
\adsurl{1981ApOpt..20.1094F}.
\end{barticle}
\endbibitem

\bibitem[\protect\citeauthoryear{{Forbes}, {Seaton}, and
  {Reeves}}{2018}]{Forbes2018}
\begin{barticle}
\bauthor{\bsnm{{Forbes}}, \binits{T.G.}},
\bauthor{\bsnm{{Seaton}}, \binits{D.B.}},
\bauthor{\bsnm{{Reeves}}, \binits{K.K.}}:
\byear{2018},
\batitle{{Reconnection in the Post-impulsive Phase of Solar Flares}}.
\bjtitle{\apj}
\bvolume{858},
\bfpage{70}.
\doiurl{https://doi.org/10.3847/1538-4357/aabad4}.
\adsurl{2018ApJ...858...70F}.
\end{barticle}
\endbibitem

\bibitem[\protect\citeauthoryear{{Fox} et~al.}{2016}]{Fox2016}
\begin{barticle}
\bauthor{\bsnm{{Fox}}, \binits{N.J.}},
\bauthor{\bsnm{{Velli}}, \binits{M.C.}},
\bauthor{\bsnm{{Bale}}, \binits{S.D.}},
\bauthor{\bsnm{{Decker}}, \binits{R.}},
\bauthor{\bsnm{{Driesman}}, \binits{A.}},
\bauthor{\bsnm{{Howard}}, \binits{R.A.}},
\bauthor{\bsnm{{Kasper}}, \binits{J.C.}},
\bauthor{\bsnm{{Kinnison}}, \binits{J.}},
\bauthor{\bsnm{{Kusterer}}, \binits{M.}},
\bauthor{\bsnm{{Lario}}, \binits{D.}},
\bauthor{\bsnm{{Lockwood}}, \binits{M.K.}},
\bauthor{\bsnm{{McComas}}, \binits{D.J.}},
\bauthor{\bsnm{{Raouafi}}, \binits{N.E.}},
\bauthor{\bsnm{{Szabo}}, \binits{A.}}:
\byear{2016},
\batitle{{The Solar Probe Plus Mission: Humanity's First Visit to Our Star}}.
\bjtitle{\ssr}
\bvolume{204},
\bfpage{7}.
\doiurl{https://doi.org/10.1007/s11214-015-0211-6}.
\adsurl{2016SSRv..204....7F}.
\end{barticle}
\endbibitem

\bibitem[\protect\citeauthoryear{{Frassati}, {Mancuso}, and
  {Bemporad}}{2020}]{Frassati2020}
\begin{barticle}
\bauthor{\bsnm{{Frassati}}, \binits{F.}},
\bauthor{\bsnm{{Mancuso}}, \binits{S.}},
\bauthor{\bsnm{{Bemporad}}, \binits{A.}}:
\byear{2020},
\batitle{{Estimate of Plasma Temperatures Across a CME-Driven Shock from a
  Comparison Between EUV and Radio Data}}.
\bjtitle{\solphys}
\bvolume{295},
\bfpage{124}.
\doiurl{https://doi.org/10.1007/s11207-020-01686-0}.
\adsurl{2020SoPh..295..124F}.
\end{barticle}
\endbibitem

\bibitem[\protect\citeauthoryear{{Frazin}, {Cranmer}, and
  {Kohl}}{2003}]{Frazin2003}
\begin{barticle}
\bauthor{\bsnm{{Frazin}}, \binits{R.A.}},
\bauthor{\bsnm{{Cranmer}}, \binits{S.R.}},
\bauthor{\bsnm{{Kohl}}, \binits{J.L.}}:
\byear{2003},
\batitle{{Empirically Determined Anisotropic Velocity Distributions and
  Outflows of O$^{5+}$ Ions in a Coronal Streamer at Solar Minimum}}.
\bjtitle{\apj}
\bvolume{597},
\bfpage{1145}.
\doiurl{https://doi.org/10.1086/378558}.
\adsurl{2003ApJ...597.1145F}.
\end{barticle}
\endbibitem

\bibitem[\protect\citeauthoryear{{Galano} et~al.}{2018}]{Galano2018}
\begin{bchapter}
\bauthor{\bsnm{{Galano}}, \binits{D.}},
\bauthor{\bsnm{{Bemporad}}, \binits{A.}},
\bauthor{\bsnm{{Buckley}}, \binits{S.}},
\bauthor{\bsnm{{Cernica}}, \binits{I.}},
\bauthor{\bsnm{{D{\'a}niel}}, \binits{V.}},
\bauthor{\bsnm{{Denis}}, \binits{F.}},
\bauthor{\bsnm{{de Vos}}, \binits{L.}},
\bauthor{\bsnm{{Fineschi}}, \binits{S.}},
\bauthor{\bsnm{{Galy}}, \binits{C.}},
\bauthor{\bsnm{{Graczyk}}, \binits{R.}},
\bauthor{\bsnm{{Horodyska}}, \binits{P.}},
\bauthor{\bsnm{{Jacob}}, \binits{J.}},
\bauthor{\bsnm{{Jansen}}, \binits{R.}},
\bauthor{\bsnm{{Kranitis}}, \binits{N.}},
\bauthor{\bsnm{{Kurowski}}, \binits{M.}},
\bauthor{\bsnm{{Ladno}}, \binits{M.}},
\bauthor{\bsnm{{Ledent}}, \binits{P.}},
\bauthor{\bsnm{{Loreggia}}, \binits{D.}},
\bauthor{\bsnm{{Melich}}, \binits{R.}},
\bauthor{\bsnm{{Mollet}}, \binits{D.}},
\bauthor{\bsnm{{Mosdorf}}, \binits{M.}},
\bauthor{\bsnm{{Paschalis}}, \binits{A.}},
\bauthor{\bsnm{{Peresty}}, \binits{R.}},
\bauthor{\bsnm{{Purica}}, \binits{M.}},
\bauthor{\bsnm{{Radzik}}, \binits{B.}},
\bauthor{\bsnm{{Rataj}}, \binits{M.}},
\bauthor{\bsnm{{Rougeot}}, \binits{R.}},
\bauthor{\bsnm{{Salvador}}, \binits{L.}},
\bauthor{\bsnm{{Thizy}}, \binits{C.}},
\bauthor{\bsnm{{Versluys}}, \binits{J.}},
\bauthor{\bsnm{{Walczak}}, \binits{T.}},
\bauthor{\bsnm{{Zarzycka}}, \binits{A.}},
\bauthor{\bsnm{{Zender}}, \binits{J.}},
\bauthor{\bsnm{{Zhukov}}, \binits{A.}}:
\byear{2018},
\bctitle{{Development of ASPIICS: a coronagraph based on Proba-3 formation
  flying mission}}.
In: \beditor{\bsnm{{Lystrup}}, \binits{M.}},
\beditor{\bsnm{{MacEwen}}, \binits{H.A.}},
\beditor{\bsnm{{Fazio}}, \binits{G.G.}},
\beditor{\bsnm{{Batalha}}, \binits{N.}},
\beditor{\bsnm{{Siegler}}, \binits{N.}},
\beditor{\bsnm{{Tong}}, \binits{E.C.}} (eds.)
\bbtitle{Space Telescopes and Instrumentation 2018: Optical, Infrared, and
  Millimeter Wave},
\bsertitle{Society of Photo-Optical Instrumentation Engineers (SPIE) Conference
  Series}
\bseriesno{10698},
\bfpage{106982Y}.
\doiurl{https://doi.org/10.1117/12.2312493}.
\adsurl{2018SPIE10698E..2YG}.
\end{bchapter}
\endbibitem

\bibitem[\protect\citeauthoryear{{Gary} and {Hurford}}{2004}]{Gary2004}
\begin{bchapter}
\bauthor{\bsnm{{Gary}}, \binits{D.E.}},
\bauthor{\bsnm{{Hurford}}, \binits{G.J.}}:
\byear{2004},
\bctitle{{Radio Spectral Diagnostics}}.
In: \beditor{\bsnm{{Gary}}, \binits{D.E.}},
\beditor{\bsnm{{Keller}}, \binits{C.U.}} (eds.)
\bbtitle{Solar and Space Weather Radiophysics},
\bsertitle{Astrophys. Space Sci. Lib.}
\bseriesno{314},
\bfpage{71}.
\doiurl{https://doi.org/10.1007/1-4020-2814-8\_4}.
\adsurl{2004ASSL..314...71G}.
\end{bchapter}
\endbibitem

\bibitem[\protect\citeauthoryear{{Gary}}{2001}]{gary2001}
\begin{barticle}
\bauthor{\bsnm{{Gary}}, \binits{G.A.}}:
\byear{2001},
\batitle{{Plasma Beta above a Solar Active Region: Rethinking the Paradigm}}.
\bjtitle{\solphys}
\bvolume{203},
\bfpage{71}.
\doiurl{https://doi.org/10.1023/A:1012722021820}.
\adsurl{2001SoPh..203...71G}.
\end{barticle}
\endbibitem

\bibitem[\protect\citeauthoryear{{Geiss}, {Gloeckler}, and {von
  Steiger}}{1995}]{Geiss1995}
\begin{barticle}
\bauthor{\bsnm{{Geiss}}, \binits{J.}},
\bauthor{\bsnm{{Gloeckler}}, \binits{G.}},
\bauthor{\bsnm{{von Steiger}}, \binits{R.}}:
\byear{1995},
\batitle{{Origin of the Solar Wind From Composition Data}}.
\bjtitle{\ssr}
\bvolume{72},
\bfpage{49}.
\doiurl{https://doi.org/10.1007/BF00768753}.
\adsurl{1995SSRv...72...49G}.
\end{barticle}
\endbibitem

\bibitem[\protect\citeauthoryear{{Gibson} et~al.}{2016}]{Gibson2016}
\begin{barticle}
\bauthor{\bsnm{{Gibson}}, \binits{S.}},
\bauthor{\bsnm{{Kucera}}, \binits{T.}},
\bauthor{\bsnm{{White}}, \binits{S.}},
\bauthor{\bsnm{{Dove}}, \binits{J.}},
\bauthor{\bsnm{{Fan}}, \binits{Y.}},
\bauthor{\bsnm{{Forland}}, \binits{B.}},
\bauthor{\bsnm{{Rachmeler}}, \binits{L.}},
\bauthor{\bsnm{{Downs}}, \binits{C.}},
\bauthor{\bsnm{{Reeves}}, \binits{K.}}:
\byear{2016},
\batitle{{FORWARD: A toolset for multiwavelength coronal magnetometry}}.
\bjtitle{Front. Astron. Space Sci.}
\bvolume{3},
\bfpage{8}.
\doiurl{https://doi.org/10.3389/fspas.2016.00008}.
\adsurl{2016FrASS...3....8G}.
\end{barticle}
\endbibitem

\bibitem[\protect\citeauthoryear{{Gilly} and {Cranmer}}{2020}]{Gilly2020}
\begin{barticle}
\bauthor{\bsnm{{Gilly}}, \binits{C.R.}},
\bauthor{\bsnm{{Cranmer}}, \binits{S.R.}}:
\byear{2020},
\batitle{{The Effect of Solar Wind Expansion and Nonequilibrium Ionization on
  the Broadening of Coronal Emission Lines}}.
\bjtitle{\apj}
\bvolume{901},
\bfpage{150}.
\doiurl{https://doi.org/10.3847/1538-4357/abb1ad}.
\adsurl{2020ApJ...901..150G}.
\end{barticle}
\endbibitem

\bibitem[\protect\citeauthoryear{{Golub} et~al.}{2007}]{Golub2007}
\begin{barticle}
\bauthor{\bsnm{{Golub}}, \binits{L.}},
\bauthor{\bsnm{{Deluca}}, \binits{E.}},
\bauthor{\bsnm{{Austin}}, \binits{G.}},
\bauthor{\bsnm{{Bookbinder}}, \binits{J.}},
\bauthor{\bsnm{{Caldwell}}, \binits{D.}},
\bauthor{\bsnm{{Cheimets}}, \binits{P.}},
\bauthor{\bsnm{{Cirtain}}, \binits{J.}},
\bauthor{\bsnm{{Cosmo}}, \binits{M.}},
\bauthor{\bsnm{{Reid}}, \binits{P.}},
\bauthor{\bsnm{{Sette}}, \binits{A.}},
\bauthor{\bsnm{{Weber}}, \binits{M.}},
\bauthor{\bsnm{{Sakao}}, \binits{T.}},
\bauthor{\bsnm{{Kano}}, \binits{R.}},
\bauthor{\bsnm{{Shibasaki}}, \binits{K.}},
\bauthor{\bsnm{{Hara}}, \binits{H.}},
\bauthor{\bsnm{{Tsuneta}}, \binits{S.}},
\bauthor{\bsnm{{Kumagai}}, \binits{K.}},
\bauthor{\bsnm{{Tamura}}, \binits{T.}},
\bauthor{\bsnm{{Shimojo}}, \binits{M.}},
\bauthor{\bsnm{{McCracken}}, \binits{J.}},
\bauthor{\bsnm{{Carpenter}}, \binits{J.}},
\bauthor{\bsnm{{Haight}}, \binits{H.}},
\bauthor{\bsnm{{Siler}}, \binits{R.}},
\bauthor{\bsnm{{Wright}}, \binits{E.}},
\bauthor{\bsnm{{Tucker}}, \binits{J.}},
\bauthor{\bsnm{{Rutledge}}, \binits{H.}},
\bauthor{\bsnm{{Barbera}}, \binits{M.}},
\bauthor{\bsnm{{Peres}}, \binits{G.}},
\bauthor{\bsnm{{Varisco}}, \binits{S.}}:
\byear{2007},
\batitle{{The X-Ray Telescope (XRT) for the Hinode Mission}}.
\bjtitle{\solphys}
\bvolume{243},
\bfpage{63}.
\doiurl{https://doi.org/10.1007/s11207-007-0182-1}.
\adsurl{2007SoPh..243...63G}.
\end{barticle}
\endbibitem

\bibitem[\protect\citeauthoryear{{Golub} et~al.}{2020}]{Golub2020}
\begin{barticle}
\bauthor{\bsnm{{Golub}}, \binits{L.}},
\bauthor{\bsnm{{Cheimets}}, \binits{P.}},
\bauthor{\bsnm{{DeLuca}}, \binits{E.E.}},
\bauthor{\bsnm{{Madsen}}, \binits{C.A.}},
\bauthor{\bsnm{{Reeves}}, \binits{K.K.}},
\bauthor{\bsnm{{Samra}}, \binits{J.}},
\bauthor{\bsnm{{Savage}}, \binits{S.}},
\bauthor{\bsnm{{Winebarger}}, \binits{A.}},
\bauthor{\bsnm{{Bruccoleri}}, \binits{A.R.}}:
\byear{2020},
\batitle{{EUV imaging and spectroscopy for improved space weather
  forecasting}}.
\bjtitle{J. Space Weather Space Clim.}
\bvolume{10},
\bfpage{37}.
\doiurl{https://doi.org/10.1051/swsc/2020040}.
\adsurl{2020JSWSC..10...37G}.
\end{barticle}
\endbibitem

\bibitem[\protect\citeauthoryear{Gombosi et~al.}{2018}]{Gombosi2018}
\begin{barticle}
\bauthor{\bsnm{Gombosi}, \binits{T.I.}},
\bauthor{\bparticle{van~der} \bsnm{Holst}, \binits{B.}},
\bauthor{\bsnm{Manchester}, \binits{W.B.}},
\bauthor{\bsnm{Sokolov}, \binits{I.V.}}:
\byear{2018},
\batitle{{Extended MHD modeling of the steady solar corona and the solar
  wind}}.
\bjtitle{Liv. Rev. Solar Phys,}
\bvolume{15},
\bfpage{4}.
\doiurl{https://doi.org/10.1007/s41116-018-0014-4}.
\end{barticle}
\endbibitem

\bibitem[\protect\citeauthoryear{{Goossens} et~al.}{2012}]{Goossens2012}
\begin{barticle}
\bauthor{\bsnm{{Goossens}}, \binits{M.}},
\bauthor{\bsnm{{Andries}}, \binits{J.}},
\bauthor{\bsnm{{Soler}}, \binits{R.}},
\bauthor{\bsnm{{Van Doorsselaere}}, \binits{T.}},
\bauthor{\bsnm{{Arregui}}, \binits{I.}},
\bauthor{\bsnm{{Terradas}}, \binits{J.}}:
\byear{2012},
\batitle{{Surface Alfv{\'e}n Waves in Solar Flux Tubes}}.
\bjtitle{\apj}
\bvolume{753},
\bfpage{111}.
\doiurl{https://doi.org/10.1088/0004-637X/753/2/111}.
\adsurl{2012ApJ...753..111G}.
\end{barticle}
\endbibitem

\bibitem[\protect\citeauthoryear{{Goryaev} et~al.}{2014}]{Goryaev2014}
\begin{barticle}
\bauthor{\bsnm{{Goryaev}}, \binits{F.}},
\bauthor{\bsnm{{Slemzin}}, \binits{V.}},
\bauthor{\bsnm{{Vainshtein}}, \binits{L.}},
\bauthor{\bsnm{{Williams}}, \binits{D.R.}}:
\byear{2014},
\batitle{{Study of Extreme-ultraviolet Emission and Properties of a Coronal
  Streamer from PROBA2/SWAP, Hinode/EIS and Mauna Loa Mk4 Observations}}.
\bjtitle{\apj}
\bvolume{781},
\bfpage{100}.
\doiurl{https://doi.org/10.1088/0004-637X/781/2/100}.
\adsurl{2014ApJ...781..100G}.
\end{barticle}
\endbibitem

\bibitem[\protect\citeauthoryear{Green et~al.}{2018}]{Green2018}
\begin{barticle}
\bauthor{\bsnm{Green}, \binits{L.M.}},
\bauthor{\bsnm{T{\"{o}}r{\"{o}}k}, \binits{T.}},
\bauthor{\bsnm{Vr{\v{s}}nak}, \binits{B.}},
\bauthor{\bsnm{Manchester}, \binits{W.}},
\bauthor{\bsnm{Veronig}, \binits{A.}}:
\byear{2018},
\batitle{{The Origin, Early Evolution and Predictability of Solar Eruptions}}.
\bjtitle{Space Sci. Rev.}
\bvolume{214},
\bfpage{46}.
\doiurl{https://doi.org/10.1007/s11214-017-0462-5}.
\end{barticle}
\endbibitem

\bibitem[\protect\citeauthoryear{{Guennou} et~al.}{2016}]{Guennou2016}
\begin{barticle}
\bauthor{\bsnm{{Guennou}}, \binits{C.}},
\bauthor{\bsnm{{Rachmeler}}, \binits{L.}},
\bauthor{\bsnm{{Seaton}}, \binits{D.}},
\bauthor{\bsnm{{Auch{\`e}re}}, \binits{F.}}:
\byear{2016},
\batitle{{Lifecycle of a large-scale polar coronal pseudostreamer/cavity
  system}}.
\bjtitle{Front. Astron. Space Sci.}
\bvolume{3},
\bfpage{14}.
\doiurl{https://doi.org/10.3389/fspas.2016.00014}.
\adsurl{2016FrASS...3...14G}.
\end{barticle}
\endbibitem

\bibitem[\protect\citeauthoryear{{Guhathakurta}
  et~al.}{1999}]{guhathakurta1999}
\begin{barticle}
\bauthor{\bsnm{{Guhathakurta}}, \binits{M.}},
\bauthor{\bsnm{{Fludra}}, \binits{A.}},
\bauthor{\bsnm{{Gibson}}, \binits{S.E.}},
\bauthor{\bsnm{{Biesecker}}, \binits{D.}},
\bauthor{\bsnm{{Fisher}}, \binits{R.}}:
\byear{1999},
\batitle{{Physical properties of a coronal hole from a coronal diagnostic
  spectrometer, Mauna Loa Coronagraph, and LASCO observations during the Whole
  Sun Month}}.
\bjtitle{\jgr}
\bvolume{104},
\bfpage{9801}.
\doiurl{https://doi.org/10.1029/1998JA900082}.
\adsurl{1999JGR...104.9801G}.
\end{barticle}
\endbibitem

\bibitem[\protect\citeauthoryear{{Habbal}, {Morgan}, and
  {Druckm{\"u}ller}}{2014}]{Habbal2014}
\begin{barticle}
\bauthor{\bsnm{{Habbal}}, \binits{S.R.}},
\bauthor{\bsnm{{Morgan}}, \binits{H.}},
\bauthor{\bsnm{{Druckm{\"u}ller}}, \binits{M.}}:
\byear{2014},
\batitle{{Exploring the Prominence-Corona Connection and its Expansion into the
  Outer Corona Using Total Solar Eclipse Observations}}.
\bjtitle{\apj}
\bvolume{793},
\bfpage{119}.
\doiurl{https://doi.org/10.1088/0004-637X/793/2/119}.
\adsurl{2014ApJ...793..119H}.
\end{barticle}
\endbibitem

\bibitem[\protect\citeauthoryear{Habbal et~al.}{2011}]{Habbal2011}
\begin{barticle}
\bauthor{\bsnm{Habbal}, \binits{S.R.}},
\bauthor{\bsnm{Druckmüller}, \binits{M.}},
\bauthor{\bsnm{Morgan}, \binits{H.}},
\bauthor{\bsnm{Ding}, \binits{A.}},
\bauthor{\bsnm{Johnson}, \binits{J.}},
\bauthor{\bsnm{Druckmüllerov{\'{a}}}, \binits{H.}},
\bauthor{\bsnm{Daw}, \binits{A.}},
\bauthor{\bsnm{Arndt}, \binits{M.B.}},
\bauthor{\bsnm{Dietzel}, \binits{M.}},
\bauthor{\bsnm{Saken}, \binits{J.}}:
\byear{2011},
\batitle{Thermodynamics of the solar corona and evolution of the solar magnetic
  field as inferred from the total solar eclipse observations of 2010 July 11}.
\bjtitle{\apj}
\bvolume{734},
\bfpage{120}.
\doiurl{https://doi.org/10.1088/0004-637x/734/2/120}.
\end{barticle}
\endbibitem

\bibitem[\protect\citeauthoryear{{Hagenaar} et~al.}{1999}]{Hagenaar1999}
\begin{barticle}
\bauthor{\bsnm{{Hagenaar}}, \binits{H.J.}},
\bauthor{\bsnm{{Schrijver}}, \binits{C.J.}},
\bauthor{\bsnm{{Title}}, \binits{A.M.}},
\bauthor{\bsnm{{Shine}}, \binits{R.A.}}:
\byear{1999},
\batitle{{Dispersal of Magnetic Flux in the Quiet Solar Photosphere}}.
\bjtitle{\apj}
\bvolume{511},
\bfpage{932}.
\doiurl{https://doi.org/10.1086/306691}.
\adsurl{1999ApJ...511..932H}.
\end{barticle}
\endbibitem

\bibitem[\protect\citeauthoryear{{Halain} et~al.}{2013}]{Halain2013}
\begin{barticle}
\bauthor{\bsnm{{Halain}}, \binits{J.-P.}},
\bauthor{\bsnm{{Berghmans}}, \binits{D.}},
\bauthor{\bsnm{{Seaton}}, \binits{D.B.}},
\bauthor{\bsnm{{Nicula}}, \binits{B.}},
\bauthor{\bsnm{{De Groof}}, \binits{A.}},
\bauthor{\bsnm{{Mierla}}, \binits{M.}},
\bauthor{\bsnm{{Mazzoli}}, \binits{A.}},
\bauthor{\bsnm{{Defise}}, \binits{J.-M.}},
\bauthor{\bsnm{{Rochus}}, \binits{P.}}:
\byear{2013},
\batitle{{The SWAP EUV Imaging Telescope. Part II: In-flight Performance and
  Calibration}}.
\bjtitle{\solphys}
\bvolume{286},
\bfpage{67}.
\doiurl{https://doi.org/10.1007/s11207-012-0183-6}.
\adsurl{2013SoPh..286...67H}.
\end{barticle}
\endbibitem

\bibitem[\protect\citeauthoryear{{Higginson}}{2016}]{Higginson2016}
\begin{botherref}
\oauthor{\bsnm{{Higginson}}, \binits{A.K.}}:
2016,
{The Dynamics of the S-Web and Implications for the Solar Wind and
  Heliosphere}.
PhD thesis,
University of Michigan.
\adsurl{2016PhDT.......184H}.
\end{botherref}
\endbibitem

\bibitem[\protect\citeauthoryear{{Hill} et~al.}{2005}]{Hill2005}
\begin{barticle}
\bauthor{\bsnm{{Hill}}, \binits{S.M.}},
\bauthor{\bsnm{{Pizzo}}, \binits{V.J.}},
\bauthor{\bsnm{{Balch}}, \binits{C.C.}},
\bauthor{\bsnm{{Biesecker}}, \binits{D.A.}},
\bauthor{\bsnm{{Bornmann}}, \binits{P.}},
\bauthor{\bsnm{{Hildner}}, \binits{E.}},
\bauthor{\bsnm{{Lewis}}, \binits{L.D.}},
\bauthor{\bsnm{{Grubb}}, \binits{R.N.}},
\bauthor{\bsnm{{Husler}}, \binits{M.P.}},
\bauthor{\bsnm{{Prendergast}}, \binits{K.}},
\bauthor{\bsnm{{Vickroy}}, \binits{J.}},
\bauthor{\bsnm{{Greer}}, \binits{S.}},
\bauthor{\bsnm{{Defoor}}, \binits{T.}},
\bauthor{\bsnm{{Wilkinson}}, \binits{D.C.}},
\bauthor{\bsnm{{Hooker}}, \binits{R.}},
\bauthor{\bsnm{{Mulligan}}, \binits{P.}},
\bauthor{\bsnm{{Chipman}}, \binits{E.}},
\bauthor{\bsnm{{Bysal}}, \binits{H.}},
\bauthor{\bsnm{{Douglas}}, \binits{J.P.}},
\bauthor{\bsnm{{Reynolds}}, \binits{R.}},
\bauthor{\bsnm{{Davis}}, \binits{J.M.}},
\bauthor{\bsnm{{Wallace}}, \binits{K.S.}},
\bauthor{\bsnm{{Russell}}, \binits{K.}},
\bauthor{\bsnm{{Freestone}}, \binits{K.}},
\bauthor{\bsnm{{Bagdigian}}, \binits{D.}},
\bauthor{\bsnm{{Page}}, \binits{T.}},
\bauthor{\bsnm{{Kerns}}, \binits{S.}},
\bauthor{\bsnm{{Hoffman}}, \binits{R.}},
\bauthor{\bsnm{{Cauffman}}, \binits{S.A.}},
\bauthor{\bsnm{{Davis}}, \binits{M.A.}},
\bauthor{\bsnm{{Studer}}, \binits{R.}},
\bauthor{\bsnm{{Berthiaume}}, \binits{F.E.}},
\bauthor{\bsnm{{Saha}}, \binits{T.T.}},
\bauthor{\bsnm{{Berthiume}}, \binits{G.D.}},
\bauthor{\bsnm{{Farthing}}, \binits{H.}},
\bauthor{\bsnm{{Zimmermann}}, \binits{F.}}:
\byear{2005},
\batitle{{The NOAA GOES-12 Solar X-Ray Imager (SXI) 1. Instrument, Operations,
  and Data}}.
\bjtitle{\solphys}
\bvolume{226},
\bfpage{255}.
\doiurl{https://doi.org/10.1007/s11207-005-7416-x}.
\adsurl{2005SoPh..226..255H}.
\end{barticle}
\endbibitem

\bibitem[\protect\citeauthoryear{{Hofmeister} et~al.}{2017}]{hofmeister2017}
\begin{barticle}
\bauthor{\bsnm{{Hofmeister}}, \binits{S.J.}},
\bauthor{\bsnm{{Veronig}}, \binits{A.}},
\bauthor{\bsnm{{Reiss}}, \binits{M.A.}},
\bauthor{\bsnm{{Temmer}}, \binits{M.}},
\bauthor{\bsnm{{Vennerstrom}}, \binits{S.}},
\bauthor{\bsnm{{Vr{\v{s}}nak}}, \binits{B.}},
\bauthor{\bsnm{{Heber}}, \binits{B.}}:
\byear{2017},
\batitle{{Characteristics of Low-latitude Coronal Holes near the Maximum of
  Solar Cycle 24}}.
\bjtitle{\apj}
\bvolume{835},
\bfpage{268}.
\doiurl{https://doi.org/10.3847/1538-4357/835/2/268}.
\adsurl{2017ApJ...835..268H}.
\end{barticle}
\endbibitem

\bibitem[\protect\citeauthoryear{{Holden}}{1894}]{Holden1894}
\begin{barticle}
\bauthor{\bsnm{{Holden}}, \binits{E.S.}}:
\byear{1894},
\batitle{{The LICK Observatory Eclipse Expeditions of January, 1889, December,
  1889, and of April, 1893}}.
\bjtitle{\pasp}
\bvolume{6},
\bfpage{245}.
\doiurl{https://doi.org/10.1086/120866}.
\adsurl{1894PASP....6..245H}.
\end{barticle}
\endbibitem

\bibitem[\protect\citeauthoryear{{Howard} et~al.}{2008}]{Howard2008}
\begin{barticle}
\bauthor{\bsnm{{Howard}}, \binits{R.A.}},
\bauthor{\bsnm{{Moses}}, \binits{J.D.}},
\bauthor{\bsnm{{Vourlidas}}, \binits{A.}},
\bauthor{\bsnm{{Newmark}}, \binits{J.S.}},
\bauthor{\bsnm{{Socker}}, \binits{D.G.}},
\bauthor{\bsnm{{Plunkett}}, \binits{S.P.}},
\bauthor{\bsnm{{Korendyke}}, \binits{C.M.}},
\bauthor{\bsnm{{Cook}}, \binits{J.W.}},
\bauthor{\bsnm{{Hurley}}, \binits{A.}},
\bauthor{\bsnm{{Davila}}, \binits{J.M.}},
\bauthor{\bsnm{{Thompson}}, \binits{W.T.}},
\bauthor{\bsnm{{St Cyr}}, \binits{O.C.}},
\bauthor{\bsnm{{Mentzell}}, \binits{E.}},
\bauthor{\bsnm{{Mehalick}}, \binits{K.}},
\bauthor{\bsnm{{Lemen}}, \binits{J.R.}},
\bauthor{\bsnm{{Wuelser}}, \binits{J.P.}},
\bauthor{\bsnm{{Duncan}}, \binits{D.W.}},
\bauthor{\bsnm{{Tarbell}}, \binits{T.D.}},
\bauthor{\bsnm{{Wolfson}}, \binits{C.J.}},
\bauthor{\bsnm{{Moore}}, \binits{A.}},
\bauthor{\bsnm{{Harrison}}, \binits{R.A.}},
\bauthor{\bsnm{{Waltham}}, \binits{N.R.}},
\bauthor{\bsnm{{Lang}}, \binits{J.}},
\bauthor{\bsnm{{Davis}}, \binits{C.J.}},
\bauthor{\bsnm{{Eyles}}, \binits{C.J.}},
\bauthor{\bsnm{{Mapson-Menard}}, \binits{H.}},
\bauthor{\bsnm{{Simnett}}, \binits{G.M.}},
\bauthor{\bsnm{{Halain}}, \binits{J.P.}},
\bauthor{\bsnm{{Defise}}, \binits{J.M.}},
\bauthor{\bsnm{{Mazy}}, \binits{E.}},
\bauthor{\bsnm{{Rochus}}, \binits{P.}},
\bauthor{\bsnm{{Mercier}}, \binits{R.}},
\bauthor{\bsnm{{Ravet}}, \binits{M.F.}},
\bauthor{\bsnm{{Delmotte}}, \binits{F.}},
\bauthor{\bsnm{{Auchere}}, \binits{F.}},
\bauthor{\bsnm{{Delaboudiniere}}, \binits{J.P.}},
\bauthor{\bsnm{{Bothmer}}, \binits{V.}},
\bauthor{\bsnm{{Deutsch}}, \binits{W.}},
\bauthor{\bsnm{{Wang}}, \binits{D.}},
\bauthor{\bsnm{{Rich}}, \binits{N.}},
\bauthor{\bsnm{{Cooper}}, \binits{S.}},
\bauthor{\bsnm{{Stephens}}, \binits{V.}},
\bauthor{\bsnm{{Maahs}}, \binits{G.}},
\bauthor{\bsnm{{Baugh}}, \binits{R.}},
\bauthor{\bsnm{{McMullin}}, \binits{D.}},
\bauthor{\bsnm{{Carter}}, \binits{T.}}:
\byear{2008},
\batitle{{Sun Earth Connection Coronal and Heliospheric Investigation
  (SECCHI)}}.
\bjtitle{\ssr}
\bvolume{136},
\bfpage{67}.
\doiurl{https://doi.org/10.1007/s11214-008-9341-4}.
\adsurl{2008SSRv..136...67H}.
\end{barticle}
\endbibitem

\bibitem[\protect\citeauthoryear{{Hundhausen}, {Gilbert}, and
  {Bame}}{1968}]{hundhausen1968}
\begin{barticle}
\bauthor{\bsnm{{Hundhausen}}, \binits{A.J.}},
\bauthor{\bsnm{{Gilbert}}, \binits{H.E.}},
\bauthor{\bsnm{{Bame}}, \binits{S.J.}}:
\byear{1968},
\batitle{{Ionization state of the interplanetary plasma}}.
\bjtitle{\jgr}
\bvolume{73},
\bfpage{5485}.
\doiurl{https://doi.org/10.1029/JA073i017p05485}.
\adsurl{1968JGR....73.5485H}.
\end{barticle}
\endbibitem

\bibitem[\protect\citeauthoryear{{Imamura} et~al.}{2014}]{Imamura2014}
\begin{barticle}
\bauthor{\bsnm{{Imamura}}, \binits{T.}},
\bauthor{\bsnm{{Tokumaru}}, \binits{M.}},
\bauthor{\bsnm{{Isobe}}, \binits{H.}},
\bauthor{\bsnm{{Shiota}}, \binits{D.}},
\bauthor{\bsnm{{Ando}}, \binits{H.}},
\bauthor{\bsnm{{Miyamoto}}, \binits{M.}},
\bauthor{\bsnm{{Toda}}, \binits{T.}},
\bauthor{\bsnm{{H{\"a}usler}}, \binits{B.}},
\bauthor{\bsnm{{P{\"a}tzold}}, \binits{M.}},
\bauthor{\bsnm{{Nabatov}}, \binits{A.}},
\bauthor{\bsnm{{Asai}}, \binits{A.}},
\bauthor{\bsnm{{Yaji}}, \binits{K.}},
\bauthor{\bsnm{{Yamada}}, \binits{M.}},
\bauthor{\bsnm{{Nakamura}}, \binits{M.}}:
\byear{2014},
\batitle{{Outflow Structure of the Quiet Sun Corona Probed by Spacecraft Radio
  Scintillations in Strong Scattering}}.
\bjtitle{\apj}
\bvolume{788},
\bfpage{117}.
\doiurl{https://doi.org/10.1088/0004-637X/788/2/117}.
\adsurl{2014ApJ...788..117I}.
\end{barticle}
\endbibitem

\bibitem[\protect\citeauthoryear{{Inhester}}{2015}]{Inhester2015}
\begin{botherref}
\oauthor{\bsnm{{Inhester}}, \binits{B.}}:
2015,
{Thomson Scattering in the Solar Corona}.
\adsurl{2015arXiv151200651I}.
\arxivurl{1512.00651}.
\end{botherref}
\endbibitem

\bibitem[\protect\citeauthoryear{{Jones} et~al.}{2018}]{Jones2018}
\begin{barticle}
\bauthor{\bsnm{{Jones}}, \binits{G.H.}},
\bauthor{\bsnm{{Knight}}, \binits{M.M.}},
\bauthor{\bsnm{{Battams}}, \binits{K.}},
\bauthor{\bsnm{{Boice}}, \binits{D.C.}},
\bauthor{\bsnm{{Brown}}, \binits{J.}},
\bauthor{\bsnm{{Giordano}}, \binits{S.}},
\bauthor{\bsnm{{Raymond}}, \binits{J.}},
\bauthor{\bsnm{{Snodgrass}}, \binits{C.}},
\bauthor{\bsnm{{Steckloff}}, \binits{J.K.}},
\bauthor{\bsnm{{Weissman}}, \binits{P.}},
\bauthor{\bsnm{{Fitzsimmons}}, \binits{A.}},
\bauthor{\bsnm{{Lisse}}, \binits{C.}},
\bauthor{\bsnm{{Opitom}}, \binits{C.}},
\bauthor{\bsnm{{Birkett}}, \binits{K.S.}},
\bauthor{\bsnm{{Bzowski}}, \binits{M.}},
\bauthor{\bsnm{{Decock}}, \binits{A.}},
\bauthor{\bsnm{{Mann}}, \binits{I.}},
\bauthor{\bsnm{{Ramanjooloo}}, \binits{Y.}},
\bauthor{\bsnm{{McCauley}}, \binits{P.}}:
\byear{2018},
\batitle{{The Science of Sungrazers, Sunskirters, and Other Near-Sun Comets}}.
\bjtitle{\ssr}
\bvolume{214},
\bfpage{20}.
\doiurl{https://doi.org/10.1007/s11214-017-0446-5}.
\adsurl{2018SSRv..214...20J}.
\end{barticle}
\endbibitem

\bibitem[\protect\citeauthoryear{{J{\"o}nsson}
  et~al.}{2017}]{jonsson_etal:2017}
\begin{barticle}
\bauthor{\bsnm{{J{\"o}nsson}}, \binits{P.}},
\bauthor{\bsnm{{Gaigalas}}, \binits{G.}},
\bauthor{\bsnm{{Rynkun}}, \binits{P.}},
\bauthor{\bsnm{{Rad{\v z}i{\= u}t{\.e}}}, \binits{L.}},
\bauthor{\bsnm{{Ekman}}, \binits{J.}},
\bauthor{\bsnm{{Gustafsson}}, \binits{S.}},
\bauthor{\bsnm{{Hartman}}, \binits{H.}},
\bauthor{\bsnm{{Wang}}, \binits{K.}},
\bauthor{\bsnm{{Godefroid}}, \binits{M.}},
\bauthor{\bsnm{{Froese Fischer}}, \binits{C.}},
\bauthor{\bsnm{{Grant}}, \binits{I.}},
\bauthor{\bsnm{{Brage}}, \binits{T.}},
\bauthor{\bsnm{{Del Zanna}}, \binits{G.}}:
\byear{2017},
\batitle{{Multiconfiguration Dirac-Hartree-Fock Calculations with Spectroscopic
  Accuracy: Applications to Astrophysics}}.
\bjtitle{Atoms}
\bvolume{5},
\bfpage{16}.
\doiurl{https://doi.org/10.3390/atoms5020016}.
\adsurl{2017Atoms...5...16J}.
\end{barticle}
\endbibitem

\bibitem[\protect\citeauthoryear{{Karpen} et~al.}{2017}]{Karpen2017}
\begin{barticle}
\bauthor{\bsnm{{Karpen}}, \binits{J.T.}},
\bauthor{\bsnm{{DeVore}}, \binits{C.R.}},
\bauthor{\bsnm{{Antiochos}}, \binits{S.K.}},
\bauthor{\bsnm{{Pariat}}, \binits{E.}}:
\byear{2017},
\batitle{{Reconnection-Driven Coronal-Hole Jets with Gravity and Solar Wind}}.
\bjtitle{\apj}
\bvolume{834},
\bfpage{62}.
\doiurl{https://doi.org/10.3847/1538-4357/834/1/62}.
\adsurl{2017ApJ...834...62K}.
\end{barticle}
\endbibitem

\bibitem[\protect\citeauthoryear{Kliem and T{\"{o}}r{\"{o}}k}{2006}]{Kliem2006}
\begin{barticle}
\bauthor{\bsnm{Kliem}, \binits{B.}},
\bauthor{\bsnm{T{\"{o}}r{\"{o}}k}, \binits{T.}}:
\byear{2006},
\batitle{{Torus Instability}}.
\bjtitle{\prl}
\bvolume{96},
\bfpage{4}.
\doiurl{https://doi.org/10.1103/PhysRevLett.96.255002}.
\end{barticle}
\endbibitem

\bibitem[\protect\citeauthoryear{{Ko} et~al.}{1997}]{ko1997}
\begin{barticle}
\bauthor{\bsnm{{Ko}}, \binits{Y.-K.}},
\bauthor{\bsnm{{Fisk}}, \binits{L.A.}},
\bauthor{\bsnm{{Geiss}}, \binits{J.}},
\bauthor{\bsnm{{Gloeckler}}, \binits{G.}},
\bauthor{\bsnm{{Guhathakurta}}, \binits{M.}}:
\byear{1997},
\batitle{{An Empirical Study of the Electron Temperature and Heavy Ion
  Velocities in the South Polar Coronal Hole}}.
\bjtitle{\solphys}
\bvolume{171},
\bfpage{345}.
\adsurl{1997SoPh..171..345K}.
\end{barticle}
\endbibitem

\bibitem[\protect\citeauthoryear{Ko et~al.}{2016}]{Ko_etal._2016}
\begin{barticle}
\bauthor{\bsnm{Ko}, \binits{Y.-K.}},
\bauthor{\bsnm{Moses}, \binits{J.D.}},
\bauthor{\bsnm{Laming}, \binits{J.M.}},
\bauthor{\bsnm{Strachan}, \binits{L.}},
\bauthor{\bsnm{Tun~Beltran}, \binits{S.}},
\bauthor{\bsnm{Tomczyk}, \binits{S.}},
\bauthor{\bsnm{Gibson}, \binits{S.E.}},
\bauthor{\bsnm{Auchère}, \binits{F.}},
\bauthor{\bsnm{Casini}, \binits{R.}},
\bauthor{\bsnm{Fineschi}, \binits{S.}},
\bauthor{\bsnm{Knoelker}, \binits{M.}},
\bauthor{\bsnm{Korendyke}, \binits{C.}},
\bauthor{\bsnm{McIntosh}, \binits{S.W.}},
\bauthor{\bsnm{Romoli}, \binits{M.}},
\bauthor{\bsnm{Rybak}, \binits{J.}},
\bauthor{\bsnm{Socker}, \binits{D.G.}},
\bauthor{\bsnm{Vourlidas}, \binits{A.}},
\bauthor{\bsnm{Wu}, \binits{Q.}}:
\byear{2016},
\batitle{Waves and Magnetism in the Solar Atmosphere (WAMIS)}.
\bjtitle{Front. Astron. Space Sci.}
\bvolume{3}.
\doiurl{https://doi.org/10.3389/fspas.2016.00001}.
\end{barticle}
\endbibitem

\bibitem[\protect\citeauthoryear{{Kocher} et~al.}{2017}]{Kocher2017}
\begin{barticle}
\bauthor{\bsnm{{Kocher}}, \binits{M.}},
\bauthor{\bsnm{{Lepri}}, \binits{S.T.}},
\bauthor{\bsnm{{Landi}}, \binits{E.}},
\bauthor{\bsnm{{Zhao}}, \binits{L.}},
\bauthor{\bsnm{{Manchester}}, \binits{I.} \bsuffix{W.~B.}}:
\byear{2017},
\batitle{{Anatomy of Depleted Interplanetary Coronal Mass Ejections}}.
\bjtitle{\apj}
\bvolume{834},
\bfpage{147}.
\doiurl{https://doi.org/10.3847/1538-4357/834/2/147}.
\adsurl{2017ApJ...834..147K}.
\end{barticle}
\endbibitem

\bibitem[\protect\citeauthoryear{{Kohl} et~al.}{1995}]{Kohl1995}
\begin{barticle}
\bauthor{\bsnm{{Kohl}}, \binits{J.L.}},
\bauthor{\bsnm{{Esser}}, \binits{R.}},
\bauthor{\bsnm{{Gardner}}, \binits{L.D.}},
\bauthor{\bsnm{{Habbal}}, \binits{S.}},
\bauthor{\bsnm{{Daigneau}}, \binits{P.S.}},
\bauthor{\bsnm{{Dennis}}, \binits{E.F.}},
\bauthor{\bsnm{{Nystrom}}, \binits{G.U.}},
\bauthor{\bsnm{{Panasyuk}}, \binits{A.}},
\bauthor{\bsnm{{Raymond}}, \binits{J.C.}},
\bauthor{\bsnm{{Smith}}, \binits{P.L.}},
\bauthor{\bsnm{{Strachan}}, \binits{L.}},
\bauthor{\bsnm{{van Ballegooijen}}, \binits{A.A.}},
\bauthor{\bsnm{{Noci}}, \binits{G.}},
\bauthor{\bsnm{{Fineschi}}, \binits{S.}},
\bauthor{\bsnm{{Romoli}}, \binits{M.}},
\bauthor{\bsnm{{Ciaravella}}, \binits{A.}},
\bauthor{\bsnm{{Modigliani}}, \binits{A.}},
\bauthor{\bsnm{{Huber}}, \binits{M.C.E.}},
\bauthor{\bsnm{{Antonucci}}, \binits{E.}},
\bauthor{\bsnm{{Benna}}, \binits{C.}},
\bauthor{\bsnm{{Giordano}}, \binits{S.}},
\bauthor{\bsnm{{Tondello}}, \binits{G.}},
\bauthor{\bsnm{{Nicolosi}}, \binits{P.}},
\bauthor{\bsnm{{Naletto}}, \binits{G.}},
\bauthor{\bsnm{{Pernechele}}, \binits{C.}},
\bauthor{\bsnm{{Spadaro}}, \binits{D.}},
\bauthor{\bsnm{{Poletto}}, \binits{G.}},
\bauthor{\bsnm{{Livi}}, \binits{S.}},
\bauthor{\bsnm{{von der L{\"u}he}}, \binits{O.}},
\bauthor{\bsnm{{Geiss}}, \binits{J.}},
\bauthor{\bsnm{{Timothy}}, \binits{J.G.}},
\bauthor{\bsnm{{Gloeckler}}, \binits{G.}},
\bauthor{\bsnm{{Allegra}}, \binits{A.}},
\bauthor{\bsnm{{Basile}}, \binits{G.}},
\bauthor{\bsnm{{Brusa}}, \binits{R.}},
\bauthor{\bsnm{{Wood}}, \binits{B.}},
\bauthor{\bsnm{{Siegmund}}, \binits{O.H.W.}},
\bauthor{\bsnm{{Fowler}}, \binits{W.}},
\bauthor{\bsnm{{Fisher}}, \binits{R.}},
\bauthor{\bsnm{{Jhabvala}}, \binits{M.}}:
\byear{1995},
\batitle{{The Ultraviolet Coronagraph Spectrometer for the Solar and
  Heliospheric Observatory}}.
\bjtitle{\solphys}
\bvolume{162},
\bfpage{313}.
\doiurl{https://doi.org/10.1007/BF00733433}.
\adsurl{1995SoPh..162..313K}.
\end{barticle}
\endbibitem

\bibitem[\protect\citeauthoryear{{Kohl} et~al.}{2006}]{Kohl2006}
\begin{barticle}
\bauthor{\bsnm{{Kohl}}, \binits{J.L.}},
\bauthor{\bsnm{{Noci}}, \binits{G.}},
\bauthor{\bsnm{{Cranmer}}, \binits{S.R.}},
\bauthor{\bsnm{{Raymond}}, \binits{J.C.}}:
\byear{2006},
\batitle{{Ultraviolet spectroscopy of the extended solar corona}}.
\bjtitle{\aapr}
\bvolume{13},
\bfpage{31}.
\doiurl{https://doi.org/10.1007/s00159-005-0026-7}.
\adsurl{2006A&ARv..13...31K}.
\end{barticle}
\endbibitem

\bibitem[\protect\citeauthoryear{{Kong} et~al.}{2017}]{Kong2017}
\begin{barticle}
\bauthor{\bsnm{{Kong}}, \binits{X.}},
\bauthor{\bsnm{{Guo}}, \binits{F.}},
\bauthor{\bsnm{{Giacalone}}, \binits{J.}},
\bauthor{\bsnm{{Li}}, \binits{H.}},
\bauthor{\bsnm{{Chen}}, \binits{Y.}}:
\byear{2017},
\batitle{{The Acceleration of High-energy Protons at Coronal Shocks: The Effect
  of Large-scale Streamer-like Magnetic Field Structures}}.
\bjtitle{\apj}
\bvolume{851},
\bfpage{38}.
\doiurl{https://doi.org/10.3847/1538-4357/aa97d7}.
\adsurl{2017ApJ...851...38K}.
\end{barticle}
\endbibitem

\bibitem[\protect\citeauthoryear{{Kontar} et~al.}{2017}]{Kontar2017}
\begin{barticle}
\bauthor{\bsnm{{Kontar}}, \binits{E.P.}},
\bauthor{\bsnm{{Yu}}, \binits{S.}},
\bauthor{\bsnm{{Kuznetsov}}, \binits{A.A.}},
\bauthor{\bsnm{{Emslie}}, \binits{A.G.}},
\bauthor{\bsnm{{Alcock}}, \binits{B.}},
\bauthor{\bsnm{{Jeffrey}}, \binits{N.L.S.}},
\bauthor{\bsnm{{Melnik}}, \binits{V.N.}},
\bauthor{\bsnm{{Bian}}, \binits{N.H.}},
\bauthor{\bsnm{{Subramanian}}, \binits{P.}}:
\byear{2017},
\batitle{{Imaging spectroscopy of solar radio burst fine structures}}.
\bjtitle{Nat. Comm.}
\bvolume{8},
\bfpage{1515}.
\doiurl{https://doi.org/10.1038/s41467-017-01307-8}.
\adsurl{2017NatCo...8.1515K}.
\end{barticle}
\endbibitem

\bibitem[\protect\citeauthoryear{{Kooi} et~al.}{2014}]{Kooi2014}
\begin{barticle}
\bauthor{\bsnm{{Kooi}}, \binits{J.E.}},
\bauthor{\bsnm{{Fischer}}, \binits{P.D.}},
\bauthor{\bsnm{{Buffo}}, \binits{J.J.}},
\bauthor{\bsnm{{Spangler}}, \binits{S.R.}}:
\byear{2014},
\batitle{{Measurements of Coronal Faraday Rotation at 4.6 R
  $_{{\ensuremath{\odot}}}$}}.
\bjtitle{\apj}
\bvolume{784},
\bfpage{68}.
\doiurl{https://doi.org/10.1088/0004-637X/784/1/68}.
\adsurl{2014ApJ...784...68K}.
\end{barticle}
\endbibitem

\bibitem[\protect\citeauthoryear{{Kooi} et~al.}{2022}]{Kooi2022}
\begin{barticle}
\bauthor{\bsnm{{Kooi}}, \binits{J.E.}},
\bauthor{\bsnm{{Wexler}}, \binits{D.B.}},
\bauthor{\bsnm{{Jensen}}, \binits{E.A.}},
\bauthor{\bsnm{{Kenny}}, \binits{M.N.}},
\bauthor{\bsnm{{Nieves-Chinchilla}}, \binits{T.}},
\bauthor{\bsnm{{Wilson}}, \binits{I.} \bsuffix{Lynn~B.}},
\bauthor{\bsnm{{Wood}}, \binits{B.E.}},
\bauthor{\bsnm{{Jian}}, \binits{L.K.}},
\bauthor{\bsnm{{Fung}}, \binits{S.F.}},
\bauthor{\bsnm{{Pevtsov}}, \binits{A.}},
\bauthor{\bsnm{{Gopalswamy}}, \binits{N.}},
\bauthor{\bsnm{{Manchester}}, \binits{W.B.}}:
\byear{2022},
\batitle{{Modern Faraday Rotation Studies to Probe the Solar Wind}}.
\bjtitle{Front. Astron. Space Sci.}
\bvolume{9},
\bfpage{841866}.
\doiurl{https://doi.org/10.3389/fspas.2022.841866}.
\adsurl{2022FrASS...941866K}.
\end{barticle}
\endbibitem

\bibitem[\protect\citeauthoryear{{Koutchmy}}{2004}]{Koutchmy2004}
\begin{bchapter}
\bauthor{\bsnm{{Koutchmy}}, \binits{S.}}:
\byear{2004},
\bctitle{{Structure and Dynamics of the Solar Corona}}.
In: \beditor{\bsnm{{Stepanov}}, \binits{A.V.}},
\beditor{\bsnm{{Benevolenskaya}}, \binits{E.E.}},
\beditor{\bsnm{{Kosovichev}}, \binits{A.G.}} (eds.)
\bbtitle{Multi-Wavelength Investigations of Solar Activity, IAU Symp.}
\bseriesno{223},
\bpublisher{Cambridge Univ. Press},
\blocation{Cambridge, UK},
\bfpage{509}.
\doiurl{https://doi.org/10.1017/S1743921304006702}.
\adsurl{2004IAUS..223..509K}.
\end{bchapter}
\endbibitem

\bibitem[\protect\citeauthoryear{{Koutchmy} and
  {Livshits}}{1992}]{Koutchmy1992}
\begin{barticle}
\bauthor{\bsnm{{Koutchmy}}, \binits{S.}},
\bauthor{\bsnm{{Livshits}}, \binits{M.}}:
\byear{1992},
\batitle{{Coronal Streamers}}.
\bjtitle{\ssr}
\bvolume{61},
\bfpage{393}.
\doiurl{https://doi.org/10.1007/BF00222313}.
\adsurl{1992SSRv...61..393K}.
\end{barticle}
\endbibitem

\bibitem[\protect\citeauthoryear{{Koutchmy} and
  {Molodensky}}{1994}]{Koutchmy1994}
\begin{bchapter}
\bauthor{\bsnm{{Koutchmy}}, \binits{S.}},
\bauthor{\bsnm{{Molodensky}}, \binits{M.}}:
\byear{1994},
\bctitle{{Magnetic structures of the intermediate corona}}.
In: \beditor{\bsnm{{Belvedere}}, \binits{G.}},
\beditor{\bsnm{{Rodono}}, \binits{M.}},
\beditor{\bsnm{{Simnett}}, \binits{G.M.}} (eds.)
\bbtitle{Advances in Solar Physics}
\bseriesno{432},
\bfpage{167}.
\doiurl{https://doi.org/10.1007/3-540-58041-7_214}.
\adsurl{1994LNP...432..167K}.
\end{bchapter}
\endbibitem

\bibitem[\protect\citeauthoryear{{Koutchmy} and
  {Nikoghossian}}{2002}]{Koutchmy2002}
\begin{barticle}
\bauthor{\bsnm{{Koutchmy}}, \binits{S.}},
\bauthor{\bsnm{{Nikoghossian}}, \binits{A.G.}}:
\byear{2002},
\batitle{{Coronal linear threads: W-L radiation of supra-thermal streams}}.
\bjtitle{\aap}
\bvolume{395},
\bfpage{983}.
\doiurl{https://doi.org/10.1051/0004-6361:20021269}.
\adsurl{2002A&A...395..983K}.
\end{barticle}
\endbibitem

\bibitem[\protect\citeauthoryear{{Kumar} et~al.}{2019}]{Kumar2019}
\begin{barticle}
\bauthor{\bsnm{{Kumar}}, \binits{P.}},
\bauthor{\bsnm{{Karpen}}, \binits{J.T.}},
\bauthor{\bsnm{{Antiochos}}, \binits{S.K.}},
\bauthor{\bsnm{{Wyper}}, \binits{P.F.}},
\bauthor{\bsnm{{DeVore}}, \binits{C.R.}},
\bauthor{\bsnm{{DeForest}}, \binits{C.E.}}:
\byear{2019},
\batitle{{Multiwavelength Study of Equatorial Coronal-hole Jets}}.
\bjtitle{\apj}
\bvolume{873},
\bfpage{93}.
\doiurl{https://doi.org/10.3847/1538-4357/ab04af}.
\adsurl{2019ApJ...873...93K}.
\end{barticle}
\endbibitem

\bibitem[\protect\citeauthoryear{{Kumar} et~al.}{2021}]{Kumar2021}
\begin{barticle}
\bauthor{\bsnm{{Kumar}}, \binits{P.}},
\bauthor{\bsnm{{Karpen}}, \binits{J.T.}},
\bauthor{\bsnm{{Antiochos}}, \binits{S.K.}},
\bauthor{\bsnm{{Wyper}}, \binits{P.F.}},
\bauthor{\bsnm{{DeVore}}, \binits{C.R.}},
\bauthor{\bsnm{{Lynch}}, \binits{B.J.}}:
\byear{2021},
\batitle{{From Pseudostreamer Jets to Coronal Mass Ejections: Observations of
  the Breakout Continuum}}.
\bjtitle{\apj}
\bvolume{907},
\bfpage{41}.
\doiurl{https://doi.org/10.3847/1538-4357/abca8b}.
\adsurl{2021ApJ...907...41K}.
\end{barticle}
\endbibitem

\bibitem[\protect\citeauthoryear{{Kumar} et~al.}{2022}]{Kumar2022}
\begin{barticle}
\bauthor{\bsnm{{Kumar}}, \binits{P.}},
\bauthor{\bsnm{{Karpen}}, \binits{J.T.}},
\bauthor{\bsnm{{Uritsky}}, \binits{V.M.}},
\bauthor{\bsnm{{Deforest}}, \binits{C.E.}},
\bauthor{\bsnm{{Raouafi}}, \binits{N.E.}},
\bauthor{\bsnm{{Richard DeVore}}, \binits{C.}}:
\byear{2022},
\batitle{{Quasi-periodic Energy Release and Jets at the Base of Solar Coronal
  Plumes}}.
\bjtitle{\apj}
\bvolume{933},
\bfpage{21}.
\doiurl{https://doi.org/10.3847/1538-4357/ac6c24}.
\adsurl{2022ApJ...933...21K}.
\end{barticle}
\endbibitem

\bibitem[\protect\citeauthoryear{{Kuzin} et~al.}{2009}]{Kuzin2009}
\begin{barticle}
\bauthor{\bsnm{{Kuzin}}, \binits{S.V.}},
\bauthor{\bsnm{{Bogachev}}, \binits{S.A.}},
\bauthor{\bsnm{{Zhitnik}}, \binits{I.A.}},
\bauthor{\bsnm{{Pertsov}}, \binits{A.A.}},
\bauthor{\bsnm{{Ignatiev}}, \binits{A.P.}},
\bauthor{\bsnm{{Mitrofanov}}, \binits{A.M.}},
\bauthor{\bsnm{{Slemzin}}, \binits{V.A.}},
\bauthor{\bsnm{{Shestov}}, \binits{S.V.}},
\bauthor{\bsnm{{Sukhodrev}}, \binits{N.K.}},
\bauthor{\bsnm{{Bugaenko}}, \binits{O.I.}}:
\byear{2009},
\batitle{{TESIS experiment on EUV imaging spectroscopy of the Sun}}.
\bjtitle{Adv. Space Res.}
\bvolume{43},
\bfpage{1001}.
\doiurl{https://doi.org/10.1016/j.asr.2008.10.021}.
\adsurl{2009AdSpR..43.1001K}.
\end{barticle}
\endbibitem

\bibitem[\protect\citeauthoryear{Lahoz and Schneider}{2014}]{Lahoz2014}
\begin{barticle}
\bauthor{\bsnm{Lahoz}, \binits{W.A.}},
\bauthor{\bsnm{Schneider}, \binits{P.}}:
\byear{2014},
\batitle{Data assimilation: making sense of Earth Observation}.
\bjtitle{Front. Environ. Sci.}
\bvolume{2}.
\doiurl{https://doi.org/10.3389/fenvs.2014.00016}.
\end{barticle}
\endbibitem

\bibitem[\protect\citeauthoryear{{Laming} and
  {Vourlidas}}{2019}]{2019AGUFMSH31B..15L}
\begin{bchapter}
\bauthor{\bsnm{{Laming}}, \binits{J.M.}},
\bauthor{\bsnm{{Vourlidas}}, \binits{A.}}:
\byear{2019},
\bctitle{{LOCKYER: Large Optimized Coronagraphs for KeY Emission line
  Research}}.
In: \bbtitle{AGU Fall Meet. Abs.}
\bseriesno{2019},
\bfpage{SH31B}.
\adsurl{2019AGUFMSH31B..15L}.
\end{bchapter}
\endbibitem

\bibitem[\protect\citeauthoryear{Laming et~al.}{2019}]{Laming_etal._2019}
\begin{barticle}
\bauthor{\bsnm{Laming}, \binits{J.M.}},
\bauthor{\bsnm{Vourlidas}, \binits{A.}},
\bauthor{\bsnm{Korendyke}, \binits{C.}},
\bauthor{\bsnm{Chua}, \binits{D.}},
\bauthor{\bsnm{Cranmer}, \binits{S.R.}},
\bauthor{\bsnm{Ko}, \binits{Y.-K.}},
\bauthor{\bsnm{Kuroda}, \binits{N.}},
\bauthor{\bsnm{Provornikova}, \binits{E.}},
\bauthor{\bsnm{Raymond}, \binits{J.C.}},
\bauthor{\bsnm{Raouafi}, \binits{N.-E.}},
\bauthor{\bsnm{Strachan}, \binits{L.}},
\bauthor{\bsnm{Tun-Beltran}, \binits{S.}},
\bauthor{\bsnm{Weberg}, \binits{M.}},
\bauthor{\bsnm{Wood}, \binits{B.E.}}:
\byear{2019},
\batitle{Element Abundances: A New Diagnostic for the Solar Wind}.
\bjtitle{\apj}
\bvolume{879},
\bfpage{124}.
\doiurl{https://doi.org/10.3847/1538-4357/ab23f1}.
\end{barticle}
\endbibitem

\bibitem[\protect\citeauthoryear{{Lamy} et~al.}{2010}]{Lamy2010}
\begin{bchapter}
\bauthor{\bsnm{{Lamy}}, \binits{P.}},
\bauthor{\bsnm{{Dam{\'e}}}, \binits{L.}},
\bauthor{\bsnm{{Viv{\`e}s}}, \binits{S.}},
\bauthor{\bsnm{{Zhukov}}, \binits{A.}}:
\byear{2010},
\bctitle{{ASPIICS: a giant coronagraph for the ESA/PROBA-3 Formation Flying
  Mission}}.
In: \beditor{\bsnm{{Oschmann}}, \binits{J.} \bsuffix{Jacobus~M.}},
\beditor{\bsnm{{Clampin}}, \binits{M.C.}},
\beditor{\bsnm{{MacEwen}}, \binits{H.A.}} (eds.)
\bbtitle{Space Telescopes and Instrumentation 2010: Optical, Infrared, and
  Millimeter Wave},
\bsertitle{Society of Photo-Optical Instrumentation Engineers (SPIE) Conference
  Series}
\bseriesno{7731},
\bfpage{773118}.
\doiurl{https://doi.org/10.1117/12.858247}.
\adsurl{2010SPIE.7731E..18L}.
\end{bchapter}
\endbibitem

\bibitem[\protect\citeauthoryear{{Landi} and {Lepri}}{2015}]{Landi2015}
\begin{barticle}
\bauthor{\bsnm{{Landi}}, \binits{E.}},
\bauthor{\bsnm{{Lepri}}, \binits{S.T.}}:
\byear{2015},
\batitle{{Photoionization in the Solar Wind}}.
\bjtitle{\apjl}
\bvolume{812},
\bfpage{L28}.
\doiurl{https://doi.org/10.1088/2041-8205/812/2/L28}.
\adsurl{2015ApJ...812L..28L}.
\end{barticle}
\endbibitem

\bibitem[\protect\citeauthoryear{{Landi}, {Habbal}, and
  {Tomczyk}}{2016}]{Landi2016}
\begin{barticle}
\bauthor{\bsnm{{Landi}}, \binits{E.}},
\bauthor{\bsnm{{Habbal}}, \binits{S.R.}},
\bauthor{\bsnm{{Tomczyk}}, \binits{S.}}:
\byear{2016},
\batitle{{Coronal plasma diagnostics from ground-based observations}}.
\bjtitle{J. Geophys. Res. (Space Phys.)}
\bvolume{121},
\bfpage{8237}.
\doiurl{https://doi.org/10.1002/2016JA022598}.
\adsurl{2016JGRA..121.8237L}.
\end{barticle}
\endbibitem

\bibitem[\protect\citeauthoryear{{Landi} et~al.}{2012}]{landi2012}
\begin{barticle}
\bauthor{\bsnm{{Landi}}, \binits{E.}},
\bauthor{\bsnm{{Gruesbeck}}, \binits{J.R.}},
\bauthor{\bsnm{{Lepri}}, \binits{S.T.}},
\bauthor{\bsnm{{Zurbuchen}}, \binits{T.H.}}:
\byear{2012},
\batitle{{New Solar Wind Diagnostic Using Both in Situ and Spectroscopic
  Measurements}}.
\bjtitle{\apj}
\bvolume{750},
\bfpage{159}.
\doiurl{https://doi.org/10.1088/0004-637X/750/2/159}.
\adsurl{2012ApJ...750..159L}.
\end{barticle}
\endbibitem

\bibitem[\protect\citeauthoryear{{Lee} et~al.}{2011}]{Lee2011}
\begin{barticle}
\bauthor{\bsnm{{Lee}}, \binits{C.O.}},
\bauthor{\bsnm{{Luhmann}}, \binits{J.G.}},
\bauthor{\bsnm{{Hoeksema}}, \binits{J.T.}},
\bauthor{\bsnm{{Sun}}, \binits{X.}},
\bauthor{\bsnm{{Arge}}, \binits{C.N.}},
\bauthor{\bsnm{{de Pater}}, \binits{I.}}:
\byear{2011},
\batitle{{Coronal Field Opens at Lower Height During the Solar Cycles 22 and 23
  Minimum Periods: IMF Comparison Suggests the Source Surface Should Be
  Lowered}}.
\bjtitle{\solphys}
\bvolume{269},
\bfpage{367}.
\doiurl{https://doi.org/10.1007/s11207-010-9699-9}.
\adsurl{2011SoPh..269..367L}.
\end{barticle}
\endbibitem

\bibitem[\protect\citeauthoryear{{Leer} and {Holzer}}{1980}]{Leer1980}
\begin{barticle}
\bauthor{\bsnm{{Leer}}, \binits{E.}},
\bauthor{\bsnm{{Holzer}}, \binits{T.E.}}:
\byear{1980},
\batitle{{Energy addition in the solar wind.}}
\bjtitle{\jgr}
\bvolume{85},
\bfpage{4681}.
\doiurl{https://doi.org/10.1029/JA085iA09p04681}.
\adsurl{1980JGR....85.4681L}.
\end{barticle}
\endbibitem

\bibitem[\protect\citeauthoryear{{Lenz}, {Lou}, and {Rosner}}{1998}]{lenz1998}
\begin{barticle}
\bauthor{\bsnm{{Lenz}}, \binits{D.D.}},
\bauthor{\bsnm{{Lou}}, \binits{Y.-Q.}},
\bauthor{\bsnm{{Rosner}}, \binits{R.}}:
\byear{1998},
\batitle{{Density Structure in a Multicomponent Coronal Loop}}.
\bjtitle{\apj}
\bvolume{504},
\bfpage{1020}.
\doiurl{https://doi.org/10.1086/306111}.
\adsurl{1998ApJ...504.1020L}.
\end{barticle}
\endbibitem

\bibitem[\protect\citeauthoryear{{Linker} et~al.}{2017}]{Linker2017}
\begin{barticle}
\bauthor{\bsnm{{Linker}}, \binits{J.A.}},
\bauthor{\bsnm{{Caplan}}, \binits{R.M.}},
\bauthor{\bsnm{{Downs}}, \binits{C.}},
\bauthor{\bsnm{{Riley}}, \binits{P.}},
\bauthor{\bsnm{{Mikic}}, \binits{Z.}},
\bauthor{\bsnm{{Lionello}}, \binits{R.}},
\bauthor{\bsnm{{Henney}}, \binits{C.J.}},
\bauthor{\bsnm{{Arge}}, \binits{C.N.}},
\bauthor{\bsnm{{Liu}}, \binits{Y.}},
\bauthor{\bsnm{{Derosa}}, \binits{M.L.}},
\bauthor{\bsnm{{Yeates}}, \binits{A.}},
\bauthor{\bsnm{{Owens}}, \binits{M.J.}}:
\byear{2017},
\batitle{{The Open Flux Problem}}.
\bjtitle{\apj}
\bvolume{848},
\bfpage{70}.
\doiurl{https://doi.org/10.3847/1538-4357/aa8a70}.
\adsurl{2017ApJ...848...70L}.
\end{barticle}
\endbibitem

\bibitem[\protect\citeauthoryear{{Lynch}}{2020}]{Lynch2020}
\begin{barticle}
\bauthor{\bsnm{{Lynch}}, \binits{B.J.}}:
\byear{2020},
\batitle{{A Model for Coronal Inflows and In/Out Pairs}}.
\bjtitle{\apj}
\bvolume{905},
\bfpage{139}.
\doiurl{https://doi.org/10.3847/1538-4357/abc5b3}.
\adsurl{2020ApJ...905..139L}.
\end{barticle}
\endbibitem

\bibitem[\protect\citeauthoryear{{Lynch} and {Edmondson}}{2013}]{Lynch2013}
\begin{barticle}
\bauthor{\bsnm{{Lynch}}, \binits{B.J.}},
\bauthor{\bsnm{{Edmondson}}, \binits{J.K.}}:
\byear{2013},
\batitle{{Sympathetic Magnetic Breakout Coronal Mass Ejections from
  Pseudostreamers}}.
\bjtitle{\apj}
\bvolume{764},
\bfpage{87}.
\doiurl{https://doi.org/10.1088/0004-637X/764/1/87}.
\adsurl{2013ApJ...764...87L}.
\end{barticle}
\endbibitem

\bibitem[\protect\citeauthoryear{{Lynch} et~al.}{2010}]{Lynch2010}
\begin{barticle}
\bauthor{\bsnm{{Lynch}}, \binits{B.J.}},
\bauthor{\bsnm{{Li}}, \binits{Y.}},
\bauthor{\bsnm{{Thernisien}}, \binits{A.F.R.}},
\bauthor{\bsnm{{Robbrecht}}, \binits{E.}},
\bauthor{\bsnm{{Fisher}}, \binits{G.H.}},
\bauthor{\bsnm{{Luhmann}}, \binits{J.G.}},
\bauthor{\bsnm{{Vourlidas}}, \binits{A.}}:
\byear{2010},
\batitle{{Sun to 1 AU propagation and evolution of a slow streamer-blowout
  coronal mass ejection}}.
\bjtitle{J. Geophys. Res. (Space Phys.)}
\bvolume{115},
\bfpage{A07106}.
\doiurl{https://doi.org/10.1029/2009JA015099}.
\adsurl{2010JGRA..115.7106L}.
\end{barticle}
\endbibitem

\bibitem[\protect\citeauthoryear{{Lyot}}{1939}]{Lyot1939}
\begin{barticle}
\bauthor{\bsnm{{Lyot}}, \binits{B.}}:
\byear{1939},
\batitle{{The study of the solar corona and prominences without eclipses
  (George Darwin Lecture, 1939)}}.
\bjtitle{\mnras}
\bvolume{99},
\bfpage{580}.
\doiurl{https://doi.org/10.1093/mnras/99.8.580}.
\adsurl{1939MNRAS..99..580L}.
\end{barticle}
\endbibitem

\bibitem[\protect\citeauthoryear{Mackay and van
  Ballegooijen}{2006}]{Mackay2006}
\begin{barticle}
\bauthor{\bsnm{Mackay}, \binits{D.H.}},
\bauthor{\bparticle{van} \bsnm{Ballegooijen}, \binits{A.A.}}:
\byear{2006},
\batitle{Models of the Large-Scale Corona. I. Formation, Evolution, and Liftoff
  of Magnetic Flux Ropes}.
\bjtitle{\apj}
\bvolume{641},
\bfpage{577}.
\doiurl{https://doi.org/10.1086/500425}.
\burl{https://doi.org/10.1086/500425}.
\end{barticle}
\endbibitem

\bibitem[\protect\citeauthoryear{{MacQueen} et~al.}{1974}]{MacQueen1974}
\begin{barticle}
\bauthor{\bsnm{{MacQueen}}, \binits{R.M.}},
\bauthor{\bsnm{{Eddy}}, \binits{J.A.}},
\bauthor{\bsnm{{Gosling}}, \binits{J.T.}},
\bauthor{\bsnm{{Hildner}}, \binits{E.}},
\bauthor{\bsnm{{Munro}}, \binits{R.H.}},
\bauthor{\bsnm{{Newkirk}}, \binits{G.A.} \bsuffix{Jr.}},
\bauthor{\bsnm{{Poland}}, \binits{A.I.}},
\bauthor{\bsnm{{Ross}}, \binits{C.L.}}:
\byear{1974},
\batitle{{The Outer Solar Corona as Observed from Skylab: Preliminary
  Results}}.
\bjtitle{\apjl}
\bvolume{187},
\bfpage{L85}.
\doiurl{https://doi.org/10.1086/181402}.
\adsurl{1974ApJ...187L..85M}.
\end{barticle}
\endbibitem

\bibitem[\protect\citeauthoryear{{MacQueen} et~al.}{1980}]{MacQueen1980}
\begin{barticle}
\bauthor{\bsnm{{MacQueen}}, \binits{R.M.}},
\bauthor{\bsnm{{Csoeke-Poeckh}}, \binits{A.}},
\bauthor{\bsnm{{Hildner}}, \binits{E.}},
\bauthor{\bsnm{{House}}, \binits{L.}},
\bauthor{\bsnm{{Reynolds}}, \binits{R.}},
\bauthor{\bsnm{{Stanger}}, \binits{A.}},
\bauthor{\bsnm{{Tepoel}}, \binits{H.}},
\bauthor{\bsnm{{Wagner}}, \binits{W.}}:
\byear{1980},
\batitle{{The High Altitude Observatory coronagraph/polarimeter on the Solar
  Maximum Mission.}}
\bjtitle{\solphys}
\bvolume{65},
\bfpage{91}.
\doiurl{https://doi.org/10.1007/BF00151386}.
\adsurl{1980SoPh...65...91M}.
\end{barticle}
\endbibitem

\bibitem[\protect\citeauthoryear{{Madsen} et~al.}{2019}]{madsen2019}
\begin{barticle}
\bauthor{\bsnm{{Madsen}}, \binits{C.A.}},
\bauthor{\bsnm{{Samra}}, \binits{J.E.}},
\bauthor{\bsnm{{Del Zanna}}, \binits{G.}},
\bauthor{\bsnm{{DeLuca}}, \binits{E.E.}}:
\byear{2019},
\batitle{{Coronal Plasma Characterization via Coordinated Infrared and Extreme
  Ultraviolet Observations of a Total Solar Eclipse}}.
\bjtitle{\apj}
\bvolume{880},
\bfpage{102}.
\doiurl{https://doi.org/10.3847/1538-4357/ab2b3c}.
\end{barticle}
\endbibitem

\bibitem[\protect\citeauthoryear{{Magyar} and {Van
  Doorsselaere}}{2022}]{Magyar2022}
\begin{barticle}
\bauthor{\bsnm{{Magyar}}, \binits{N.}},
\bauthor{\bsnm{{Van Doorsselaere}}, \binits{T.}}:
\byear{2022},
\batitle{{Phase Mixing and the 1/f Spectrum in the Solar Wind}}.
\bjtitle{\apj}
\bvolume{938},
\bfpage{98}.
\doiurl{https://doi.org/10.3847/1538-4357/ac8b81}.
\adsurl{2022ApJ...938...98M}.
\end{barticle}
\endbibitem

\bibitem[\protect\citeauthoryear{{Mahrous} et~al.}{2018}]{Mahrous2018}
\begin{barticle}
\bauthor{\bsnm{{Mahrous}}, \binits{A.}},
\bauthor{\bsnm{{Alielden}}, \binits{K.}},
\bauthor{\bsnm{{Vr{\v{s}}nak}}, \binits{B.}},
\bauthor{\bsnm{{Youssef}}, \binits{M.}}:
\byear{2018},
\batitle{{Type II solar radio burst band-splitting: Measure of coronal magnetic
  field strength}}.
\bjtitle{J. Atmos. Solar-Terr. Phys.}
\bvolume{172},
\bfpage{75}.
\doiurl{https://doi.org/10.1016/j.jastp.2018.03.018}.
\adsurl{2018JASTP.172...75M}.
\end{barticle}
\endbibitem

\bibitem[\protect\citeauthoryear{{Maia} et~al.}{2007}]{Maia2007}
\begin{barticle}
\bauthor{\bsnm{{Maia}}, \binits{D.J.F.}},
\bauthor{\bsnm{{Gama}}, \binits{R.}},
\bauthor{\bsnm{{Mercier}}, \binits{C.}},
\bauthor{\bsnm{{Pick}}, \binits{M.}},
\bauthor{\bsnm{{Kerdraon}}, \binits{A.}},
\bauthor{\bsnm{{Karlick{\'y}}}, \binits{M.}}:
\byear{2007},
\batitle{{The Radio-Coronal Mass Ejection Event on 2001 April 15}}.
\bjtitle{\apj}
\bvolume{660},
\bfpage{874}.
\doiurl{https://doi.org/10.1086/508011}.
\adsurl{2007ApJ...660..874M}.
\end{barticle}
\endbibitem

\bibitem[\protect\citeauthoryear{{Majumdar}, {Patel}, and
  {Pant}}{2022}]{Majumdar2022}
\begin{barticle}
\bauthor{\bsnm{{Majumdar}}, \binits{S.}},
\bauthor{\bsnm{{Patel}}, \binits{R.}},
\bauthor{\bsnm{{Pant}}, \binits{V.}}:
\byear{2022},
\batitle{{On the Variation in the Volumetric Evolution of CMEs from the Inner
  to Outer Corona}}.
\bjtitle{\apj}
\bvolume{929},
\bfpage{11}.
\doiurl{https://doi.org/10.3847/1538-4357/ac5909}.
\adsurl{2022ApJ...929...11M}.
\end{barticle}
\endbibitem

\bibitem[\protect\citeauthoryear{{Majumdar} et~al.}{2020}]{Majumdar2020ApJM}
\begin{barticle}
\bauthor{\bsnm{{Majumdar}}, \binits{S.}},
\bauthor{\bsnm{{Pant}}, \binits{V.}},
\bauthor{\bsnm{{Patel}}, \binits{R.}},
\bauthor{\bsnm{{Banerjee}}, \binits{D.}}:
\byear{2020},
\batitle{{Connecting 3D Evolution of Coronal Mass Ejections to Their Source
  Regions}}.
\bjtitle{\apj}
\bvolume{899},
\bfpage{6}.
\doiurl{https://doi.org/10.3847/1538-4357/aba1f2}.
\adsurl{2020ApJ...899....6M}.
\end{barticle}
\endbibitem

\bibitem[\protect\citeauthoryear{Malanushenko et~al.}{2022}]{Malanushenko2022}
\begin{barticle}
\bauthor{\bsnm{Malanushenko}, \binits{A.}},
\bauthor{\bsnm{Cheung}, \binits{M.C.M.}},
\bauthor{\bsnm{DeForest}, \binits{C.E.}},
\bauthor{\bsnm{Klimchuk}, \binits{J.A.}},
\bauthor{\bsnm{Rempel}, \binits{M.}}:
\byear{2022},
\batitle{{The Coronal Veil}}.
\bjtitle{\apj}
\bvolume{927},
\bfpage{1}.
\doiurl{https://doi.org/10.3847/1538-4357/ac3df9}.
\end{barticle}
\endbibitem

\bibitem[\protect\citeauthoryear{{Mann} et~al.}{2018}]{Mann2018}
\begin{barticle}
\bauthor{\bsnm{{Mann}}, \binits{G.}},
\bauthor{\bsnm{{Breitling}}, \binits{F.}},
\bauthor{\bsnm{{Vocks}}, \binits{C.}},
\bauthor{\bsnm{{Aurass}}, \binits{H.}},
\bauthor{\bsnm{{Steinmetz}}, \binits{M.}},
\bauthor{\bsnm{{Strassmeier}}, \binits{K.G.}},
\bauthor{\bsnm{{Bisi}}, \binits{M.M.}},
\bauthor{\bsnm{{Fallows}}, \binits{R.A.}},
\bauthor{\bsnm{{Gallagher}}, \binits{P.}},
\bauthor{\bsnm{{Kerdraon}}, \binits{A.}},
\bauthor{\bsnm{{Mackinnon}}, \binits{A.}},
\bauthor{\bsnm{{Magdalenic}}, \binits{J.}},
\bauthor{\bsnm{{Rucker}}, \binits{H.}},
\bauthor{\bsnm{{Anderson}}, \binits{J.}},
\bauthor{\bsnm{{Asgekar}}, \binits{A.}},
\bauthor{\bsnm{{Avruch}}, \binits{I.M.}},
\bauthor{\bsnm{{Bell}}, \binits{M.E.}},
\bauthor{\bsnm{{Bentum}}, \binits{M.J.}},
\bauthor{\bsnm{{Bernardi}}, \binits{G.}},
\bauthor{\bsnm{{Best}}, \binits{P.}},
\bauthor{\bsnm{{B{\^\i}rzan}}, \binits{L.}},
\bauthor{\bsnm{{Bonafede}}, \binits{A.}},
\bauthor{\bsnm{{Broderick}}, \binits{J.W.}},
\bauthor{\bsnm{{Br{\"u}ggen}}, \binits{M.}},
\bauthor{\bsnm{{Butcher}}, \binits{H.R.}},
\bauthor{\bsnm{{Ciardi}}, \binits{B.}},
\bauthor{\bsnm{{Corstanje}}, \binits{A.}},
\bauthor{\bsnm{{de Gasperin}}, \binits{F.}},
\bauthor{\bsnm{{de Geus}}, \binits{E.}},
\bauthor{\bsnm{{Deller}}, \binits{A.}},
\bauthor{\bsnm{{Duscha}}, \binits{S.}},
\bauthor{\bsnm{{Eisl{\"o}ffel}}, \binits{J.}},
\bauthor{\bsnm{{Engels}}, \binits{D.}},
\bauthor{\bsnm{{Falcke}}, \binits{H.}},
\bauthor{\bsnm{{Fender}}, \binits{R.}},
\bauthor{\bsnm{{Ferrari}}, \binits{C.}},
\bauthor{\bsnm{{Frieswijk}}, \binits{W.}},
\bauthor{\bsnm{{Garrett}}, \binits{M.A.}},
\bauthor{\bsnm{{Grie{\ss}meier}}, \binits{J.}},
\bauthor{\bsnm{{Gunst}}, \binits{A.W.}},
\bauthor{\bsnm{{van Haarlem}}, \binits{M.}},
\bauthor{\bsnm{{Hassall}}, \binits{T.E.}},
\bauthor{\bsnm{{Heald}}, \binits{G.}},
\bauthor{\bsnm{{Hessels}}, \binits{J.W.T.}},
\bauthor{\bsnm{{Hoeft}}, \binits{M.}},
\bauthor{\bsnm{{H{\"o}randel}}, \binits{J.}},
\bauthor{\bsnm{{Horneffer}}, \binits{A.}},
\bauthor{\bsnm{{Juette}}, \binits{E.}},
\bauthor{\bsnm{{Karastergiou}}, \binits{A.}},
\bauthor{\bsnm{{Klijn}}, \binits{W.F.A.}},
\bauthor{\bsnm{{Kondratiev}}, \binits{V.I.}},
\bauthor{\bsnm{{Kramer}}, \binits{M.}},
\bauthor{\bsnm{{Kuniyoshi}}, \binits{M.}},
\bauthor{\bsnm{{Kuper}}, \binits{G.}},
\bauthor{\bsnm{{Maat}}, \binits{P.}},
\bauthor{\bsnm{{Markoff}}, \binits{S.}},
\bauthor{\bsnm{{McFadden}}, \binits{R.}},
\bauthor{\bsnm{{McKay-Bukowski}}, \binits{D.}},
\bauthor{\bsnm{{McKean}}, \binits{J.P.}},
\bauthor{\bsnm{{Mulcahy}}, \binits{D.D.}},
\bauthor{\bsnm{{Munk}}, \binits{H.}},
\bauthor{\bsnm{{Nelles}}, \binits{A.}},
\bauthor{\bsnm{{Norden}}, \binits{M.J.}},
\bauthor{\bsnm{{Orru}}, \binits{E.}},
\bauthor{\bsnm{{Paas}}, \binits{H.}},
\bauthor{\bsnm{{Pandey-Pommier}}, \binits{M.}},
\bauthor{\bsnm{{Pandey}}, \binits{V.N.}},
\bauthor{\bsnm{{Pizzo}}, \binits{R.}},
\bauthor{\bsnm{{Polatidis}}, \binits{A.G.}},
\bauthor{\bsnm{{Rafferty}}, \binits{D.}},
\bauthor{\bsnm{{Reich}}, \binits{W.}},
\bauthor{\bsnm{{R{\"o}ttgering}}, \binits{H.}},
\bauthor{\bsnm{{Scaife}}, \binits{A.M.M.}},
\bauthor{\bsnm{{Schwarz}}, \binits{D.J.}},
\bauthor{\bsnm{{Serylak}}, \binits{M.}},
\bauthor{\bsnm{{Sluman}}, \binits{J.}},
\bauthor{\bsnm{{Smirnov}}, \binits{O.}},
\bauthor{\bsnm{{Stappers}}, \binits{B.W.}},
\bauthor{\bsnm{{Tagger}}, \binits{M.}},
\bauthor{\bsnm{{Tang}}, \binits{Y.}},
\bauthor{\bsnm{{Tasse}}, \binits{C.}},
\bauthor{\bsnm{{ter Veen}}, \binits{S.}},
\bauthor{\bsnm{{Thoudam}}, \binits{S.}},
\bauthor{\bsnm{{Toribio}}, \binits{M.C.}},
\bauthor{\bsnm{{Vermeulen}}, \binits{R.}},
\bauthor{\bsnm{{van Weeren}}, \binits{R.J.}},
\bauthor{\bsnm{{Wise}}, \binits{M.W.}},
\bauthor{\bsnm{{Wucknitz}}, \binits{O.}},
\bauthor{\bsnm{{Yatawatta}}, \binits{S.}},
\bauthor{\bsnm{{Zarka}}, \binits{P.}},
\bauthor{\bsnm{{Zensus}}, \binits{J.A.}}:
\byear{2018},
\batitle{{Tracking of an electron beam through the solar corona with LOFAR}}.
\bjtitle{\aap}
\bvolume{611},
\bfpage{A57}.
\doiurl{https://doi.org/10.1051/0004-6361/201629017}.
\adsurl{2018A&A...611A..57M}.
\end{barticle}
\endbibitem

\bibitem[\protect\citeauthoryear{Marsch}{2006}]{Marsch2006}
\begin{barticle}
\bauthor{\bsnm{Marsch}, \binits{E.}}:
\byear{2006},
\batitle{{Kinetic physics of the solar corona and solar wind}}.
\bjtitle{Liv. Rev. Solar Phys,}
\bvolume{3},
\bfpage{1}.
\doiurl{https://doi.org/10.12942/lrsp-2006-1}.
\end{barticle}
\endbibitem

\bibitem[\protect\citeauthoryear{{Martinez}}{1978}]{Martinez1978}
\begin{barticle}
\bauthor{\bsnm{{Martinez}}, \binits{J.}}:
\byear{1978},
\batitle{{A Better Photograph of Coronal Streamers}}.
\bjtitle{Sky Tel.}
\bvolume{56},
\bfpage{92}.
\adsurl{1978S&T....56...92M}.
\end{barticle}
\endbibitem

\bibitem[\protect\citeauthoryear{{Mason}, {Antiochos}, and
  {Vourlidas}}{2021}]{EMason2021}
\begin{barticle}
\bauthor{\bsnm{{Mason}}, \binits{E.I.}},
\bauthor{\bsnm{{Antiochos}}, \binits{S.K.}},
\bauthor{\bsnm{{Vourlidas}}, \binits{A.}}:
\byear{2021},
\batitle{{An Observational Study of a ``Rosetta Stone'' Solar Eruption}}.
\bjtitle{\apjl}
\bvolume{914},
\bfpage{L8}.
\doiurl{https://doi.org/10.3847/2041-8213/ac0259}.
\adsurl{2021ApJ...914L...8M}.
\end{barticle}
\endbibitem

\bibitem[\protect\citeauthoryear{{Mason} et~al.}{2021}]{JMason2021}
\begin{barticle}
\bauthor{\bsnm{{Mason}}, \binits{J.P.}},
\bauthor{\bsnm{{Chamberlin}}, \binits{P.C.}},
\bauthor{\bsnm{{Seaton}}, \binits{D.}},
\bauthor{\bsnm{{Burkepile}}, \binits{J.}},
\bauthor{\bsnm{{Colaninno}}, \binits{R.}},
\bauthor{\bsnm{{Dissauer}}, \binits{K.}},
\bauthor{\bsnm{{Eparvier}}, \binits{F.G.}},
\bauthor{\bsnm{{Fan}}, \binits{Y.}},
\bauthor{\bsnm{{Gibson}}, \binits{S.}},
\bauthor{\bsnm{{Jones}}, \binits{A.R.}},
\bauthor{\bsnm{{Kay}}, \binits{C.}},
\bauthor{\bsnm{{Kirk}}, \binits{M.}},
\bauthor{\bsnm{{Kohnert}}, \binits{R.}},
\bauthor{\bsnm{{Pesnell}}, \binits{W.D.}},
\bauthor{\bsnm{{Thompson}}, \binits{B.J.}},
\bauthor{\bsnm{{Veronig}}, \binits{A.M.}},
\bauthor{\bsnm{{West}}, \binits{M.J.}},
\bauthor{\bsnm{{Windt}}, \binits{D.}},
\bauthor{\bsnm{{Woods}}, \binits{T.N.}}:
\byear{2021},
\batitle{{SunCET: The Sun Coronal Ejection Tracker Concept}}.
\bjtitle{J. Space Weather Space Clim.}
\bvolume{11},
\bfpage{20}.
\doiurl{https://doi.org/10.1051/swsc/2021004}.
\adsurl{2021JSWSC..11...20M}.
\end{barticle}
\endbibitem

\bibitem[\protect\citeauthoryear{{Mason} et~al.}{2022}]{Mason2022}
\begin{barticle}
\bauthor{\bsnm{{Mason}}, \binits{J.P.}},
\bauthor{\bsnm{{Seaton}}, \binits{D.B.}},
\bauthor{\bsnm{{Jones}}, \binits{A.R.}},
\bauthor{\bsnm{{Jin}}, \binits{M.}},
\bauthor{\bsnm{{Chamberlin}}, \binits{P.C.}},
\bauthor{\bsnm{{Sims}}, \binits{A.}},
\bauthor{\bsnm{{Woods}}, \binits{T.N.}}:
\byear{2022},
\batitle{{Simultaneous High Dynamic Range Algorithm, Testing, and Instrument
  Simulation}}.
\bjtitle{\apj}
\bvolume{924},
\bfpage{63}.
\doiurl{https://doi.org/10.3847/1538-4357/ac33a1}.
\adsurl{2022ApJ...924...63M}.
\end{barticle}
\endbibitem

\bibitem[\protect\citeauthoryear{{Masson} et~al.}{2014}]{Masson2014}
\begin{barticle}
\bauthor{\bsnm{{Masson}}, \binits{S.}},
\bauthor{\bsnm{{McCauley}}, \binits{P.}},
\bauthor{\bsnm{{Golub}}, \binits{L.}},
\bauthor{\bsnm{{Reeves}}, \binits{K.K.}},
\bauthor{\bsnm{{DeLuca}}, \binits{E.E.}}:
\byear{2014},
\batitle{{Dynamics of the Transition Corona}}.
\bjtitle{\apj}
\bvolume{787},
\bfpage{145}.
\doiurl{https://doi.org/10.1088/0004-637X/787/2/145}.
\adsurl{2014ApJ...787..145M}.
\end{barticle}
\endbibitem

\bibitem[\protect\citeauthoryear{{McCauley}, {Cairns}, and
  {Morgan}}{2018}]{McCauley2018b}
\begin{barticle}
\bauthor{\bsnm{{McCauley}}, \binits{P.I.}},
\bauthor{\bsnm{{Cairns}}, \binits{I.H.}},
\bauthor{\bsnm{{Morgan}}, \binits{J.}}:
\byear{2018},
\batitle{{Densities Probed by Coronal Type III Radio Burst Imaging}}.
\bjtitle{\solphys}
\bvolume{293},
\bfpage{132}.
\doiurl{https://doi.org/10.1007/s11207-018-1353-y}.
\adsurl{2018SoPh..293..132M}.
\end{barticle}
\endbibitem

\bibitem[\protect\citeauthoryear{{McCauley} et~al.}{2017}]{McCauley2017}
\begin{barticle}
\bauthor{\bsnm{{McCauley}}, \binits{P.I.}},
\bauthor{\bsnm{{Cairns}}, \binits{I.H.}},
\bauthor{\bsnm{{Morgan}}, \binits{J.}},
\bauthor{\bsnm{{Gibson}}, \binits{S.E.}},
\bauthor{\bsnm{{Harding}}, \binits{J.C.}},
\bauthor{\bsnm{{Lonsdale}}, \binits{C.}},
\bauthor{\bsnm{{Oberoi}}, \binits{D.}}:
\byear{2017},
\batitle{{Type III Solar Radio Burst Source Region Splitting due to a
  Quasi-separatrix Layer}}.
\bjtitle{\apj}
\bvolume{851},
\bfpage{151}.
\doiurl{https://doi.org/10.3847/1538-4357/aa9cee}.
\adsurl{2017ApJ...851..151M}.
\end{barticle}
\endbibitem

\bibitem[\protect\citeauthoryear{{McComas} et~al.}{2007}]{McComas2007}
\begin{barticle}
\bauthor{\bsnm{{McComas}}, \binits{D.J.}},
\bauthor{\bsnm{{Velli}}, \binits{M.}},
\bauthor{\bsnm{{Lewis}}, \binits{W.S.}},
\bauthor{\bsnm{{Acton}}, \binits{L.W.}},
\bauthor{\bsnm{{Balat-Pichelin}}, \binits{M.}},
\bauthor{\bsnm{{Bothmer}}, \binits{V.}},
\bauthor{\bsnm{{Dirling}}, \binits{R.B.}},
\bauthor{\bsnm{{Feldman}}, \binits{W.C.}},
\bauthor{\bsnm{{Gloeckler}}, \binits{G.}},
\bauthor{\bsnm{{Habbal}}, \binits{S.R.}},
\bauthor{\bsnm{{Hassler}}, \binits{D.M.}},
\bauthor{\bsnm{{Mann}}, \binits{I.}},
\bauthor{\bsnm{{Matthaeus}}, \binits{W.H.}},
\bauthor{\bsnm{{McNutt}}, \binits{R.L.}},
\bauthor{\bsnm{{Mewaldt}}, \binits{R.A.}},
\bauthor{\bsnm{{Murphy}}, \binits{N.}},
\bauthor{\bsnm{{Ofman}}, \binits{L.}},
\bauthor{\bsnm{{Sittler}}, \binits{E.C.}},
\bauthor{\bsnm{{Smith}}, \binits{C.W.}},
\bauthor{\bsnm{{Zurbuchen}}, \binits{T.H.}}:
\byear{2007},
\batitle{{Understanding coronal heating and solar wind acceleration: Case for
  in situ near-Sun measurements}}.
\bjtitle{Rev. Geophys.}
\bvolume{45},
\bfpage{RG1004}.
\doiurl{https://doi.org/10.1029/2006RG000195}.
\adsurl{2007RvGeo..45.1004M}.
\end{barticle}
\endbibitem

\bibitem[\protect\citeauthoryear{{McGregor} et~al.}{2008}]{McGregor2008}
\begin{barticle}
\bauthor{\bsnm{{McGregor}}, \binits{S.L.}},
\bauthor{\bsnm{{Hughes}}, \binits{W.J.}},
\bauthor{\bsnm{{Arge}}, \binits{C.N.}},
\bauthor{\bsnm{{Owens}}, \binits{M.J.}}:
\byear{2008},
\batitle{{Analysis of the magnetic field discontinuity at the potential field
  source surface and Schatten Current Sheet interface in the Wang-Sheeley-Arge
  model}}.
\bjtitle{J. Geophys. Res. (Space Phys.)}
\bvolume{113},
\bfpage{A08112}.
\doiurl{https://doi.org/10.1029/2007JA012330}.
\adsurl{2008JGRA..113.8112M}.
\end{barticle}
\endbibitem

\bibitem[\protect\citeauthoryear{{Meyer} et~al.}{2020}]{Meyer2020}
\begin{barticle}
\bauthor{\bsnm{{Meyer}}, \binits{K.A.}},
\bauthor{\bsnm{{Mackay}}, \binits{D.H.}},
\bauthor{\bsnm{{Talpeanu}}, \binits{D.-C.}},
\bauthor{\bsnm{{Upton}}, \binits{L.A.}},
\bauthor{\bsnm{{West}}, \binits{M.J.}}:
\byear{2020},
\batitle{{Investigation of the Middle Corona with SWAP and a Data-Driven
  Non-Potential Coronal Magnetic Field Model}}.
\bjtitle{\solphys}
\bvolume{295},
\bfpage{101}.
\doiurl{https://doi.org/10.1007/s11207-020-01668-2}.
\adsurl{2020SoPh..295..101M}.
\end{barticle}
\endbibitem

\bibitem[\protect\citeauthoryear{{Mierla} et~al.}{2013}]{Mierla2013}
\begin{barticle}
\bauthor{\bsnm{{Mierla}}, \binits{M.}},
\bauthor{\bsnm{{Seaton}}, \binits{D.B.}},
\bauthor{\bsnm{{Berghmans}}, \binits{D.}},
\bauthor{\bsnm{{Chifu}}, \binits{I.}},
\bauthor{\bsnm{{De Groof}}, \binits{A.}},
\bauthor{\bsnm{{Inhester}}, \binits{B.}},
\bauthor{\bsnm{{Rodriguez}}, \binits{L.}},
\bauthor{\bsnm{{Stenborg}}, \binits{G.}},
\bauthor{\bsnm{{Zhukov}}, \binits{A.N.}}:
\byear{2013},
\batitle{{Study of a Prominence Eruption using PROBA2/SWAP and STEREO/EUVI
  Data}}.
\bjtitle{\solphys}
\bvolume{286},
\bfpage{241}.
\doiurl{https://doi.org/10.1007/s11207-012-9965-0}.
\adsurl{2013SoPh..286..241M}.
\end{barticle}
\endbibitem

\bibitem[\protect\citeauthoryear{{Mierla} et~al.}{2020}]{Mierla2020}
\begin{barticle}
\bauthor{\bsnm{{Mierla}}, \binits{M.}},
\bauthor{\bsnm{{Janssens}}, \binits{J.}},
\bauthor{\bsnm{{D'Huys}}, \binits{E.}},
\bauthor{\bsnm{{Wauters}}, \binits{L.}},
\bauthor{\bsnm{{West}}, \binits{M.J.}},
\bauthor{\bsnm{{Seaton}}, \binits{D.B.}},
\bauthor{\bsnm{{Berghmans}}, \binits{D.}},
\bauthor{\bsnm{{Podladchikova}}, \binits{E.}}:
\byear{2020},
\batitle{{Long-Term Evolution of the Solar Corona Using PROBA2 Data}}.
\bjtitle{\solphys}
\bvolume{295},
\bfpage{66}.
\doiurl{https://doi.org/10.1007/s11207-020-01635-x}.
\adsurl{2020SoPh..295...66M}.
\end{barticle}
\endbibitem

\bibitem[\protect\citeauthoryear{{Mierla} et~al.}{2022}]{Mierla2022}
\begin{barticle}
\bauthor{\bsnm{{Mierla}}, \binits{M.}},
\bauthor{\bsnm{{Zhukov}}, \binits{A.N.}},
\bauthor{\bsnm{{Berghmans}}, \binits{D.}},
\bauthor{\bsnm{{Parenti}}, \binits{S.}},
\bauthor{\bsnm{{Auch{\`e}re}}, \binits{F.}},
\bauthor{\bsnm{{Heinzel}}, \binits{P.}},
\bauthor{\bsnm{{Seaton}}, \binits{D.B.}},
\bauthor{\bsnm{{Palmerio}}, \binits{E.}},
\bauthor{\bsnm{{Jej{\v{c}}i{\v{c}}}}, \binits{S.}},
\bauthor{\bsnm{{Janssens}}, \binits{J.}},
\bauthor{\bsnm{{Kraaikamp}}, \binits{E.}},
\bauthor{\bsnm{{Nicula}}, \binits{B.}},
\bauthor{\bsnm{{Long}}, \binits{D.M.}},
\bauthor{\bsnm{{Hayes}}, \binits{L.A.}},
\bauthor{\bsnm{{Jebaraj}}, \binits{I.C.}},
\bauthor{\bsnm{{Talpeanu}}, \binits{D.-C.}},
\bauthor{\bsnm{{D'Huys}}, \binits{E.}},
\bauthor{\bsnm{{Dolla}}, \binits{L.}},
\bauthor{\bsnm{{Gissot}}, \binits{S.}},
\bauthor{\bsnm{{Magdaleni{\'c}}}, \binits{J.}},
\bauthor{\bsnm{{Rodriguez}}, \binits{L.}},
\bauthor{\bsnm{{Shestov}}, \binits{S.}},
\bauthor{\bsnm{{Stegen}}, \binits{K.}},
\bauthor{\bsnm{{Verbeeck}}, \binits{C.}},
\bauthor{\bsnm{{Sasso}}, \binits{C.}},
\bauthor{\bsnm{{Romoli}}, \binits{M.}},
\bauthor{\bsnm{{Andretta}}, \binits{V.}}:
\byear{2022},
\batitle{{Prominence eruption observed in He II 304 {\r{A}} up to $>6$
  R$_{{\ensuremath{\odot}}}$ by EUI/FSI aboard Solar Orbiter}}.
\bjtitle{\aap}
\bvolume{662},
\bfpage{L5}.
\doiurl{https://doi.org/10.1051/0004-6361/202244020}.
\adsurl{2022A&A...662L...5M}.
\end{barticle}
\endbibitem

\bibitem[\protect\citeauthoryear{{Miki{\'c}} et~al.}{1999}]{Mikic1999}
\begin{barticle}
\bauthor{\bsnm{{Miki{\'c}}}, \binits{Z.}},
\bauthor{\bsnm{{Linker}}, \binits{J.A.}},
\bauthor{\bsnm{{Schnack}}, \binits{D.D.}},
\bauthor{\bsnm{{Lionello}}, \binits{R.}},
\bauthor{\bsnm{{Tarditi}}, \binits{A.}}:
\byear{1999},
\batitle{{Magnetohydrodynamic modeling of the global solar corona}}.
\bjtitle{Phys. Plasmas}
\bvolume{6},
\bfpage{2217}.
\doiurl{https://doi.org/10.1063/1.873474}.
\adsurl{1999PhPl....6.2217M}.
\end{barticle}
\endbibitem

\bibitem[\protect\citeauthoryear{{Miki{\'c}} et~al.}{2018}]{Mikic2018}
\begin{barticle}
\bauthor{\bsnm{{Miki{\'c}}}},
\bauthor{\bsnm{{}}, \binits{Z.}},
\bauthor{\bsnm{{Downs}}, \binits{C.}},
\bauthor{\bsnm{{Linker}}, \binits{J.A.}},
\bauthor{\bsnm{{Caplan}}, \binits{R.M.}},
\bauthor{\bsnm{{Mackay}}, \binits{D.H.}},
\bauthor{\bsnm{{Upton}}, \binits{L.A.}},
\bauthor{\bsnm{{Riley}}, \binits{P.}},
\bauthor{\bsnm{{Lionello}}, \binits{R.}},
\bauthor{\bsnm{{T{\"o}r{\"o}k}}, \binits{T.}},
\bauthor{\bsnm{{Titov}}, \binits{V.S.}},
\bauthor{\bsnm{{Wijaya}}, \binits{J.}},
\bauthor{\bsnm{{Druckm{\"u}ller}}, \binits{M.}},
\bauthor{\bsnm{{Pasachoff}}, \binits{J.M.}},
\bauthor{\bsnm{{Carlos}}, \binits{W.}}:
\byear{2018},
\batitle{{Predicting the corona for the 21 August 2017 total solar eclipse}}.
\bjtitle{Nat. Astron.}
\bvolume{2},
\bfpage{913}.
\doiurl{https://doi.org/10.1038/s41550-018-0562-5}.
\adsurl{2018NatAs...2..913M}.
\end{barticle}
\endbibitem

\bibitem[\protect\citeauthoryear{Mondal, Oberoi, and
  Vourlidas}{2020}]{mondal2020}
\begin{barticle}
\bauthor{\bsnm{Mondal}, \binits{S.}},
\bauthor{\bsnm{Oberoi}, \binits{D.}},
\bauthor{\bsnm{Vourlidas}, \binits{A.}}:
\byear{2020},
\batitle{Estimation of the {{Physical Parameters}} of a {{CME}} at {{High
  Coronal Heights Using Low-frequency Radio Observations}}}.
\bjtitle{\apj}
\bvolume{893},
\bfpage{28}.
\doiurl{https://doi.org/10.3847/1538-4357/ab7fab}.
\end{barticle}
\endbibitem

\bibitem[\protect\citeauthoryear{{Morgan} and
  {Druckm{\"u}ller}}{2014}]{Morgan2014}
\begin{barticle}
\bauthor{\bsnm{{Morgan}}, \binits{H.}},
\bauthor{\bsnm{{Druckm{\"u}ller}}, \binits{M.}}:
\byear{2014},
\batitle{{Multi-Scale Gaussian Normalization for Solar Image Processing}}.
\bjtitle{\solphys}
\bvolume{289},
\bfpage{2945}.
\doiurl{https://doi.org/10.1007/s11207-014-0523-9}.
\adsurl{2014SoPh..289.2945M}.
\end{barticle}
\endbibitem

\bibitem[\protect\citeauthoryear{{Morgan} and {Habbal}}{2007}]{Morgan2007}
\begin{barticle}
\bauthor{\bsnm{{Morgan}}, \binits{H.}},
\bauthor{\bsnm{{Habbal}}, \binits{S.R.}}:
\byear{2007},
\batitle{{Are solar maximum fan streamers a consequence of twisting sheet
  structures?}}
\bjtitle{\aap}
\bvolume{465},
\bfpage{L47}.
\doiurl{https://doi.org/10.1051/0004-6361:20077126}.
\adsurl{2007A&A...465L..47M}.
\end{barticle}
\endbibitem

\bibitem[\protect\citeauthoryear{Morosan et~al.}{2019}]{morosan2019}
\begin{barticle}
\bauthor{\bsnm{Morosan}, \binits{D.E.}},
\bauthor{\bsnm{Carley}, \binits{E.P.}},
\bauthor{\bsnm{Hayes}, \binits{L.A.}},
\bauthor{\bsnm{Murray}, \binits{S.A.}},
\bauthor{\bsnm{Zucca}, \binits{P.}},
\bauthor{\bsnm{Fallows}, \binits{R.A.}},
\bauthor{\bsnm{McCauley}, \binits{J.}},
\bauthor{\bsnm{Kilpua}, \binits{E.K.J.}},
\bauthor{\bsnm{Mann}, \binits{G.}},
\bauthor{\bsnm{Vocks}, \binits{C.}},
\bauthor{\bsnm{Gallagher}, \binits{P.T.}}:
\byear{2019},
\batitle{Multiple Regions of Shock-Accelerated Particles during a Solar Coronal
  Mass Ejection}.
\bjtitle{Nat. Astron.}
\bvolume{3},
\bfpage{452}.
\doiurl{https://doi.org/10.1038/s41550-019-0689-z}.
\end{barticle}
\endbibitem

\bibitem[\protect\citeauthoryear{{Morton}, {Tomczyk}, and
  {Pinto}}{2015}]{Morton2015}
\begin{barticle}
\bauthor{\bsnm{{Morton}}, \binits{R.J.}},
\bauthor{\bsnm{{Tomczyk}}, \binits{S.}},
\bauthor{\bsnm{{Pinto}}, \binits{R.}}:
\byear{2015},
\batitle{{Investigating Alfv{\'e}nic wave propagation in coronal open-field
  regions}}.
\bjtitle{Nat. Comm.}
\bvolume{6},
\bfpage{7813}.
\doiurl{https://doi.org/10.1038/ncomms8813}.
\adsurl{2015NatCo...6.7813M}.
\end{barticle}
\endbibitem

\bibitem[\protect\citeauthoryear{{Morton}, {Tomczyk}, and
  {Pinto}}{2016}]{Morton2016ApJ}
\begin{barticle}
\bauthor{\bsnm{{Morton}}, \binits{R.J.}},
\bauthor{\bsnm{{Tomczyk}}, \binits{S.}},
\bauthor{\bsnm{{Pinto}}, \binits{R.F.}}:
\byear{2016},
\batitle{{A Global View of Velocity Fluctuations in the Corona below 1.3 R
  $_{{\ensuremath{\odot}}}$ with CoMP}}.
\bjtitle{\apj}
\bvolume{828},
\bfpage{89}.
\doiurl{https://doi.org/10.3847/0004-637X/828/2/89}.
\adsurl{2016ApJ...828...89M}.
\end{barticle}
\endbibitem

\bibitem[\protect\citeauthoryear{{Morton}, {Weberg}, and
  {McLaughlin}}{2019}]{Morton2019}
\begin{barticle}
\bauthor{\bsnm{{Morton}}, \binits{R.J.}},
\bauthor{\bsnm{{Weberg}}, \binits{M.J.}},
\bauthor{\bsnm{{McLaughlin}}, \binits{J.A.}}:
\byear{2019},
\batitle{{A basal contribution from p-modes to the Alfv{\'e}nic wave flux in
  the Sun's corona}}.
\bjtitle{Nat. Astron.}
\bvolume{3},
\bfpage{223}.
\doiurl{https://doi.org/10.1038/s41550-018-0668-9}.
\adsurl{2019NatAs...3..223M}.
\end{barticle}
\endbibitem

\bibitem[\protect\citeauthoryear{{Moses} et~al.}{2020}]{moses2020}
\begin{barticle}
\bauthor{\bsnm{{Moses}}, \binits{J.D.}},
\bauthor{\bsnm{{Antonucci}}, \binits{E.}},
\bauthor{\bsnm{{Newmark}}, \binits{J.}},
\bauthor{\bsnm{{Auch{\`e}re}}, \binits{F.}},
\bauthor{\bsnm{{Fineschi}}, \binits{S.}},
\bauthor{\bsnm{{Romoli}}, \binits{M.}},
\bauthor{\bsnm{{Telloni}}, \binits{D.}},
\bauthor{\bsnm{{Massone}}, \binits{G.}},
\bauthor{\bsnm{{Zangrilli}}, \binits{L.}},
\bauthor{\bsnm{{Focardi}}, \binits{M.}},
\bauthor{\bsnm{{Landini}}, \binits{F.}},
\bauthor{\bsnm{{Pancrazzi}}, \binits{M.}},
\bauthor{\bsnm{{Rossi}}, \binits{G.}},
\bauthor{\bsnm{{Malvezzi}}, \binits{A.M.}},
\bauthor{\bsnm{{Wang}}, \binits{D.}},
\bauthor{\bsnm{{Leclec'h}}, \binits{J.-C.}},
\bauthor{\bsnm{{Moalic}}, \binits{J.-P.}},
\bauthor{\bsnm{{Rouesnel}}, \binits{F.}},
\bauthor{\bsnm{{Abbo}}, \binits{L.}},
\bauthor{\bsnm{{Canou}}, \binits{A.}},
\bauthor{\bsnm{{Barbey}}, \binits{N.}},
\bauthor{\bsnm{{Guennou}}, \binits{C.}},
\bauthor{\bsnm{{Laming}}, \binits{J.M.}},
\bauthor{\bsnm{{Lemen}}, \binits{J.}},
\bauthor{\bsnm{{Wuelser}}, \binits{J.-P.}},
\bauthor{\bsnm{{Kohl}}, \binits{J.L.}},
\bauthor{\bsnm{{Gardner}}, \binits{L.D.}}:
\byear{2020},
\batitle{{Global helium abundance measurements in the solar corona}}.
\bjtitle{Nat. Astron.}
\bvolume{4},
\bfpage{1134}.
\doiurl{https://doi.org/10.1038/s41550-020-1156-6}.
\adsurl{2020NatAs...4.1134M}.
\end{barticle}
\endbibitem

\bibitem[\protect\citeauthoryear{{Murphy}, {Raymond}, and
  {Korreck}}{2011}]{Murphy2011}
\begin{barticle}
\bauthor{\bsnm{{Murphy}}, \binits{N.A.}},
\bauthor{\bsnm{{Raymond}}, \binits{J.C.}},
\bauthor{\bsnm{{Korreck}}, \binits{K.E.}}:
\byear{2011},
\batitle{{Plasma Heating During a Coronal Mass Ejection Observed By the Solar
  and Heliospheric Observatory}}.
\bjtitle{\apj}
\bvolume{735},
\bfpage{17}.
\doiurl{https://doi.org/10.1088/0004-637X/735/1/17}.
\adsurl{2011ApJ...735...17M}.
\end{barticle}
\endbibitem

\bibitem[\protect\citeauthoryear{{Newkirk} and {Lacey}}{1970}]{Newkirk1970}
\begin{barticle}
\bauthor{\bsnm{{Newkirk}}, \binits{G.}},
\bauthor{\bsnm{{Lacey}}, \binits{L.}}:
\byear{1970},
\batitle{{The Corona during the March 7, 1970, Eclipse}}.
\bjtitle{\nat}
\bvolume{226},
\bfpage{1098}.
\doiurl{https://doi.org/10.1038/2261098a0}.
\adsurl{1970Natur.226.1098N}.
\end{barticle}
\endbibitem

\bibitem[\protect\citeauthoryear{{Newkirk}, {Altschuler}, and
  {Harvey}}{1968}]{Newkirk1968}
\begin{bchapter}
\bauthor{\bsnm{{Newkirk}}, \binits{G.}},
\bauthor{\bsnm{{Altschuler}}, \binits{M.D.}},
\bauthor{\bsnm{{Harvey}}, \binits{J.}}:
\byear{1968},
\bctitle{{Influence of Magnetic Fields on the Structure of the Solar Corona}}.
In: \beditor{\bsnm{{Kiepenheuer}}, \binits{K.O.}} (ed.)
\bbtitle{Structure and Development of Solar Active Regions, Proc. IAU Symp}
\bseriesno{35},
\bpublisher{Reidel},
\blocation{Dordrecht},
\bfpage{379}.
\adsurl{1968IAUS...35..379N}.
\end{bchapter}
\endbibitem

\bibitem[\protect\citeauthoryear{{Noci}, {Kohl}, and
  {Withbroe}}{1987}]{noci1987}
\begin{barticle}
\bauthor{\bsnm{{Noci}}, \binits{G.}},
\bauthor{\bsnm{{Kohl}}, \binits{J.L.}},
\bauthor{\bsnm{{Withbroe}}, \binits{G.L.}}:
\byear{1987},
\batitle{{Solar wind diagnostics from Doppler-enhanced scattering}}.
\bjtitle{\apj}
\bvolume{315},
\bfpage{706}.
\doiurl{https://doi.org/10.1086/165172}.
\adsurl{1987ApJ...315..706N}.
\end{barticle}
\endbibitem

\bibitem[\protect\citeauthoryear{{O'Hara} et~al.}{2019}]{OHara2019}
\begin{barticle}
\bauthor{\bsnm{{O'Hara}}, \binits{J.P.}},
\bauthor{\bsnm{{Mierla}}, \binits{M.}},
\bauthor{\bsnm{{Podladchikova}}, \binits{O.}},
\bauthor{\bsnm{{D'Huys}}, \binits{E.}},
\bauthor{\bsnm{{West}}, \binits{M.J.}}:
\byear{2019},
\batitle{{Exceptional Extended Field-of-view Observations by PROBA2/SWAP on
  2017 April 1 and 3}}.
\bjtitle{\apj}
\bvolume{883},
\bfpage{59}.
\doiurl{https://doi.org/10.3847/1538-4357/ab3b08}.
\adsurl{2019ApJ...883...59O}.
\end{barticle}
\endbibitem

\bibitem[\protect\citeauthoryear{{Owocki}, {Holzer}, and
  {Hundhausen}}{1983}]{owocki1983}
\begin{barticle}
\bauthor{\bsnm{{Owocki}}, \binits{S.P.}},
\bauthor{\bsnm{{Holzer}}, \binits{T.E.}},
\bauthor{\bsnm{{Hundhausen}}, \binits{A.J.}}:
\byear{1983},
\batitle{{The solar wind ionization state as a coronal temperature
  diagnostic}}.
\bjtitle{\apj}
\bvolume{275},
\bfpage{354}.
\doiurl{https://doi.org/10.1086/161538}.
\adsurl{1983ApJ...275..354O}.
\end{barticle}
\endbibitem

\bibitem[\protect\citeauthoryear{{Parker}}{1958}]{Parker1958}
\begin{barticle}
\bauthor{\bsnm{{Parker}}, \binits{E.N.}}:
\byear{1958},
\batitle{{Dynamics of the Interplanetary Gas and Magnetic Fields.}}
\bjtitle{\apj}
\bvolume{128},
\bfpage{664}.
\doiurl{https://doi.org/10.1086/146579}.
\adsurl{1958ApJ...128..664P}.
\end{barticle}
\endbibitem

\bibitem[\protect\citeauthoryear{{Pasachoff} et~al.}{2011}]{Pasachoff2011}
\begin{barticle}
\bauthor{\bsnm{{Pasachoff}}, \binits{J.M.}},
\bauthor{\bsnm{{Ru{\v{s}}in}}, \binits{V.}},
\bauthor{\bsnm{{Druckm{\"u}llerov{\'a}}}, \binits{H.}},
\bauthor{\bsnm{{Saniga}}, \binits{M.}},
\bauthor{\bsnm{{Lu}}, \binits{M.}},
\bauthor{\bsnm{{Malamut}}, \binits{C.}},
\bauthor{\bsnm{{Seaton}}, \binits{D.B.}},
\bauthor{\bsnm{{Golub}}, \binits{L.}},
\bauthor{\bsnm{{Engell}}, \binits{A.J.}},
\bauthor{\bsnm{{Hill}}, \binits{S.W.}},
\bauthor{\bsnm{{Lucas}}, \binits{R.}}:
\byear{2011},
\batitle{{Structure and Dynamics of the 2010 July 11 Eclipse White-light
  Corona}}.
\bjtitle{\apj}
\bvolume{734},
\bfpage{114}.
\doiurl{https://doi.org/10.1088/0004-637X/734/2/114}.
\adsurl{2011ApJ...734..114P}.
\end{barticle}
\endbibitem

\bibitem[\protect\citeauthoryear{{Pascoe}, {Wright}, and {De
  Moortel}}{2010}]{Pascoe2010}
\begin{barticle}
\bauthor{\bsnm{{Pascoe}}, \binits{D.J.}},
\bauthor{\bsnm{{Wright}}, \binits{A.N.}},
\bauthor{\bsnm{{De Moortel}}, \binits{I.}}:
\byear{2010},
\batitle{{Coupled Alfv{\'e}n and Kink Oscillations in Coronal Loops}}.
\bjtitle{\apj}
\bvolume{711},
\bfpage{990}.
\doiurl{https://doi.org/10.1088/0004-637X/711/2/990}.
\adsurl{2010ApJ...711..990P}.
\end{barticle}
\endbibitem

\bibitem[\protect\citeauthoryear{{Patel} et~al.}{2020}]{patel2020A&A}
\begin{barticle}
\bauthor{\bsnm{{Patel}}, \binits{R.}},
\bauthor{\bsnm{{Pant}}, \binits{V.}},
\bauthor{\bsnm{{Chandrashekhar}}, \binits{K.}},
\bauthor{\bsnm{{Banerjee}}, \binits{D.}}:
\byear{2020},
\batitle{{A statistical study of plasmoids associated with a post-CME current
  sheet}}.
\bjtitle{\aap}
\bvolume{644},
\bfpage{A158}.
\doiurl{https://doi.org/10.1051/0004-6361/202039000}.
\adsurl{2020A&A...644A.158P}.
\end{barticle}
\endbibitem

\bibitem[\protect\citeauthoryear{{Pizzo} et~al.}{2005}]{Pizzo2005}
\begin{barticle}
\bauthor{\bsnm{{Pizzo}}, \binits{V.J.}},
\bauthor{\bsnm{{Hill}}, \binits{S.M.}},
\bauthor{\bsnm{{Balch}}, \binits{C.C.}},
\bauthor{\bsnm{{Biesecker}}, \binits{D.A.}},
\bauthor{\bsnm{{Bornmann}}, \binits{P.}},
\bauthor{\bsnm{{Hildner}}, \binits{E.}},
\bauthor{\bsnm{{Grubb}}, \binits{R.N.}},
\bauthor{\bsnm{{Chipman}}, \binits{E.G.}},
\bauthor{\bsnm{{Davis}}, \binits{J.M.}},
\bauthor{\bsnm{{Wallace}}, \binits{K.S.}},
\bauthor{\bsnm{{Russell}}, \binits{K.}},
\bauthor{\bsnm{{Cauffman}}, \binits{S.A.}},
\bauthor{\bsnm{{Saha}}, \binits{T.T.}},
\bauthor{\bsnm{{Berthiume}}, \binits{G.D.}}:
\byear{2005},
\batitle{{The NOAA GOES-12 Solar X-Ray Imager (SXI) 2. Performance}}.
\bjtitle{\solphys}
\bvolume{226},
\bfpage{283}.
\doiurl{https://doi.org/10.1007/s11207-005-7417-9}.
\adsurl{2005SoPh..226..283P}.
\end{barticle}
\endbibitem

\bibitem[\protect\citeauthoryear{{Plowman}}{2021}]{Plowman2021}
\begin{barticle}
\bauthor{\bsnm{{Plowman}}, \binits{J.}}:
\byear{2021},
\batitle{{Three-dimensional Reconstruction of Coronal Plasma Properties from a
  Single Perspective}}.
\bjtitle{\apj}
\bvolume{922},
\bfpage{109}.
\doiurl{https://doi.org/10.3847/1538-4357/ac2664}.
\adsurl{2021ApJ...922..109P}.
\end{barticle}
\endbibitem

\bibitem[\protect\citeauthoryear{{Pneuman} and {Kopp}}{1971}]{Pneuman1971}
\begin{barticle}
\bauthor{\bsnm{{Pneuman}}, \binits{G.W.}},
\bauthor{\bsnm{{Kopp}}, \binits{R.A.}}:
\byear{1971},
\batitle{{Gas-Magnetic Field Interactions in the Solar Corona}}.
\bjtitle{\solphys}
\bvolume{18},
\bfpage{258}.
\doiurl{https://doi.org/10.1007/BF00145940}.
\adsurl{1971SoPh...18..258P}.
\end{barticle}
\endbibitem

\bibitem[\protect\citeauthoryear{{Prasad} et~al.}{2017}]{Prasad2017}
\begin{barticle}
\bauthor{\bsnm{{Prasad}}, \binits{B.R.}},
\bauthor{\bsnm{Banerjee}, \binits{D.}},
\bauthor{\bsnm{Singh}, \binits{J.}},
\bauthor{\bsnm{Nagabhushana}, \binits{S.}},
\bauthor{\bsnm{Kumar}, \binits{A.}},
\bauthor{\bsnm{Kamath}, \binits{P.U.}},
\bauthor{\bsnm{Kathiravan}, \binits{S.}},
\bauthor{\bsnm{Venkata}, \binits{S.}},
\bauthor{\bsnm{Rajkumar}, \binits{N.}},
\bauthor{\bsnm{Natarajan}, \binits{V.}},
\bauthor{\bsnm{Juneja}, \binits{M.}},
\bauthor{\bsnm{Somu}, \binits{P.}},
\bauthor{\bsnm{Pant}, \binits{V.}},
\bauthor{\bsnm{Shaji}, \binits{N.}},
\bauthor{\bsnm{Sankarsubramanian}, \binits{K.}},
\bauthor{\bsnm{Patra}, \binits{A.}},
\bauthor{\bsnm{Venkateswaran}, \binits{R.}},
\bauthor{\bsnm{Adoni}, \binits{A.A.}},
\bauthor{\bsnm{Narendra}, \binits{S.}},
\bauthor{\bsnm{Haridas}, \binits{T.R.}},
\bauthor{\bsnm{Mathew}, \binits{S.K.}},
\bauthor{\bsnm{Krishna}, \binits{R.M.}},
\bauthor{\bsnm{Amareswari}, \binits{K.}},
\bauthor{\bsnm{Jaiswal}, \binits{B.}}:
\byear{2017},
\batitle{{Visible Emission Line Coronagraph on Aditya-L1}}.
\bjtitle{Curr. Science}
\bvolume{113},
\bfpage{613}.
\doiurl{https://doi.org/10.18520/cs/v113/i04/613-615}.
\end{barticle}
\endbibitem

\bibitem[\protect\citeauthoryear{{Rachmeler} et~al.}{2014}]{Rachmeler2014}
\begin{barticle}
\bauthor{\bsnm{{Rachmeler}}, \binits{L.A.}},
\bauthor{\bsnm{{Platten}}, \binits{S.J.}},
\bauthor{\bsnm{{Bethge}}, \binits{C.}},
\bauthor{\bsnm{{Seaton}}, \binits{D.B.}},
\bauthor{\bsnm{{Yeates}}, \binits{A.R.}}:
\byear{2014},
\batitle{{Observations of a Hybrid Double-streamer/Pseudostreamer in the Solar
  Corona}}.
\bjtitle{\apjl}
\bvolume{787},
\bfpage{L3}.
\doiurl{https://doi.org/10.1088/2041-8205/787/1/L3}.
\adsurl{2014ApJ...787L...3R}.
\end{barticle}
\endbibitem

\bibitem[\protect\citeauthoryear{{Rakowski}, {Laming}, and
  {Lepri}}{2007}]{rakowski2007}
\begin{barticle}
\bauthor{\bsnm{{Rakowski}}, \binits{C.E.}},
\bauthor{\bsnm{{Laming}}, \binits{J.M.}},
\bauthor{\bsnm{{Lepri}}, \binits{S.T.}}:
\byear{2007},
\batitle{{Ion Charge States in Halo Coronal Mass Ejections: What Can We Learn
  about the Explosion?}}
\bjtitle{\apj}
\bvolume{667},
\bfpage{602}.
\doiurl{https://doi.org/10.1086/520914}.
\adsurl{2007ApJ...667..602R}.
\end{barticle}
\endbibitem

\bibitem[\protect\citeauthoryear{{Raouafi} et~al.}{2016}]{Raouafi2016b}
\begin{barticle}
\bauthor{\bsnm{{Raouafi}}, \binits{N.E.}},
\bauthor{\bsnm{{Riley}}, \binits{P.}},
\bauthor{\bsnm{{Gibson}}, \binits{S.}},
\bauthor{\bsnm{{Fineschi}}, \binits{S.}},
\bauthor{\bsnm{{Solanki}}, \binits{S.K.}}:
\byear{2016},
\batitle{{Diagnostics of Coronal Magnetic Fields Through the Hanle Effect in UV
  and IR Lines}}.
\bjtitle{Front. Astron. Space Sci.}
\bvolume{3},
\bfpage{20}.
\doiurl{https://doi.org/10.3389/fspas.2016.00020}.
\adsurl{2016FrASS...3...20R}.
\end{barticle}
\endbibitem

\bibitem[\protect\citeauthoryear{Raouafi et~al.}{2016}]{Raouafi2016a}
\begin{barticle}
\bauthor{\bsnm{Raouafi}, \binits{N.E.}},
\bauthor{\bsnm{Patsourakos}, \binits{S.}},
\bauthor{\bsnm{Pariat}, \binits{E.}},
\bauthor{\bsnm{Young}, \binits{P.R.}},
\bauthor{\bsnm{Sterling}, \binits{A.C.}},
\bauthor{\bsnm{Savcheva}, \binits{A.}},
\bauthor{\bsnm{Shimojo}, \binits{M.}},
\bauthor{\bsnm{Moreno-Insertis}, \binits{F.}},
\bauthor{\bsnm{DeVore}, \binits{C.R.}},
\bauthor{\bsnm{Archontis}, \binits{V.}},
\bauthor{\bsnm{T{\"{o}}r{\"{o}}k}, \binits{T.}},
\bauthor{\bsnm{Mason}, \binits{H.}},
\bauthor{\bsnm{Curdt}, \binits{W.}},
\bauthor{\bsnm{Meyer}, \binits{K.}},
\bauthor{\bsnm{Dalmasse}, \binits{K.}},
\bauthor{\bsnm{Matsui}, \binits{Y.}}:
\byear{2016},
\batitle{{Solar Coronal Jets: Observations, Theory, and Modeling}}.
\bjtitle{Space Sci. Rev.}
\bvolume{201},
\bfpage{1}.
\doiurl{https://doi.org/10.1007/s11214-016-0260-5}.
\end{barticle}
\endbibitem

\bibitem[\protect\citeauthoryear{{Raouafi} et~al.}{2023}]{Raouafi2023}
\begin{botherref}
\oauthor{\bsnm{{Raouafi}}, \binits{N.E.}},
\oauthor{\bsnm{{Stenborg}}, \binits{G.}},
\oauthor{\bsnm{{Seaton}}, \binits{D.B.}},
\oauthor{\bsnm{{Wang}}, \binits{H.}},
\oauthor{\bsnm{{Wang}}, \binits{J.}},
\oauthor{\bsnm{{DeForest}}, \binits{C.E.}},
\oauthor{\bsnm{{Bale}}, \binits{S.D.}},
\oauthor{\bsnm{{Drake}}, \binits{J.F.}},
\oauthor{\bsnm{{Uritsky}}, \binits{V.M.}},
\oauthor{\bsnm{{Karpen}}, \binits{J.T.}},
\oauthor{\bsnm{{DeVore}}, \binits{C.R.}},
\oauthor{\bsnm{{Sterling}}, \binits{A.C.}},
\oauthor{\bsnm{{Horbury}}, \binits{T.S.}},
\oauthor{\bsnm{{Harra}}, \binits{L.K.}},
\oauthor{\bsnm{{Bourouaine}}, \binits{S.}},
\oauthor{\bsnm{{Kasper}}, \binits{J.C.}},
\oauthor{\bsnm{{Kumar}}, \binits{P.}},
\oauthor{\bsnm{{Phan}}, \binits{T.D.}},
\oauthor{\bsnm{{Velli}}, \binits{M.}}:
2023,
{Magnetic Reconnection as the Driver of the Solar Wind}.
\textit{arXiv e-prints},
arXiv:2301.00903.
\doiurl{https://doi.org/10.48550/arXiv.2301.00903}.
\adsurl{2023arXiv230100903R}.
\end{botherref}
\endbibitem

\bibitem[\protect\citeauthoryear{{Raymond} et~al.}{1997}]{Raymond1997}
\begin{barticle}
\bauthor{\bsnm{{Raymond}}, \binits{J.C.}},
\bauthor{\bsnm{{Kohl}}, \binits{J.L.}},
\bauthor{\bsnm{{Noci}}, \binits{G.}},
\bauthor{\bsnm{{Antonucci}}, \binits{E.}},
\bauthor{\bsnm{{Tondello}}, \binits{G.}},
\bauthor{\bsnm{{Huber}}, \binits{M.C.E.}},
\bauthor{\bsnm{{Gardner}}, \binits{L.D.}},
\bauthor{\bsnm{{Nicolosi}}, \binits{P.}},
\bauthor{\bsnm{{Fineschi}}, \binits{S.}},
\bauthor{\bsnm{{Romoli}}, \binits{M.}},
\bauthor{\bsnm{{Spadaro}}, \binits{D.}},
\bauthor{\bsnm{{Siegmund}}, \binits{O.H.W.}},
\bauthor{\bsnm{{Benna}}, \binits{C.}},
\bauthor{\bsnm{{Ciaravella}}, \binits{A.}},
\bauthor{\bsnm{{Cranmer}}, \binits{S.}},
\bauthor{\bsnm{{Giordano}}, \binits{S.}},
\bauthor{\bsnm{{Karovska}}, \binits{M.}},
\bauthor{\bsnm{{Martin}}, \binits{R.}},
\bauthor{\bsnm{{Michels}}, \binits{J.}},
\bauthor{\bsnm{{Modigliani}}, \binits{A.}},
\bauthor{\bsnm{{Naletto}}, \binits{G.}},
\bauthor{\bsnm{{Panasyuk}}, \binits{A.}},
\bauthor{\bsnm{{Pernechele}}, \binits{C.}},
\bauthor{\bsnm{{Poletto}}, \binits{G.}},
\bauthor{\bsnm{{Smith}}, \binits{P.L.}},
\bauthor{\bsnm{{Suleiman}}, \binits{R.M.}},
\bauthor{\bsnm{{Strachan}}, \binits{L.}}:
\byear{1997},
\batitle{{Composition of Coronal Streamers from the SOHO Ultraviolet
  Coronagraph Spectrometer}}.
\bjtitle{\solphys}
\bvolume{175},
\bfpage{645}.
\doiurl{https://doi.org/10.1023/A:1004948423169}.
\adsurl{1997SoPh..175..645R}.
\end{barticle}
\endbibitem

\bibitem[\protect\citeauthoryear{{Raymond} et~al.}{2014}]{Raymond2014}
\begin{barticle}
\bauthor{\bsnm{{Raymond}}, \binits{J.C.}},
\bauthor{\bsnm{{McCauley}}, \binits{P.I.}},
\bauthor{\bsnm{{Cranmer}}, \binits{S.R.}},
\bauthor{\bsnm{{Downs}}, \binits{C.}}:
\byear{2014},
\batitle{{The Solar Corona as Probed by Comet Lovejoy (C/2011 W3)}}.
\bjtitle{\apj}
\bvolume{788},
\bfpage{152}.
\doiurl{https://doi.org/10.1088/0004-637X/788/2/152}.
\adsurl{2014ApJ...788..152R}.
\end{barticle}
\endbibitem

\bibitem[\protect\citeauthoryear{{Raymond} et~al.}{2018}]{Raymond2018}
\begin{barticle}
\bauthor{\bsnm{{Raymond}}, \binits{J.C.}},
\bauthor{\bsnm{{Downs}}, \binits{C.}},
\bauthor{\bsnm{{Knight}}, \binits{M.M.}},
\bauthor{\bsnm{{Battams}}, \binits{K.}},
\bauthor{\bsnm{{Giordano}}, \binits{S.}},
\bauthor{\bsnm{{Rosati}}, \binits{R.}}:
\byear{2018},
\batitle{{Comet C/2011 W3 (Lovejoy) between 2 and 10 Solar Radii: Physical
  Parameters of the Comet and the Corona}}.
\bjtitle{\apj}
\bvolume{858},
\bfpage{19}.
\doiurl{https://doi.org/10.3847/1538-4357/aabade}.
\adsurl{2018ApJ...858...19R}.
\end{barticle}
\endbibitem

\bibitem[\protect\citeauthoryear{{Reames}}{1999}]{Reames1999}
\begin{barticle}
\bauthor{\bsnm{{Reames}}, \binits{D.V.}}:
\byear{1999},
\batitle{{Particle acceleration at the Sun and in the heliosphere}}.
\bjtitle{\ssr}
\bvolume{90},
\bfpage{413}.
\doiurl{https://doi.org/10.1023/A:1005105831781}.
\adsurl{1999SSRv...90..413R}.
\end{barticle}
\endbibitem

\bibitem[\protect\citeauthoryear{{Reid}}{2020}]{Reid2020}
\begin{barticle}
\bauthor{\bsnm{{Reid}}, \binits{H.A.S.}}:
\byear{2020},
\batitle{{A review of recent type III imaging spectroscopy}}.
\bjtitle{Front. Astron. Space Sci.}
\bvolume{7},
\bfpage{56}.
\doiurl{https://doi.org/10.3389/fspas.2020.00056}.
\adsurl{2020FrASS...7...56R}.
\end{barticle}
\endbibitem

\bibitem[\protect\citeauthoryear{{Reid} and {Kontar}}{2021}]{Reid2021}
\begin{barticle}
\bauthor{\bsnm{{Reid}}, \binits{H.A.S.}},
\bauthor{\bsnm{{Kontar}}, \binits{E.P.}}:
\byear{2021},
\batitle{{Fine structure of type III solar radio bursts from Langmuir wave
  motion in turbulent plasma}}.
\bjtitle{Nat. Astron.}
\bvolume{5},
\bfpage{796}.
\doiurl{https://doi.org/10.1038/s41550-021-01370-8}.
\adsurl{2021NatAs...5..796R}.
\end{barticle}
\endbibitem

\bibitem[\protect\citeauthoryear{Reid and Ratcliffe}{2014}]{reid2014}
\begin{barticle}
\bauthor{\bsnm{Reid}, \binits{H.A.S.}},
\bauthor{\bsnm{Ratcliffe}, \binits{H.}}:
\byear{2014},
\batitle{A Review of Solar Type {{III}} Radio Bursts}.
\bjtitle{Res. Astron. Astrophys.}
\bvolume{14},
\bfpage{773}.
\doiurl{https://doi.org/10.1088/1674-4527/14/7/003}.
\end{barticle}
\endbibitem

\bibitem[\protect\citeauthoryear{{Reva} et~al.}{2017}]{Reva2017}
\begin{barticle}
\bauthor{\bsnm{{Reva}}, \binits{A.A.}},
\bauthor{\bsnm{{Kirichenko}}, \binits{A.S.}},
\bauthor{\bsnm{{Ulyanov}}, \binits{A.S.}},
\bauthor{\bsnm{{Kuzin}}, \binits{S.V.}}:
\byear{2017},
\batitle{{Observations of the Coronal Mass Ejection with a Complex Acceleration
  Profile}}.
\bjtitle{\apj}
\bvolume{851},
\bfpage{108}.
\doiurl{https://doi.org/10.3847/1538-4357/aa9986}.
\adsurl{2017ApJ...851..108R}.
\end{barticle}
\endbibitem

\bibitem[\protect\citeauthoryear{{Rickett}}{1990}]{Rickett1990}
\begin{barticle}
\bauthor{\bsnm{{Rickett}}, \binits{B.J.}}:
\byear{1990},
\batitle{{Radio propagation through the turbulent interstellar plasma.}}
\bjtitle{Annual~Rev.~Astron.~Astrophys.}
\bvolume{28},
\bfpage{561}.
\doiurl{https://doi.org/10.1146/annurev.aa.28.090190.003021}.
\adsurl{1990ARA&A..28..561R}.
\end{barticle}
\endbibitem

\bibitem[\protect\citeauthoryear{{Riley} et~al.}{2019}]{Riley2019}
\begin{barticle}
\bauthor{\bsnm{{Riley}}, \binits{P.}},
\bauthor{\bsnm{{Linker}}, \binits{J.A.}},
\bauthor{\bsnm{{Mikic}}, \binits{Z.}},
\bauthor{\bsnm{{Caplan}}, \binits{R.M.}},
\bauthor{\bsnm{{Downs}}, \binits{C.}},
\bauthor{\bsnm{{Thumm}}, \binits{J.-L.}}:
\byear{2019},
\batitle{{Can an Unobserved Concentration of Magnetic Flux Above the Poles of
  the Sun Resolve the Open Flux Problem?}}
\bjtitle{\apj}
\bvolume{884},
\bfpage{18}.
\doiurl{https://doi.org/10.3847/1538-4357/ab3a98}.
\adsurl{2019ApJ...884...18R}.
\end{barticle}
\endbibitem

\bibitem[\protect\citeauthoryear{{Rivera} et~al.}{2019}]{rivera2019}
\begin{barticle}
\bauthor{\bsnm{{Rivera}}, \binits{Y.J.}},
\bauthor{\bsnm{{Landi}}, \binits{E.}},
\bauthor{\bsnm{{Lepri}}, \binits{S.T.}},
\bauthor{\bsnm{{Gilbert}}, \binits{J.A.}}:
\byear{2019},
\batitle{{Empirical Modeling of CME Evolution Constrained to ACE/SWICS Charge
  State Distributions}}.
\bjtitle{\apj}
\bvolume{874},
\bfpage{164}.
\doiurl{https://doi.org/10.3847/1538-4357/ab0e11}.
\adsurl{2019ApJ...874..164R}.
\end{barticle}
\endbibitem

\bibitem[\protect\citeauthoryear{{Rivera} et~al.}{2021}]{Rivera2021}
\begin{barticle}
\bauthor{\bsnm{{Rivera}}, \binits{Y.J.}},
\bauthor{\bsnm{{Lepri}}, \binits{S.T.}},
\bauthor{\bsnm{{Raymond}}, \binits{J.C.}},
\bauthor{\bsnm{{Reeves}}, \binits{K.K.}},
\bauthor{\bsnm{{Stevens}}, \binits{M.L.}},
\bauthor{\bsnm{{Zhao}}, \binits{L.}}:
\byear{2021},
\batitle{{Solar Origin of Bare Ion Anomalies in the Solar Wind and
  Interplanetary Coronal Mass Ejections}}.
\bjtitle{\apj}
\bvolume{921},
\bfpage{93}.
\doiurl{https://doi.org/10.3847/1538-4357/ac1676}.
\adsurl{2021ApJ...921...93R}.
\end{barticle}
\endbibitem

\bibitem[\protect\citeauthoryear{{Rivera} et~al.}{2022a}]{Rivera2022b}
\begin{barticle}
\bauthor{\bsnm{{Rivera}}, \binits{Y.J.}},
\bauthor{\bsnm{{Higginson}}, \binits{A.}},
\bauthor{\bsnm{{Lepri}}, \binits{S.T.}},
\bauthor{\bsnm{{Viall}}, \binits{N.M.}},
\bauthor{\bsnm{{Alterman}}, \binits{B.L.}},
\bauthor{\bsnm{{Landi}}, \binits{E.}},
\bauthor{\bsnm{{Spitzer}}, \binits{S.A.}},
\bauthor{\bsnm{{Raines}}, \binits{J.M.}},
\bauthor{\bsnm{{Cranmer}}, \binits{S.R.}},
\bauthor{\bsnm{{Laming}}, \binits{J.M.}},
\bauthor{\bsnm{{Mason}}, \binits{E.I.}},
\bauthor{\bsnm{{Wallace}}, \binits{S.}},
\bauthor{\bsnm{{Raymond}}, \binits{J.C.}},
\bauthor{\bsnm{{Lynch}}, \binits{B.J.}},
\bauthor{\bsnm{{Gilly}}, \binits{C.R.}},
\bauthor{\bsnm{{Chen}}, \binits{T.Y.}},
\bauthor{\bsnm{{Dewey}}, \binits{R.M.}}:
\byear{2022}a,
\batitle{{Deciphering the birth region, formation, and evolution of ambient and
  transient solar wind using heavy ion observations}}.
\bjtitle{Frontiers in Astronomy and Space Sciences}
\bvolume{9},
\bfpage{1056347}.
\doiurl{https://doi.org/10.3389/fspas.2022.1056347}.
\adsurl{2022FrASS...956347R}.
\end{barticle}
\endbibitem

\bibitem[\protect\citeauthoryear{{Rivera} et~al.}{2022b}]{Rivera2022}
\begin{barticle}
\bauthor{\bsnm{{Rivera}}, \binits{Y.J.}},
\bauthor{\bsnm{{Raymond}}, \binits{J.C.}},
\bauthor{\bsnm{{Landi}}, \binits{E.}},
\bauthor{\bsnm{{Lepri}}, \binits{S.T.}},
\bauthor{\bsnm{{Reeves}}, \binits{K.K.}},
\bauthor{\bsnm{{Stevens}}, \binits{M.L.}},
\bauthor{\bsnm{{Alterman}}, \binits{B.L.}}:
\byear{2022}b,
\batitle{{Manifestation of Gravitational Settling in Coronal Mass Ejections
  Measured in the Heliosphere}}.
\bjtitle{\apj}
\bvolume{936},
\bfpage{83}.
\doiurl{https://doi.org/10.3847/1538-4357/ac8873}.
\adsurl{2022ApJ...936...83R}.
\end{barticle}
\endbibitem

\bibitem[\protect\citeauthoryear{{Robbrecht}, {Berghmans}, and {Van der
  Linden}}{2009}]{Robbrecht2009}
\begin{barticle}
\bauthor{\bsnm{{Robbrecht}}, \binits{E.}},
\bauthor{\bsnm{{Berghmans}}, \binits{D.}},
\bauthor{\bsnm{{Van der Linden}}, \binits{R.A.M.}}:
\byear{2009},
\batitle{{Automated LASCO CME Catalog for Solar Cycle 23: Are CMEs Scale
  Invariant?}}
\bjtitle{\apj}
\bvolume{691},
\bfpage{1222}.
\doiurl{https://doi.org/10.1088/0004-637X/691/2/1222}.
\adsurl{2009ApJ...691.1222R}.
\end{barticle}
\endbibitem

\bibitem[\protect\citeauthoryear{{Rochus} et~al.}{2020}]{Rochus2020}
\begin{barticle}
\bauthor{\bsnm{{Rochus}}, \binits{P.}},
\bauthor{\bsnm{{Auch{\`e}re}}, \binits{F.}},
\bauthor{\bsnm{{Berghmans}}, \binits{D.}},
\bauthor{\bsnm{{Harra}}, \binits{L.}},
\bauthor{\bsnm{{Schmutz}}, \binits{W.}},
\bauthor{\bsnm{{Sch{\"u}hle}}, \binits{U.}},
\bauthor{\bsnm{{Addison}}, \binits{P.}},
\bauthor{\bsnm{{Appourchaux}}, \binits{T.}},
\bauthor{\bsnm{{Aznar Cuadrado}}, \binits{R.}},
\bauthor{\bsnm{{Baker}}, \binits{D.}},
\bauthor{\bsnm{{Barbay}}, \binits{J.}},
\bauthor{\bsnm{{Bates}}, \binits{D.}},
\bauthor{\bsnm{{BenMoussa}}, \binits{A.}},
\bauthor{\bsnm{{Bergmann}}, \binits{M.}},
\bauthor{\bsnm{{Beurthe}}, \binits{C.}},
\bauthor{\bsnm{{Borgo}}, \binits{B.}},
\bauthor{\bsnm{{Bonte}}, \binits{K.}},
\bauthor{\bsnm{{Bouzit}}, \binits{M.}},
\bauthor{\bsnm{{Bradley}}, \binits{L.}},
\bauthor{\bsnm{{B{\"u}chel}}, \binits{V.}},
\bauthor{\bsnm{{Buchlin}}, \binits{E.}},
\bauthor{\bsnm{{B{\"u}chner}}, \binits{J.}},
\bauthor{\bsnm{{Cab{\'e}}}, \binits{F.}},
\bauthor{\bsnm{{Cadiergues}}, \binits{L.}},
\bauthor{\bsnm{{Chaigneau}}, \binits{M.}},
\bauthor{\bsnm{{Chares}}, \binits{B.}},
\bauthor{\bsnm{{Choque Cortez}}, \binits{C.}},
\bauthor{\bsnm{{Coker}}, \binits{P.}},
\bauthor{\bsnm{{Condamin}}, \binits{M.}},
\bauthor{\bsnm{{Coumar}}, \binits{S.}},
\bauthor{\bsnm{{Curdt}}, \binits{W.}},
\bauthor{\bsnm{{Cutler}}, \binits{J.}},
\bauthor{\bsnm{{Davies}}, \binits{D.}},
\bauthor{\bsnm{{Davison}}, \binits{G.}},
\bauthor{\bsnm{{Defise}}, \binits{J.-M.}},
\bauthor{\bsnm{{Del Zanna}}, \binits{G.}},
\bauthor{\bsnm{{Delmotte}}, \binits{F.}},
\bauthor{\bsnm{{Delouille}}, \binits{V.}},
\bauthor{\bsnm{{Dolla}}, \binits{L.}},
\bauthor{\bsnm{{Dumesnil}}, \binits{C.}},
\bauthor{\bsnm{{D{\"u}rig}}, \binits{F.}},
\bauthor{\bsnm{{Enge}}, \binits{R.}},
\bauthor{\bsnm{{Fran{\c{c}}ois}}, \binits{S.}},
\bauthor{\bsnm{{Fourmond}}, \binits{J.-J.}},
\bauthor{\bsnm{{Gillis}}, \binits{J.-M.}},
\bauthor{\bsnm{{Giordanengo}}, \binits{B.}},
\bauthor{\bsnm{{Gissot}}, \binits{S.}},
\bauthor{\bsnm{{Green}}, \binits{L.M.}},
\bauthor{\bsnm{{Guerreiro}}, \binits{N.}},
\bauthor{\bsnm{{Guilbaud}}, \binits{A.}},
\bauthor{\bsnm{{Gyo}}, \binits{M.}},
\bauthor{\bsnm{{Haberreiter}}, \binits{M.}},
\bauthor{\bsnm{{Hafiz}}, \binits{A.}},
\bauthor{\bsnm{{Hailey}}, \binits{M.}},
\bauthor{\bsnm{{Halain}}, \binits{J.-P.}},
\bauthor{\bsnm{{Hansotte}}, \binits{J.}},
\bauthor{\bsnm{{Hecquet}}, \binits{C.}},
\bauthor{\bsnm{{Heerlein}}, \binits{K.}},
\bauthor{\bsnm{{Hellin}}, \binits{M.-L.}},
\bauthor{\bsnm{{Hemsley}}, \binits{S.}},
\bauthor{\bsnm{{Hermans}}, \binits{A.}},
\bauthor{\bsnm{{Hervier}}, \binits{V.}},
\bauthor{\bsnm{{Hochedez}}, \binits{J.-F.}},
\bauthor{\bsnm{{Houbrechts}}, \binits{Y.}},
\bauthor{\bsnm{{Ihsan}}, \binits{K.}},
\bauthor{\bsnm{{Jacques}}, \binits{L.}},
\bauthor{\bsnm{{J{\'e}r{\^o}me}}, \binits{A.}},
\bauthor{\bsnm{{Jones}}, \binits{J.}},
\bauthor{\bsnm{{Kahle}}, \binits{M.}},
\bauthor{\bsnm{{Kennedy}}, \binits{T.}},
\bauthor{\bsnm{{Klaproth}}, \binits{M.}},
\bauthor{\bsnm{{Kolleck}}, \binits{M.}},
\bauthor{\bsnm{{Koller}}, \binits{S.}},
\bauthor{\bsnm{{Kotsialos}}, \binits{E.}},
\bauthor{\bsnm{{Kraaikamp}}, \binits{E.}},
\bauthor{\bsnm{{Langer}}, \binits{P.}},
\bauthor{\bsnm{{Lawrenson}}, \binits{A.}},
\bauthor{\bsnm{{Le Clech'}}, \binits{J.-C.}},
\bauthor{\bsnm{{Lenaerts}}, \binits{C.}},
\bauthor{\bsnm{{Liebecq}}, \binits{S.}},
\bauthor{\bsnm{{Linder}}, \binits{D.}},
\bauthor{\bsnm{{Long}}, \binits{D.M.}},
\bauthor{\bsnm{{Mampaey}}, \binits{B.}},
\bauthor{\bsnm{{Markiewicz-Innes}}, \binits{D.}},
\bauthor{\bsnm{{Marquet}}, \binits{B.}},
\bauthor{\bsnm{{Marsch}}, \binits{E.}},
\bauthor{\bsnm{{Matthews}}, \binits{S.}},
\bauthor{\bsnm{{Mazy}}, \binits{E.}},
\bauthor{\bsnm{{Mazzoli}}, \binits{A.}},
\bauthor{\bsnm{{Meining}}, \binits{S.}},
\bauthor{\bsnm{{Meltchakov}}, \binits{E.}},
\bauthor{\bsnm{{Mercier}}, \binits{R.}},
\bauthor{\bsnm{{Meyer}}, \binits{S.}},
\bauthor{\bsnm{{Monecke}}, \binits{M.}},
\bauthor{\bsnm{{Monfort}}, \binits{F.}},
\bauthor{\bsnm{{Morinaud}}, \binits{G.}},
\bauthor{\bsnm{{Moron}}, \binits{F.}},
\bauthor{\bsnm{{Mountney}}, \binits{L.}},
\bauthor{\bsnm{{M{\"u}ller}}, \binits{R.}},
\bauthor{\bsnm{{Nicula}}, \binits{B.}},
\bauthor{\bsnm{{Parenti}}, \binits{S.}},
\bauthor{\bsnm{{Peter}}, \binits{H.}},
\bauthor{\bsnm{{Pfiffner}}, \binits{D.}},
\bauthor{\bsnm{{Philippon}}, \binits{A.}},
\bauthor{\bsnm{{Phillips}}, \binits{I.}},
\bauthor{\bsnm{{Plesseria}}, \binits{J.-Y.}},
\bauthor{\bsnm{{Pylyser}}, \binits{E.}},
\bauthor{\bsnm{{Rabecki}}, \binits{F.}},
\bauthor{\bsnm{{Ravet-Krill}}, \binits{M.-F.}},
\bauthor{\bsnm{{Rebellato}}, \binits{J.}},
\bauthor{\bsnm{{Renotte}}, \binits{E.}},
\bauthor{\bsnm{{Rodriguez}}, \binits{L.}},
\bauthor{\bsnm{{Roose}}, \binits{S.}},
\bauthor{\bsnm{{Rosin}}, \binits{J.}},
\bauthor{\bsnm{{Rossi}}, \binits{L.}},
\bauthor{\bsnm{{Roth}}, \binits{P.}},
\bauthor{\bsnm{{Rouesnel}}, \binits{F.}},
\bauthor{\bsnm{{Roulliay}}, \binits{M.}},
\bauthor{\bsnm{{Rousseau}}, \binits{A.}},
\bauthor{\bsnm{{Ruane}}, \binits{K.}},
\bauthor{\bsnm{{Scanlan}}, \binits{J.}},
\bauthor{\bsnm{{Schlatter}}, \binits{P.}},
\bauthor{\bsnm{{Seaton}}, \binits{D.B.}},
\bauthor{\bsnm{{Silliman}}, \binits{K.}},
\bauthor{\bsnm{{Smit}}, \binits{S.}},
\bauthor{\bsnm{{Smith}}, \binits{P.J.}},
\bauthor{\bsnm{{Solanki}}, \binits{S.K.}},
\bauthor{\bsnm{{Spescha}}, \binits{M.}},
\bauthor{\bsnm{{Spencer}}, \binits{A.}},
\bauthor{\bsnm{{Stegen}}, \binits{K.}},
\bauthor{\bsnm{{Stockman}}, \binits{Y.}},
\bauthor{\bsnm{{Szwec}}, \binits{N.}},
\bauthor{\bsnm{{Tamiatto}}, \binits{C.}},
\bauthor{\bsnm{{Tandy}}, \binits{J.}},
\bauthor{\bsnm{{Teriaca}}, \binits{L.}},
\bauthor{\bsnm{{Theobald}}, \binits{C.}},
\bauthor{\bsnm{{Tychon}}, \binits{I.}},
\bauthor{\bsnm{{van Driel-Gesztelyi}}, \binits{L.}},
\bauthor{\bsnm{{Verbeeck}}, \binits{C.}},
\bauthor{\bsnm{{Vial}}, \binits{J.-C.}},
\bauthor{\bsnm{{Werner}}, \binits{S.}},
\bauthor{\bsnm{{West}}, \binits{M.J.}},
\bauthor{\bsnm{{Westwood}}, \binits{D.}},
\bauthor{\bsnm{{Wiegelmann}}, \binits{T.}},
\bauthor{\bsnm{{Willis}}, \binits{G.}},
\bauthor{\bsnm{{Winter}}, \binits{B.}},
\bauthor{\bsnm{{Zerr}}, \binits{A.}},
\bauthor{\bsnm{{Zhang}}, \binits{X.}},
\bauthor{\bsnm{{Zhukov}}, \binits{A.N.}}:
\byear{2020},
\batitle{{The Solar Orbiter EUI instrument: The Extreme Ultraviolet Imager}}.
\bjtitle{\aap}
\bvolume{642},
\bfpage{A8}.
\doiurl{https://doi.org/10.1051/0004-6361/201936663}.
\adsurl{2020A&A...642A...8R}.
\end{barticle}
\endbibitem

\bibitem[\protect\citeauthoryear{{Romoli} et~al.}{2021}]{romoli2021}
\begin{barticle}
\bauthor{\bsnm{{Romoli}}, \binits{M.}},
\bauthor{\bsnm{{Antonucci}}, \binits{E.}},
\bauthor{\bsnm{{Andretta}}, \binits{V.}},
\bauthor{\bsnm{{Capuano}}, \binits{G.E.}},
\bauthor{\bsnm{{Da Deppo}}, \binits{V.}},
\bauthor{\bsnm{{De Leo}}, \binits{Y.}},
\bauthor{\bsnm{{Downs}}, \binits{C.}},
\bauthor{\bsnm{{Fineschi}}, \binits{S.}},
\bauthor{\bsnm{{Heinzel}}, \binits{P.}},
\bauthor{\bsnm{{Landini}}, \binits{F.}},
\bauthor{\bsnm{{Liberatore}}, \binits{A.}},
\bauthor{\bsnm{{Naletto}}, \binits{G.}},
\bauthor{\bsnm{{Nicolini}}, \binits{G.}},
\bauthor{\bsnm{{Pancrazzi}}, \binits{M.}},
\bauthor{\bsnm{{Sasso}}, \binits{C.}},
\bauthor{\bsnm{{Spadaro}}, \binits{D.}},
\bauthor{\bsnm{{Susino}}, \binits{R.}},
\bauthor{\bsnm{{Telloni}}, \binits{D.}},
\bauthor{\bsnm{{Teriaca}}, \binits{L.}},
\bauthor{\bsnm{{Uslenghi}}, \binits{M.}},
\bauthor{\bsnm{{Wang}}, \binits{Y.-M.}},
\bauthor{\bsnm{{Bemporad}}, \binits{A.}},
\bauthor{\bsnm{{Capobianco}}, \binits{G.}},
\bauthor{\bsnm{{Casti}}, \binits{M.}},
\bauthor{\bsnm{{Fabi}}, \binits{M.}},
\bauthor{\bsnm{{Frassati}}, \binits{F.}},
\bauthor{\bsnm{{Frassetto}}, \binits{F.}},
\bauthor{\bsnm{{Giordano}}, \binits{S.}},
\bauthor{\bsnm{{Grimani}}, \binits{C.}},
\bauthor{\bsnm{{Jerse}}, \binits{G.}},
\bauthor{\bsnm{{Magli}}, \binits{E.}},
\bauthor{\bsnm{{Massone}}, \binits{G.}},
\bauthor{\bsnm{{Messerotti}}, \binits{M.}},
\bauthor{\bsnm{{Moses}}, \binits{D.}},
\bauthor{\bsnm{{Pelizzo}}, \binits{M.-G.}},
\bauthor{\bsnm{{Romano}}, \binits{P.}},
\bauthor{\bsnm{{Sch{\"u}hle}}, \binits{U.}},
\bauthor{\bsnm{{Slemer}}, \binits{A.}},
\bauthor{\bsnm{{Stangalini}}, \binits{M.}},
\bauthor{\bsnm{{Straus}}, \binits{T.}},
\bauthor{\bsnm{{Volpicelli}}, \binits{C.A.}},
\bauthor{\bsnm{{Zangrilli}}, \binits{L.}},
\bauthor{\bsnm{{Zuppella}}, \binits{P.}},
\bauthor{\bsnm{{Abbo}}, \binits{L.}},
\bauthor{\bsnm{{Auch{\`e}re}}, \binits{F.}},
\bauthor{\bsnm{{Aznar Cuadrado}}, \binits{R.}},
\bauthor{\bsnm{{Berlicki}}, \binits{A.}},
\bauthor{\bsnm{{Bruno}}, \binits{R.}},
\bauthor{\bsnm{{Ciaravella}}, \binits{A.}},
\bauthor{\bsnm{{D'Amicis}}, \binits{R.}},
\bauthor{\bsnm{{Lamy}}, \binits{P.}},
\bauthor{\bsnm{{Lanzafame}}, \binits{A.}},
\bauthor{\bsnm{{Malvezzi}}, \binits{A.M.}},
\bauthor{\bsnm{{Nicolosi}}, \binits{P.}},
\bauthor{\bsnm{{Nistic{\`o}}}, \binits{G.}},
\bauthor{\bsnm{{Peter}}, \binits{H.}},
\bauthor{\bsnm{{Plainaki}}, \binits{C.}},
\bauthor{\bsnm{{Poletto}}, \binits{L.}},
\bauthor{\bsnm{{Reale}}, \binits{F.}},
\bauthor{\bsnm{{Solanki}}, \binits{S.K.}},
\bauthor{\bsnm{{Strachan}}, \binits{L.}},
\bauthor{\bsnm{{Tondello}}, \binits{G.}},
\bauthor{\bsnm{{Tsinganos}}, \binits{K.}},
\bauthor{\bsnm{{Velli}}, \binits{M.}},
\bauthor{\bsnm{{Ventura}}, \binits{R.}},
\bauthor{\bsnm{{Vial}}, \binits{J.-C.}},
\bauthor{\bsnm{{Woch}}, \binits{J.}},
\bauthor{\bsnm{{Zimbardo}}, \binits{G.}}:
\byear{2021},
\batitle{{First light observations of the solar wind in the outer corona with
  the Metis coronagraph}}.
\bjtitle{\aap}
\bvolume{656},
\bfpage{A32}.
\doiurl{https://doi.org/10.1051/0004-6361/202140980}.
\adsurl{2021A&A...656A..32R}.
\end{barticle}
\endbibitem

\bibitem[\protect\citeauthoryear{Samra et~al.}{2021}]{Samra2021}
\begin{barticle}
\bauthor{\bsnm{Samra}, \binits{J.E.}},
\bauthor{\bsnm{Marquez}, \binits{V.}},
\bauthor{\bsnm{Cheimets}, \binits{P.}},
\bauthor{\bsnm{DeLuca}, \binits{E.E.}},
\bauthor{\bsnm{Golub}, \binits{L.}},
\bauthor{\bsnm{Hannigan}, \binits{J.W.}},
\bauthor{\bsnm{Madsen}, \binits{C.A.}},
\bauthor{\bsnm{Vira}, \binits{A.}}:
\byear{2021},
\batitle{{The Airborne Infrared Spectrometer: Development, Characterization,
  and the 21 August 2017 Eclipse Observation}}.
\bjtitle{\aj}
\bvolume{164},
\bfpage{39}.
\doiurl{https://doi.org/10.3847/1538-3881/ac7218}.
\adsurl{2022AJ....164...39S}.
\end{barticle}
\endbibitem

\bibitem[\protect\citeauthoryear{{Sanchez-Diaz} et~al.}{2017}]{SanchezDiaz2017}
\begin{barticle}
\bauthor{\bsnm{{Sanchez-Diaz}}, \binits{E.}},
\bauthor{\bsnm{{Rouillard}}, \binits{A.P.}},
\bauthor{\bsnm{{Davies}}, \binits{J.A.}},
\bauthor{\bsnm{{Lavraud}}, \binits{B.}},
\bauthor{\bsnm{{Sheeley}}, \binits{N.R.}},
\bauthor{\bsnm{{Pinto}}, \binits{R.F.}},
\bauthor{\bsnm{{Kilpua}}, \binits{E.}},
\bauthor{\bsnm{{Plotnikov}}, \binits{I.}},
\bauthor{\bsnm{{Genot}}, \binits{V.}}:
\byear{2017},
\batitle{{Observational Evidence for the Associated Formation of Blobs and
  Raining Inflows in the Solar Corona}}.
\bjtitle{\apjl}
\bvolume{835},
\bfpage{L7}.
\doiurl{https://doi.org/10.3847/2041-8213/835/1/L7}.
\adsurl{2017ApJ...835L...7S}.
\end{barticle}
\endbibitem

\bibitem[\protect\citeauthoryear{{Sarkar} et~al.}{2019}]{Sarkar2019}
\begin{barticle}
\bauthor{\bsnm{{Sarkar}}, \binits{R.}},
\bauthor{\bsnm{{Srivastava}}, \binits{N.}},
\bauthor{\bsnm{{Mierla}}, \binits{M.}},
\bauthor{\bsnm{{West}}, \binits{M.J.}},
\bauthor{\bsnm{{D'Huys}}, \binits{E.}}:
\byear{2019},
\batitle{{Evolution of the Coronal Cavity From the Quiescent to Eruptive Phase
  Associated with Coronal Mass Ejection}}.
\bjtitle{\apj}
\bvolume{875},
\bfpage{101}.
\doiurl{https://doi.org/10.3847/1538-4357/ab11c5}.
\adsurl{2019ApJ...875..101S}.
\end{barticle}
\endbibitem

\bibitem[\protect\citeauthoryear{{Savage} and {McKenzie}}{2011}]{Savage2011}
\begin{barticle}
\bauthor{\bsnm{{Savage}}, \binits{S.L.}},
\bauthor{\bsnm{{McKenzie}}, \binits{D.E.}}:
\byear{2011},
\batitle{{Quantitative Examination of a Large Sample of Supra-arcade Downflows
  in Eruptive Solar Flares}}.
\bjtitle{\apj}
\bvolume{730},
\bfpage{98}.
\doiurl{https://doi.org/10.1088/0004-637X/730/2/98}.
\adsurl{2011ApJ...730...98S}.
\end{barticle}
\endbibitem

\bibitem[\protect\citeauthoryear{{Savage}, {McKenzie}, and
  {Reeves}}{2012}]{Savage2012}
\begin{barticle}
\bauthor{\bsnm{{Savage}}, \binits{S.L.}},
\bauthor{\bsnm{{McKenzie}}, \binits{D.E.}},
\bauthor{\bsnm{{Reeves}}, \binits{K.K.}}:
\byear{2012},
\batitle{{Re-interpretation of Supra-arcade Downflows in Solar Flares}}.
\bjtitle{\apjl}
\bvolume{747},
\bfpage{L40}.
\doiurl{https://doi.org/10.1088/2041-8205/747/2/L40}.
\adsurl{2012ApJ...747L..40S}.
\end{barticle}
\endbibitem

\bibitem[\protect\citeauthoryear{{Schatten}, {Wilcox}, and
  {Ness}}{1969}]{Schatten1969}
\begin{barticle}
\bauthor{\bsnm{{Schatten}}, \binits{K.H.}},
\bauthor{\bsnm{{Wilcox}}, \binits{J.M.}},
\bauthor{\bsnm{{Ness}}, \binits{N.F.}}:
\byear{1969},
\batitle{{A model of interplanetary and coronal magnetic fields}}.
\bjtitle{\solphys}
\bvolume{6},
\bfpage{442}.
\doiurl{https://doi.org/10.1007/BF00146478}.
\adsurl{1969SoPh....6..442S}.
\end{barticle}
\endbibitem

\bibitem[\protect\citeauthoryear{{Schlenker} et~al.}{2021}]{Schlenker2021}
\begin{barticle}
\bauthor{\bsnm{{Schlenker}}, \binits{M.J.}},
\bauthor{\bsnm{{Antiochos}}, \binits{S.K.}},
\bauthor{\bsnm{{MacNeice}}, \binits{P.J.}},
\bauthor{\bsnm{{Mason}}, \binits{E.I.}}:
\byear{2021},
\batitle{{The Effect of Thermal Nonequilibrium on Helmet Streamers}}.
\bjtitle{\apj}
\bvolume{916},
\bfpage{115}.
\doiurl{https://doi.org/10.3847/1538-4357/ac069d}.
\adsurl{2021ApJ...916..115S}.
\end{barticle}
\endbibitem

\bibitem[\protect\citeauthoryear{{Schrijver} et~al.}{2008}]{Schrijver2008}
\begin{barticle}
\bauthor{\bsnm{{Schrijver}}, \binits{C.J.}},
\bauthor{\bsnm{{Elmore}}, \binits{C.}},
\bauthor{\bsnm{{Kliem}}, \binits{B.}},
\bauthor{\bsnm{{T{\"o}r{\"o}k}}, \binits{T.}},
\bauthor{\bsnm{{Title}}, \binits{A.M.}}:
\byear{2008},
\batitle{{Observations and Modeling of the Early Acceleration Phase of Erupting
  Filaments Involved in Coronal Mass Ejections}}.
\bjtitle{\apj}
\bvolume{674},
\bfpage{586}.
\doiurl{https://doi.org/10.1086/524294}.
\adsurl{2008ApJ...674..586S}.
\end{barticle}
\endbibitem

\bibitem[\protect\citeauthoryear{{Seaton} et~al.}{2013a}]{Seaton2013b}
\begin{barticle}
\bauthor{\bsnm{{Seaton}}, \binits{D.B.}},
\bauthor{\bsnm{{De Groof}}, \binits{A.}},
\bauthor{\bsnm{{Shearer}}, \binits{P.}},
\bauthor{\bsnm{{Berghmans}}, \binits{D.}},
\bauthor{\bsnm{{Nicula}}, \binits{B.}}:
\byear{2013}a,
\batitle{{SWAP Observations of the Long-term, Large-scale Evolution of the
  Extreme-ultraviolet Solar Corona}}.
\bjtitle{\apj}
\bvolume{777},
\bfpage{72}.
\doiurl{https://doi.org/10.1088/0004-637X/777/1/72}.
\adsurl{2013ApJ...777...72S}.
\end{barticle}
\endbibitem

\bibitem[\protect\citeauthoryear{{Seaton} et~al.}{2013b}]{Seaton2013}
\begin{barticle}
\bauthor{\bsnm{{Seaton}}, \binits{D.B.}},
\bauthor{\bsnm{{Berghmans}}, \binits{D.}},
\bauthor{\bsnm{{Nicula}}, \binits{B.}},
\bauthor{\bsnm{{Halain}}, \binits{J.-P.}},
\bauthor{\bsnm{{De Groof}}, \binits{A.}},
\bauthor{\bsnm{{Thibert}}, \binits{T.}},
\bauthor{\bsnm{{Bloomfield}}, \binits{D.S.}},
\bauthor{\bsnm{{Raftery}}, \binits{C.L.}},
\bauthor{\bsnm{{Gallagher}}, \binits{P.T.}},
\bauthor{\bsnm{{Auch{\`e}re}}, \binits{F.}},
\bauthor{\bsnm{{Defise}}, \binits{J.-M.}},
\bauthor{\bsnm{{D'Huys}}, \binits{E.}},
\bauthor{\bsnm{{Lecat}}, \binits{J.-H.}},
\bauthor{\bsnm{{Mazy}}, \binits{E.}},
\bauthor{\bsnm{{Rochus}}, \binits{P.}},
\bauthor{\bsnm{{Rossi}}, \binits{L.}},
\bauthor{\bsnm{{Sch{\"u}hle}}, \binits{U.}},
\bauthor{\bsnm{{Slemzin}}, \binits{V.}},
\bauthor{\bsnm{{Yalim}}, \binits{M.S.}},
\bauthor{\bsnm{{Zender}}, \binits{J.}}:
\byear{2013}b,
\batitle{{The SWAP EUV Imaging Telescope Part I: Instrument Overview and
  Pre-Flight Testing}}.
\bjtitle{\solphys}
\bvolume{286},
\bfpage{43}.
\doiurl{https://doi.org/10.1007/s11207-012-0114-6}.
\adsurl{2013SoPh..286...43S}.
\end{barticle}
\endbibitem

\bibitem[\protect\citeauthoryear{{Seaton} et~al.}{2021}]{Seaton2021}
\begin{barticle}
\bauthor{\bsnm{{Seaton}}, \binits{D.B.}},
\bauthor{\bsnm{{Hughes}}, \binits{J.M.}},
\bauthor{\bsnm{{Tadikonda}}, \binits{S.K.}},
\bauthor{\bsnm{{Caspi}}, \binits{A.}},
\bauthor{\bsnm{{DeForest}}, \binits{C.E.}},
\bauthor{\bsnm{{Krimchansky}}, \binits{A.}},
\bauthor{\bsnm{{Hurlburt}}, \binits{N.E.}},
\bauthor{\bsnm{{Seguin}}, \binits{R.}},
\bauthor{\bsnm{{Slater}}, \binits{G.}}:
\byear{2021},
\batitle{{The Sun's dynamic extended corona observed in extreme ultraviolet}}.
\bjtitle{Nat. Astron.}
\bvolume{5},
\bfpage{1029}.
\doiurl{https://doi.org/10.1038/s41550-021-01427-8}.
\adsurl{2021NatAs...5.1029S}.
\end{barticle}
\endbibitem

\bibitem[\protect\citeauthoryear{{Seaton} et~al.}{2023}]{Seaton2023}
\begin{botherref}
\oauthor{\bsnm{{Seaton}}, \binits{D.B.}},
\oauthor{\bsnm{{Berghmans}}, \binits{D.}},
\oauthor{\bsnm{{de Groof}}, \binits{A.}},
\oauthor{\bsnm{{D'Huys}}, \binits{E.}},
\oauthor{\bsnm{{Nicula}}, \binits{B.}},
\oauthor{\bsnm{{Rachmeler}}, \binits{L.A.}},
\oauthor{\bsnm{{West}}, \binits{M.J.}}:
2023,
{A Simple Azimuthally-Varying Radial Filter for Wide-Field EUV Solar Images}.
\textit{Solar Physics}
\textbf{In Prep.}.
\end{botherref}
\endbibitem

\bibitem[\protect\citeauthoryear{Sheeley and Wang}{2002}]{Sheeley2002}
\begin{barticle}
\bauthor{\bsnm{Sheeley}, \binits{J.} \bsuffix{N.~R.}},
\bauthor{\bsnm{Wang}, \binits{Y.-M.}}:
\byear{2002},
\batitle{Characteristics of Coronal Inflows}.
\bjtitle{\apj}
\bvolume{579},
\bfpage{874–887}.
\doiurl{https://doi.org/10.1086/342923}.
\end{barticle}
\endbibitem

\bibitem[\protect\citeauthoryear{{Sheeley} and {Wang}}{2014}]{Sheeley2014}
\begin{barticle}
\bauthor{\bsnm{{Sheeley}}, \binits{J.} \bsuffix{N.~R.}},
\bauthor{\bsnm{{Wang}}, \binits{Y.-M.}}:
\byear{2014},
\batitle{{Coronal Inflows during the Interval 1996-2014}}.
\bjtitle{\apj}
\bvolume{797},
\bfpage{10}.
\doiurl{https://doi.org/10.1088/0004-637X/797/1/10}.
\adsurl{2014ApJ...797...10S}.
\end{barticle}
\endbibitem

\bibitem[\protect\citeauthoryear{{Sheeley} et~al.}{1997}]{Sheeley1997}
\begin{barticle}
\bauthor{\bsnm{{Sheeley}}, \binits{N.R.}},
\bauthor{\bsnm{{Wang}}, \binits{Y.-M.}},
\bauthor{\bsnm{{Hawley}}, \binits{S.H.}},
\bauthor{\bsnm{{Brueckner}}, \binits{G.E.}},
\bauthor{\bsnm{{Dere}}, \binits{K.P.}},
\bauthor{\bsnm{{Howard}}, \binits{R.A.}},
\bauthor{\bsnm{{Koomen}}, \binits{M.J.}},
\bauthor{\bsnm{{Korendyke}}, \binits{C.M.}},
\bauthor{\bsnm{{Michels}}, \binits{D.J.}},
\bauthor{\bsnm{{Paswaters}}, \binits{S.E.}},
\bauthor{\bsnm{{Socker}}, \binits{D.G.}},
\bauthor{\bsnm{{St. Cyr}}, \binits{O.C.}},
\bauthor{\bsnm{{Wang}}, \binits{D.}},
\bauthor{\bsnm{{Lamy}}, \binits{P.L.}},
\bauthor{\bsnm{{Llebaria}}, \binits{A.}},
\bauthor{\bsnm{{Schwenn}}, \binits{R.}},
\bauthor{\bsnm{{Simnett}}, \binits{G.M.}},
\bauthor{\bsnm{{Plunkett}}, \binits{S.}},
\bauthor{\bsnm{{Biesecker}}, \binits{D.A.}}:
\byear{1997},
\batitle{{Measurements of Flow Speeds in the Corona Between 2 and 30
  R$_{{\ensuremath{\odot}}}$}}.
\bjtitle{\apj}
\bvolume{484},
\bfpage{472}.
\doiurl{https://doi.org/10.1086/304338}.
\adsurl{1997ApJ...484..472S}.
\end{barticle}
\endbibitem

\bibitem[\protect\citeauthoryear{{Shen} et~al.}{2018}]{shen2018}
\begin{barticle}
\bauthor{\bsnm{{Shen}}, \binits{C.}},
\bauthor{\bsnm{{Kong}}, \binits{X.}},
\bauthor{\bsnm{{Guo}}, \binits{F.}},
\bauthor{\bsnm{{Raymond}}, \binits{J.C.}},
\bauthor{\bsnm{{Chen}}, \binits{B.}}:
\byear{2018},
\batitle{{The Dynamical Behavior of Reconnection-driven Termination Shocks in
  Solar Flares: Magnetohydrodynamic Simulations}}.
\bjtitle{\apj}
\bvolume{869},
\bfpage{116}.
\doiurl{https://doi.org/10.3847/1538-4357/aaeed3}.
\adsurl{2018ApJ...869..116S}.
\end{barticle}
\endbibitem

\bibitem[\protect\citeauthoryear{{Shen} et~al.}{2022}]{Shen2022}
\begin{barticle}
\bauthor{\bsnm{{Shen}}, \binits{C.}},
\bauthor{\bsnm{{Chen}}, \binits{B.}},
\bauthor{\bsnm{{Reeves}}, \binits{K.K.}},
\bauthor{\bsnm{{Yu}}, \binits{S.}},
\bauthor{\bsnm{{Polito}}, \binits{V.}},
\bauthor{\bsnm{{Xie}}, \binits{X.}}:
\byear{2022},
\batitle{{The origin of underdense plasma downflows associated with magnetic
  reconnection in solar flares}}.
\bjtitle{Nat. Astron.}
\bvolume{6},
\bfpage{317}.
\doiurl{https://doi.org/10.1038/s41550-021-01570-2}.
\adsurl{2022NatAs...6..317S}.
\end{barticle}
\endbibitem

\bibitem[\protect\citeauthoryear{{Shestov} et~al.}{2021}]{Shestov2021}
\begin{barticle}
\bauthor{\bsnm{{Shestov}}, \binits{S.V.}},
\bauthor{\bsnm{{Zhukov}}, \binits{A.N.}},
\bauthor{\bsnm{{Inhester}}, \binits{B.}},
\bauthor{\bsnm{{Dolla}}, \binits{L.}},
\bauthor{\bsnm{{Mierla}}, \binits{M.}}:
\byear{2021},
\batitle{{Expected performances of the PROBA-3/ASPIICS solar coronagraph:
  Simulated data}}.
\bjtitle{\aap}
\bvolume{652},
\bfpage{A4}.
\doiurl{https://doi.org/10.1051/0004-6361/202140467}.
\adsurl{2021A&A...652A...4S}.
\end{barticle}
\endbibitem

\bibitem[\protect\citeauthoryear{{Shoda}, {Yokoyama}, and
  {Suzuki}}{2018}]{Shoda2018}
\begin{barticle}
\bauthor{\bsnm{{Shoda}}, \binits{M.}},
\bauthor{\bsnm{{Yokoyama}}, \binits{T.}},
\bauthor{\bsnm{{Suzuki}}, \binits{T.K.}}:
\byear{2018},
\batitle{{A Self-consistent Model of the Coronal Heating and Solar Wind
  Acceleration Including Compressible and Incompressible Heating Processes}}.
\bjtitle{\apj}
\bvolume{853},
\bfpage{190}.
\doiurl{https://doi.org/10.3847/1538-4357/aaa3e1}.
\adsurl{2018ApJ...853..190S}.
\end{barticle}
\endbibitem

\bibitem[\protect\citeauthoryear{{Sieyra} et~al.}{2020}]{Sieyra2020}
\begin{barticle}
\bauthor{\bsnm{{Sieyra}}, \binits{M.V.}},
\bauthor{\bsnm{{C{\'e}cere}}, \binits{M.}},
\bauthor{\bsnm{{Cremades}}, \binits{H.}},
\bauthor{\bsnm{{Iglesias}}, \binits{F.A.}},
\bauthor{\bsnm{{Sahade}}, \binits{A.}},
\bauthor{\bsnm{{Mierla}}, \binits{M.}},
\bauthor{\bsnm{{Stenborg}}, \binits{G.}},
\bauthor{\bsnm{{Costa}}, \binits{A.}},
\bauthor{\bsnm{{West}}, \binits{M.J.}},
\bauthor{\bsnm{{D'Huys}}, \binits{E.}}:
\byear{2020},
\batitle{{Analysis of Large Deflections of Prominence-CME Events during the
  Rising Phase of Solar Cycle 24}}.
\bjtitle{\solphys}
\bvolume{295},
\bfpage{126}.
\doiurl{https://doi.org/10.1007/s11207-020-01694-0}.
\adsurl{2020SoPh..295..126S}.
\end{barticle}
\endbibitem

\bibitem[\protect\citeauthoryear{{Simon}, {Title}, and
  {Weiss}}{2001}]{SimonTitleWeiss2001}
\begin{barticle}
\bauthor{\bsnm{{Simon}}, \binits{G.W.}},
\bauthor{\bsnm{{Title}}, \binits{A.M.}},
\bauthor{\bsnm{{Weiss}}, \binits{N.O.}}:
\byear{2001},
\batitle{{Sustaining the Sun's Magnetic Network with Emerging Bipoles}}.
\bjtitle{\apj}
\bvolume{561},
\bfpage{427}.
\doiurl{https://doi.org/10.1086/322243}.
\adsurl{2001ApJ...561..427S}.
\end{barticle}
\endbibitem

\bibitem[\protect\citeauthoryear{{Soler} et~al.}{2019}]{Soler2019}
\begin{barticle}
\bauthor{\bsnm{{Soler}}, \binits{R.}},
\bauthor{\bsnm{{Terradas}}, \binits{J.}},
\bauthor{\bsnm{{Oliver}}, \binits{R.}},
\bauthor{\bsnm{{Ballester}}, \binits{J.L.}}:
\byear{2019},
\batitle{{Energy Transport and Heating by Torsional Alfv{\'e}n Waves
  Propagating from the Photosphere to the Corona in the Quiet Sun}}.
\bjtitle{\apj}
\bvolume{871},
\bfpage{3}.
\doiurl{https://doi.org/10.3847/1538-4357/aaf64c}.
\adsurl{2019ApJ...871....3S}.
\end{barticle}
\endbibitem

\bibitem[\protect\citeauthoryear{Stenborg, Vourlidas, and
  Howard}{2008}]{Stenborg2008}
\begin{barticle}
\bauthor{\bsnm{Stenborg}, \binits{G.}},
\bauthor{\bsnm{Vourlidas}, \binits{A.}},
\bauthor{\bsnm{Howard}, \binits{R.A.}}:
\byear{2008},
\batitle{A Fresh View of the Extreme-Ultraviolet Corona from the Application of
  a New Image-Processing Technique}.
\bjtitle{\apj}
\bvolume{674},
\bfpage{1201–1206}.
\doiurl{https://doi.org/10.1086/525556}.
\end{barticle}
\endbibitem

\bibitem[\protect\citeauthoryear{{Sterling} et~al.}{2015}]{Sterling2015}
\begin{barticle}
\bauthor{\bsnm{{Sterling}}, \binits{A.C.}},
\bauthor{\bsnm{{Moore}}, \binits{R.L.}},
\bauthor{\bsnm{{Falconer}}, \binits{D.A.}},
\bauthor{\bsnm{{Adams}}, \binits{M.}}:
\byear{2015},
\batitle{{Small-scale filament eruptions as the driver of X-ray jets in solar
  coronal holes}}.
\bjtitle{\nat}
\bvolume{523},
\bfpage{437}.
\doiurl{https://doi.org/10.1038/nature14556}.
\adsurl{2015Natur.523..437S}.
\end{barticle}
\endbibitem

\bibitem[\protect\citeauthoryear{{Strachan} et~al.}{1993}]{strachan1993}
\begin{barticle}
\bauthor{\bsnm{{Strachan}}, \binits{L.}},
\bauthor{\bsnm{{Kohl}}, \binits{J.L.}},
\bauthor{\bsnm{{Weiser}}, \binits{H.}},
\bauthor{\bsnm{{Withbroe}}, \binits{G.L.}},
\bauthor{\bsnm{{Munro}}, \binits{R.H.}}:
\byear{1993},
\batitle{{A Doppler Dimming Determination of Coronal Outflow Velocity}}.
\bjtitle{\apj}
\bvolume{412},
\bfpage{410}.
\doiurl{https://doi.org/10.1086/172930}.
\adsurl{1993ApJ...412..410S}.
\end{barticle}
\endbibitem

\bibitem[\protect\citeauthoryear{{Strachan} et~al.}{2002}]{Strachan2002}
\begin{barticle}
\bauthor{\bsnm{{Strachan}}, \binits{L.}},
\bauthor{\bsnm{{Suleiman}}, \binits{R.}},
\bauthor{\bsnm{{Panasyuk}}, \binits{A.V.}},
\bauthor{\bsnm{{Biesecker}}, \binits{D.A.}},
\bauthor{\bsnm{{Kohl}}, \binits{J.L.}}:
\byear{2002},
\batitle{{Empirical Densities, Kinetic Temperatures, and Outflow Velocities in
  the Equatorial Streamer Belt at Solar Minimum}}.
\bjtitle{\apj}
\bvolume{571},
\bfpage{1008}.
\doiurl{https://doi.org/10.1086/339984}.
\adsurl{2002ApJ...571.1008S}.
\end{barticle}
\endbibitem

\bibitem[\protect\citeauthoryear{{Strachan} et~al.}{2017}]{Strachan2017}
\begin{bchapter}
\bauthor{\bsnm{{Strachan}}, \binits{L.}},
\bauthor{\bsnm{{Laming}}, \binits{J.M.}},
\bauthor{\bsnm{{Ko}}, \binits{Y.-K.}},
\bauthor{\bsnm{{Tun Beltran}}, \binits{S.}},
\bauthor{\bsnm{{Korendyke}}, \binits{C.M.}},
\bauthor{\bsnm{{Brown}}, \binits{C.M.}},
\bauthor{\bsnm{{Socker}}, \binits{D.G.}},
\bauthor{\bsnm{{Galysh}}, \binits{I.J.}},
\bauthor{\bsnm{{Finne}}, \binits{T.T.}},
\bauthor{\bsnm{{Eisenhower}}, \binits{K.C.}},
\bauthor{\bsnm{{Brechbiel}}, \binits{D.J.}},
\bauthor{\bsnm{{Noya}}, \binits{M.}},
\bauthor{\bsnm{{Provornikova}}, \binits{E.}},
\bauthor{\bsnm{{Gardner}}, \binits{L.D.}}:
\byear{2017},
\bctitle{{The Ultraviolet Spectro-Coronagraph (UVSC) Pathfinder Experiment for
  the Remote Detection of Suprathermal Seed Particles: Instrument Status}}.
In: \bbtitle{AAS/Solar Physics Division Abstracts \#48},
\bsertitle{AAS/Solar Physics Division Meeting}
\bseriesno{48},
\bfpage{110.07}.
\adsurl{2017SPD....4811007S}.
\end{bchapter}
\endbibitem

\bibitem[\protect\citeauthoryear{{Telloni}, {Giordano}, and
  {Antonucci}}{2019}]{Telloni2019}
\begin{barticle}
\bauthor{\bsnm{{Telloni}}, \binits{D.}},
\bauthor{\bsnm{{Giordano}}, \binits{S.}},
\bauthor{\bsnm{{Antonucci}}, \binits{E.}}:
\byear{2019},
\batitle{{On the Fast Solar Wind Heating and Acceleration Processes: A
  Statistical Study Based on the UVCS Survey Data}}.
\bjtitle{\apjl}
\bvolume{881},
\bfpage{L36}.
\doiurl{https://doi.org/10.3847/2041-8213/ab3731}.
\adsurl{2019ApJ...881L..36T}.
\end{barticle}
\endbibitem

\bibitem[\protect\citeauthoryear{{Terradas}, {Goossens}, and
  {Verth}}{2010}]{Terradas2010}
\begin{barticle}
\bauthor{\bsnm{{Terradas}}, \binits{J.}},
\bauthor{\bsnm{{Goossens}}, \binits{M.}},
\bauthor{\bsnm{{Verth}}, \binits{G.}}:
\byear{2010},
\batitle{{Selective spatial damping of propagating kink waves due to resonant
  absorption}}.
\bjtitle{\aap}
\bvolume{524},
\bfpage{A23}.
\doiurl{https://doi.org/10.1051/0004-6361/201014845}.
\adsurl{2010A&A...524A..23T}.
\end{barticle}
\endbibitem

\bibitem[\protect\citeauthoryear{Thalmann et~al.}{2015}]{Thalmann2015}
\begin{barticle}
\bauthor{\bsnm{Thalmann}, \binits{J.K.}},
\bauthor{\bsnm{Su}, \binits{Y.}},
\bauthor{\bsnm{Temmer}, \binits{M.}},
\bauthor{\bsnm{Veronig}, \binits{A.M.}}:
\byear{2015},
\batitle{{The Confined X-Class Flares of Solar Active Region 2192}}.
\bjtitle{\apj}
\bvolume{801},
\bfpage{L23}.
\doiurl{https://doi.org/10.1088/2041-8205/801/2/L23}.
\end{barticle}
\endbibitem

\bibitem[\protect\citeauthoryear{{Thernisien}, {Vourlidas}, and
  {Howard}}{2009}]{Thernisien2009}
\begin{barticle}
\bauthor{\bsnm{{Thernisien}}, \binits{A.}},
\bauthor{\bsnm{{Vourlidas}}, \binits{A.}},
\bauthor{\bsnm{{Howard}}, \binits{R.A.}}:
\byear{2009},
\batitle{{Forward Modeling of Coronal Mass Ejections Using STEREO/SECCHI
  Data}}.
\bjtitle{\solphys}
\bvolume{256},
\bfpage{111}.
\doiurl{https://doi.org/10.1007/s11207-009-9346-5}.
\adsurl{2009SoPh..256..111T}.
\end{barticle}
\endbibitem

\bibitem[\protect\citeauthoryear{{Thernisien}, {Howard}, and
  {Vourlidas}}{2006}]{Thernisien2006}
\begin{barticle}
\bauthor{\bsnm{{Thernisien}}, \binits{A.F.R.}},
\bauthor{\bsnm{{Howard}}, \binits{R.A.}},
\bauthor{\bsnm{{Vourlidas}}, \binits{A.}}:
\byear{2006},
\batitle{{Modeling of Flux Rope Coronal Mass Ejections}}.
\bjtitle{\apj}
\bvolume{652},
\bfpage{763}.
\doiurl{https://doi.org/10.1086/508254}.
\adsurl{2006ApJ...652..763T}.
\end{barticle}
\endbibitem

\bibitem[\protect\citeauthoryear{{Thurgood}, {Morton}, and
  {McLaughlin}}{2014}]{Thurgood2014}
\begin{barticle}
\bauthor{\bsnm{{Thurgood}}, \binits{J.O.}},
\bauthor{\bsnm{{Morton}}, \binits{R.J.}},
\bauthor{\bsnm{{McLaughlin}}, \binits{J.A.}}:
\byear{2014},
\batitle{{First Direct Measurements of Transverse Waves in Solar Polar Plumes
  Using SDO/AIA}}.
\bjtitle{\apjl}
\bvolume{790},
\bfpage{L2}.
\doiurl{https://doi.org/10.1088/2041-8205/790/1/L2}.
\adsurl{2014ApJ...790L...2T}.
\end{barticle}
\endbibitem

\bibitem[\protect\citeauthoryear{Titov et~al.}{}]{Titov2011}
\begin{botherref}
\oauthor{\bsnm{Titov}, \binits{V.S.}},
\oauthor{\bsnm{Miki{\'{c}}}, \binits{Z.}},
\oauthor{\bsnm{Linker}, \binits{J.A.}},
\oauthor{\bsnm{Lionello}, \binits{R.}},
\oauthor{\bsnm{Antiochos}, \binits{S.K.}}:
Magnetic Topology of Coronal Hole Linkages.
\textit{\apj},
111.
\doiurl{https://doi.org/10.1088/0004-637X/731/2/111}.
\end{botherref}
\endbibitem

\bibitem[\protect\citeauthoryear{{Tomczyk} et~al.}{2007}]{Tomczyk2007}
\begin{barticle}
\bauthor{\bsnm{{Tomczyk}}, \binits{S.}},
\bauthor{\bsnm{{McIntosh}}, \binits{S.W.}},
\bauthor{\bsnm{{Keil}}, \binits{S.L.}},
\bauthor{\bsnm{{Judge}}, \binits{P.G.}},
\bauthor{\bsnm{{Schad}}, \binits{T.}},
\bauthor{\bsnm{{Seeley}}, \binits{D.H.}},
\bauthor{\bsnm{{Edmondson}}, \binits{J.}}:
\byear{2007},
\batitle{{Alfv{\'e}n Waves in the Solar Corona}}.
\bjtitle{Science}
\bvolume{317},
\bfpage{1192}.
\doiurl{https://doi.org/10.1126/science.1143304}.
\adsurl{2007Sci...317.1192T}.
\end{barticle}
\endbibitem

\bibitem[\protect\citeauthoryear{{Tomczyk} et~al.}{2008}]{Tomczyk2008}
\begin{barticle}
\bauthor{\bsnm{{Tomczyk}}, \binits{S.}},
\bauthor{\bsnm{{Card}}, \binits{G.L.}},
\bauthor{\bsnm{{Darnell}}, \binits{T.}},
\bauthor{\bsnm{{Elmore}}, \binits{D.F.}},
\bauthor{\bsnm{{Lull}}, \binits{R.}},
\bauthor{\bsnm{{Nelson}}, \binits{P.G.}},
\bauthor{\bsnm{{Streander}}, \binits{K.V.}},
\bauthor{\bsnm{{Burkepile}}, \binits{J.}},
\bauthor{\bsnm{{Casini}}, \binits{R.}},
\bauthor{\bsnm{{Judge}}, \binits{P.G.}}:
\byear{2008},
\batitle{{An Instrument to Measure Coronal Emission Line Polarization}}.
\bjtitle{\solphys}
\bvolume{247},
\bfpage{411}.
\doiurl{https://doi.org/10.1007/s11207-007-9103-6}.
\adsurl{2008SoPh..247..411T}.
\end{barticle}
\endbibitem

\bibitem[\protect\citeauthoryear{{Tomczyk} et~al.}{2016}]{Tomczyk2016}
\begin{barticle}
\bauthor{\bsnm{{Tomczyk}}, \binits{S.}},
\bauthor{\bsnm{{Landi}}, \binits{E.}},
\bauthor{\bsnm{{Burkepile}}, \binits{J.T.}},
\bauthor{\bsnm{{Casini}}, \binits{R.}},
\bauthor{\bsnm{{DeLuca}}, \binits{E.E.}},
\bauthor{\bsnm{{Fan}}, \binits{Y.}},
\bauthor{\bsnm{{Gibson}}, \binits{S.E.}},
\bauthor{\bsnm{{Lin}}, \binits{H.}},
\bauthor{\bsnm{{McIntosh}}, \binits{S.W.}},
\bauthor{\bsnm{{Solomon}}, \binits{S.C.}},
\bauthor{\bsnm{{Toma}}, \binits{G.}},
\bauthor{\bsnm{{Wijn}}, \binits{A.G.}},
\bauthor{\bsnm{{Zhang}}, \binits{J.}}:
\byear{2016},
\batitle{{Scientific objectives and capabilities of the Coronal Solar Magnetism
  Observatory}}.
\bjtitle{J. Geophys. Res. (Space Phys.)}
\bvolume{121},
\bfpage{7470}.
\doiurl{https://doi.org/10.1002/2016JA022871}.
\adsurl{2016JGRA..121.7470T}.
\end{barticle}
\endbibitem

\bibitem[\protect\citeauthoryear{{T{\"o}r{\"o}k} et~al.}{2011}]{Torok2011}
\begin{barticle}
\bauthor{\bsnm{{T{\"o}r{\"o}k}}, \binits{T.}},
\bauthor{\bsnm{{Panasenco}}, \binits{O.}},
\bauthor{\bsnm{{Titov}}, \binits{V.S.}},
\bauthor{\bsnm{{Miki{\'c}}}, \binits{Z.}},
\bauthor{\bsnm{{Reeves}}, \binits{K.K.}},
\bauthor{\bsnm{{Velli}}, \binits{M.}},
\bauthor{\bsnm{{Linker}}, \binits{J.A.}},
\bauthor{\bsnm{{De Toma}}, \binits{G.}}:
\byear{2011},
\batitle{{A Model for Magnetically Coupled Sympathetic Eruptions}}.
\bjtitle{\apjl}
\bvolume{739},
\bfpage{L63}.
\doiurl{https://doi.org/10.1088/2041-8205/739/2/L63}.
\adsurl{2011ApJ...739L..63T}.
\end{barticle}
\endbibitem

\bibitem[\protect\citeauthoryear{{Tousey} et~al.}{1973}]{Tousey1973}
\begin{barticle}
\bauthor{\bsnm{{Tousey}}, \binits{R.}},
\bauthor{\bsnm{{Bartoe}}, \binits{J.D.F.}},
\bauthor{\bsnm{{Bohlin}}, \binits{J.D.}},
\bauthor{\bsnm{{Brueckner}}, \binits{G.E.}},
\bauthor{\bsnm{{Purcell}}, \binits{J.D.}},
\bauthor{\bsnm{{Scherrer}}, \binits{V.E.}},
\bauthor{\bsnm{{Sheeley}}, \binits{J.} \bsuffix{N.~R.}},
\bauthor{\bsnm{{Schumacher}}, \binits{R.J.}},
\bauthor{\bsnm{{Vanhoosier}}, \binits{M.E.}}:
\byear{1973},
\batitle{{A Preliminary Study of the Extreme Ultraviolet Spectroheliograms from
  Skylab}}.
\bjtitle{\solphys}
\bvolume{33},
\bfpage{265}.
\doiurl{https://doi.org/10.1007/BF00152418}.
\adsurl{1973SoPh...33..265T}.
\end{barticle}
\endbibitem

\bibitem[\protect\citeauthoryear{{Tousey} et~al.}{1977}]{Tousey1977}
\begin{barticle}
\bauthor{\bsnm{{Tousey}}, \binits{R.}},
\bauthor{\bsnm{{Bartoe}}, \binits{J.-D.F.}},
\bauthor{\bsnm{{Brueckner}}, \binits{G.E.}},
\bauthor{\bsnm{{Purcell}}, \binits{J.D.}}:
\byear{1977},
\batitle{{Extreme ultraviolet spectroheliograph ATM experiment S082A.}}
\bjtitle{\applopt}
\bvolume{16},
\bfpage{870}.
\doiurl{https://doi.org/10.1364/AO.16.000870}.
\adsurl{1977ApOpt..16..870T}.
\end{barticle}
\endbibitem

\bibitem[\protect\citeauthoryear{Upton and Hathaway}{2014}]{Upton2014}
\begin{barticle}
\bauthor{\bsnm{Upton}, \binits{L.}},
\bauthor{\bsnm{Hathaway}, \binits{D.H.}}:
\byear{2014},
\batitle{Effects of Meridional Flow Variations on Solar Cycles 23 and 24}.
\bjtitle{\apj}
\bvolume{792},
\bfpage{142}.
\doiurl{https://doi.org/10.1088/0004-637x/792/2/142}.
\burl{https://doi.org/10.1088/0004-637x/792/2/142}.
\end{barticle}
\endbibitem

\bibitem[\protect\citeauthoryear{{Uritsky} et~al.}{2021}]{Uritsky2021}
\begin{barticle}
\bauthor{\bsnm{{Uritsky}}, \binits{V.M.}},
\bauthor{\bsnm{{DeForest}}, \binits{C.E.}},
\bauthor{\bsnm{{Karpen}}, \binits{J.T.}},
\bauthor{\bsnm{{DeVore}}, \binits{C.R.}},
\bauthor{\bsnm{{Kumar}}, \binits{P.}},
\bauthor{\bsnm{{Raouafi}}, \binits{N.E.}},
\bauthor{\bsnm{{Wyper}}, \binits{P.F.}}:
\byear{2021},
\batitle{{Plumelets: Dynamic Filamentary Structures in Solar Coronal Plumes}}.
\bjtitle{\apj}
\bvolume{907},
\bfpage{1}.
\doiurl{https://doi.org/10.3847/1538-4357/abd186}.
\adsurl{2021ApJ...907....1U}.
\end{barticle}
\endbibitem

\bibitem[\protect\citeauthoryear{{Uzzo}, {Ko}, and {Raymond}}{2004}]{uzzo2004}
\begin{barticle}
\bauthor{\bsnm{{Uzzo}}, \binits{M.}},
\bauthor{\bsnm{{Ko}}, \binits{Y.-K.}},
\bauthor{\bsnm{{Raymond}}, \binits{J.C.}}:
\byear{2004},
\batitle{{Active Region Streamer Diagnostics 2001 September 14-16}}.
\bjtitle{\apj}
\bvolume{603},
\bfpage{760}.
\doiurl{https://doi.org/10.1086/381525}.
\adsurl{2004ApJ...603..760U}.
\end{barticle}
\endbibitem

\bibitem[\protect\citeauthoryear{{Uzzo} et~al.}{2003}]{uzzo2003}
\begin{barticle}
\bauthor{\bsnm{{Uzzo}}, \binits{M.}},
\bauthor{\bsnm{{Ko}}, \binits{Y.-K.}},
\bauthor{\bsnm{{Raymond}}, \binits{J.C.}},
\bauthor{\bsnm{{Wurz}}, \binits{P.}},
\bauthor{\bsnm{{Ipavich}}, \binits{F.M.}}:
\byear{2003},
\batitle{{Elemental Abundances for the 1996 Streamer Belt}}.
\bjtitle{\apj}
\bvolume{585},
\bfpage{1062}.
\doiurl{https://doi.org/10.1086/346132}.
\adsurl{2003ApJ...585.1062U}.
\end{barticle}
\endbibitem

\bibitem[\protect\citeauthoryear{{Uzzo} et~al.}{2006}]{Uzzo2006}
\begin{barticle}
\bauthor{\bsnm{{Uzzo}}, \binits{M.}},
\bauthor{\bsnm{{Strachan}}, \binits{L.}},
\bauthor{\bsnm{{Vourlidas}}, \binits{A.}},
\bauthor{\bsnm{{Ko}}, \binits{Y.-K.}},
\bauthor{\bsnm{{Raymond}}, \binits{J.C.}}:
\byear{2006},
\batitle{{Physical Properties of a 2003 April Quiescent Streamer}}.
\bjtitle{\apj}
\bvolume{645},
\bfpage{720}.
\doiurl{https://doi.org/10.1086/504286}.
\adsurl{2006ApJ...645..720U}.
\end{barticle}
\endbibitem

\bibitem[\protect\citeauthoryear{{van Ballegooijen}, {Priest}, and
  {Mackay}}{2000}]{vanBallegooijen2000}
\begin{barticle}
\bauthor{\bsnm{{van Ballegooijen}}, \binits{A.A.}},
\bauthor{\bsnm{{Priest}}, \binits{E.R.}},
\bauthor{\bsnm{{Mackay}}, \binits{D.H.}}:
\byear{2000},
\batitle{{Mean Field Model for the Formation of Filament Channels on the Sun}}.
\bjtitle{\apj}
\bvolume{539},
\bfpage{983}.
\doiurl{https://doi.org/10.1086/309265}.
\adsurl{2000ApJ...539..983V}.
\end{barticle}
\endbibitem

\bibitem[\protect\citeauthoryear{{van de Hulst}}{1950}]{vandehulst:1950}
\begin{barticle}
\bauthor{\bsnm{{van de Hulst}}, \binits{H.C.}}:
\byear{1950},
\batitle{{The electron density of the solar corona}}.
\bjtitle{Bull.~Astron.~Inst.~Netherlands}
\bvolume{11},
\bfpage{135}.
\adsurl{1950BAN....11..135V}.
\end{barticle}
\endbibitem

\bibitem[\protect\citeauthoryear{{V{\'a}squez}, {van Ballegooijen}, and
  {Raymond}}{2003}]{Vasquez2003}
\begin{barticle}
\bauthor{\bsnm{{V{\'a}squez}}, \binits{A.M.}},
\bauthor{\bsnm{{van Ballegooijen}}, \binits{A.A.}},
\bauthor{\bsnm{{Raymond}}, \binits{J.C.}}:
\byear{2003},
\batitle{{The Effect of Proton Temperature Anisotropy on the Solar Minimum
  Corona and Wind}}.
\bjtitle{\apj}
\bvolume{598},
\bfpage{1361}.
\doiurl{https://doi.org/10.1086/379008}.
\adsurl{2003ApJ...598.1361V}.
\end{barticle}
\endbibitem

\bibitem[\protect\citeauthoryear{{Vernazza} and {Reeves}}{1978}]{Vernazza1978}
\begin{barticle}
\bauthor{\bsnm{{Vernazza}}, \binits{J.E.}},
\bauthor{\bsnm{{Reeves}}, \binits{E.M.}}:
\byear{1978},
\batitle{{Extreme ultraviolet composite spectra of representative solar
  features.}}
\bjtitle{\apjs}
\bvolume{37},
\bfpage{485}.
\doiurl{https://doi.org/10.1086/190539}.
\adsurl{1978ApJS...37..485V}.
\end{barticle}
\endbibitem

\bibitem[\protect\citeauthoryear{{Verscharen}, {Klein}, and
  {Maruca}}{2019}]{Verscharen2019}
\begin{barticle}
\bauthor{\bsnm{{Verscharen}}, \binits{D.}},
\bauthor{\bsnm{{Klein}}, \binits{K.G.}},
\bauthor{\bsnm{{Maruca}}, \binits{B.A.}}:
\byear{2019},
\batitle{{The multi-scale nature of the solar wind}}.
\bjtitle{Liv. Rev. Solar Phys,}
\bvolume{16},
\bfpage{5}.
\doiurl{https://doi.org/10.1007/s41116-019-0021-0}.
\adsurl{2019LRSP...16....5V}.
\end{barticle}
\endbibitem

\bibitem[\protect\citeauthoryear{{Viall} and {Borovsky}}{2020}]{Viall2020}
\begin{barticle}
\bauthor{\bsnm{{Viall}}, \binits{N.M.}},
\bauthor{\bsnm{{Borovsky}}, \binits{J.E.}}:
\byear{2020},
\batitle{{Nine Outstanding Questions of Solar Wind Physics}}.
\bjtitle{J. Geophys. Res. (Space Phys.)}
\bvolume{125},
\bfpage{e26005}.
\doiurl{https://doi.org/10.1029/2018JA026005}.
\adsurl{2020JGRA..12526005V}.
\end{barticle}
\endbibitem

\bibitem[\protect\citeauthoryear{{von Steiger} et~al.}{2000}]{vonSteiger2000}
\begin{barticle}
\bauthor{\bsnm{{von Steiger}}, \binits{R.}},
\bauthor{\bsnm{{Schwadron}}, \binits{N.A.}},
\bauthor{\bsnm{{Fisk}}, \binits{L.A.}},
\bauthor{\bsnm{{Geiss}}, \binits{J.}},
\bauthor{\bsnm{{Gloeckler}}, \binits{G.}},
\bauthor{\bsnm{{Hefti}}, \binits{S.}},
\bauthor{\bsnm{{Wilken}}, \binits{B.}},
\bauthor{\bsnm{{Wimmer-Schweingruber}}, \binits{R.F.}},
\bauthor{\bsnm{{Zurbuchen}}, \binits{T.H.}}:
\byear{2000},
\batitle{{Composition of quasi-stationary solar wind flows from Ulysses/Solar
  Wind Ion Composition Spectrometer}}.
\bjtitle{\jgr}
\bvolume{105},
\bfpage{27217}.
\doiurl{https://doi.org/10.1029/1999JA000358}.
\adsurl{2000JGR...10527217V}.
\end{barticle}
\endbibitem

\bibitem[\protect\citeauthoryear{Vourlidas and Webb}{2018}]{Vourlidas2018}
\begin{barticle}
\bauthor{\bsnm{Vourlidas}, \binits{A.}},
\bauthor{\bsnm{Webb}, \binits{D.F.}}:
\byear{2018},
\batitle{{Streamer-blowout Coronal Mass Ejections: Their Properties and
  Relation to the Coronal Magnetic Field Structure}}.
\bjtitle{\apj}
\bvolume{861},
\bfpage{103}.
\doiurl{https://doi.org/10.3847/1538-4357/aaca3e}.
\end{barticle}
\endbibitem

\bibitem[\protect\citeauthoryear{Vourlidas, Carley, and
  Vilmer}{2020}]{Vourlidas_Carley_Vilmer_2020}
\begin{barticle}
\bauthor{\bsnm{Vourlidas}, \binits{A.}},
\bauthor{\bsnm{Carley}, \binits{E.P.}},
\bauthor{\bsnm{Vilmer}, \binits{N.}}:
\byear{2020},
\batitle{Radio Observations of Coronal Mass Ejections: Space Weather Aspects}.
\bjtitle{Front. Astron. Space Sci.}
\bvolume{7}.
\doiurl{https://doi.org/10.3389/fspas.2020.00043}.
\end{barticle}
\endbibitem

\bibitem[\protect\citeauthoryear{Vourlidas et~al.}{2017}]{Vourlidas2017}
\begin{barticle}
\bauthor{\bsnm{Vourlidas}, \binits{A.}},
\bauthor{\bsnm{Balmaceda}, \binits{L.A.}},
\bauthor{\bsnm{Stenborg}, \binits{G.}},
\bauthor{\bsnm{Lago}, \binits{A.D.}}:
\byear{2017},
\batitle{{ Multi-viewpoint Coronal Mass Ejection Catalog Based on STEREO COR2
  Observations }}.
\bjtitle{\apj}
\bvolume{838},
\bfpage{141}.
\doiurl{https://doi.org/10.3847/1538-4357/aa67f0}.
\end{barticle}
\endbibitem

\bibitem[\protect\citeauthoryear{Vourlidas et~al.}{2020}]{Vourlidas2020}
\begin{botherref}
\oauthor{\bsnm{Vourlidas}, \binits{A.}},
\oauthor{\bsnm{Viall}, \binits{N.}},
\oauthor{\bsnm{Laming}, \binits{M.}},
\oauthor{\bsnm{Cranmer}, \binits{S.}},
\oauthor{\bsnm{Arge}, \binits{C.}},
\oauthor{\bsnm{DeForest}, \binits{C.}},
\oauthor{\bparticle{de} \bsnm{Toma}, \binits{G.}},
\oauthor{\bsnm{Caspi}, \binits{A.}},
\oauthor{\bsnm{Raouafi}, \binits{N.}}:
2020,
Exploring the Critical Coronal Transition Region: The Key to Uncovering the
  Genesis of the Solar Wind and Solar Eruptions.
\textit{Earth Space Sci. Open Archive},
3.
\doiurl{https://doi.org/10.1002/essoar.10504451.1}.
\end{botherref}
\endbibitem

\bibitem[\protect\citeauthoryear{{Wang}}{1996}]{Wang1996}
\begin{barticle}
\bauthor{\bsnm{{Wang}}, \binits{Y.-M.}}:
\byear{1996},
\batitle{{Nonradial Coronal Streamers}}.
\bjtitle{\apjl}
\bvolume{456},
\bfpage{L119}.
\doiurl{https://doi.org/10.1086/309871}.
\adsurl{1996ApJ...456L.119W}.
\end{barticle}
\endbibitem

\bibitem[\protect\citeauthoryear{{Wang}}{2015}]{Wang2015}
\begin{barticle}
\bauthor{\bsnm{{Wang}}, \binits{Y.-M.}}:
\byear{2015},
\batitle{{Pseudostreamers as the Source of a Separate Class of Solar Coronal
  Mass Ejections}}.
\bjtitle{\apjl}
\bvolume{803},
\bfpage{L12}.
\doiurl{https://doi.org/10.1088/2041-8205/803/1/L12}.
\adsurl{2015ApJ...803L..12W}.
\end{barticle}
\endbibitem

\bibitem[\protect\citeauthoryear{Wang and Hess}{2018}]{Wang2018}
\begin{barticle}
\bauthor{\bsnm{Wang}, \binits{Y.-M.}},
\bauthor{\bsnm{Hess}, \binits{P.}}:
\byear{2018},
\batitle{{Gradual Streamer Expansions and the Relationship between Blobs and
  Inflows}}.
\bjtitle{\apj}
\bvolume{859},
\bfpage{135}.
\doiurl{https://doi.org/10.3847/1538-4357/aabfd5}.
\end{barticle}
\endbibitem

\bibitem[\protect\citeauthoryear{{Wang} and {Sheeley}}{1992}]{Wang1992}
\begin{barticle}
\bauthor{\bsnm{{Wang}}, \binits{Y.-M.}},
\bauthor{\bsnm{{Sheeley}}, \binits{J.} \bsuffix{N.~R.}}:
\byear{1992},
\batitle{{On Potential Field Models of the Solar Corona}}.
\bjtitle{\apj}
\bvolume{392},
\bfpage{310}.
\doiurl{https://doi.org/10.1086/171430}.
\adsurl{1992ApJ...392..310W}.
\end{barticle}
\endbibitem

\bibitem[\protect\citeauthoryear{{Wang}, {Sheeley}, and
  {Rich}}{2007}]{Wang2007}
\begin{barticle}
\bauthor{\bsnm{{Wang}}, \binits{Y.-M.}},
\bauthor{\bsnm{{Sheeley}}, \binits{N.R.} \bsuffix{Jr.}},
\bauthor{\bsnm{{Rich}}, \binits{N.B.}}:
\byear{2007},
\batitle{{Coronal Pseudostreamers}}.
\bjtitle{\apj}
\bvolume{658},
\bfpage{1340}.
\doiurl{https://doi.org/10.1086/511416}.
\adsurl{2007ApJ...658.1340W}.
\end{barticle}
\endbibitem

\bibitem[\protect\citeauthoryear{{Weberg}, {Lepri}, and
  {Zurbuchen}}{2015}]{Weberg2015}
\begin{barticle}
\bauthor{\bsnm{{Weberg}}, \binits{M.J.}},
\bauthor{\bsnm{{Lepri}}, \binits{S.T.}},
\bauthor{\bsnm{{Zurbuchen}}, \binits{T.H.}}:
\byear{2015},
\batitle{{Coronal Sources, Elemental Fractionation, and Release Mechanisms of
  Heavy Ion Dropouts in the Solar Wind}}.
\bjtitle{\apj}
\bvolume{801},
\bfpage{99}.
\doiurl{https://doi.org/10.1088/0004-637X/801/2/99}.
\adsurl{2015ApJ...801...99W}.
\end{barticle}
\endbibitem

\bibitem[\protect\citeauthoryear{{Weberg}, {Zurbuchen}, and
  {Lepri}}{2012}]{Weberg2012}
\begin{barticle}
\bauthor{\bsnm{{Weberg}}, \binits{M.J.}},
\bauthor{\bsnm{{Zurbuchen}}, \binits{T.H.}},
\bauthor{\bsnm{{Lepri}}, \binits{S.T.}}:
\byear{2012},
\batitle{{ACE/SWICS Observations of Heavy Ion Dropouts within the Solar Wind}}.
\bjtitle{\apj}
\bvolume{760},
\bfpage{30}.
\doiurl{https://doi.org/10.1088/0004-637X/760/1/30}.
\adsurl{2012ApJ...760...30W}.
\end{barticle}
\endbibitem

\bibitem[\protect\citeauthoryear{{West} and {Seaton}}{2015}]{West2015}
\begin{barticle}
\bauthor{\bsnm{{West}}, \binits{M.J.}},
\bauthor{\bsnm{{Seaton}}, \binits{D.B.}}:
\byear{2015},
\batitle{{SWAP Observations of Post-flare Giant Arches in the Long-Duration 14
  October 2014 Solar Eruption}}.
\bjtitle{\apjl}
\bvolume{801},
\bfpage{L6}.
\doiurl{https://doi.org/10.1088/2041-8205/801/1/L6}.
\adsurl{2015ApJ...801L...6W}.
\end{barticle}
\endbibitem

\bibitem[\protect\citeauthoryear{{West} et~al.}{2020}]{West2020}
\begin{barticle}
\bauthor{\bsnm{{West}}, \binits{M.J.}},
\bauthor{\bsnm{{Kintziger}}, \binits{C.}},
\bauthor{\bsnm{{Haberreiter}}, \binits{M.}},
\bauthor{\bsnm{{Gyo}}, \binits{M.}},
\bauthor{\bsnm{{Berghmans}}, \binits{D.}},
\bauthor{\bsnm{{Gissot}}, \binits{S.}},
\bauthor{\bsnm{{B{\"u}chel}}, \binits{V.}},
\bauthor{\bsnm{{Golub}}, \binits{L.}},
\bauthor{\bsnm{{Shestov}}, \binits{S.}},
\bauthor{\bsnm{{Davies}}, \binits{J.A.}}:
\byear{2020},
\batitle{{LUCI onboard Lagrange, the next generation of EUV space weather
  monitoring}}.
\bjtitle{J. Space Weather Space Clim.}
\bvolume{10},
\bfpage{49}.
\doiurl{https://doi.org/10.1051/swsc/2020052}.
\adsurl{2020JSWSC..10...49W}.
\end{barticle}
\endbibitem

\bibitem[\protect\citeauthoryear{{West} et~al.}{2022}]{West2022}
\begin{barticle}
\bauthor{\bsnm{{West}}, \binits{M.J.}},
\bauthor{\bsnm{{Seaton}}, \binits{D.B.}},
\bauthor{\bsnm{{D'Huys}}, \binits{E.}},
\bauthor{\bsnm{{Mierla}}, \binits{M.}},
\bauthor{\bsnm{{Laurenza}}, \binits{M.}},
\bauthor{\bsnm{{Meyer}}, \binits{K.A.}},
\bauthor{\bsnm{{Berghmans}}, \binits{D.}},
\bauthor{\bsnm{{Rachmeler}}, \binits{L.R.}},
\bauthor{\bsnm{{Rodriguez}}, \binits{L.}},
\bauthor{\bsnm{{Stegen}}, \binits{K.}}:
\byear{2022},
\batitle{{A Review of the Extended EUV Corona Observed by the Sun Watcher with
  Active Pixels and Image Processing (SWAP) Instrument}}.
\bjtitle{\solphys}
\bvolume{297},
\bfpage{136}.
\doiurl{https://doi.org/10.1007/s11207-022-02063-9}.
\adsurl{2022SoPh..297..136W}.
\end{barticle}
\endbibitem

\bibitem[\protect\citeauthoryear{{Wexler}, {Jensen}, and
  {Heiles}}{2021}]{Wexler2021a}
\begin{barticle}
\bauthor{\bsnm{{Wexler}}, \binits{D.B.}},
\bauthor{\bsnm{{Jensen}}, \binits{E.A.}},
\bauthor{\bsnm{{Heiles}}, \binits{C.}}:
\byear{2021},
\batitle{{Middle Corona Magnetic Field Strength Determined by Spacecraft Radio
  Faraday Rotation}}.
\bjtitle{Res. Notes Am. Astron. Soc.}
\bvolume{5},
\bfpage{165}.
\doiurl{https://doi.org/10.3847/2515-5172/ac1521}.
\adsurl{2021RNAAS...5..165W}.
\end{barticle}
\endbibitem

\bibitem[\protect\citeauthoryear{{Wexler} et~al.}{2017}]{Wexler2017}
\begin{barticle}
\bauthor{\bsnm{{Wexler}}, \binits{D.B.}},
\bauthor{\bsnm{{Jensen}}, \binits{E.A.}},
\bauthor{\bsnm{{Hollweg}}, \binits{J.V.}},
\bauthor{\bsnm{{Heiles}}, \binits{C.}},
\bauthor{\bsnm{{Efimov}}, \binits{A.I.}},
\bauthor{\bsnm{{Vierinen}}, \binits{J.}},
\bauthor{\bsnm{{Coster}}, \binits{A.J.}}:
\byear{2017},
\batitle{{Faraday rotation fluctuations of MESSENGER radio signals through the
  equatorial lower corona near solar minimum}}.
\bjtitle{Space Weather}
\bvolume{15},
\bfpage{310}.
\doiurl{https://doi.org/10.1002/2016SW001558}.
\adsurl{2017SpWea..15..310W}.
\end{barticle}
\endbibitem

\bibitem[\protect\citeauthoryear{{Wexler} et~al.}{2019}]{Wexler2019}
\begin{barticle}
\bauthor{\bsnm{{Wexler}}, \binits{D.B.}},
\bauthor{\bsnm{{Hollweg}}, \binits{J.V.}},
\bauthor{\bsnm{{Efimov}}, \binits{A.I.}},
\bauthor{\bsnm{{Lukanina}}, \binits{L.A.}},
\bauthor{\bsnm{{Coster}}, \binits{A.J.}},
\bauthor{\bsnm{{Vierinen}}, \binits{J.}},
\bauthor{\bsnm{{Jensen}}, \binits{E.A.}}:
\byear{2019},
\batitle{{Spacecraft Radio Frequency Fluctuations in the Solar Corona: A
  MESSENGER-HELIOS Composite Study}}.
\bjtitle{\apj}
\bvolume{871},
\bfpage{202}.
\doiurl{https://doi.org/10.3847/1538-4357/aaf6a8}.
\adsurl{2019ApJ...871..202W}.
\end{barticle}
\endbibitem

\bibitem[\protect\citeauthoryear{{Wexler} et~al.}{2021}]{Wexler2021b}
\begin{barticle}
\bauthor{\bsnm{{Wexler}}, \binits{D.B.}},
\bauthor{\bsnm{{Stevens}}, \binits{M.L.}},
\bauthor{\bsnm{{Case}}, \binits{A.W.}},
\bauthor{\bsnm{{Song}}, \binits{P.}}:
\byear{2021},
\batitle{{Alfv{\'e}n Speed Transition Zone in the Solar Corona}}.
\bjtitle{\apjl}
\bvolume{919},
\bfpage{L33}.
\doiurl{https://doi.org/10.3847/2041-8213/ac25fa}.
\adsurl{2021ApJ...919L..33W}.
\end{barticle}
\endbibitem

\bibitem[\protect\citeauthoryear{{Wexler} et~al.}{2020}]{Wexler2020}
\begin{barticle}
\bauthor{\bsnm{{Wexler}}, \binits{D.}},
\bauthor{\bsnm{{Imamura}}, \binits{T.}},
\bauthor{\bsnm{{Efimov}}, \binits{A.}},
\bauthor{\bsnm{{Song}}, \binits{P.}},
\bauthor{\bsnm{{Lukanina}}, \binits{L.}},
\bauthor{\bsnm{{Ando}}, \binits{H.}},
\bauthor{\bsnm{{Jensen}}, \binits{E.}},
\bauthor{\bsnm{{Vierinen}}, \binits{J.}},
\bauthor{\bsnm{{Coster}}, \binits{A.}}:
\byear{2020},
\batitle{{Coronal Electron Density Fluctuations Inferred from Akatsuki
  Spacecraft Radio Observations}}.
\bjtitle{\solphys}
\bvolume{295},
\bfpage{111}.
\doiurl{https://doi.org/10.1007/s11207-020-01677-1}.
\adsurl{2020SoPh..295..111W}.
\end{barticle}
\endbibitem

\bibitem[\protect\citeauthoryear{Wiegelmann}{2007}]{Wiegelmann2007}
\begin{barticle}
\bauthor{\bsnm{Wiegelmann}, \binits{T.}}:
\byear{2007},
\batitle{Computing Nonlinear Force-Free Coronal Magnetic Fields in Spherical
  Geometry}.
\bjtitle{\solphys}
\bvolume{240},
\bfpage{227}.
\doiurl{https://doi.org/10.1007/s11207-006-0266-3}.
\end{barticle}
\endbibitem

\bibitem[\protect\citeauthoryear{{Wilhelm} et~al.}{1995}]{Wilhelm1995}
\begin{barticle}
\bauthor{\bsnm{{Wilhelm}}, \binits{K.}},
\bauthor{\bsnm{{Curdt}}, \binits{W.}},
\bauthor{\bsnm{{Marsch}}, \binits{E.}},
\bauthor{\bsnm{{Sch{\"u}hle}}, \binits{U.}},
\bauthor{\bsnm{{Lemaire}}, \binits{P.}},
\bauthor{\bsnm{{Gabriel}}, \binits{A.}},
\bauthor{\bsnm{{Vial}}, \binits{J.-C.}},
\bauthor{\bsnm{{Grewing}}, \binits{M.}},
\bauthor{\bsnm{{Huber}}, \binits{M.C.E.}},
\bauthor{\bsnm{{Jordan}}, \binits{S.D.}},
\bauthor{\bsnm{{Poland}}, \binits{A.I.}},
\bauthor{\bsnm{{Thomas}}, \binits{R.J.}},
\bauthor{\bsnm{{K{\"u}hne}}, \binits{M.}},
\bauthor{\bsnm{{Timothy}}, \binits{J.G.}},
\bauthor{\bsnm{{Hassler}}, \binits{D.M.}},
\bauthor{\bsnm{{Siegmund}}, \binits{O.H.W.}}:
\byear{1995},
\batitle{{SUMER - Solar Ultraviolet Measurements of Emitted Radiation}}.
\bjtitle{\solphys}
\bvolume{162},
\bfpage{189}.
\doiurl{https://doi.org/10.1007/BF00733430}.
\adsurl{1995SoPh..162..189W}.
\end{barticle}
\endbibitem

\bibitem[\protect\citeauthoryear{{Wilson} et~al.}{2022}]{Wilson2022}
\begin{barticle}
\bauthor{\bsnm{{Wilson}}, \binits{M.L.}},
\bauthor{\bsnm{{Raymond}}, \binits{J.C.}},
\bauthor{\bsnm{{Lepri}}, \binits{S.T.}},
\bauthor{\bsnm{{Lionello}}, \binits{R.}},
\bauthor{\bsnm{{Murphy}}, \binits{N.A.}},
\bauthor{\bsnm{{Reeves}}, \binits{K.K.}},
\bauthor{\bsnm{{Shen}}, \binits{C.}}:
\byear{2022},
\batitle{{Constraining the CME Core Heating and Energy Budget with SOHO/UVCS}}.
\bjtitle{\apj}
\bvolume{927},
\bfpage{27}.
\doiurl{https://doi.org/10.3847/1538-4357/ac4d35}.
\adsurl{2022ApJ...927...27W}.
\end{barticle}
\endbibitem

\bibitem[\protect\citeauthoryear{{Winebarger} et~al.}{2002}]{Winebarger2002}
\begin{barticle}
\bauthor{\bsnm{{Winebarger}}, \binits{A.R.}},
\bauthor{\bsnm{{Warren}}, \binits{H.}},
\bauthor{\bsnm{{van Ballegooijen}}, \binits{A.}},
\bauthor{\bsnm{{DeLuca}}, \binits{E.E.}},
\bauthor{\bsnm{{Golub}}, \binits{L.}}:
\byear{2002},
\batitle{{Steady Flows Detected in Extreme-Ultraviolet Loops}}.
\bjtitle{\apjl}
\bvolume{567},
\bfpage{L89}.
\doiurl{https://doi.org/10.1086/339796}.
\adsurl{2002ApJ...567L..89W}.
\end{barticle}
\endbibitem

\bibitem[\protect\citeauthoryear{{Wl{\'e}rick} and
  {Axtell}}{1957}]{Wlerick1957}
\begin{barticle}
\bauthor{\bsnm{{Wl{\'e}rick}}, \binits{G.}},
\bauthor{\bsnm{{Axtell}}, \binits{J.}}:
\byear{1957},
\batitle{{A New Instrument for Observing the Electron Corona.}}
\bjtitle{\apj}
\bvolume{126},
\bfpage{253}.
\doiurl{https://doi.org/10.1086/146396}.
\adsurl{1957ApJ...126..253W}.
\end{barticle}
\endbibitem

\bibitem[\protect\citeauthoryear{{Woo}}{1978}]{woo1978}
\begin{barticle}
\bauthor{\bsnm{{Woo}}, \binits{R.}}:
\byear{1978},
\batitle{{Errata: Radial Dependence of Solar Wind Properties Deduced from
  HELIOS 1/2 and Pioneer 10/11 Radio Scattering Observations}}.
\bjtitle{\apj}
\bvolume{223},
\bfpage{704}.
\doiurl{https://doi.org/10.1086/156304}.
\adsurl{1978ApJ...223..704W}.
\end{barticle}
\endbibitem

\bibitem[\protect\citeauthoryear{{Wyper} et~al.}{2021}]{Wyper2021}
\begin{barticle}
\bauthor{\bsnm{{Wyper}}, \binits{P.F.}},
\bauthor{\bsnm{{Antiochos}}, \binits{S.K.}},
\bauthor{\bsnm{{DeVore}}, \binits{C.R.}},
\bauthor{\bsnm{{Lynch}}, \binits{B.J.}},
\bauthor{\bsnm{{Karpen}}, \binits{J.T.}},
\bauthor{\bsnm{{Kumar}}, \binits{P.}}:
\byear{2021},
\batitle{{A Model for the Coupled Eruption of a Pseudostreamer and Helmet
  Streamer}}.
\bjtitle{\apj}
\bvolume{909},
\bfpage{54}.
\doiurl{https://doi.org/10.3847/1538-4357/abd9ca}.
\adsurl{2021ApJ...909...54W}.
\end{barticle}
\endbibitem

\bibitem[\protect\citeauthoryear{{Yang} et~al.}{2020}]{yang2020}
\begin{barticle}
\bauthor{\bsnm{{Yang}}, \binits{Z.}},
\bauthor{\bsnm{{Bethge}}, \binits{C.}},
\bauthor{\bsnm{{Tian}}, \binits{H.}},
\bauthor{\bsnm{{Tomczyk}}, \binits{S.}},
\bauthor{\bsnm{{Morton}}, \binits{R.}},
\bauthor{\bsnm{{Del Zanna}}, \binits{G.}},
\bauthor{\bsnm{{McIntosh}}, \binits{S.W.}},
\bauthor{\bsnm{{Karak}}, \binits{B.B.}},
\bauthor{\bsnm{{Gibson}}, \binits{S.}},
\bauthor{\bsnm{{Samanta}}, \binits{T.}},
\bauthor{\bsnm{{He}}, \binits{J.}},
\bauthor{\bsnm{{Chen}}, \binits{Y.}},
\bauthor{\bsnm{{Wang}}, \binits{L.}}:
\byear{2020},
\batitle{{Global maps of the magnetic field in the solar corona}}.
\bjtitle{Science}
\bvolume{369},
\bfpage{694}.
\doiurl{https://doi.org/10.1126/science.abb4462}.
\end{barticle}
\endbibitem

\bibitem[\protect\citeauthoryear{{Yashiro} et~al.}{2004}]{Yashiro2004}
\begin{barticle}
\bauthor{\bsnm{{Yashiro}}, \binits{S.}},
\bauthor{\bsnm{{Gopalswamy}}, \binits{N.}},
\bauthor{\bsnm{{Michalek}}, \binits{G.}},
\bauthor{\bsnm{{St. Cyr}}, \binits{O.C.}},
\bauthor{\bsnm{{Plunkett}}, \binits{S.P.}},
\bauthor{\bsnm{{Rich}}, \binits{N.B.}},
\bauthor{\bsnm{{Howard}}, \binits{R.A.}}:
\byear{2004},
\batitle{{A catalog of white light coronal mass ejections observed by the SOHO
  spacecraft}}.
\bjtitle{J. Geophys. Res. (Space Phys.)}
\bvolume{109},
\bfpage{A07105}.
\doiurl{https://doi.org/10.1029/2003JA010282}.
\adsurl{2004JGRA..109.7105Y}.
\end{barticle}
\endbibitem

\bibitem[\protect\citeauthoryear{{Yeates}, {Mackay}, and {van
  Ballegooijen}}{2008}]{Yeates2008}
\begin{barticle}
\bauthor{\bsnm{{Yeates}}, \binits{A.R.}},
\bauthor{\bsnm{{Mackay}}, \binits{D.H.}},
\bauthor{\bsnm{{van Ballegooijen}}, \binits{A.A.}}:
\byear{2008},
\batitle{{Modelling the Global Solar Corona II: Coronal Evolution and Filament
  Chirality Comparison}}.
\bjtitle{\solphys}
\bvolume{247},
\bfpage{103}.
\doiurl{https://doi.org/10.1007/s11207-007-9097-0}.
\adsurl{2008SoPh..247..103Y}.
\end{barticle}
\endbibitem

\bibitem[\protect\citeauthoryear{{Yeates} et~al.}{2018}]{Yeates2018}
\begin{barticle}
\bauthor{\bsnm{{Yeates}}, \binits{A.R.}},
\bauthor{\bsnm{{Amari}}, \binits{T.}},
\bauthor{\bsnm{{Contopoulos}}, \binits{I.}},
\bauthor{\bsnm{{Feng}}, \binits{X.}},
\bauthor{\bsnm{{Mackay}}, \binits{D.H.}},
\bauthor{\bsnm{{Miki{\'c}}}, \binits{Z.}},
\bauthor{\bsnm{{Wiegelmann}}, \binits{T.}},
\bauthor{\bsnm{{Hutton}}, \binits{J.}},
\bauthor{\bsnm{{Lowder}}, \binits{C.A.}},
\bauthor{\bsnm{{Morgan}}, \binits{H.}},
\bauthor{\bsnm{{Petrie}}, \binits{G.}},
\bauthor{\bsnm{{Rachmeler}}, \binits{L.A.}},
\bauthor{\bsnm{{Upton}}, \binits{L.A.}},
\bauthor{\bsnm{{Canou}}, \binits{A.}},
\bauthor{\bsnm{{Chopin}}, \binits{P.}},
\bauthor{\bsnm{{Downs}}, \binits{C.}},
\bauthor{\bsnm{{Druckm{\"u}ller}}, \binits{M.}},
\bauthor{\bsnm{{Linker}}, \binits{J.A.}},
\bauthor{\bsnm{{Seaton}}, \binits{D.B.}},
\bauthor{\bsnm{{T{\"o}r{\"o}k}}, \binits{T.}}:
\byear{2018},
\batitle{{Global Non-Potential Magnetic Models of the Solar Corona During the
  March 2015 Eclipse}}.
\bjtitle{\ssr}
\bvolume{214},
\bfpage{99}.
\doiurl{https://doi.org/10.1007/s11214-018-0534-1}.
\adsurl{2018SSRv..214...99Y}.
\end{barticle}
\endbibitem

\bibitem[\protect\citeauthoryear{{Young}, {Klimchuk}, and
  {Mason}}{1999}]{young1999}
\begin{barticle}
\bauthor{\bsnm{{Young}}, \binits{P.R.}},
\bauthor{\bsnm{{Klimchuk}}, \binits{J.A.}},
\bauthor{\bsnm{{Mason}}, \binits{H.E.}}:
\byear{1999},
\batitle{{Temperature and density in a polar plume - measurements from
  CDS/SOHO}}.
\bjtitle{\aap}
\bvolume{350},
\bfpage{286}.
\adsurl{1999A&A...350..286Y}.
\end{barticle}
\endbibitem

\bibitem[\protect\citeauthoryear{{Yu} et~al.}{2020}]{SYu2020}
\begin{barticle}
\bauthor{\bsnm{{Yu}}, \binits{S.}},
\bauthor{\bsnm{{Chen}}, \binits{B.}},
\bauthor{\bsnm{{Reeves}}, \binits{K.K.}},
\bauthor{\bsnm{{Gary}}, \binits{D.E.}},
\bauthor{\bsnm{{Musset}}, \binits{S.}},
\bauthor{\bsnm{{Fleishman}}, \binits{G.D.}},
\bauthor{\bsnm{{Nita}}, \binits{G.M.}},
\bauthor{\bsnm{{Glesener}}, \binits{L.}}:
\byear{2020},
\batitle{{Magnetic Reconnection during the Post-impulsive Phase of a
  Long-duration Solar Flare: Bidirectional Outflows as a Cause of Microwave and
  X-Ray Bursts}}.
\bjtitle{\apj}
\bvolume{900},
\bfpage{17}.
\doiurl{https://doi.org/10.3847/1538-4357/aba8a6}.
\adsurl{2020ApJ...900...17Y}.
\end{barticle}
\endbibitem

\bibitem[\protect\citeauthoryear{{Zhang} and {Dere}}{2006}]{Zhang2006}
\begin{barticle}
\bauthor{\bsnm{{Zhang}}, \binits{J.}},
\bauthor{\bsnm{{Dere}}, \binits{K.P.}}:
\byear{2006},
\batitle{{A Statistical Study of Main and Residual Accelerations of Coronal
  Mass Ejections}}.
\bjtitle{\apj}
\bvolume{649},
\bfpage{1100}.
\doiurl{https://doi.org/10.1086/506903}.
\adsurl{2006ApJ...649.1100Z}.
\end{barticle}
\endbibitem

\bibitem[\protect\citeauthoryear{{Zhang} et~al.}{2021}]{Zhang2021}
\begin{barticle}
\bauthor{\bsnm{{Zhang}}, \binits{J.}},
\bauthor{\bsnm{{Temmer}}, \binits{M.}},
\bauthor{\bsnm{{Gopalswamy}}, \binits{N.}},
\bauthor{\bsnm{{Malandraki}}, \binits{O.}},
\bauthor{\bsnm{{Nitta}}, \binits{N.V.}},
\bauthor{\bsnm{{Patsourakos}}, \binits{S.}},
\bauthor{\bsnm{{Shen}}, \binits{F.}},
\bauthor{\bsnm{{Vr{\v{s}}nak}}, \binits{B.}},
\bauthor{\bsnm{{Wang}}, \binits{Y.}},
\bauthor{\bsnm{{Webb}}, \binits{D.}},
\bauthor{\bsnm{{Desai}}, \binits{M.I.}},
\bauthor{\bsnm{{Dissauer}}, \binits{K.}},
\bauthor{\bsnm{{Dresing}}, \binits{N.}},
\bauthor{\bsnm{{Dumbovi{\'c}}}, \binits{M.}},
\bauthor{\bsnm{{Feng}}, \binits{X.}},
\bauthor{\bsnm{{Heinemann}}, \binits{S.G.}},
\bauthor{\bsnm{{Laurenza}}, \binits{M.}},
\bauthor{\bsnm{{Lugaz}}, \binits{N.}},
\bauthor{\bsnm{{Zhuang}}, \binits{B.}}:
\byear{2021},
\batitle{{Earth-affecting solar transients: a review of progresses in solar
  cycle 24}}.
\bjtitle{Prog. Earth Planet. Sci.}
\bvolume{8},
\bfpage{56}.
\doiurl{https://doi.org/10.1186/s40645-021-00426-7}.
\adsurl{2021PEPS....8...56Z}.
\end{barticle}
\endbibitem

\bibitem[\protect\citeauthoryear{{Zhao} et~al.}{2017}]{Zhao2017}
\begin{barticle}
\bauthor{\bsnm{{Zhao}}, \binits{L.}},
\bauthor{\bsnm{{Landi}}, \binits{E.}},
\bauthor{\bsnm{{Lepri}}, \binits{S.T.}},
\bauthor{\bsnm{{Kocher}}, \binits{M.}},
\bauthor{\bsnm{{Zurbuchen}}, \binits{T.H.}},
\bauthor{\bsnm{{Fisk}}, \binits{L.A.}},
\bauthor{\bsnm{{Raines}}, \binits{J.M.}}:
\byear{2017},
\batitle{{An Anomalous Composition in Slow Solar Wind as a Signature of
  Magnetic Reconnection in its Source Region}}.
\bjtitle{\apjs}
\bvolume{228},
\bfpage{4}.
\doiurl{https://doi.org/10.3847/1538-4365/228/1/4}.
\adsurl{2017ApJS..228....4Z}.
\end{barticle}
\endbibitem

\bibitem[\protect\citeauthoryear{{Zhitnik} et~al.}{2002}]{Zhitnik2002}
\begin{bchapter}
\bauthor{\bsnm{{Zhitnik}}, \binits{I.A.}},
\bauthor{\bsnm{{Bougaenko}}, \binits{O.I.}},
\bauthor{\bsnm{{Delaboudiniere}}, \binits{J.-P.}},
\bauthor{\bsnm{{Ignatiev}}, \binits{A.P.}},
\bauthor{\bsnm{{Korneev}}, \binits{V.V.}},
\bauthor{\bsnm{{Krutov}}, \binits{V.V.}},
\bauthor{\bsnm{{Kuzin}}, \binits{S.V.}},
\bauthor{\bsnm{{Lisin}}, \binits{D.V.}},
\bauthor{\bsnm{{Mitrofanov}}, \binits{A.V.}},
\bauthor{\bsnm{{Oparin}}, \binits{S.N.}},
\bauthor{\bsnm{{Oraevsky}}, \binits{V.N.}},
\bauthor{\bsnm{{Pertsov}}, \binits{A.A.}},
\bauthor{\bsnm{{Slemzin}}, \binits{V.A.}},
\bauthor{\bsnm{{Sobelman}}, \binits{I.I.}},
\bauthor{\bsnm{{Stepanov}}, \binits{A.I.}},
\bauthor{\bsnm{{Schwarz}}, \binits{J.}}:
\byear{2002},
\bctitle{{SPIRIT X-ray telescope/spectroheliometer results}}.
In: \beditor{\bsnm{{Wilson}}, \binits{A.}} (ed.)
\bbtitle{Solar Variability: From Core to Outer Frontiers}
\bseriesno{SP-2},
\bpublisher{ESA, Noordwijk},
\bfpage{915}.
\adsurl{2002ESASP.506..915Z}.
\end{bchapter}
\endbibitem

\bibitem[\protect\citeauthoryear{{Zhukov} et~al.}{2008}]{Zhukov2008}
\begin{barticle}
\bauthor{\bsnm{{Zhukov}}, \binits{A.N.}},
\bauthor{\bsnm{{Saez}}, \binits{F.}},
\bauthor{\bsnm{{Lamy}}, \binits{P.}},
\bauthor{\bsnm{{Llebaria}}, \binits{A.}},
\bauthor{\bsnm{{Stenborg}}, \binits{G.}}:
\byear{2008},
\batitle{{The Origin of Polar Streamers in the Solar Corona}}.
\bjtitle{\apj}
\bvolume{680},
\bfpage{1532}.
\doiurl{https://doi.org/10.1086/587924}.
\adsurl{2008ApJ...680.1532Z}.
\end{barticle}
\endbibitem

\bibitem[\protect\citeauthoryear{{Zimovets} et~al.}{2012}]{Zimovets2012}
\begin{barticle}
\bauthor{\bsnm{{Zimovets}}, \binits{I.}},
\bauthor{\bsnm{{Vilmer}}, \binits{N.}},
\bauthor{\bsnm{{Chian}}, \binits{A.C.-L.}},
\bauthor{\bsnm{{Sharykin}}, \binits{I.}},
\bauthor{\bsnm{{Struminsky}}, \binits{A.}}:
\byear{2012},
\batitle{{Spatially resolved observations of a split-band coronal type II radio
  burst}}.
\bjtitle{\aap}
\bvolume{547},
\bfpage{A6}.
\doiurl{https://doi.org/10.1051/0004-6361/201219454}.
\adsurl{2012A&A...547A...6Z}.
\end{barticle}
\endbibitem

\end{thebibliography}

\end{article} 
\end{document}